\documentclass[12pt]{article}
\usepackage[english]{babel}%

\usepackage[utf8]{inputenc}

\usepackage{color}

\usepackage{epsfig}
\usepackage{graphics}
\usepackage{amssymb,amsfonts,amsmath,latexsym}

\usepackage[T2A]{fontenc}

\newcommand{\N}{N\raise.7ex\hbox{\underline{$\circ $}}$\;$}

\numberwithin{equation}{section}

\usepackage{setspace}
\begin{document}
\begin{center}
{\bf
 RELATIVISTIC WAVE EQUATIONS WITH EXTENDED \newline A
SET OF THE LORENTZ GROUP REPRESENTATIONS}
\\ [5mm]
V.A. Pletyukhov
\\[3mm]
Doctor of Phys.-Math. Sciences\\
Professor of Department of
General and Theoretical Physics
\\
Brest State University named after A.S. Pushkin

\author{{\bf V.A. Pletyukhov}\\
erphys@brsu.brest.by; Brest State University, Brest, Belarus}

\date{}

%\maketitle
\end{center}

\begin{abstract}
It is shown that the use of extended sets of irreducible
representations of the Lorentz group opens new possibilities for
the theory of relativistic wave equations from the point of view
of the space-time description of both the internal structure and
the isospin degrees of freedom of elementary particles. The
approach developed in this work also makes it possible to apply
the methods of the theory of relativistic wave equations in
superstring and gauge models of fundamental interactions.
\end{abstract}

\section*{Introduction}
\addcontentsline{toc}{chapter}{ \bf{Introduction} }

\begin{spacing}{0.95}
\hspace{0.5cm}
Equations of Newton, Maxwell, Einstein, Schr\"{o}dinger, Dirac,
Yang-Mills ...

Each of them was an epoch-making event in physics. Newton's
equations marked the beginning of theoretical physics. Maxwell's
equations involve the introduction of a fundamentally new physical
concept of the field, the unification of electrical and magnetic
phenomena, the prediction of the existence of electromagnetic
waves.
Einstein's equations combined the properties of matter and
space-time, created the basis for describing the universe as a
 physical object.
The Schrodinger equation led to an understanding of the
probabilistic nature of physical processes in the micro-world.
The Dirac equation is a corner stone of the quantum-mechanical
description of physical phenomena,  it created the basis for
quantum field theory and  predicted the existence of a new kind of
matter (antiparticles).
The Yang-Mills equations lie in the foundation of the theory of
gauge fields, the unified theory of electromagnetic and weak
interactions, the Standard Model.

The Dirac equation served simultaneously as the starting  model
for the creation of the general theory of relativistic wave
equations (RWE) -- first-order relativistic quantum-mechanical
equations written in matrix-differential form.
The fundamental idea  of this theory is  the governing of amny
 RWE by with a  corresponding
 set of irreducible representations of the group of geometric
(space-time) symmetries of the Minkowski space.
Besides  Dirac [1], the most significant contributions to
development of the theory of RWE ware made by Pauli and  Fierz [2;
3], Baba [4; 5], Harish--Chandra [6; 7], Gelfand and  Yaglom [8;
9],  and the Belorussian theoretical physicist F.I. Fedorov
[10--13].

We can formulate the following postulate basis for this theory:

\begin{quotation}

 1) any RWE must satisfies the invariance requirements with respect
to the transformations of the proper Lorentz group and  operation
of the spatial reflection,  also  possibility of the Lagrangian
formulation of the theory is assumed;

 2) RWE describing a single physical micro-object should not be
decaying in the sense of the full Lorentz group;

 3) among the states of the micro-object, there cannot be
those  which  correspond  to zero energy;

 4) the correct RWE  must lead to a positive definite density of
energy (charge) in the case of a whole (half-integer) spin;

5) fields with  integral (and half-integral) spins are  described
on the basis of tensor (spinor) representations of the Lorentz
group;

 6) usually when constructing for a particle with spin a corresponding  RWE
  we   restrict ourselves to the
minimally necessary set of irreducible representations of the
Lorentz group.

\end{quotation}

 The listed provisions of the theory of RWE were
formulated in the 20s--50s of the last century. They were based
on  idea  that  elementary particles are  nonstructural
point-like micro objects with a single internal degree of freedom
(spin), the last  has a spatio-temporal interpretation. However, with
the establishment of new experimental facts (the existence of internal
structure for some  particles, the presence of additional
internal degrees of freedom besides spin, etc.), the above
ideas  have undergone significant changes. The very
concept of "elementary particle" \hspace{1mm} has also changed.

 There arose the idea of
existence of fundamentally new physical objects that unify the
qualities of micro-particles (fields) with nonzero and zero mass
(for example, the electroweak fields) and the properties of
massless micro-objects with different helicity values (fields
interacting with non-closed strings). The "old" \hspace{1mm} problems of the
theory of RWE also remained unresolved. On the one hand, its
postulate basis seems insufficiently complete, since it does not
restrict the  spectrum of possible elementary particles.
On the other hand, the only use of   symmetries with  space-time origin
in the RWE theory makes it difficult to describe additional
internal degrees of freedom besides spin (isospin). The question
of the origin of the mass remained unsolved in
the RWE theory.

Thus, the theory of RWE has faced challenges, to which it, as would
seem, is not capable of giving adequate answers. Therefore,
interest in this theory in the last two or three decades has significantly
decreased.

 However, even in 1955--1957 czech physicists
Petrash and Ulegla constructed and investigated a RWE for a
micro-object with spin on the basis of the use of a set of
irreducible representations of the Lorentz group extended in
comparison with the Dirac equation. It was shown that in an
external magnetic field this micro-object exhibits an internal
electromagnetic structure in the form of an anomalous magnetic
moment. Even earlier, in 1928, the English physicist Darwin
proposed, as an alternative to the Dirac equation,  the RWE
which is nor  disintegrated in the full Lorentz group, it  (in our opinion
was not quite fair) in the literature, is  called the
Dirac--K\"{a}hler equation. The Dirac--K\"{a}hler equation contains
two-fold (twice repeating) scalar and vector representations of
the proper Lorentz group. Because of this, it has an internal
symmetry, described by a noncompact group, which forms a
semi-direct product; It does not commute with the group of Lorentz
symmetry. The latter circumstance makes it possible to interpret
the Dirac--K\"{a}hler equation, in spite of the tensor character of the
components of its wave function, as the RWE for a micro-object with
spin and some isospin degree of freedom with space-time origin.

 The Dirac--K\"{a}hler and Petrash--Ulegla equations can be
regarded as the first successful attempts to go beyond the
postulates 1) -- 6) and thereby greatly expanded the capabilities of
the theory of RWE. They showed, in particular, that if the
postulate 6) is abandoned, the possibility of a spatio-temporal
description of both the internal structure and the isospin degrees
of freedom of particles appears.
 Since the middle of 1960s,
this direction began to develop actively in a number of scientific
centers of the Republic of Belarus on the initiative and under the
leadership of Academician of the Academy of Sciences of Belarus
F.I. Fedorov. Over the past
decade, a wealth of results in the theory of RWE with an extended
set of representations of the Lorentz group had been accumulated.
In this paper, we present some significant results that, in our
view, can not only be adapted to modern experimental achievements
and theoretical trends in high-energy physics, but also to some
extent they may help in resolve  the problems that remain unsolved.

In Chapter
1, we consider in detail those restrictions that conditions (1) --
(5) imply on the algebraic and group structure of
matrix-differential RWE-s. The analysis is carried out in the
so-called canonical basis, or the Gel'fand--Yaglom basis, which is
most convenient for establishing the spin and mass characteristics
of a micro-object. Chapter 2 shows how the Gelfand--Yaglom approach
works specifically with respect to the simplest known equations
for particles with lower spins. In addition to the Dirac equation,
these are the Duffin--Kemmer equations for scalar and vector
particles, the Fierz--Pauli equation for spin 3/2 and 2 particles.

In Chapter 3, we present RWE-s for micro-objects with spins, which are  based on extended
sets of representations of the Lorentz group, including repeated
ones. It is shown that such an extension allows us to describe t the
internal electromagnetic structure of the micro-object in the
framework of the conventional theory of the RWE theory. All the
"extended" \hspace{1mm} RVE-s considered in this chapter are renormalizable and
do not contain unprincipled solutions.

In Chapter 4, a RWE is constructed for a vector particle with a nonzero mass
and an additional internal quantum number, chirality. As shown below
in Chapter 7, such a particle in a certain sense can be regarded as a massive
analog of the unified massless Maxwell--Kalb--Ramond field.

Chapters 5 and 6 give a matrix formulation of the Dirac--K\"{a}hler equation
and its maximal tensor generalizations, bounded by the dimension of
space-time. The possibility of consistent quantization of these equations
from Fermi--Dirac statistics is shown both from the point of view of the
correct corpuscular picture of the second-quantized field and in relation to
the probabilistic interpretation of the theory.

Chapter 7 shows the possibility of a joint description of fields of the
Maxwellian type and the Kalba--Ramond type (the Notof of
Ogievetskii and Polubarinov) in the approach of the RWE theory. This allows us
to speak of a single massless vector field with three helicity values, which
can claim to play the role of the carrier of the interaction of open strings
in string theory, reduced to the four-dimensional Minkowski space. A matrix
interpretation of the mechanism of mass generation is also given by
gauge-invariant mixing of massless fields with different helicity values.
This mechanism does not lead to the appearance of an additional massive
scalar field and can act as an alternative to the generally accepted Higgs
mechanism.

In Chapter 8, the status of the so-called massive gauge-invariant
fields is studied in detail. It is established what is the
similarity and difference between massive and massless
gauge-invariant fields from the point of view of the formalism of
the RWE theory.

Finally, in Chapter 9 we propose tensor and matrix formulations of
a massively massless vector field with three massive and one
massless quanta, which can be treated as an electroweak fields. The
requirement of the non-disintegration of the corresponding RWE
leads to the necessity of introducing a massive scalar field as an
integral component of the electroweak fields. This implies a
fundamentally different treatment of the origin of the Higgs
boson, which is not related to the mechanism of spontaneous
symmetry breaking and mass generation. \newline ~

\end{spacing}

\section{Relativistic Wave Equations. The Gel'fand--Yaglom approach}
\addcontentsline{toc}{chapter}{ \bf{Relativistic Wave Equations. The Gel'fand--Yaglom approach} }
\label{intro}

\hspace{0.5cm}
A relativistic quantum mechanical description of free elementary microobjects can be ever reduced to a system ofthe first-order linear differential equations with constant coefficients. In case of microobjects with a nonzero mass such a system can be expressed in the matrix differential form \footnote{Here and elsewhere we use the metrics $g_{\mu \nu} = \mathrm{diag} (1,1,1,1)$, and therefore there is no necessity to distinguish between covariant and contravariant Lorentz indices. We also assume a summation over repeated indices, in accordance with the  Einstein's rule.}
\begin{align}
(\Gamma_{\mu} \partial_{\mu} + m) \Psi (x) =0,
\label{eq:nonzero_mass}
\end{align}
where $\Psi$ is a multicomponent wavefunction, $\Gamma_{\mu}$ are square matrices, $m$ is a scalar parameter associated with a mass. In turm, for zero-mass microobjects we have
\begin{align}
(\Gamma_{\mu} \partial_{\mu} + \Gamma_0) \Psi (x) =0,
\label{eq:zero_mass}
\end{align}
where $\Gamma_0$ is a singular matrix ($\det \Gamma_0 =0$), which, in particular, can be zero. These two forms of expressions are precisely the ones which are associated with the term "relativistic wave equations". Note that for a matrix $\Gamma_0$ in \eqref{eq:zero_mass} being non-singular, this equation can be ever transformed to the form \eqref{eq:nonzero_mass}. Therefore,
in the following we always assume $\Gamma_0$ to be singular, unless it is explicitly stated otherwise.

The main and indisputable requirement imposed on the equations \eqref{eq:nonzero_mass} and \eqref{eq:zero_mass} is their invariance under transformations of the restricted Lorentz group. From this  follow the two conditions. Firstly, the function $\Psi$ is being transformed with some representation $T$ of the restricted Lorentz group (it will be also shown later that the representation $T$ must be reducible). Secondly, the matrices $\Gamma_{\mu}$ and $\Gamma_0$ must satisfy the following conditions
\begin{align}
T^{-1} \Gamma_{\mu} T &= L_{\mu \nu} \Gamma_{\nu}, \label{nonsing_trans} \\
T^{-1} \Gamma_{0} T &= \Gamma_{0},  \label{sing_trans}
\end{align}
where $L_{\mu \nu}$ is the Lorentz matrix. Applying \eqref{nonsing_trans} to infinitesimally small Lorentz transformations
\begin{align}
T = 1 + \delta \omega_{[\mu \nu]} J^{[\mu \nu]},
\label{InfLoTra}
\end{align}
we arrive at the following relation
\begin{align}
[J^{[\mu \nu]}, \Gamma_{\alpha}]_- = \delta_{\nu \alpha} \Gamma_{\mu} - \delta_{\mu \alpha} \Gamma_{\nu}.
\label{Jcond}
\end{align}
Setting $\mu = i$ and $\nu = \alpha =4$ in \eqref{Jcond}, one can express the matrices $\Gamma_i \, (i=1,2,3)$ via $\Gamma_4$ and the boost operators $J^{[i 4]}$ of the Lorentz transformations:
\begin{align}
\Gamma_i = [J^{[i 4]}, \Gamma_4 ]_- .
\label{gam_i_gen}
\end{align}
Thus it appears that among the all matrices $\Gamma_{\mu}$ the matrix $\Gamma_4$ plays the main role.

Next we recall that every irreducible finite-dimensional representation $\tau$ of the Lorentz group is given by a pair of numbers $l_1 , l_2$ which can take, either simultaneously or separately, either integer (including zero) or half-integer positive values. A representation $\tau$ acting in the representation space $R^{\tau}$ generates in general a reducible representation of its subgroup -- the rotation group. In the other words, the representation space $R^{\tau}$ can be decomposed into a direct sum of the invariant subspaces $R^{\tau}_s$. In each of them a representation of the rotation group induced  by the representation $\tau \sim (l_1 , l_2)$ of the Lorentz group is irreducible and given by either integer of half-integer weight $s$, the representation $\tau$ containing all
possible weights from $|l_1 - l_2|$ to $l_1 + l_2$. Thus, the dimension of $R^{\tau}$ equals $(2 l_1 +1) (2 l_2 +1)$.

Depending on values of $l_1$ and $l_2$ (integer or half-integer), all irreducible finite-dimensional representation of the Lorentz group fall into four classes:
\begin{align}
\begin{split}
\mathrm{class} \,\, +1: & \quad  l_1, l_2 \,\, \mathrm{both \,\, integer}; \\
 \mathrm{class} \,\, -1: & \quad  l_1, l_2 \,\, \mathrm{both \,\, half \,\, integer}; \\
 \mathrm{class} \,\, +\varepsilon : & \quad  l_1 \,\, \mathrm{integer}, l_2 \,\, \mathrm{half\,\, integer}; \\
 \mathrm{class} \,\, -\varepsilon : & \quad  l_1 \,\, \mathrm{half \,\, integer}, l_2 \,\, \mathrm{integer}.
\end{split}
\end{align}
Irreducible representations $\tau \sim (l_1 , l_2)$ and $\tau' \sim (l'_1, l'_2)$ are called linking, if the following conditions are simultaneously fulfilled
\begin{align}
l'_ 1 = l_1 \pm \frac12 , \quad l'_2 = l_2 \pm \frac12 ,
\end{align}
the signs $+$ and $-$ are being uncorrelated.

A pictorial representation of linking irreducible representations of the Lorentz group is conveniently given by the so called linking scheme in which the linking components are connected with each other by the bar. It is obvious that the only allowed linkings can occur between representations of the classes $+1$ and $-1$, and between representations of the classes $+\varepsilon$ and $-\varepsilon$. Thus, there exist two general types of linking schemes:
\begin{align}
\begin{tabular}{ccccccccc}
& & &   & $(0,0)$ &  & & & \\
& & &   &   $|$       &  & & & \\
 & & $(0,1)$   & --- & $( \frac12, \frac12)$  & --- & $(1,0)$ & &  \\
 &  & $|$ &  & $|$ &  & $|$ &  &  \\
$(0,2)$ &  --- & $ (\frac12, \frac32) $ & --- & $ (1,1) $ & --- & $(\frac32, \frac12)$ & --- & $ (2,0)$ \\
$|$ &  & $|$ &  & $|$ &  & $|$ &  & $|$
\end{tabular}
\label{type_1}
\end{align}
in the first case, and
\begin{align}
\begin{tabular}{ccccccccccc}
& & &   & $(0,\frac12)$ &  --- & $(\frac12 , 0)$ & & & & \\
& & &   &   $|$       &  & $|$ & & \\
 & & $(0,\frac32)$   & --- & $( \frac12, 1)$  & --- & $(1,\frac12)$ & --- &  $(\frac32,0)$ & & \\
 &  & $|$ &  & $|$ &  & $|$ &  & $|$ & &  \\
$(0,\frac52)$ &  --- & $ (\frac12, 2) $ & --- & $ (1,\frac32) $ & --- & $(\frac32, 1)$ & --- & $ (2,\frac12)$ & --- & $(\frac52,0)$ \\
$|$ &  & $|$ &  & $|$ &  & $|$ &  & $|$ & & $|$
\end{tabular}
\label{type_eps}
\end{align}
in the second case.

It is important to note that in the depicted linking schemes each irreducible component can occur more than once. Then it is said about the RWE with multiple (repeated) Lorentz group representations. It is obvious that any linking scheme of the type \eqref{type_1} containes only integer-valued weights $s$ of irreducible representations of the rotation group, and hence the RWE corresponding to it describes microobjects with integer spin. Analogously, linking schemes of the type \eqref{type_eps} serve for a description of microobjects with half-integer spin.

Let us first consider which constraints on the representation $T$ acting in the space $R$ of the wavefunction $\Psi$ are imposed by the invariance condition of the RWE \eqref{eq:nonzero_mass} under the restricted Lorentz group transformations.

Suppose that the representation $T$ consists of a single irreducible representation $\tau \sim (l_1 , l_2)$ of the restricted Lorentz group. Then the term $m \Psi$ in \eqref{eq:nonzero_mass} is being transformed by means of the representation $\tau$. In turn, the term $\Gamma_{\mu} \partial_{\mu} \Psi$ is being transformed by a representation, which fully or partially consists of irreducible representations contained in a direct product of the representations
\begin{align}
(l_1 , l_2) & \otimes (\frac12 , \frac12)= \nonumber \\
&= (l_1 + \frac12 , l_2 + \frac12) \oplus (l_1 + \frac12 , l_2 - \frac12) \oplus (l_1 - \frac12 , l_2 + \frac12) \oplus (l_1 - \frac12 , l_2 - \frac12).
\label{dp_decomp}
\end{align}
But since neither of the representations occurring in the decomposition \eqref{dp_decomp} coincides with the representation $(l_1, l_2)$, the terms $\Gamma_{\mu} \partial_{\mu} \Psi$ and $ m \Psi$ can not be transformed in the same way under the Lorentz transformations. It means that the RWE \eqref{eq:nonzero_mass} can not be based on a single irreducible Lorentz group transformation. Arguing similarly, we come to a conclusion that the representation $T$ must be reducible, and it must consist of linking irreducible representations.

For instance, the well known and most simple RWEs -- the Dirac equation (spin $\frac12$), the Duffin--Kemmer equation (spins $0$ and $1$), and the Fierz--Pauli equation (spin $\frac32$) -- are based on the following linking schemes
\begin{align}
(0, \frac12) - (\frac12 , 0),
\label{ls1}
\end{align}
\begin{align}
\begin{tabular}{c}
$(0,0)$ \\
$|$ \\
$(\frac12,\frac12)$
\end{tabular},
\label{ls2}
\end{align}
\begin{align}
(0, 1) - (\frac12 , \frac12) - (1,0),
\label{ls3}
\end{align}
and
\begin{align}
\begin{tabular}{ccc}
$(0,\frac12)$ & --- & $(\frac12,0 )$ \\
$|$ & & $|$ \\
$(\frac12,1)$ & --- & $(1, \frac12)$
\end{tabular},
\label{ls4}
\end{align}
respectively.

There is yet another possible variant of a linking scheme, when it consists of separate fragments, each of which fulfils the above stated conditions, but stays unlinked from the others. In this case the corresponding RWE will be falling apart with respect to the restricted Lorentz group transformations.

Usually, on the RWE \eqref{eq:nonzero_mass}  also imposed a condition of its invariance under spatial reflections, which together with the pure Lorentz transformations form the full Lorentz group. This condition leads to the fact that for each irreducible representation $\tau \sim (l_1 , l_2)$ with $l_1 \neq l_2$   occurring in a linking scheme there must be also present the  representation $\dot{\tau} \sim (l_2 , l_1)$, which is called conjugate to $\tau$.

A conceptual aspect of the RWE theory is the statement asserting that an entire physical microobject must be described by an equation which does not fall apart with respect to the full Lorentz group transformations.

All said above about the RWE \eqref{eq:nonzero_mass} also holds for the RWE \eqref{eq:zero_mass}, except for the case $\Gamma_0 = 0$. This exceptional case won't be considered in the present paper, and we refer an interested reader to the other papers, e.g. Ref.

A physical interest represent those RWE which can be derived by virtue of the variational principle from a Lorentz invariant Lagrangian function (Lagrangian density). For its construction it is necessary to introduce invariant quadratic combinations (quadratic forms) consisting of field functions, their first derivatives,  and the matrices $\Gamma_{\mu}$ and $\Gamma_0$. There is, however, a problem that matrices $T$ of the finite-dimensional representations of the Lorentz group are not unitary. Therefore, a simple quadratic form
\begin{align}
\Psi^{\dagger} \Psi =( \Psi^T)^* \Psi
\label{QuFo}
\end{align}
does not appear to be a  Lorentz invariant. Instead of the quadratic form \eqref{QuFo} one should introduce the so called Lorentz invariant bilinear form
\begin{align}
\overline{\Psi} \Psi = \Psi^{\dagger}\eta  \Psi, \quad \overline{\Psi} =  \Psi^{\dagger}\eta ,
\label{InBiFo}
\end{align}
where $\eta$ is some number matrix. To enforce the relativistic invariance of the form \eqref{InBiFo} one must claim that the matrix $\eta$ obeys the condition
\begin{align}
T^{\dagger} \eta T = \eta .
\label{InvCond}
\end{align}
Applying \eqref{InvCond} to the  infinitesimal transformations \eqref{InfLoTra} we obtain an equivalent formulation of the relativistic  invariance condition on $\eta$
\begin{align}
\eta J^{[ij]} & =  J^{[ij]} \eta , \label{inv1} \\
\eta J^{[i4]} & = - J^{[i4]} \eta . \label{inv2}
\end{align}
A matrix $\eta$ obeying these conditions is called the matrix of the Lorentz invariant bilinear form.

Under the fulfilment of the conditions \eqref{nonsing_trans}, \eqref{sing_trans}, \eqref{inv1}, \eqref{inv2}, the following combinations also appear to be the Lorentz invariants
\begin{align}
\partial_{\mu} (\overline{\Psi} \Gamma_{\mu} \Psi), \quad (\partial_{\mu} \overline{\Psi}) \Gamma_{\mu} \Psi, \quad \overline{\Psi} \Gamma_{\mu} (\partial_{\mu} \Psi).
\end{align}

A Lagrangian density  leading to the equation \eqref{eq:nonzero_mass} can be chosen, e.g., in the form
\begin{align}
\mathcal{L} (x) = - \frac12 \overline{\Psi} (\Gamma_{\mu} \partial_{\mu} + m) \Psi + \frac12( (\partial_{\mu} \overline{\Psi}) \Gamma_{\mu} -m \overline{\Psi} )\Psi.
\label{lagr_dens}
\end{align}
Varying it in accordance with the least action principle, we arrive at the equation \eqref{eq:nonzero_mass} for the function $\Psi$, and at the equation
\begin{align}
- (\partial_{\mu} \overline{\Psi}) \Gamma_{\mu} + m \overline{\Psi} = 0
\label{eqOPsi}
\end{align}
for the function $\overline{\Psi}$. The dynamical variables like the energy and the  momentum of the field are obtained from the Lagrangian density \eqref{lagr_dens} according to the general formulas
\begin{align}
E = - \int \overline{\Psi}\,  \Gamma_4 \, \partial_4 \Psi\,  d^3 x , \quad P=  - \frac{i}{c} \int \overline{\Psi}\,  \Gamma_4 \, \partial_i \Psi \, d^3 x ,
\end{align}
following from the Noether's theorem.

The following simplified form of the Lagrangian density is often  used
\begin{align}
\mathcal{L} (x) = - \overline{\Psi} (\Gamma_{\mu} \partial_{\mu}+ m) \Psi ,
\label{eq126}
\end{align}
which also allows us to obtain the equation \eqref{eq:nonzero_mass} and the correct expression for the dynamical variables. The equation \eqref{eq:zero_mass} can be derived from the Lagrangian density
\begin{align}
\mathcal {L} (x) =  - \overline{\Psi} (\Gamma_{\mu} \partial_{\mu} +\Gamma_0 )\Psi .
\end{align}

From the physical arguments it follows that the bilinear form \eqref{InBiFo} must be real valued
\begin{align}
( \Psi^{\dagger} \eta \Psi)^{\dagger} = \Psi^{\dagger} \eta^{\dagger} \Psi = \Psi^{\dagger} \eta \Psi,
\nonumber
\end{align}
implying
\begin{align}
\eta^{\dagger}= \eta .
\label{etaHerm}
\end{align}
Moreover, a normalization of the wavefunction $\Psi$ can be chosen in such a way that the following condition is fulfilled
\begin{align}
\eta^2 =1, \quad \mathrm{or} \quad \eta^{-1} = \eta .
\nonumber
\end{align}
Being combined with \eqref{etaHerm}, this leads to the relations
\begin{align}
\eta = \eta^T = \eta^* = \eta^{\dagger} = \eta^{-1} .
\label{etaCond}
\end{align}

The conditions \eqref{etaHerm} and \eqref{etaCond} do not yet fully define the matrix $\eta$, since they do not take into account that the function $\Psi (x)$ is a solution of the equation \eqref{eqOPsi}. To establish additional constraints imposed on the matrix $\eta$  due to the latter property , let us perform a hermitian conjugation of the equation \eqref{eq:nonzero_mass} and multiply the result by $\eta$. Thus, we obtain
\begin{align}
(\partial_i \Psi^{\dagger} \Gamma_i^{\dagger} - \partial_4 \Psi^{\dagger} \Gamma_4^{\dagger} +\Psi^{\dagger} m) =0 .
\label{eq:Psi_spec}
\end{align}
Let us require a possibility to proceed from the equation \eqref{eq:Psi_spec} to the equation \eqref{eqOPsi}. Obviously, to enable this the following commutation relations must hold
\begin{align}
\Gamma_i^{\dagger} \eta = - \eta \Gamma_i, \quad \Gamma_4^{\dagger} \eta = \eta \Gamma_4 .
\label{cond1}
\end{align}

In case of the RWE \eqref{eq:zero_mass} it is necessary to add to the relations \eqref{cond1} the condition
\begin{align}
 \Gamma_0^{\dagger} \eta = \eta \Gamma_0 ,
\label{cond2}
\end{align}
which along with \eqref{cond1} leads to the equation
\begin{align}
- (\partial_{\mu} \overline{\Psi}) \Gamma_{\mu} + \overline{\Psi} \Gamma_0 =0 .
\end{align}
for the function $\overline{\Psi}$.

On the choice of the  possible RWEs it is also imposed the following condition: among  states of a microobject there should be no states with zero energy.
This condition essentially restricts an allowed form of minimal polynomials for the matrices $\Gamma_{\mu}$.

As it is shown in [11], the minimal polynomial for the matrix $\Gamma_4$ (and, hence, for all the matrices $\Gamma_{\mu}$) must have the structure
\begin{align}
\Gamma_4^n (\Gamma_4^2 -\lambda_1^2) (\Gamma_4^2 - \lambda_2^2) \ldots =0,
\label{min_pol}
\end{align}
where $\lambda_i$ are distinct real valued numbers, $n$ is either an integer positive number or zero.

The definiteness conditions for the  energy and the charge for microobjects with a single mass value can be represented in the form of  inequalities
\begin{align}
(-1)^{n+1} \left[ (\mathrm{Sp} (\Gamma_4^{n+1} \eta))^2 - (\mathrm{Sp} (\Gamma_4^{n} \eta))^2 \right] & >0, \label{ineq1} \\
(-1)^n \left[[ (\mathrm{Sp} (\Gamma_4^{n+1} \eta))^2 - (\mathrm{Sp} (\Gamma_4^{n} \eta))^2 \right] & >0 , \label{ineq2}
\end{align}
respectively. Moreover, in accordance with the Pauli theorem, it is assumed that the condition \eqref{ineq1} takes place for integer case, while the condition \eqref{ineq2} takes place for half-integer spin.

Finally, what concerns the matrix $\Gamma_0$ in the equation \eqref{eq:zero_mass}, upon a certain choice of the basis in the wavefunction space the matrix $\Gamma_0$ can be casted to a form, in which it consists of independent scalar blocks $a_{\tau}$ put in correspondence to irreducible representations
 $\tau \in T$. A part of these blocks is zero valued. From the invariance condition of the RWE \eqref{eq:zero_mass} under transformation of the full Lorentz group it follows that nonzero blocks $a_{\tau}$ obey the equality
\begin{align}
a_{\tau} = a_{\dot{\tau}}.
\end{align}
In case of finite-dimensional RWEs this condition can be also derived from a requirement of a possibility to formulate the RWE \eqref{eq:zero_mass} in terms of the Lagrangian density.

For a construction of a RWE with a given spectrum of mass and spin states it is convenient to use the so called canonical basis, or the Gel'fand--Yaglom basis. In this basis the components of the wavefunction $\psi^{\tau}_{sk}$ describe the "pure" states, i.e. the states with a definite spine value $s$ and its projection $k$, the upper index $\tau$ pointing to the corresponding irreducible representation. In the Gel'fand--Yaglom basis the matrix $\Gamma_4$ has a quasi-diagonal form
\begin{align}
\Gamma_4 = \bigoplus_s C^s \otimes I_{2s+1},
\label{osumG4}
\end{align}
where $I_{2s+1}$ is a unit matrix of the dimension $2 s+1$; $C^s$ is a matrix block corresponding to the spin $s$ in the following sense: if the matrix $C^s$ has nozero roots, then a particle possesses the spin $s$. Possible mass values of a microobject in case of the RWE \eqref{eq:nonzero_mass} are expressed via nonzero roots $\lambda_i$ of the block $C^s$ by
\begin{align}
m_i^{(s)} = \frac{m}{|\lambda_i^{(s)}|}.
\end{align}

The spin block $C^s$ is constructed in the following way. From a linking scheme, on which a RWE is based, one selects all irreducible components $\tau \sim (l_1 , l_2)$ satisfying the condition
\begin{align}
|l_1 - l_2 | \leq s\leq l_1 + l_2 .
\label{s_range}
\end{align}
(It is said that these representations form the block $C^s$.) Then, these components are enumerated for convenience. The matrix $C^s$ consists of the elements $c^s_{\tau \tau'}$, the representation labels $\tau, \tau'$ playing the role of matrix indices. To the nonlinking components correspond zero valued elements  $c^s_{\tau \tau'}$. It follows that in a linking scheme intended for a description of a spin $s$ there must be at least two linking irreducible representations, which satisfy the condition \eqref{s_range}.

The requirement of the relativistic invariance of the RWEs \eqref{eq:nonzero_mass}, \eqref{eq:zero_mass} imposes the following constraints on the elements $c_{\tau \tau'}^s$:
\begin{align}
c^s_{\tau \tau'} &=c_{\tau \tau'} \sqrt{(s+l_+ +2)(s - l_+ - 1)}  \quad \mathrm{if} \quad l'_+ = l_+ + 1, \quad l'_- = l_-  , \nonumber \\
c^s_{\tau \tau'} &=c_{\tau \tau'} \sqrt{(s+l_- +1)(s - l_-)} \quad \mathrm{if} \quad l'_+ = l_+ , \quad l'_- = l_- + 1 ,\label{c_cond} \\
c^s_{\tau \tau'} &=c_{\tau \tau'} (s+ \frac12)\quad \mathrm{if} \quad l'_+ = l_+ , \quad l'_- = l_-  , \nonumber
\end{align}
where $l_+ = l_1 + l_2$, $l_- = |l_1 - l_2|$, $l'_+ = l'_1 + l'_2$, $l'_- = |l'_1 - l'_2|$ ; and $c_{\tau \tau'}$ are arbitrary nonzero complex numbers for linking representations, and zeroes otherwise.

The invariance of the RWEs with respect to spatial reflections, realized by a matrix $P$, imposes on the numbers $c_{\tau \tau'}$  the following constraints
\begin{align}
c_{\tau \tau'} &= c_{\dot{\tau} \dot{\tau}'} \quad \mathrm{if} \quad \dot{\tau} \neq \tau, \,\, \dot{\tau}' \neq \tau' ; \label{c1_cond}  \\
c_{\tau \tau'} &=\pm  c_{\dot{\tau} \dot{\tau}'} \quad \mathrm{if} \quad \dot{\tau} = \tau, \,\, \dot{\tau}' \neq \tau' \quad  \mathrm{or} \quad  \dot{\tau} \neq \tau, \,\, \dot{\tau}' = \tau'  . \label{c2_cond}
\end{align}
In the second line for the first case,  the sign "+" is chosen if the operator $P$ acts in the subspaces $R^{\tau}$, $R^{\tau'}$  equally, that is
\begin{align}
P \psi_{sk}^{\tau} = (-1)^{[s]} \psi_{sk}^{\tau}, \quad P \psi_{sk}^{\tau'}= (-1)^{[s]} \psi_{sk}^{\dot{\tau}'};
\label{Pc1_cond}
\end{align}
 the sign "$-$" is chosen if the operator$P$ acts in the subspaces $R^{\tau}$, $R^{\tau'}$  differently, that is
\begin{align}
P \psi_{sk}^{\tau} = (-1)^{[s]+1} \psi_{sk}^{\tau}, \quad P \psi_{sk}^{\tau'}= (-1)^{[s]} \psi_{sk}^{\dot{\tau}'}.
\label{Pc2_cond}
\end{align}

The matrix $\eta$ of the bilinear form \eqref{QuFo} in the Gel'fand--Yaglom basis has the structure which is analogous to \eqref{osumG4}. Namely,
\begin{align}
\eta =\bigoplus_s \eta^s \otimes I_{2 s+1}.
\end{align}
The conditions \eqref{inv1}, \eqref{inv2}, and \eqref{etaCond} yield that the only nonzero elements are $\eta^s_{\tau \dot{\tau}}$; moreover,
\begin{align}
\eta_{\tau  \dot{\tau}}^s = \eta_{\dot{\tau} \tau}^s = - \eta_{\tau \dot{\tau}}^{s+1}.
\label{eta_cond}
\end{align}

The conditions \eqref{cond1} lead to the relation
\begin{align}
c_{\tau \tau'}^s \eta^s_{\tau' \dot{\tau}'} = (c^s_{\dot{\tau}' \dot{\tau}} )^* \eta^s_{\tau \dot{\tau}}.
\label{c_eta_cond}
\end{align}

After imposing on the elements of the matrices $\Gamma_4$, $\Gamma_0$, and $\eta$ the constraints \eqref{c_cond}--\eqref{c2_cond}, \eqref{eta_cond}, \eqref{c_eta_cond}, there remains, as a rule, an arbitrariness which can be used to satisfy the conditions \eqref{min_pol} and \eqref{ineq1} (or \eqref{ineq2}). If this turns out impossible, then one can not construct a RWE for description of a microobject with given spectrum of mass and spin states on the basis of a considered linking scheme.

\section{The simplest relativistic wave equations for particles  with lowest spins}
\hspace{0.5cm}
In the following we will apply the Gel'fand--Yaglom approach whose main features are described in the previous section. We will demonstrate how to use it in practice on examples of the well known and most simple RWEs with lowest spins.

Let us first consider the linking scheme \eqref{ls2} in which the irreducible representation $(0,0)$ corresponds to the scalar function $\psi_0$, and the representation $(\frac12, \frac12)$ corresponds to the vector function $\psi_{\mu}$. The matrix $\Gamma_4$ of the RWE \eqref{eq:nonzero_mass}, which is based on the given set of irreducible representations of the Lorentz group,  has the following form in the Gel'fand--Yaglom basis
\begin{align}
\Gamma_4 = \left( \begin{tabular}{cc}
$C^0$ &  $0$ \\
$0$ & $C^1 \otimes I_3$
\end{tabular} \right),
\label{gam4_s2}
\end{align}
where the blocks $C^0$ and $C^1$ correspond to spins $0$ and $1$, respectively. Labelling the representations under consideration by the numbers
\begin{align}
(0,0) \sim 1 , \quad (\frac12 , \frac12) \sim 2,
\end{align}
which play the role of matrix indices, we obtain the following expressions for the matrices $C^0$ and $C^1$:
\begin{align}
C^0 = \left( \begin{tabular}{cc}
$0$ &  $c_{12}^0$ \\
 $c_{21}^0$ & $0$
\end{tabular} \right), \qquad C^1 =0 .
\end{align}
It is also easy to see that the relations \eqref{c_cond}--\eqref{c2_cond} do not impose any constraints on the numbers $c_{12}^0$ and $c_{21}^0$.

Matrix elements of the Lorentz invariant bilinear form
\begin{align}
\eta =  \left( \begin{tabular}{cc}
$\eta^0$ &  $0$  \\
$0$  & $\eta^1 \otimes I_3$
\end{tabular} \right), \quad \eta^0 =   \left( \begin{tabular}{cc}
$\eta^0_{11}$ &   $0$ \\
$0$  & $\eta^0_{22}$
\end{tabular} \right), \quad \eta^1 = \eta_{22}^1
\label{eq24}
\end{align}
can be chosen as
\begin{align}
\eta^0_{11} = \eta^0_{22} = - \eta_{21}^1 =1.
\end{align}
At this choice the condition \eqref{c_eta_cond} leads to the equality
\begin{align}
c_{21}^0 = (c_{12}^0)^* .
\end{align}
Setting $c_{12}^0 =1$, we obtain the final expressions for the matrices $C^0$ and $\Gamma_4$
\begin{align}
C^0 = \left(  \begin{tabular}{cc} $0$ & $1$ \\ $1$ & $0$  \end{tabular} \right) , \quad
\Gamma_4 = \left( \begin{tabular}{ccccc}
$ 0$ & $1$ & $ 0$ & $0$ & $0$ \\
$ 1$ & $0$ & $ 0$ & $0$ & $0$ \\
$ 0$ & $0$ & $ 0$ & $0$ & $0$ \\
$ 0$ & $0$ & $ 0$ & $0$ & $0$ \\
$ 0$ & $0$ & $ 0$ & $0$ & $0$
\end{tabular}\right).
\label{gam4_2}
\end{align}
It is easy to check that the minimal equation for the matrix $\Gamma_4$ is
\begin{align}
\Gamma_4 (\Gamma_4^2 -1) =0.
\label{gam4_2min}
\end{align}
The form of the other matrices $\Gamma_i$ can be found on the basis of Eq.~\eqref{gam_i_gen}.

For the definiteness of the energy in this case $(n=1)$ the following inequality must hold
\begin{align}
(-1)^2 \left[ \left(\mathrm{Sp} (\Gamma_4^2 \eta ) \right)^2 - \left(\mathrm{Sp} (\Gamma_4 \eta ) \right)^2 \right] >0 .
\label{def_en_s2}
\end{align}
Its validity is easily confirmed by a direct evaluation.

From \eqref{gam4_2} and \eqref{gam4_2min} it follows that to the state with spin $s=0$ corresponds a single mass value, while the state with spin $s=1$ is absent. In the literature the RWE with the matrix $\Gamma_4$ defined by Eq.~\eqref{gam4_2}  is usually called the Duffin--Kemmer equation for a scalar particle. In the tensor formulation it reads
\begin{align}
\partial_{\mu} \psi_{\mu} + m\psi_0 = 0 , \quad \partial_{\mu} \psi_0 + m \psi_{\mu} = 0.
\end{align}
It is easy to derive from it a second order equation for a scalar particle
\begin{align}
(\square -m^2) \psi_0 = 0.
\end{align}
The latter equation means that we indeed achieve a description of a particle with a nonzero mass and spin $s=0$.

For constructing the simplest RWE for a particle with a nonzero mass and spin $s=1$ serves the linking scheme \eqref{ls3}, where a combination of the representations $(0,1)$ and $(1,0)$ corresponds to the second-rank antisymmetric tensor $\psi_{[\mu \nu]}$. In this case the matrix $\Gamma_4$ in the Gel'fand--Yaglom basis has still the block structure \eqref{gam4_s2}. Introducing the labels $(\frac12, \frac12) \sim 1$, $(0,1) \sim 2$, $(1,0) \sim 3$, we establish the following expressions for the spin blocks
\begin{align}
C^1  =  \left(
\begin{tabular}{ccc}
$0$ & $c_{12}^1$ & $c_{13}^1$ \\
$c_{21}^1$ & $0$ & $0$ \\
$c_{31}^1$ & $0$ & $0$
\end{tabular} \right), \quad C^0 = 0.
\label{eq212}
\end{align}
Like in case of spin $s=0$, the relations \eqref{c_cond} do not impose here any constraints on the numbers $c_{ij}^1$. The invariance under spatial reflections
yields the relations (see equation \eqref{c2_cond})
\begin{align}
c_{12}^1 = \pm c_{13}^1 , \quad c_{21}^1  = \pm c_{31}^1 .
\label{Pcond_s2}
\end{align}
A possibility of the Lagrangian formulation leads to the condition
\begin{align}
c_{12}^1 = \frac{\eta_{11}^1}{\eta_{23}^1} (c_{31}^1)^* .
\label{c_eta_s2}
\end{align}
Nonvanishing matrix elements of the invariant bilinear form
\begin{align}
 \eta &= \left(\begin{tabular}{cc} $\eta^0$ & $0$  \\ $0$ & $\eta^1 \otimes I_3$ \end{tabular}  \right), \quad \eta^0 = \eta_{11}^0 , \nonumber  \\
\eta^1 &= \left(\begin{tabular}{ccc}
$\eta^1$ & $0$ & $0$ \\
$0$ & $0$ & $\eta_{23}^1$ \\
$0$ & $\eta_{32}^1$ & $0$
\end{tabular} \right)
\end{align}
can be chosen as
\begin{align}
-\eta_{11}^0 = \eta_{11}^1 = \pm \eta_{23}^1 = \pm \eta_{32}^1 = 1 .
\end{align}
Then, accounting \eqref{Pcond_s2} we deduce from \eqref{c_eta_s2} the relations
\begin{align}
c_{21}^1 = (c_{12}^1)^* , \quad c_{31}^1 = (c_{13}^1)^* .
\end{align}

Choosing $c_{12}^1 =\frac{1}{\sqrt{2}} $, we obtain
\begin{align}
C^1  = \frac{1}{\sqrt{2}} \left(
\begin{tabular}{ccc}
$0$ & $1$ & $\pm 1$ \\
$1$ & $0$ & $0$ \\
$ \pm 1$ & $0$ & $0$
\end{tabular} \right),
\label{C1_1}
\end{align}
where the signs are correlated. The block $C^1$ in \eqref{C1_1} has the only nonzero eigenvalue $1$  (up to a sign), that is the constructed RWE describes
a particle with a nonzero mass and spin $s=1$. The relations \eqref{gam4_2min} and \eqref{def_en_s2} are also fulfilled. In the literature this equation is called the Duffin--Kemmer equation for a vector particle. In the tensor formulation it reads
\begin{align}
\partial_{\nu} \psi_{[\mu \nu]} + m \psi_{\mu} = 0 , \quad - \partial_{\mu} \psi_{\nu} + \partial_{\nu} \psi_{\mu} + m \psi_{[\mu \nu]} = 0 .
\label{proca1}
\end{align}
From this system of the first order equations one can derive a second order equation
\begin{align}
(\square -m^2) \psi_{\mu} =0 , \quad \partial_{\mu} \psi_{\mu} =0 ,
\end{align}
which is called the Proc\'a equation.

The simplest equation for a spin-$\frac12$ particle -- the Dirac equation -- follows from the linking scheme \eqref{ls1}, corresponding to a first-rank bispinor. The matrix $\Gamma_4$ of this equation in its standard form \eqref{eq:nonzero_mass} is written in the Gel'fand--Yaglom basis as
\begin{align}
\Gamma_4 = C^{1/2} \otimes I_2 = \left( \begin{tabular}{cccc}
$0$ & $0$ & $c_{12}^{1/2}$ & $0$ \\
$0$ & $0$ & $0$ & $c_{12}^{1/2}$ \\
$c_{21}^{1/2}$ & $0$ & $0$ & $0$ \\
$0$ & $c_{21}^{1/2}$ & $0$ & $0$ \\
\end{tabular} \right),
\end{align}
where the following labeling $(0, \frac12) \sim 1$, $(\frac12 ,0) \sim 2$ of the lower indices is introduced. Applying the conditions \eqref{c_cond}, \eqref{c1_cond}, and \eqref{c_eta_cond} to the matrix elements $c_{ij}^{1/2}$, we obtain
\begin{align}
\Gamma_4 = \left( \begin{tabular}{cc} $0$ &  $I_2$ \\  $I_2$ &  $0$  \end{tabular} \right) = \sigma_1 \otimes I_2 ,
\end{align}
the other matrices $\Gamma_i$ acquiring the form
\begin{align}
\Gamma_1 = \sigma_2 \otimes \sigma_1 , \quad \Gamma_2 = \sigma_2 \otimes \sigma_2 , \quad \Gamma_3 = \sigma_2 \otimes \sigma_3 ,
\end{align}
expressed via the Pauli matrices $\sigma_i$. The matrix of the bilinear form coincides with $\Gamma_4$:
\begin{align}
\eta = \Gamma_4 .
\end{align}
The Dirac matrices satisfy the algebra
\begin{align}
\Gamma_{\mu} \Gamma_{\nu} + \Gamma_{\nu} \Gamma_{\mu} = 2 \delta_{\mu \nu}
\end{align}
and the minimal equation
\begin{align}
\Gamma_{\mu}^2 - 1 = 0 \quad (\mathrm{no \,\, summation \,\, over \,\, }\mu).
\label{min_pol_dir}
\end{align}
Note that the unitary transformation
\begin{align}
U = \frac{1}{\sqrt{2}} \left( \begin{tabular}{cc} $I_2$ &  $I_2$ \\  $I_2$ &  $-I_2$  \end{tabular} \right)
\end{align}
transforms the matrix $\Gamma_4$ into the form
\begin{align}
\Gamma_4 = \left( \begin{tabular}{cc} $I_2$ &  $0$ \\  $0$ &  $-I_2$  \end{tabular}  \right) = \sigma_3 \otimes I_2,
\end{align}
while leaving the other matrices $\Gamma_i$ unchanged.

The condition of the charge positive definiteness \eqref{ineq2} acquires in the considered case ($n=0$) the form
\begin{align}
\left(\mathrm{Sp} (\Gamma_4 \eta) \right)^2 -\left (\mathrm{Sp} \, \eta \right)^2 >0.
\label{pd_ch1}
\end{align}
Since
\begin{align}
\mathrm{Sp} \, \eta = \mathrm{Sp} \,  \Gamma_4= 0 , \quad \mathrm{Sp} (\Gamma_4 \eta) = 4,
\nonumber
\end{align}
the inequality \eqref{pd_ch1} is fulfilled.

It is necessary to remark an important property which is common for the RWEs discussed above: the matrix $\Gamma_4$ in the Duffin--Kemmer equations for both scalar and vector particles as well as in the Dirac equation can be ever diagonalized. At the same time  it is known from Refs. [8; 9] that among all finite-dimensional equations with a diagonalizable matrix $\Gamma_4$ only the Duffin--Kemmer equation has the positive-definite energy, and only the Dirac equation has a positive-definite charge. It is, however, possible to construct equations with either a positive-definite energy or a positive-definite charge, which feature a non-diagonalizable matrix $\Gamma_4$. Below we consider examples of such equations.

The simplest and the most known equation among the RWEs of this type is the Fierz--Pauli equation for a spin-$\frac32$ particle. Let us give its formulation in the Gel'fand--Yaglom approach. To this end, we take the linking scheme \eqref{ls4} of Lorentz group irreducible representations
and analyze all possibilities to construct on its basis admissible theories for spin $s=\frac32$.

The matrix $\Gamma_4$ of the equation \eqref{eq:zero_mass}, corresponding to the set of representations contained in the linking scheme \eqref{ls4},
has the following form in the Gel'fand--Yaglom basis
\begin{align}
\Gamma_4 &= \left(\begin{tabular}{cc} $C^{1/2} \otimes I_2$ & $0$ \\
$0$ & $C^{3/2} \otimes I_4$
\end{tabular} \right), \label{gam4_s2init} \\
C^{3/2} &=  \left( \begin{tabular}{cc}
$0$ & $c_{34}^{3/2}$  \\
 $c_{43}^{3/2}$  &  $0$
\end{tabular} \right), \quad
C^{1/2} = \left( \begin{tabular}{cccc}
$0$ & $c_{12}^{1/2}$ & $c_{13}^{1/2}$ & $0$ \\
$c_{21}^{1/2}$ & $0$ & $0$ & $c_{24}^{1/2}$ \\
$c_{31}^{1/2}$ & $0$ & $0$ & $c_{34}^{1/2}$ \\
$0$ & $c_{42}^{1/2} $ & $c_{43}^{1/2} $ & $0$
 \end{tabular} \right),
\end{align}
where the labeling $(0, \frac12) \sim 1$, $(\frac12, 0 ) \sim 2$ , $(\frac12 , 1) \sim 3$, $(1, \frac12) \sim 4$ is used. The matrix $\eta$ has an analogous structure
\begin{align}
\eta = (\eta^{1/2} \otimes I_2 ) \oplus (\eta^{3/2} \otimes I_4 ),
\nonumber
\end{align}
the following matrix elements being nonzero
\begin{align}
\eta_{12}^{1/2} = \eta_{21}^{1/2}, \quad \eta_{34}^{1/2} = \eta_{43}^{1/2} =- \eta_{34}^{3/2} =- \eta_{43}^{3/2}.
\nonumber
\end{align}

The invariance of the RWE theory under the reduced Lorentz group transformations yields in this case the relations
\begin{align}
c_{34}^{3/2} = 2 c_{34}^{1/2}, \quad c_{43}^{3/2} = 2 c_{43}^{1/2}.
\label{invLor}
\end{align}
For the $P$-invariance it is necessary to demand (see equation \eqref{c1_cond}--\eqref{Pc2_cond})
\begin{align}
\begin{split}
c_{12}^{1/2} & = c_{21}^{1/2}, \quad c_{34}^{1/2} = c_{43}^{1/2}, \\
c_{13}^{1/2} & = c_{24}^{1/2}, \quad c_{31}^{1/2} = c_{42}^{1/2}, \quad c_{34}^{3/2} = c_{43}^{3/2}.
\end{split}
\label{invP}
\end{align}
Finally, the condition of a derivability of a sought equation from an invariant Lagrangian function in accordance with \eqref{c_eta_cond} imposes on the elements $c_{ij}^s$ of the matrix $\Gamma_4$ the constraints
\begin{align}
c_{12}^{1/2},c_{21}^{1/2}, c_{34}^{1/2},c_{43}^{1/2} \in \mathbb{R}
\label{invLag1}
\end{align}
and
\begin{align}
\begin{split}
c_{42}^{1/2} \eta_{21}^{1/2} &= (c_{13}^{1/2})^* \eta_{43}^{1/2} , \\
c_{31}^{1/2} \eta_{12}^{1/2} &= (c_{24}^{1/2})^* \eta_{34}^{1/2} .
\end{split}
\label{c_eta_cond2}
\end{align}

For the sake of convenience let us rewrite the relations \eqref{c_eta_cond2} as
\begin{align}
c_{42}^{1/2} =( c_{13}^{1/2})^* f , \quad c_{31}^{1/2} =( c_{24}^{1/2})^* f , \quad f = \frac{\eta_{34}^{1/2}}{\eta_{12}^{1/2}} =\frac{\eta_{43}^{1/2}}{\eta_{21}^{1/2}}.
\end{align}
Without loss of generality, the parameter $f$ can be chosen to be either $+1$ or $-1$. This is equivalent to the two essentially different ways of defining the matrix $\eta$:
\begin{align}
- \eta_{12}^{1/2} &= - \eta_{34}^{1/2} = \eta_{34}^{3/2} = 1, \label{def_eta1} \\
\eta_{12}^{1/2} &= - \eta_{34}^{1/2} = \eta_{34}^{3/2} = 1 . \label{def_eta2}
\end{align}

In the first place, we consider all possible RWEs for spin $s=\frac32$ imposing only the constraints \eqref{invLor}, \eqref{invLag1}, \eqref{c_eta_cond2}, that is including into the set of admissible RWEs both $P$-invariant and $P$-noninvariant ones. For the spin block $C^{3/2}$ there exists~-- up to equivalence~-- the unique choice of its elements: $c_{34}^{3/2}= c_{43}^{3/2}=1$. According to \eqref{invLor} it follows $c_{34}^{1/2}=c_{43}^{1/2} = \frac12$, and thus the blocks $C^{1/2}$ and $C^{3/2}$ acquire the form
\begin{align}
C^{1/2} = \left(  \begin{tabular}{cccc}
$0$ & $a$ & $c$ & $0$ \\
$b$ & $0$ & $0$ & $d$ \\
$f d^*$ & $0$ & $0$ & $1/2$ \\
$0$ & $ f c^* $ & $1/2 $ & $0$
 \end{tabular}\right) , \quad C^{3/2} = \left(  \begin{tabular}{cc} $0$ & $1$ \\ $1$ & $0$ \end{tabular} \right),
\label{c12c32}
\end{align}
where the notations
\begin{align}
c_{12}^{1/2} = a, \quad c_{21}^{1/2} = b , \quad c_{13}^{1/2} = c , \quad c_{24}^{1/2} =d
\nonumber
\end{align}
are introduced for brevity.

In order to obtain on the basis of \eqref{c12c32} a RWE describing the pure spin $s = \frac32$, it is necessary to require turning into zero of all roots of the characteristic equations for the spin block $C^{1/2}$
\begin{align}
\lambda^4 - \left( f c^* d + f c d^* + a b+ \frac14\right) \lambda^2+ \left( \frac14 a b - \frac12 fa |d|^2  - \frac12 f b |c|^2 + |c|^2 |d|^2 \right) = 0.
\end{align}
The roots of this equation equal zero  under fulfillment of the conditions
\begin{align}
\begin{split}
&  \frac14 a b - \frac12 fa |d|^2  - \frac12 f b |c|^2 + |c|^2 |d|^2  = 0 , \\
&   f c^* d + f c d^* + a b+ \frac14  = 0 .
\end{split}
\label{zero_cond}
\end{align}

The structure \eqref{c12c32} of the spin blocks $C^{1/2}$, $C^{3/2}$ of the matrix $\Gamma_4$ is associated with some generalized RWE, from which under certain choices of the parameters $a$, $b$, $c$, $d$  follow formulations of concrete equations for a spin-$\frac32$ particle. Taking into account the possibilities \eqref{def_eta1} and \eqref{def_eta2} of defining the matrix of the bilinear form $\eta$, it is meaningful to introduce RWEs of the first type with the matrices
\begin{align}
C^{1/2} =  \left( \begin{tabular}{cccc}
$0$ & $a$ & $c$ & $0$ \\
$b$ & $0$ & $0$ & $d$ \\
$d^*$ & $0$ & $0$ & $1/2$ \\
$0$ & $c^* $ & $1/2 $ & $0$
 \end{tabular} \right), \quad C^{3/2} = \left(  \begin{tabular}{cc} $0$ & $1$ \\ $1$ & $0$ \end{tabular} \right) \quad \mathrm{for} \quad f=+1,
\label{C_type1}
\end{align}
and RWEs of the second type with the matrices
\begin{align}
C^{1/2} =  \left(  \begin{tabular}{cccc}
$0$ & $a$ & $c$ & $0$ \\
$b$ & $0$ & $0$ & $d$ \\
$-d^*$ & $0$ & $0$ & $1/2$ \\
$0$ & $-c^* $ & $1/2 $ & $0$
 \end{tabular}\right), \quad C^{3/2} = \left(  \begin{tabular}{cc} $0$ & $1$ \\ $1$ & $0$ \end{tabular} \right)  \quad \mathrm{for} \quad f=-1.
\label{C_type2}
\end{align}

In case of RWEs of the second type the system \eqref{zero_cond} admits the solution
\begin{align}
a=b= - \frac12 , \quad c=d= \frac12
\nonumber
\end{align}
which leads to the matrix $\Gamma_4$ with the blocks
\begin{align}
C^{1/2} = \frac12 \left(  \begin{tabular}{cccc}
$0$ & $-1$ & $1$ & $0$ \\
$-1$ & $0$ & $0$ & $1$ \\
$-1$ & $0$ & $0$ & $1$ \\
$0$ & $-1$ & $1$ & $0$
 \end{tabular}\right), \quad C^{3/2} = \left( \begin{tabular}{cc} $0$ & $1$ \\ $1$ & $0$ \end{tabular} \right).
\end{align}
It is easy to check that the condition \eqref{invP} is also fulfilled. Thus, we obtain a $P$-invariant RWE, which appears to be the Fierz--Pauli equation in the Gel'fand--Yaglom formalism.

The matrix $\Gamma_4$ \eqref{gam4_s2init}, \eqref{C_type2} of the Fierz--Pauli equation satisfies the minimal equation
\begin{align}
\Gamma_4^2 (\Gamma_4^2 - 1) = 0,
\end{align}
and therefore it can not be diagonalized. The charge definiteness condition \eqref{ineq2} in this case acquires the form
\begin{align}
\left(\mathrm{Sp} (\Gamma_4^3 \eta) \right)^2 - \left(\mathrm{Sp} (\Gamma_4^2 \eta) \right)^2 >0.
\label{ch_def}
\end{align}

Taking into account the explicit expression for the blocks $\eta^{1/2}$, $\eta^{3/2}$ of the matrix $\eta$
\begin{align}
\eta^{1/2} = \left(  \begin{tabular}{cccc}
$0$ & $1$ & $0$ & $0$ \\
$1$ & $0$ & $0$ & $0$ \\
$0$ & $0$ & $0$ & $-1$ \\
$0$ & $0$ & $-1$ & $0$
 \end{tabular}\right), \quad \eta^{3/2} = \left( \begin{tabular}{cc} $0$ & $1$ \\ $1$ & $0$ \end{tabular} \right),
\end{align}
we obtain
\begin{align}
\mathrm{Sp} (\Gamma_4^2 \eta) =0, \quad \mathrm{Sp} (\Gamma_4^3 \eta)  \in \mathbb{R}\backslash \{  0 \}.
\end{align}
Thus, the condition \eqref{ch_def} is fulfilled.

In the framework of the first type RWEs it turns out impossible to construct a $P$-invariant RWE, admitting the Lagrangian formulation. In fact, the conditions \eqref{invP} mean (for $f=+1$) that $a=b$, $c=d$, and then it follows from \eqref{zero_cond} that $2 |c|^2 + a^2 + \frac14 =0$. But the latter condition can not be fulfilled, since according to \eqref{invLag1} the parameter $a=c_{12}^{1/2}$ is real-valued.

If we relax the condition \eqref{invP}, then the system \eqref{zero_cond} has the following solutions for $f=-1$:
\begin{align}
a=b=c=-d = \frac12 \quad \mathrm{and} \quad a=b=-c=d = \frac12,
\nonumber
\end{align}
which lead to the spin blocks
\begin{align}
C^{1/2} = \frac12 \left(  \begin{tabular}{cccc}
$0$ & $1$ & $\pm 1$ & $0$ \\
$1$ & $0$ & $0$ & $\mp 1$ \\
$\mp 1$ & $0$ & $0$ & $1$ \\
$0$ & $\pm 1 $ & $1$ & $0$
 \end{tabular}\right), \quad C^{3/2} = \left(  \begin{tabular}{cc} $0$ & $1$ \\ $1$ & $0$ \end{tabular} \right),
\end{align}
where the signs "$+$" and "$-$" correspond to the two above stated solutions, respectively. The RWEs obtained in this way describe a particle with spin $s=\frac32$ and admit the Lagrangian formulation, but do not appear invariant with respect to spatial reflections.

The analysis of the linking scheme \eqref{ls4} in the Gel'fand--Yaglom approach allows us to make an important conclusion. Setting
\begin{align}
a=b= \frac12 , \quad c=d= \frac{\sqrt{3}}{2}
\end{align}
in the framwork of the first RWEs, we arrive at the matrix $\Gamma_4$ whose minimal polynom  has the form \eqref{min_pol_dir}. This choice agrees with the conditions \eqref{invLor}--\eqref{c_eta_cond2} and leads to the RWE of the Dirac type, describing a variable spin $\frac12, \frac32$. In the spin-tensor formulation this equation acquires a familiar shape
\begin{align}
(\gamma_{\mu} \partial_{\mu} + m)_{\alpha \beta} \psi_{\nu}^{\beta} =0,
\end{align}
being  the starting point for deriving the Rarita--Schwinger equation by means of imposing the constraints
\begin{align}
\gamma_{\nu} \psi_{\nu}^{\beta} = 0 , \quad \partial_{\nu} \psi_{\nu}^{\beta} = 0.
\end{align}
Since these conditions projecting out the redundant spin $\frac12$ are $P$-invariant and do not change the matrix structure of the bilinear form \eqref{def_eta1}, it becomes obvious that the Rarita--Schwinger equation belongs to the first type RWEs. In turn, the Fiertz-Pauli equation belongs to the second type. Therefore, the identification of these equations usually made in the literature requires a critical rethinking.

\section{RWEs with an extended set of the Lorentz group representations and internal structure of microobjects}
\hspace{0.5cm}
A characteristic feature of all RWE considered above consists in the fact that they are based on the sets of the Lorentz group irreducible representations which are minimally necessary for constructing theory for a given spin. Along with that, in accordance with the ideology of the relativistic quantum mechanics,
which interprets elementary particles as point-like structureless objects, such RWEs take into account only spin properties of particles. A possibility of describing other internal properties of particles in the orthodox version of the RWE theory is not provided.

Relaxing the requirement of minimality in usage of sets of the Lorentz group irreducible representations opens new possibilities of the RWE theory approach for a spatio-temporal (geometrized) description of internal properties of particles. To obtain equations which do not disintegrate with respect to the Lorentz group and which are capable of reflecting an internal structure of a particle with spin $s$, one can use the following possibilities: Either to include into a linking scheme representations with higher weights or to employ multiple representations of the Lorentz group. In the present chapter we show how to describe an internal electromagnetic structure of particles with lowest spins in the framework of the RWE theory with extended sets of the Lorentz group representations.

For the first time, the RWE for a particle with spin $s = \frac12$, which arises after involving additional -- with respect to the bispinor -- irreducible components in the representation space of a wavefunction, was proposed by Petras [17]. Here we give a brief description of the theory of the Petras equation in the Gel'fand--Yaglom approach. To this end, we consider the following linking scheme
\begin{align}
\begin{tabular}{ccc}
$ (0,\frac12)'$  & --- &  $(\frac12 ,0)'$ \\
$|$ & & $|$ \\
$(\frac12 ,1)$ & --- &  $(1,\frac12 )$ \\
$|$ & & $|$ \\
$ (0,\frac12 )$  & --- &  $(\frac12 ,0)$
\end{tabular} .
\label{scheme31}
\end{align}

Let us enumerate the irreducible representations contained in \eqref{scheme31}:
\begin{align}
\begin{split}
& (0, \frac12) \sim 1, \quad  (0, \frac12)' \sim 2 , \quad (1, \frac12) \sim 3 , \\
& (\frac12 ,0) \sim 4 , \quad (\frac12 , 0)' \sim 5 , \quad (\frac12 ,1) \sim 6 .
\end{split}
\end{align}
Then, we for the spin blocks $C^{1/2}$, $C^{3/2}$ of  the matrix
\begin{align}
\Gamma_4 = (C^{1/2} \otimes I_2) \oplus (C^{3/2} \otimes I_4)
\end{align}
we obtain the following expressions
\begin{align}
C^{1/2} = \left( \begin{tabular}{cccccc}
$0$ & $0$ & $0$ & $c_{14}^{1/2}$ & $0$ & $c_{16}^{1/2}$ \\
$0$ & $0$ & $0$ & $0$ & $c_{25}^{1/2}$ & $c_{26}^{1/2}$ \\
$0$ & $0$ & $0$ & $c_{34}^{1/2}$ & $c_{35}^{1/2}$ & $c_{36}^{1/2}$ \\
$c_{41}^{1/2}$ & $0$ & $c_{43}^{1/2}$ & $0$ & $0$ & $0$ \\
$0$ & $c_{52}^{1/2}$ & $c_{53}^{1/2}$ & $0$ & $0$ & $0$ \\
$c_{61}^{1/2}$ & $c_{62}^{1/2}$ & $c_{63}^{1/2}$ & $0$ & $0$ & $0$
\end{tabular} \right), \quad
C^{3/2} = \left( \begin{tabular}{cc}
$0$ & $c_{36}^{3/2}$ \\
$c_{63}^{3/2} $ & $0$ \end{tabular} \right).
\label{eq34}
\end{align}
To exclude spin $s = \frac32$, we impose constraints
\begin{align}
c_{36}^{3/2} = c_{63}^{3/2} =0, \quad \mathrm{or} \quad C^{3/2} =0,
\label{eq35}
\end{align}
from whence it follows
\begin{align}
c_{36}^{1/2} = c_{63}^{1/2} =0
\label{eq36}
\end{align}
 by virtue of \eqref{c_cond}. The condition \eqref{c1_cond} of the $P$-invariance leads to the relations
\begin{align}
\begin{split}
& c_{14}^{1/2} =c_{14}^{1/2} , \quad  c_{25}^{1/2} =c_{52}^{1/2} , \quad  c_{16}^{1/2} =c_{43}^{1/2} , \\
& c_{26}^{1/2} =c_{53}^{1/2} , \quad  c_{34}^{1/2} =c_{61}^{1/2} , \quad  c_{35}^{1/2} =c_{62}^{1/2} .
\end{split}
\label{eq37}
\end{align}
A possibility of the Lagrangian formulation (formula \eqref{c_eta_cond}) implies
\begin{align}
c_{14}^{1/2}, \,\, c_{25}^{1/2} \in \mathbb{R}; \quad
c_{34}^{1/2} = \frac{\eta_{63}^{1/2}}{\eta_{14}^{1/2}} (c_{16}^{1/2})^*,  \quad
c_{35}^{1/2} = \frac{\eta_{63}^{1/2}}{\eta_{25}^{1/2}} (c_{26}^{1/2})^*.
\label{eq38}
\end{align}

With account of the constraints \eqref{eq35}--\eqref{eq38} the spin block $C^{1/2}$ acquires the form
\begin{align}
C^{1/2} = \left( \begin{tabular}{cc} $0$ & $C$ \\ $C$ & $0$ \end{tabular} \right), \quad
C = \left( \begin{tabular}{ccc}
$c_1$  & $0$ & $c_3$ \\
$0$  & $c_2$ & $c_4$ \\
$f_1 c_3^*$  & $f_2 c_4^*$ & $0$
 \end{tabular} \right),
 \label{eq39}
\end{align}
where the notations
\begin{align}
\begin{split}
c_1 = c_{14}^{1/2}, \quad c_2 & = c_{25}^{1/2}, \quad c_3 =c_{16}^{1/2} , \quad c_4 =c_{26}^{1/2} , \\
f_1 &= \frac{\eta_{63}^{1/2}}{\eta_{14}^{1/2}}, \quad f_2 = \frac{\eta_{63}^{1/2}}{\eta_{25}^{1/2}}
\end{split}
\label{eq310}
\end{align}
are introduced for brevity.

A characteristic equation for the block $C$ reads
\begin{align}
\lambda^3 - (c_1 + c_2) \lambda^2 + (c_1 c_2 - f_1 |c_3|^2 - f_2 |c_4|^2) \lambda + f_1 c_2 |c_3|^2 + f_2 c_1 |c_4|^2 =0.
\end{align}
To obtain a single mass value, it is necessary to impose
\begin{align}
\begin{split}
c_1 c_2 - f_1 |c_3|^2 - f_2 |c_4|^2 &=0 , \\
f_1 c_2 |c_3|^2 + f_2 c_1 |c_4|^2 &=0.
\end{split}
\label{eq312}
\end{align}
Without loss of generality, we choose the only nonzero eigenvalue of the block $C$ to be
\begin{align}
\lambda = c_1 + c_2 =1.
\label{eq313}
\end{align}
Such a choice yields the following minimal polynomials for the $C^{1/2}$ and the matrix $\Gamma_4$:
\begin{align}
(C^{1/2})^2 [(C^{1/2})^2-1]=0, \quad \Gamma_4^2 (\Gamma_4^2 -1) =0.
\end{align}

It remains to impose the condition of the charge definiteness \eqref{ineq2}, which in the present case ($C^{3/2} = 0$, $n = 2$) acquires the form
\begin{align}
\mathrm{Sp} \, \, \left( (C^{1/2} )^3 \eta^{1/2} \right) \neq 0 ,
\end{align}
where
\begin{align}
\eta^{1/2} = \left( \begin{tabular}{cc} $0$ & $\eta'$ \\ $\eta'$ & $0$ \end{tabular} \right), \quad
\eta' = \left( \begin{tabular}{ccc}
$\eta_{14}^{1/2}$  & $0$ & $0$ \\
$0$  & $\eta_{25}^{1/2}$ & $0$ \\
$0$  & $0$ & $\eta_{36}^{1/2}$
 \end{tabular} \right).
\end{align}
It follows with account of \eqref{eq39} that
\begin{align}
\eta_{14}^{1/2} c_1^3 + \eta_{25}^{1/2} c_2^3 + (2 \eta_{63}^{1/2} + \eta_{14}^{1/2}) c_1 |c_3|^2 +(2 \eta_{63}^{1/2} + \eta_{25}^{1/2})  c_2 |c_4|^2 \neq 0 .
\label{eq317}
\end{align}

A simultaneous fulfilment of the conditions \eqref{eq312}, \eqref{eq313}, \eqref{eq317} is ensured, e.g., by the choice
\begin{align}
& c_1 = \frac13, \quad c_2 = \frac23 , \quad  c_3 = \frac{\sqrt{2}}{3}, \quad  c_4 =\frac23 , \\
& \eta_{14}^{1/2} =-1, \quad \eta_{25}^{1/2} = 1, \quad \eta_{36}^{1/2} = 1,
\end{align}
giving the following matrices $C$ and $\eta'$
\begin{align}
C = \left( \begin{tabular}{ccc}
$1/3$  & $0$ & $\sqrt{2}/3$ \\
$0$  & $2/3$ & $2/3$ \\
$-\sqrt{2}/3$  & $2/3$ & $0$
 \end{tabular} \right), \quad
\eta' = \left( \begin{tabular}{ccc}
$-1$  & $0$ & $0$ \\
$0$  & $1$ & $0$ \\
$0$  & $0$ & $1$
 \end{tabular} \right).
\end{align}

Thus, we obtained the 20-component RWE with the linking scheme \eqref{scheme31} featuring the nondiagonalizable matrix $\Gamma_4$. This equation describes a spin-$\frac12$ particle and obeys all physical requirements.

Let us now show that on the basis of the linking scheme \eqref{scheme31} it is also possible to construct a RWE for a particle with spin $s= \frac32$ [18]. Sticking with the previous labelling of the irreducible representations contained in \eqref{scheme31}, we again arrive at the general form \eqref{eq34} of the spin blocks $C^{1/2}$, $C^{3/2}$ of the matrix $\Gamma_4$ in the Gel'fand--Yaglom basis.

The condition  \eqref{c_cond} of the relativistic invariance leads to the following constraints
\begin{align}
c_{36}^{3/2} = 2 c_{36}^{1/2} , \quad c_{63}^{3/2}= 2 c_{63}^{1/2} .
\label{eq321}
\end{align}
The RWE invariance with respect to spatial reflections gives besides \eqref{eq37} the relations
\begin{align}
c_{36}^{1/2} = c_{63}^{1/2}, \quad
c_{36}^{3/2} = c_{63}^{3/2}.
\label{eq322}
\end{align}
A possibility of the Lagrangian formulation of the theory completes the relations \eqref{eq37} by the condition
\begin{align}
c_{36}^{1/2} \in \mathbb{R} .
\label{eq323}
\end{align}
Since in the present case we are interested in spin $\frac32$, we can set without loss of generality
\begin{align}
c_{36}^{3/2} = c_{63}^{3/2} = 1 .
\label{eq324}
\end{align}
Under this choice the conditions \eqref{eq321}--\eqref{eq323} yield the following spin blocks $C^{1/2}$ and $C^{3/2}$:
\begin{align}
C^{1/2} &= \left(\begin{tabular}{cc} $0$ & $C$ \\ $C$ & $0$ \end{tabular} \right) , \quad
C = \left(\begin{tabular}{ccc}
$c_1$ & $0$ & $c_3$ \\
$0$ & $c_2$ & $c_4$  \\
$f_1 c_3^*$ & $f_2 c_4^*$ & $1/2$
\end{tabular} \right), \label{eq325}  \\
C^{3/2} &= \left(\begin{tabular}{cc} $0$ & $1$ \\ $1$ & $0$ \end{tabular} \right),
\label{eq326}
\end{align}
where the notations \eqref{eq310} are used.

A characteristic equation for the spin block $C$ has the form
\begin{align}
\lambda^3 - \left(c_1 +c_2 +\frac12 \right) \lambda^2 + \left(\frac{c_1 +c_2}{2}+ c_1 c_2 - f_1 |c_3|^2 - f_2 |c_4|^2 \right) \lambda & \nonumber \\
-\frac{c_1 c_2}{2} +f_1 c_2 |c_3|^2 + f_2 c_1 |c_4|^2 &=0  .
\label{eq327}
\end{align}
To exclude states with spin $\frac12$ we must claim that all eigenvalues of the block $C^{1/2}$ are equal to zero. This requirement leads to the conditions
\begin{align}
\begin{split}
c_1 +c_2 +\frac12  &=0, \\
\frac{c_1 +c_2}{2}+ c_1 c_2 - f_1 |c_3|^2 - f_2 |c_4|^2 &=0, \\
-\frac{c_1 c_2}{2} +f_1 c_2 |c_3|^2 + f_2 c_1 |c_4|^2 &=0,
\end{split}
\label{eq328}
\end{align}
where the numbers $f_1$ and $f_2$ may independently of each other take values either $+1$ or $-1$.

It is easy to see that the conditions \eqref{eq328} are compatible with each other. Moreover, there is still much flexibility in choice of parameters $c_1$, $c_2$, $c_3$, $c_4$, $f_1$, $f_2$. Nevertheless, the minimal equations for the spin blocks $C^{1/2}$ \eqref{eq325} and $C^{3/2}$ \eqref{eq326} and the matrix $\Gamma_4$ always have the same forms:
\begin{align}
& (C^{1/2})^3 =0, \quad (C^{3/2})^2-1 =0 , \label{eq329} \\
& \qquad \qquad \Gamma_4^3 (\Gamma_4^2 -1) =0.
\label{eq330}
\end{align}

For example, let us choose
\begin{align}
f_1 &= -1, \quad f_2 = -1 , \\
c_1 &= \frac12, \quad c_2 = -1, \quad |c_3| = \frac{1}{2\sqrt{3}}, \quad |c_4| = \sqrt{\frac23},
\end{align}
realizing  one of the admissible possibilities. This choice leads to the matrix of the bilinear form having blocks
\begin{align}
\eta^{1/2} =\left(\begin{tabular}{cccccc}
$0$ & $0$ & $0$ & $1$ & $0$ & $0$ \\
$0$ & $0$ & $0$ & $0$ & $1$ & $0$ \\
$0$ & $0$ & $0$ & $0$ & $0$ & $-1$ \\
$1$ & $0$ & $0$ & $0$ & $0$ & $0$ \\
$0$ & $1$ & $0$ & $0$ & $0$ & $0$ \\
$0$ & $0$ & $-1$ & $0$ & $0$ & $0$
\end{tabular} \right) , \quad
\eta^{3/2} = \left(\begin{tabular}{cc} $0$ & $1$ \\ $1$ & $0$ \end{tabular} \right),
\end{align}
as well as to the spin block $C^{1/2}$ of the matrix $\Gamma_4$
\begin{align}
C^{1/2} =\left(\begin{tabular}{cccccc}
$0$ & $0$ & $0$ & $\frac12$ & $0$ & $\frac{1}{2 \sqrt{3}}$ \\
$0$ & $0$ & $0$ & $0$ & $-1$ & $\sqrt{\frac23}$ \\
$0$ & $0$ & $0$ & $-\frac{1}{2 \sqrt{3}}$ & $-\sqrt{\frac23}$ & $\frac12$ \\
$\frac12$ & $0$ & $\frac{1}{2 \sqrt{3}}$ & $0$ & $0$ & $0$ \\
$0$ & $-1$ & $\sqrt{\frac23}$ & $0$ & $0$ & $0$ \\
$-\frac{1}{2 \sqrt{3}}$ & $-\sqrt{\frac23}$ & $\frac12$ & $0$ & $0$ & $0$
\end{tabular} \right).
\end{align}

By virtue of \eqref{eq330} the charge definiteness condition for $n=3$ acquires the form
\begin{align}
\left( \mathrm{Sp} \, (\Gamma_4^4 \eta )\right)^2 -\left(  \mathrm{Sp} \, (\Gamma_4^3 \eta ) \right)^2 <0 .
\label{ineq43}
\end{align}
Using the explicit form of the matrices $\Gamma_4$ and $\eta$, we obtain the relations
\begin{align}
  \mathrm{Sp} \, (\Gamma_4^4 \eta ) =0, \quad  \mathrm{Sp} \, (\Gamma_4^3 \eta ) =8,
\end{align}
which ensure a fulfilment of  the inequality \eqref{ineq43}.

To construct RWEs describing microobjects with spins $s=0$ and $s=1$ and differing from the well-known Duffin--Kemmer equations, we consider the following set of the Lorentz group irreducible representations [19,20]
\begin{align}
(0,0) \oplus 2 (\frac12, \frac12) \oplus (0,1) \oplus (1,0),
\label{eq337}
\end{align}
which constitute the linking scheme
\begin{align}
\begin{tabular}{ccccc}
&  & $(0,0)$ &  & \\
& & $|$ & & \\
$(0,1)$ & --- & $2 (\frac12, \frac12)$ & --- & $(1,0)$.
\end{tabular}
\label{eq338}
\end{align}
Here the vector representation $(\frac12 , \frac12)$ has the multiplicity two. In the following, we label one of them with the prime to distinguish the two vector representations from each other.

The block structure of the matrix $\Gamma_4$ corresponding to the RWE based on the scheme \eqref{eq337} has the form \eqref{gam4_s2} in the Gel'fand--Yaglom basis. Labelling the irreducible components contained in \eqref{eq337}  by
\begin{align}
(0,0) \sim 1, \quad  (\frac12 , \frac12)' \sim 2 , \quad  (\frac12 , \frac12) \sim 3  ,\quad (0,1) \sim 4, \quad (1,0) \sim 5 ,
\end{align}
we obtained after applying the conditions of the relativistic and $P$-invariance conditions the following spin blocks
\begin{align}
C^0 = \left(\begin{tabular}{ccc}
$0$ & $c_{12}^0$ & $c_{13}^0$ \\
$c_{21}^0$ & $0$ & $0$ \\
$c_{31}^0$ & $0$ & $0$
\end{tabular}\right) , \quad
C^1 = \left( \begin{tabular}{cccc}
$0$ & $0$ & $c_{24}^1$ & $c_{24}^1$ \\
$0$ & $0$ & $c_{34}^1$ & $c_{34}^1$ \\
$c_{42}^1$ & $c_{43}^1$ & $0$ & $0$ \\
$c_{42}^1$ & $c_{43}^1$ & $0$ & $0$
\end{tabular} \right).
\label{eq340}
\end{align}

The matrix of the bilinear form in the Gel'fand--Yaglom basis reads
\begin{align}
\eta &= \left( \begin{tabular}{cc} $\eta^0$ & $0$ \\ $0$ & $\eta^1 \otimes I_3$ \end{tabular} \right)  , \label{eq341} \\
\eta^0 &= \left( \begin{tabular}{ccc}
$\eta_{11}^0$ & $0$ & $0$ \\
$0$ & $\eta_{22}^0$ & $0$ \\
$0$ & $0$ & $\eta_{33}^0$
\end{tabular} \right)  , \quad
\eta^1 = \left( \begin{tabular}{cccc}
$\eta_{22}^1$ & $0$ & $0$ & $0$ \\
$0$ & $\eta_{33}^1$ & $0$ & $0$ \\
$0$ & $0$ & $0$ & $\eta_{45}^1$ \\
$0$ & $0$ & $\eta_{54}^1$ & $0$
\end{tabular} \right). \nonumber
\end{align}
Choosing its nonzero elements to be
\begin{align}
\eta_{11}^0 = - \eta_{22}^0 = \eta_{33}^0 = \eta_{22}^1 = -\eta_{33}^1 =- \eta_{45}^1 = - \eta_{54}^1 = 1,
\label{eq342}
\end{align}
we reduce the condition  \eqref{c_eta_cond} of a possibility of the Lagrangian formulation to the relations
\begin{align}
c_{21}^0 = -(c_{12}^0)^* , \quad c_{31}^0 = (c_{13}^0)^* , \quad c_{42}^1 = -(c_{24}^1)^* , \quad c_{43}^1 = (c_{34}^1)^* .
\end{align}

The remaining arbitrariness in choice of values for the matrix $\Gamma_4$ elements we can use for obtaining RWE with a desired value of spin.

Thus, setting
\begin{align}
c_{12}^0 = 0, \quad c_{13}^0 = c_{24}^1 = c_{34}^1 = 1,
\end{align}
we obtain the spin blocks
\begin{align}
C^0 = \left(\begin{tabular}{ccc}
$0$ & $0$ & $1$ \\
$0$ & $0$ & $0$ \\
$1$ & $0$ & $0$
\end{tabular}\right) , \quad
C^1 = \left( \begin{tabular}{cccc}
$0$ & $0$ & $1$ & $1$ \\
$0$ & $0$ & $1$ & $1$ \\
$-1$ & $1$ & $0$ & $0$ \\
$-1$ & $1$ & $0$ & $0$
\end{tabular} \right).
\label{eq345}
\end{align}
It is easy to check the spin blocks $C^{1}$ and $C^0$ and the matrix $\Gamma_4$ obey the following minimal equations
\begin{align}
& (C^{1})^3 =0, \quad C^0 [ (C^{0})^2-1] =0 , \label{eq346} \\
& \qquad \qquad \Gamma_4^3 (\Gamma_4^2 -1) =0. \label{eq347}
\end{align}

From \eqref{eq346} it follows that to the state with spin $s=0$ corresponds the only mass value, while all eigenvalues of the block $C^1$ are zero, i.e. states with $s=1$ are absent. Thus, we obtain the RWE for a microobject with spin $s=0$ and a single mass value.

Using  the relations \eqref{eq341}, \eqref{eq342}, and \eqref{eq345} for matrices $\Gamma_4$ and $\eta$ it is easy to check the identities
\begin{align}
  \mathrm{Sp} \, (\Gamma_4^3 \eta ) =0, \quad  \mathrm{Sp} \, (\Gamma_4^4 \eta ) =2.
\end{align}
They ensure the fulfilment of the energy definiteness condition \eqref{ineq1} which for the present case ($n=3$) is expressed by the inequality
\begin{align}
(-1)^4 \left[ \left( \mathrm{Sp} \, (\Gamma_4^4 \eta )\right)^2 -\left(  \mathrm{Sp} \, (\Gamma_4^3 \eta ) \right)^2  \right] > 0 .
\label{eq349}
\end{align}

In the tensor formulation the constructed RWE reads
\begin{align}
\begin{split}
\partial_{\mu} \psi_{\mu} + m \psi_0 = 0, \\
\partial_{\nu} \psi_{[\mu \nu ]} + \partial_{\mu} \psi_0 + m \psi_{\mu} =0, \\
-\partial_{\nu} \psi_{[\mu \nu ]} + m \psi'_{\mu} =0, \\
-\partial_{\mu} \psi_{\nu}+ \partial_{\nu} \psi_{\mu} -\partial_{\mu} \psi'_{\nu}+ \partial_{\nu} \psi'_{\mu} + m \psi_{[\mu \nu]} =0 ,
\end{split}
\label{eq350}
\end{align}
where $\psi_0$ is a scalar, $\psi_{\mu}$ and $\psi'_{\mu}$ are 4-vectors, $\psi_{[\mu \nu]}$ is an antisymmetric tensor of the second rank. From \eqref{eq350}
it is easy to derive the second-order equation
\begin{align}
(\square- m^2) \psi_0 = 0 ,
\end{align}
which means that the system \eqref{eq350} indeed describes a particle with nonzero mass and spin $s=0$.

To construct a RWE for a microparticle with spin $s=1$ on the basis of the linking scheme \eqref{eq338}, we choose the following values for the matrix elements of the matrix \eqref{eq341}:
\begin{align}
\eta_{11}^0 =\eta_{22}^0 =-\eta_{33}^0 = -\eta_{22}^1 =\eta_{33}^1=\eta_{45}^1 =\eta_{54}^1 =  1.
\end{align}
Then, according to  the condition \eqref{c_eta_cond} we obtain
\begin{align}
c_{21}^0 = (c_{12}^0)^* , \quad c_{31}^0 =-  (c_{13}^0)^* , \quad c_{42}^1 =- (c_{24}^1)^* , \quad c_{43}^1 =  (c_{34}^1)^* .
\end{align}

Within the remaining freedom in choice of the matrix elements of the spin blocks $C^0$ and $C^1$ \eqref{eq340} we make a particular selection of values
\begin{align}
\begin{split}
c_{12}^0 &= c_{13}^0 = c_{21}^0 = - c_{31}^0 = 1,\\
c_{24}^1 &= c_{42}^1 =0 ,\quad c_{34}^1 = c_{43}^1 = \frac{1}{\sqrt{2}}.
\end{split}
\end{align}
Thus, we arrive at the final form of the matrices $\eta^0$, $\eta^1$, $C^0$, and $C^1$
\begin{align}
\eta^0 &= \left( \begin{tabular}{ccc}
$1$ & $0$ & $0$ \\
$0$ & $1$ & $0$ \\
$0$ & $0$ & $-1$
\end{tabular} \right)  , \quad
\eta^1 = \left( \begin{tabular}{cccc}
$-1$ & $0$ & $0$ & $0$ \\
$0$ & $1$ & $0$ & $0$ \\
$0$ & $0$ & $0$ & $1$ \\
$0$ & $0$ & $1$ & $0$
\end{tabular} \right),
\label{eq355}
\end{align}
\begin{align}
C^0 = \left(\begin{tabular}{ccc}
$0$ & $1$ & $1$ \\
$1$ & $0$ & $0$ \\
$-1$ & $0$ & $0$
\end{tabular}\right) , \quad
C^1 = \frac{1}{\sqrt{2}} \left( \begin{tabular}{cccc}
$0$ & $0$ & $0$ & $0$ \\
$0$ & $0$ & $1$ & $\pm 1$ \\
$0$ & $1$ & $0$ & $0$ \\
$0$ & $\pm 1$ & $0$ & $0$
\end{tabular} \right).
\label{eq356}
\end{align}

It is easy to check that the minimal equations for the spin blocks \eqref{eq356} of the matrix $\Gamma_4$ have the form
\begin{align}
(C^0)^3 = 0 , \quad C^1 [(C^1)^2 - 1] =0.
\label{eq357}
\end{align}
This means that the corresponding RWE indeed describes a microparticle with spin $s=1$.

From \eqref{eq357} it follows that the minimal equation for the matrix $\Gamma_4$ coincides with its analog \eqref{eq347} for a scalar particle. Therefore the condition of the energy definiteness in the present case should coincide with \eqref{eq349}. Using the definitions \eqref{eq355} and \eqref{eq356}, one can verify that the condition \eqref{eq349} is valid for a vector particle as well.

A tensor formulation of the obtained RWE with the extended set of representations for a particle with spin $s=1$ reads
\begin{align}
\begin{split}
\partial_{\mu} \psi_{\mu} +\partial_{\mu} \psi'_{\mu} +m \psi_0 = 0, \\
\partial_{\lambda} \psi_{[\mu \lambda ]} -\partial_{\mu} \psi_0 +m \psi_{\mu} = 0, \\
\partial_{\mu} \psi_0 +m \psi'_{\mu} = 0, \\
-\partial_{\mu} \psi_{\nu} +\partial_{\nu} \psi'_{\mu} +m \psi_{[\mu \nu]} = 0.
\end{split}
\label{eq358}
\end{align}
From this system one can derive the equations
\begin{align}
(\square - m^2) (\psi_{\mu} + \psi'_{\mu}) =0 , \quad \partial_{\mu} (\psi_{\mu} + \psi'_{\mu}) = 0 ,
\end{align}
which unambiguously indicate that the system \eqref{eq358} does describe a vector particle with nonzero mass.

Other versions of the extended RWEs for particles with lowest spins are proposed in the papers [21] (spin $\frac12$), [22] (spin $0$), and [23] (spin $1$).

A question of the physical inequivalence of  RWEs with minimal and extended sets of representations of the Lorentz group has been discussed for the first time for specific equations in the papers [24, 25] (spin $\frac12$), [19, 20] (spins $0$ and $1$), and [26] (spin $\frac32$). A general study of this question for particles having arbitrary spin $s$ and a single value of mass $m$ and interacting with the electromagnetic field has been carried out in the papers [27, 28]. The essence and the main results of the latter study are the following.

First, one introduces minimal and extended equations for free particles
\begin{align}
(\Gamma_{\mu}^{(0)} \partial_{\mu} + m) \Psi_0 (x) = 0, \\
(\Gamma_{\mu}^{(1)} \partial_{\mu} + m) \Psi_1 (x) = 0,
\end{align}
which are defined in the representation spaces corresponding to the irreducible Lorentz group representations $T_0$ and $T_1 = T_0 + T'$, respectively.
Second, one finds an explicit form of operators $R$ and $K$ transforming $\Psi_0$ into $\Psi_1$ and vice versa:
\begin{align}
R & = (A, 0  ) , \quad    K= \left(\begin{tabular}{c} $F$ \\ $G$ \end{tabular} \right) , \\
R \Psi_1 &= (A, 0  )\left(\begin{tabular}{c} $\Psi_1^0$ \\ $\Psi_1^1$ \end{tabular} \right) = A \Psi_1^0 = \Psi_0 , \\
K \Psi_0 & =\left(\begin{tabular}{c} $F$ \\ $G$ \end{tabular} \right) \Psi_0 =\left(\begin{tabular}{c} $F \Psi_0$ \\ $G \Psi_0$ \end{tabular} \right)   =\left(\begin{tabular}{c} $\Psi_1^0$ \\ $\Psi_1^1$ \end{tabular} \right)  = \Psi_1 .
\end{align}
Here $A$ and $F$ are rectangular number-valued matrices, satisfying the condition
\begin{align}
A F = I,
\end{align}
while the matrix $G$ in general contains differentiation operators. Moreover, the operators $R$ and $K$ should obey the relation
\begin{align}
R \Gamma_{\mu}^{(1)} K = \Gamma_{\mu}^{(0)} + B_{\mu},
\end{align}
where the matrices $B_{\mu}$ satisfy the equation
\begin{align}
B_{\mu} \partial_{\mu} \Psi_0 (x) = 0 .
\end{align}

Next, one considers the equations
\begin{align}
(\Gamma_{\mu}^{(0)}  \mathcal{ D}_{\mu}   + m) \Phi_0 (x) =0, \label{eq368} \\
(\Gamma_{\mu}^{(1)}  \mathcal{ D}_{\mu}   + m) \Phi_1 (x) =0, \label{eq369}
\end{align}
describing particles which interact with the electromagnetic field added by means of the minimal coupling
\begin{align}
\partial_{\mu} \to \mathcal{D}_{\mu} = \partial_{\mu} - i e A_{\mu}.
\label{eq370}
\end{align}
The operator $R$, transforming $ \Phi_1 (x)$ into $ \Phi_0 (x)$ has the same form as for free particles. For the operator $K'$ which realizes the inverse transformation we obtain
\begin{align}
K' = \left( \begin{tabular}{c} $F$ \\ $G+G'$ \end{tabular} \right),
\end{align}
where the addition $G'$ is caused by the derivative extension \eqref{eq370}. In light of this, the equation \eqref{eq369} can be cast to the form
\begin{align}
(R \Gamma_{\mu}^{(1)}  \mathcal{ D}_{\mu}  K' + m) \Phi_0 (x) =0 ,
\end{align}
or
\begin{align}
(\Gamma_{\mu}^{(0)}  \mathcal{ D}_{\mu}  + Q + m) \Phi_0 (x) =0 ,
\end{align}
where
\begin{align}
Q \sim R \Gamma_{\mu}^{(1)} \mathcal{D}_{\mu} G' + B_{\mu} \mathcal{ D}_{\mu} .
\label{eq374}
\end{align}
Thus, after bringing the equation \eqref{eq369} into shape of the equation \eqref{eq368} for a wavefunction with a minimally necessary number of components, we observe in the latter an occurrence of the additional term $Q$ which is given by \eqref{eq374}. For particles with spins $s = \frac12$ and
$s=\frac32$ described by the RWEs with the extended sets of representations, which have been discussed above, this term acquires the form
\begin{align}
Q  \sim \frac{i e}{m} (\partial_{\mu} A_{\nu} - \partial_{\nu} A_{\mu}) J^{[\mu \nu]} =   \frac{i e}{m} F_{[\mu \nu]}  J^{[\mu \nu]} ,
\end{align}
where $J^{[\mu \nu]}$ are the generators of the Lorentz group representations in the representation spaces \eqref{ls1} and \eqref{ls4}, respectively. In the nonrelativistic approximation, this term describes an additional -- anomalous -- magnetic moment, and it leads to an interaction of the Pauli type in the Lagrangian.

In the cases of the extended RWEs for particles with lowest spins, which have been discussed above, this additional term has the form
\begin{align}
Q & \sim \frac{e^2}{m} F_{[ \mu \nu]} F_{[ \mu \nu]} e^{00}, \label{eq376} \\
Q & \sim \frac{e^2}{m} F_{[ \mu \nu]} F_{[ \rho \sigma]} e^{[\mu \nu], [\rho \sigma]}  , \label{eq377}
\end{align}
where $e^{AB}$ are the generalized Kronecker symbols defined [29] by the formulas
\begin{align}
(e^{AB})_{CD} = \delta_{AC} \delta_{BD}.
\end{align}
In the case of a particle with spin $s=0$ the term $Q$ given by \eqref{eq376} describes in the nonrelativistic approximation dipole electric and magnetic polarizablities of this particle which are  induced by an external electromagnetic field.  In the case of a particle with spin $s=1$ the analogous term \eqref{eq377} describes a particle's static tensor electric polarizability.

It is obvious that an additional interaction with an external electromagnetic field should influence a form of matrix elements for specific scattering processes.
Detailed calculations of some of these processes have been performed in the papers [19-28]. It has been shown that in the first order perturbation theory the mentioned interaction does not show up. For examples, scattering on the Coulomb centre happens in the same way for both types of RWEs based on the minimal and the extended sets of the Lorentz group representations. In turn, a calculation of cross-sections of the typical second order process~-- the Compton light scattering on particles described by the RWEs with extended sets of representations~-- leads in all cases to matrix elements having the form
\begin{align}
M_1 = M_0 + M',
\end{align}
where $M_0$ is a matrix element corresponding to a particle described by an RWE with a minimal set of representations, and $M'$ is a correction caused by
the presence of an internal electromagnetic structure in this particle. Explicit expressions for these corrections can be found in the above mentioned publications.

Thus, a simple extension of a set of used representations, which can also be realized by an inclusion of replicated irreducible components, allows us to reflect internal particles' structure by means of the conventional spatiotemporal description in terms of the RWE theory. Apparently, in the conceptual respect this approach is more advantageous as compared to the popular phenomenological approach, in which additional terms describing specific structural effects are introduced into the Lagrangian by hand.

In conclusion, we remark an important property that all extended RWEs for lowest spins considered in this chapter are free of difficulties which emerge in theories for higher spins upon introducing an electromagnetic field by means of the minimal coupling. In particular, they appear to be renormalizable and do not contain noncausal solutions, in spite of the nondiagonalizability of the matrix $\Gamma_4$ (in this respect, see e.g. [22, 30]).

\section{RWE for a chiral particle with spin $1$}
\hspace{0.5cm}
\label{chirRWE}

In the previous chapter we have shown that RWEs with extended sets of irreducible Lorentz group representations (including multiple ones) allow us to account an internal structure of elementary particles in the framework of equations which do not fall apart in the sense of the relativistic invariance. Moreover, a usage of multiple representations in the RWE theory allows us also to achieve a spatio-temporal description\footnote{To avoid misunderstanding, we note that by a spatio-temporal description we understand in the present context a usage of equations which do not disintegrate under the full Lorentz group transformations. In the other words, we speak about a description of internal properties of elementary particles without taking into consideration field indices of a non-Lorentzian origin.} of particles' additional (besides spin) internal degrees of freedom.

Let us demonstrate the above statements on the example of a chirality -- a degree of freedom which is related to a double degeneracy of states conjugated with each other by the spatial inversion. A notion of chirality is nowadays widely used in the hadron physics, where it arises after neglecting the masses of the light $u$- and $d$-quarks. There is a row of experiments at low energies whose results witness in favour of an existence of interactions mediated by chiral particles with spin $1$; see, e.g.,  [31] for a detailed review.

Let us consider a linking scheme of irreducible Lorentz group representations
\begin{align}
\begin{tabular}{ccccccc}
& &   & $(\frac12,\frac12)'$ &  & & \\
& &  $\diagup$  &         & $\diagdown$  & & \\
 & $(0,1)$   &  &   &  & $(1,0)$, &  \\
   &   &  $\diagdown$ &  & $\diagup$  &  &    \\
   &   &  & $ (\frac12,\frac12) $ &  & &
\end{tabular}
\label{eq41}
\end{align}
which contains a twofold representation $(\frac12, \frac12)$ (like in \eqref{eq338}, the prime is introduced to distinguish between the two of them. The matrix $\Gamma_4$ of the RWE corresponding to the scheme \eqref{eq41} has in the Gel'fand--Yaglom's basis a quasidiagonal form \eqref{gam4_s2}. As usual, we introduce a labelling of the involved representations
\begin{align}
(\frac12,\frac12) \sim 1, \quad (\frac12,\frac12)' \sim 2, \quad (0,1) \sim 3, \quad (1,0) \sim 4 .
\label{eq42}
\end{align}
Since the components $1$ and $2$ which form the spin block $C^{0}$ do not link with each other, we obtain
\begin{align}
C^0 = \left(
\begin{tabular}{cc}
$0$ & $0$ \\
$0$ & $0$
\end{tabular} \right),
\label{eq43}
\end{align}
that is the spin $s = 0$ is absent.

For the block $C^{1}$ we have the following general expression
\begin{align}
C^1 = \left(
\begin{tabular}{cccc}
$0$ & $0$ & $c_{13}^1$ & $c_{14}^1$ \\
$0$ & $0$ & $c_{23}^1$ & $c_{24}^1$ \\
$c_{31}^1$ & $c_{32}^1$ & $0$ & $0$ \\
$c_{41}^1$ & $c_{42}^1$ & $0$ & $0$
\end{tabular} \right).
\label{eq44}
\end{align}

In the present consideration, a requirement of the invariance of the RWE, which is being constructed, under transformations of the proper Lorentz group does not impose any constraints on elements $c_{ij}^1$. The invariance with respect to $P$-inversion leads to the conditions
\begin{align}
c_{14}^1 = \pm c_{13}^1, \quad c_{41}^1 = \pm c_{31}^1, \quad c_{24}^1 = \pm c_{23}^1, \quad c_{42}^1 = \pm c_{32}^1.
\label{eq45}
\end{align}
In this case the sign "$+$" ("$-$") is chosen -- according to the formulas \eqref{c2_cond}--\eqref{Pc2_cond} -- when both representations  $(\frac12, \frac12)$ and $(\frac12, \frac12)'$ in the scheme \eqref{eq41} appear to be true vector (pseudovector) ones. For these two situations we obtain the spin block $C^{1}$ in the form
\begin{align}
C^1 = \left(
\begin{tabular}{cccc}
$0$ & $0$ & $c_{13}^1$ & $\pm c_{13}^1$ \\
$0$ & $0$ & $c_{23}^1$ & $\pm c_{23}^1$ \\
$c_{31}^1$ & $c_{32}^1$ & $0$ & $0$ \\
$\pm c_{31}^1$ & $\pm c_{32}^1$ & $0$ & $0$
\end{tabular} \right),
\label{eq46}
\end{align}
where the upper and the lower signs are correlated and correspond to either of these situations.

The characteristic equation for the block $C^{1}$ reads
\begin{align}
\lambda^4 -2 \lambda^2 (c_{13}^1 c_{31}^1 + c_{23}^1 c_{32}^1) =0 .
\label{eq47}
\end{align}
It follows that this block has the only -- up to a sign -- nonzero eigenvalue
\begin{align}
\lambda = \pm \sqrt{2 (c_{13}^1 c_{31}^1 + c_{23}^1 c_{32}^1)} .
\label{eq48}
\end{align}
In the other words, the state with spin $1$ does not contain additional internal degrees of freedom. Obviously, the RWE obtained in such a way is reduced to the Dirac-K\"ahler equation for a vector particle.

The things are different, when one the representations in the scheme \eqref{eq41} is vectorial, and the other one is pseudovectorial. In this case the requirement of the $P$-invariance leads to the conditions
\begin{align}
c_{14}^1 =  c_{13}^1, \quad c_{41}^1 = c_{31}^1, \quad c_{24}^1 = - c_{23}^1, \quad c_{42}^1 = - c_{32}^1
\label{eq49}
\end{align}
and to the corresponding spin block $C^{1}$
\begin{align}
C^1 = \left(
\begin{tabular}{cccc}
$0$ & $0$ & $c_{13}^1$ & $c_{13}^1$ \\
$0$ & $0$ & $c_{23}^1$ & $- c_{23}^1$ \\
$c_{31}^1$ & $c_{32}^1$ & $0$ & $0$ \\
$c_{31}^1$ & $- c_{32}^1$ & $0$ & $0$
\end{tabular} \right).
\label{eq410}
\end{align}
Its characteristic equation reads
\begin{align}
\lambda^4 -2 \lambda^2 (c_{13}^1 c_{31}^1 + c_{23}^1 c_{32}^1) + 4 c_{13}^1 c_{31}^1 c_{23}^1 c_{32}^1=0 .
\label{eq411}
\end{align}
The solutions of the equation \eqref{eq411} $\lambda_1 = \pm \sqrt{2 c_{13}^1 c_{31}^1}$ and $\lambda_2 = \pm \sqrt{2 c_{23}^1 c_{32}^1}$ mean that we deal with a particle with spin $s = 1$ and, generally speaking, with the two possible mass values
\begin{align}
 m_1 = \frac{m}{\sqrt{2 c_{13}^1 c_{31}^1}}, \quad m_2 = \frac{m}{\sqrt{2 c_{23}^1 c_{32}^1}} .
\label{eq412}
\end{align}
However, if we set
\begin{align}
c_{13}^1 c_{31}^1 = c_{23}^1 c_{32}^1 ,
\label{eq413}
\end{align}
then we arrive at the RWE for a particle with a single mass value, but with a twofold degeneracy of states classified by some other -- additional -- quantum number.

The arbitrariness, remaining in the choice of elements $c^{1}_{ij}$ of the block $C^{1}$ \eqref{eq410} after imposing upon them the constraints \eqref{eq413}, can be used for obtaining the Lagrangian formulation of the RWE under consideration. To this end the elements $\eta^0_{ij}$, $\eta^1_{ij}$ of the matrix  $\eta$ of the bilinear form we define in the following way:
\begin{align}
-\eta^0_{11} = \eta^0_{22} = \eta^1_{11} = -\eta^1_{22} = \eta^1_{34} = \eta^1_{43} =1 .
\label{eq414}
\end{align}

Then, the requirement  \eqref{c_eta_cond} of the relativistic invariance of the Lagrangian for this theory leads to the relations
\begin{align}
c_{31}^1 = (c_{13}^1 )^* , \quad c_{32}^1 = -  (c_{23}^1 )^* .
\label{eq415}
\end{align}
Without loss of generality, we choose the remaining arbitrary elements  $c^1_{13}$, $c^1_{23}$, e.g.,  to be
\begin{align}
c_{13}^1 = c_{23}^1 = \frac{1}{\sqrt{2}},
\label{eq416}
\end{align}
the eigenvalues of the block  $C^{1}$ being  equal $\pm 1$.

Thus, for the nonzero block $C^{1}$ of the matrix $\Gamma_4$ and for the blocks $\eta^0$, $\eta^1$ of the matrix $\eta$ we establish their final form
\begin{align}
C^1 &= \frac{1}{\sqrt{2}}  \left(
\begin{tabular}{cccc}
$0$ & $0$ & $1$ & $1$ \\
$0$ & $0$ & $1$ & $- 1$ \\
$1$ & $1$ & $0$ & $0$ \\
$1$ & $- 1$ & $0$ & $0$
\end{tabular} \right),
\label{eq417} \\
\eta^0 &= \left(
\begin{tabular}{cc}
$-1$ & $0$ \\
$0$ & $1$
\end{tabular} \right),  \quad \eta^1 = \left(
\begin{tabular}{cccc}
$1$ & $0$ & $0$ & $0$ \\
$0$ & $-1$ & $0$ & $0$ \\
$0$ & $0$ & $0$ & $1$ \\
$0$ & $0$ & $1$ & $0$
\end{tabular} \right). \label{eq418}
\end{align}

For clarifying the meaning of the twofold degeneracy of the states of a microobject, which is described by the derived RWE, it is convenient to use its tensor formulation
\begin{align}
\begin{split}
& \partial_{\nu} \psi_{[\mu \nu]} + m \psi_{\mu} =0, \\
& \partial_{\nu}  \tilde{\psi}_{[\mu \nu]} + m \tilde{\psi}_{\mu}  =0, \\
-& \partial_{\mu}  \psi_{\nu} +  \partial_{\nu}  \psi_{\mu}+ \varepsilon_{\mu \nu \alpha \beta} \partial_{\alpha} \tilde{\psi}_{\beta} + m \psi_{[\mu \nu]}=0,
\end{split}
\label{eq419}
\end{align}
where the quantities $\psi_{\mu}$, $\tilde{\psi}_{\mu}$, $\psi_{[\mu \nu]} $ are assigned to the representation $(\frac12, \frac12)$, $(\frac12, \frac12)'$, $(0, 1) \oplus (1, 0)$, respectively; and
\begin{align}
\tilde{\psi}_{[\mu \nu]} = \frac12 \varepsilon_{\mu \nu \alpha \beta} \psi_{[\alpha \beta]} ,
\label{eq420}
\end{align}
with  $\varepsilon_{\mu \nu \alpha \beta}$ being the Levi-Civita tensor ($ \varepsilon_{1234}= - i$).

Making in  the system \eqref{eq419} the substitutions
\begin{align}
\begin{split}
\varphi_{\mu} &= \frac{1}{\sqrt{2}} (\psi_{\mu} - i \tilde{\psi}_{\mu}), \quad \varphi_{[\mu \nu]} = \frac{1}{\sqrt{2}} (\psi_{[\mu \nu]}  -i \tilde{\psi}_{[\mu \nu]} ), \\
\chi_{\mu} &=  \frac{1}{\sqrt{2}} (\psi_{\mu} + i \tilde{\psi}_{\mu}), \quad \chi_{[\mu \nu]} = \frac{1}{\sqrt{2}} (\psi_{[\mu \nu]}  + i \tilde{\psi}_{[\mu \nu]} ),
\end{split}
\label{eq421}
\end{align}
we transform it to a direct sum of the two seven-component subsystems:
\begin{align}
\begin{split}
& \partial_{\nu} \varphi_{[\mu  \nu]} + m \varphi_{\mu} =0 , \\
- &\partial_{\mu} \varphi_{\nu} + \partial_{\nu} \varphi_{\mu} + i \varepsilon_{\mu \nu \alpha \beta} \partial_{\alpha} \varphi_{\beta} + m \varphi_{[\mu  \nu]}  =0
\end{split}
\label{eq422}
\end{align}
and
\begin{align}
\begin{split}
& \partial_{\nu} \chi_{[\mu  \nu]} + m \chi_{\mu} =0 , \\
-&  \partial_{\mu} \chi_{\nu} + \partial_{\nu} \chi_{\mu} - i \varepsilon_{\mu \nu \alpha \beta} \partial_{\alpha} \chi_{\beta} + m \chi_{[\mu  \nu]}  =0 .
\end{split}
\label{eq423}
\end{align}

The subsystems \eqref{eq422}, \eqref{eq423} are invariant in the sense of the proper Lorentz group transformations. However, under the spatial inversion they are transformed into each other.  Therefore with respect to the full Lorentz group transformations the system \eqref{eq422}, \eqref{eq423}~-- as well as the system \eqref{eq419}~-- appears to be nondisintegrating. Moreover, we note that to the subsystems \eqref{eq422}, \eqref{eq423} alone it is impossible to assign a Lagrangian which would satisfy the standard requirements. A correct Lagrangian formulation is only possible for the system  \eqref{eq422}, \eqref{eq423} considered as the whole.

Thus, the twofold degenerate states of a vector particle, which is described by the constructed RWE, are related to each other by the $P$-inversion. Therefore, the additional quantum number which we talked about above distinguishes between the $P$-conjugated states and can be interpreted as chirality, in analogy with the notion of chirality for massless particles.

It is important to note the following. As it is known, to construct a theory of massless particles with chirality $S$ it suffices to use just the representations $(0,S)$ and $(S,0)$, the notions of chirality and spirality (spin's projection on the momentum direction) coinciding in their essence for this case. In the case of
massive particles this, however, does not take place any more. At first, the notions of chirality and spirality do not coincide for massive particles, and, at second, a theory of such particles, as follows from the above considerations, is necessarily based on a set of {\it linking} irreducible Lorentz group representations, including multiple components. A possible physical meaning of chirality in application to massive virtual particles will be discussed in Chapter 7.

\section{Tensor RWEs of the Dirac type \\
and geometrized description of internal degrees of freedom of fundamental particles}
\hspace{0.5cm}
\label{TensorRWEDirac}

Since an existence of additional internal quantum numbers for fundamental particles is by now a firmly established fact, there arises a question whether it is possible to apply the RWE theory for describing degrees of freedom associated with internal and, in particular,  gauge symmetries. The traditional gauge theories of fundamental particles and their interactions are based as a rule on the Dirac equation whose wavefunction is equipped by a free non-Lorentzian index playing the role of an internal variable. From the RWE theory point of view, such an approach actually means a usage of equations which are falling apart under the Lorentz group transformations. On this basis are constructed the renowned models of electroweak and strong interactions, and the Standard $SU(3)\times  SU(2) \times U(1)$ Model.

However, this approach is not capable of solving a number of problems. In particular, it is not very effective in including the gravitational interaction into a general scheme. It is presently hoped that a solution of this problem lies in employing symmetry groups whose transformations would affect both spatiotemporal and internal variables on equal basis. In the other words, it is a matter of an eventual geometrized introduction of internal degrees of freedom.

Let us briefly review the most known approaches in this direction:
\begin{itemize}
  \item
  theories of the Kaluza-Klein type, which operate with the space-time of a dimension greater than four, additional dimensions being treated on equal footing with the four standard~-- observable~-- ones. A compactification of extra dimensions leads to a release of internal degrees of freedom retaining their geometrical character;
  \item
  supersymmetry-supergravitation, uniting particles with different spins and statistics into entire supermultiplets. One of the premises in this approach consists in the existence of a new mathematical structure -- the supersymmetry transformations, which mix up together bosonic and fermionic fields. In analogy with the Lorentz transformations which reveal the connection between the space and the time, the supersymmetry transformations unite the space-time and internal degrees of freedom of particles into the entire entity;
  \item
  string and superstring theories, which include the ideas of Kaluza and Klein, the supersymmetries, the gauge approach, and the general relativity.
\end{itemize}

It seems however feasible to formulate yet another approach of a geometrized description of internal degrees of freedom, which is based on the usage of an extended set of the Lorentz group representations (including multiple ones) in the framework of the RWE theory. A natural possibility in this respect consists in using nondisintegrating~-- with respect to the full Lorentz group -- equations whose wavefunction possesses transformation properties of a direct product of the Dirac bispinors, and  whose matrices $\Gamma_\mu$ satisfy commutation relations of the Dirac matrices' algebra. In the following, we will call such RWEs Dirac-like RWEs, or RWEs of the Dirac type.

The most widely known RWE of the discussed type is the Dirac-K\"ahler (DK) equation, which represents itself the most general differential equation (or the system of equations) of the first order over the field of complex numbers for the full set of antisymmetric tensor fields in the Minkowski space.
On the other hand, in the appropriate basis (let us call it fermionic) a wavefunction of the DK equation possesses the Lorentz transformation properties of a direct product of the Dirac bispinor times the charge-conjugated bispinor.
In the tensor formulation the DK equation can be represented by the system
\begin{align}
\begin{split}
& \partial_{\mu} \psi_{\mu} + m \psi_0 =0 , \\
&  \partial_{\nu} \psi_{[\mu \nu]} +  \partial_{\mu}  \psi_0 + m \psi_{\mu}  =0 , \\
-&  \partial_{\mu}\psi_{\nu} + \partial_{\nu}\psi_{\mu}  + i \varepsilon_{\mu \nu \alpha \beta} \partial_{\alpha} \tilde{\psi}_{\beta}  + m \psi_{[\mu \nu]} =0 , \\
& \partial_{\mu} \tilde{\psi}_{\mu} + m \tilde{\psi}_0 =0 , \\
&  \partial_{\nu} \tilde{\psi}_{[\mu \nu]} +  \partial_{\mu}  \tilde{\psi}_0 + m \tilde{\psi}_{\mu}  =0 , \\
& \frac12 \varepsilon_{\mu \nu \alpha \beta} \partial_{\nu} \psi_{[ \alpha \beta ]}  + \partial_{\mu}  \tilde{\psi}_0  +  m \tilde{\psi}_{\mu} =0 .
\end{split}
\label{eq51}
\end{align}
Here $\psi_0$ is a scalar, $\psi_{\mu}$ is a vector, $\psi_{[\mu \nu]}$ is a second-rank antisymmetric tensor,
$\tilde{\psi}_{\mu} = \frac{1}{3!} \varepsilon_{\mu \nu \alpha \beta} \psi_{[\nu \alpha \beta]} $ is a pseudovector, which is dual conjugated to the third-rank antisymmetric tensor  $\psi_{[\nu \alpha \beta]} $, and $\tilde{\psi}_{0} = \frac{1}{4!} \varepsilon_{\mu \nu \alpha \beta} \psi_{[ \mu \nu \alpha \beta]} $ is a pseudoscalar which is dual conjugated to the fourth-rank antisymmetric tensor  $\psi_{[ \mu \nu \alpha \beta]}$.

The system \eqref{eq51} is nondisintegrating in the sense of the full Lorentz group. It can be written in the matrix-differential form \eqref{eq:nonzero_mass}, which is standard for the RWE theory, where the wavefunction  $\Psi$ is represented by the column vector with tensor-valued components
\begin{align}
\Psi = (\psi_0 , \tilde{\psi}_0 , \psi_{\mu}, \tilde{\psi}_{\mu}, \psi_{[\mu \nu]})^T ,
\label{eq52}
\end{align}
and the $16 \times 16$ matrices $\Gamma_\mu$ are expressed by
\begin{align}
\begin{split}
& \Gamma_{\mu} = \Gamma_{\mu}^{(+)}+  \Gamma_{\mu}^{(-)} , \\
& \Gamma_{\mu}^{(+)} = e^{\tilde{0} \tilde{\mu}} + e^{\tilde{\mu} \tilde{0}} + e^{\lambda, [ \lambda \mu ]} + e^{[\lambda \mu ],\lambda}, \\
& \Gamma_{\mu}^{(-)} = e^{0 \mu} + e^{\mu 0} + \frac{i}{2} \varepsilon_{\lambda \mu \alpha \beta} ( e^{\tilde{\lambda}, [ \alpha \beta ]} + e^{[\alpha \beta ], \tilde{\lambda}}).
\end{split}
\label{eq53}
\end{align}

Using the known rules of the generalized Kronecker symbols' multiplication [29], it is easy to check that the matrices  $\Gamma_\mu$ \eqref{eq53} satisfy the Dirac matrices' algebra
\begin{align}
\Gamma_{\mu} \Gamma_{\nu} + \Gamma_{\nu} \Gamma_{\mu} = 2 \delta_{\mu \nu} .
\label{eq54}
\end{align}

For establishing the group of internal symmetry of the DK field one can conveniently pass from the tensor basis  \eqref{eq52} to the fermionic one, in which the matrices $\Gamma_\mu$ and the matrix $\eta$ of the Lorentz-invariant  bilinear form read
\begin{align}
\Gamma_{\mu} &= I_4 \otimes \gamma_4 ,
\label{eq55} \\
\eta &= \gamma_4 \otimes \gamma_4 .
\label{eq56}
\end{align}

We remind that by an internal symmetry transformation of the RWE  \eqref{eq:nonzero_mass} we understand a linear transformation of the wavefunction
\begin{align}
\Psi' (x) = Q \Psi (x),
\label{eq57}
\end{align}
which does not touch spatiotemporal coordinates and which leaves invariant the equation \eqref{eq:nonzero_mass} and its Lagrangian \eqref{eq126}. For this to happen, a matrix   $Q$ must satisfy the conditions
\begin{align}
 \, [ \Gamma_\mu , Q]_- =0 , \label{eq58} \\
 Q^+ \eta Q = \eta . \label{eq59}
\end{align}

Applying the conditions \eqref{eq58}, \eqref{eq59} to the matirces $\Gamma_\mu$ and  $\eta$ leads us to a noncompact 15-parametric group  $SU (2, 2)$, whose generators may serve the Hermitian matrices
\begin{align}
\Gamma'_{\mu}, \quad \Gamma'_5 , \quad i \Gamma'_{\mu} \Gamma'_5 , \quad i \Gamma'_{[\mu} \Gamma'_{\nu ]} = \frac{i}{2} (\Gamma'_{\mu} \Gamma'_{\nu} - \Gamma'_{\nu} \Gamma'_{\mu}).
\label{eq510}
\end{align}
Here
\begin{align}
\Gamma'_5 = \Gamma'_1 \Gamma'_2 \Gamma'_3 \Gamma'_4
\label{eq511}
\end{align}
and
\begin{align}
\Gamma'_{\mu} = \Gamma^{(+)}_{\mu} -\Gamma^{(-)}_{\mu}
\label{eq512}
\end{align}
is the second set of  $16 \times 16$ matrices, satisfying  -- like  $\Gamma_\mu$ -- the algebra of Dirac matrices and commuting with the matrices  $\Gamma_\mu$. In the fermionic basis these matrices have the form
\begin{align}
\Gamma'_{\mu} = \gamma_{\mu} \otimes I_4 .
\label{eq513}
\end{align}
A characteristic feature of the internal symmetry group of the DK equation is that its generators  \eqref{eq510} do not commute with the Lorentz generators
\begin{align}
J_{[\mu \nu]} = \frac14 (\Gamma_{[ \mu} \Gamma_{\nu ]} + \Gamma'_{[ \mu} \Gamma'_{\nu ]})
\label{eq514}
\end{align}
from the representation of the wavefunction  $\Psi$. Along with this, the group  $G$ corresponding tothe  full invariance algebra of the DK equation appears to be  a semidirect product of the Lorentz group $\Lambda$ and the group of the internal symmetry $Q$: $G = \Lambda \oslash Q$. On the other hand, the group can be represented as a direct product  $G = \Lambda' \otimes Q$, where $\Lambda'$  is an overdefined Lorentz group with respect to which the wavefunction $\Psi$ is no longer a collection of the tensor-valued components, but rather a collection of four Dirac fields with a usual internal symmetry (i.e. commuting with the Lorentz group transformations).

The above statements remain valid for all interactions, including gauge ones, which do not violate an internal symmetry of the free Lagrangian. They assert
that it is generally possible to apply the DK equation for a description of particles with spin $s = \frac12$ and internal degrees of freedom, which thus have a geometric origin  (for details see, e.g., [32]). The idea that the DK equation can be exploited as a geometric model for the generations of quarks (or leptons) was first put forward in the works [33, 34].

Let us now give a matrix formulation of the DK equation in the Gel'fand--Yaglom basis, which will be useful in the following.

We begin with the next set of irreducible representations of the proper Lorentz group
\begin{align}
\begin{tabular}{ccccc}
& & $2 (0,0)$ & & \\
& & $|$ & & \\
$(0,1)$ & --- & $2 (\frac12, \frac12)$ & --- & $(1,0)$,
\end{tabular}
\label{eq515}
\end{align}
containing twofold components  $(0, 0)$ and $(\frac12, \frac12)$. The matrix $\Gamma_4$ of the corresponding RWE in the Gel'fand--Yaglom basis has the form  \eqref{gam4_s2}, where
\begin{align}
C^0 = \left(
\begin{tabular}{cccc}
$0$ & $0$ & $c_{13}^0$ & $c_{14}^0$ \\
$0$ & $0$ & $c_{23}^0$ & $c_{24}^0$ \\
$c_{31}^0$ & $c_{32}^0$ & $0$ & $0$ \\
$c_{41}^0$ & $c_{42}^0$ & $0$ & $0$
\end{tabular} \right), \quad
C^1 = \left(
\begin{tabular}{cccc}
$0$ & $0$ & $c_{35}^1$ & $c_{36}^1$ \\
$0$ & $0$ & $c_{45}^1$ & $c_{46}^1$ \\
$c_{53}^1$ & $c_{54}^1$ & $0$ & $0$ \\
$c_{63}^1$ & $c_{64}^1$ & $0$ & $0$
\end{tabular} \right),
\label{eq516}
\end{align}
and the following labelling of the irreducible representations contained in  \eqref{eq515} is adopted:
\begin{align}
(0,0) \sim 1, \quad (0,0)' \sim 2, \quad (\frac12 , \frac12) \sim 3 , \quad (\frac12 , \frac12)' \sim 4, \quad (0,1) \sim 5, \quad (1,0) \sim 6.
\label{eq517}
\end{align}
Here, like it was also the case earlier, the prime is used to distinguish between the multiple representations.

Let us first consider the spin block $C^{1}$. The conditions of the relativistic and $P$-invariance of the theory impose on elements  $c^{1}_{ij}$ in general case the constraints
\begin{align}
c^1_{35} = \pm c^1_{36}, \quad c^1_{45} = \pm c^1_{46}, \quad c^1_{53} = \pm c^1_{63}, \quad c^1_{54} = \pm c^1_{64}.
\label{eq518}
\end{align}
Here the choice of signs "$+$" or "$-$" depends on the definition of the spatial inversion operator, which is given either by  \eqref{Pc1_cond} or \eqref{Pc2_cond}. In the present context this means that the sign "$+$" ("$-$") in  \eqref{eq518} occurs for the true vectorial (pseudovectorial) character of the multiple representations $(\frac12,\frac12)$. It turns out that one can construct a RWE satisfying all necessary physical requirements, if one chooses one of the representations $(\frac12,\frac12)$ as true vectorial and the other as pseudovectorial (which is denoted in the following as  $(\frac12,\frac12)'$). Then the relations \eqref{eq518} acquire the form
\begin{align}
c^1_{35} =  c^1_{36} , \quad c^1_{45} = - c^1_{46}, \quad  c^1_{53} =  c^1_{63} , \quad c^1_{54} = - c^1_{64},
\label{eq519}
\end{align}
and for the block $C^{1}$ one gets an expression
\begin{align}
C^1 = \left(
\begin{tabular}{cccc}
$0$ & $0$ & $c_{35}^1$ & $c_{35}^1$ \\
$0$ & $0$ & $c_{45}^1$ & $-c_{45}^1$ \\
$c_{53}^1$ & $c_{54}^1$ & $0$ & $0$ \\
$c_{53}^1$ & $-c_{54}^1$ & $0$ & $0$
\end{tabular} \right).
\label{eq520}
\end{align}

Analogously, one of the representations $(0,0)$ in \eqref{eq515} we choose as scalar and the other as pseudoscalar (also labelling it in the following with the prime). And since in a $P$-invariant RWE a vector (pseudovector) representation can not link with  a pseudscalar (scalar) one, the following equalities should take place
\begin{align}
c^0_{14}=c^0_{23}=c^0_{41}=c^0_{32}=0.
\label{eq521}
\end{align}
With this in view, the block $C^{0}$ \eqref{eq518} is transformed to the form
\begin{align}
C^0 = \left(
\begin{tabular}{cccc}
$0$ & $0$ & $c_{13}^0$ & $0$ \\
$0$ & $0$ & $0$ & $c_{24}^0$ \\
$c_{31}^0$ & $0$ & $0$ & $0$ \\
$0$ & $c_{42}^0$ & $0$ & $0$
\end{tabular} \right),
\label{eq522}
\end{align}
and the representations \eqref{eq515} build the linking scheme
\begin{align}
\begin{tabular}{ccccccccc}
& & &   & $(0,1)$ &  & & & \\
& & &  $\diagup$  &         & $\diagdown$  & & & \\
$(0,0)'$ &  --- & $(\frac12,\frac12)'$   &  &   &  & $(\frac12,\frac12) $ &  --- & $(0,0)$ . \\
 &  &   &  $\diagdown$ &  & $\diagup$  &  &   & \\
&   &   &  & $ (1,0)$ &  & &  &
\end{tabular}
\label{eq523}
\end{align}

The blocks  $\eta^{0}$, $\eta^{1}$ of the matrix of the bilinear invariant form  $\eta$ \eqref{eq24} have in this case the form
\begin{align}
\begin{split}
\eta &= \eta^0 \oplus (\eta^1 \otimes I_3), \\
\eta^0 &=  \left(
\begin{tabular}{cccc}
$\eta_{11}^0$ & $0$ & $0$ & $0$ \\
$0$ & $\eta_{22}^0$ & $0$ & $0$ \\
$0$ & $0$ & $\eta_{33}^0$ & $0$ \\
$0$ & $0$ & $0$ & $\eta_{44}^0$
\end{tabular} \right), \quad
\eta^1 =  \left(
\begin{tabular}{cccc}
$\eta_{33}^1$ & $0$ & $0$ & $0$ \\
$0$ & $\eta_{44}^1$ & $0$ & $0$ \\
$0$ & $0$ & $0$ & $\eta_{56}^1$ \\
$0$ & $0$ & $\eta_{65}^1$ & $0$
\end{tabular} \right),
\end{split}
\label{eq524}
\end{align}
where according to \eqref{eta_cond} it also holds
\begin{align}
\eta_{33}^1 = - \eta_{33}^0, \quad \eta_{44}^1 = - \eta_{44}^0, \quad \eta_{65}^1 = \pm \eta_{56}^1 .
\label{eq525}
\end{align}

The conditions \eqref{c_eta_cond} along with the equalities \eqref{eq519} lead to the relations
\begin{align}
c_{31}^0 = \frac{\eta_{33}^0}{\eta_{11}^0}(c_{13}^0)^*, \quad c_{42}^0 = \frac{\eta_{44}^0}{\eta_{22}^0}(c_{24}^0)^*, \quad
c_{53}^1 = \frac{\eta_{56}^1}{\eta_{33}^1}(c_{35}^1)^*, \quad c_{54}^1 = \frac{\eta_{56}^1}{\eta_{44}^1}(c_{45}^1)^*.
\label{eq526}
\end{align}
Choosing now for the remaining free elements $c^s_{\tau \tau '}$
 and
$\eta^s_{\tau \dot{\tau}}$ the values equal, e.g., to
\begin{align}
c_{13}^0 &= c_{24}^0=1 , \quad c_{35}^1 = c_{45}^1 = \frac{1}{\sqrt{2}}, \label{eq527} \\
\eta_{11}^0 &=- \eta_{22}^0 =\eta_{33}^0 =- \eta_{44}^0 =- \eta_{56}^1 =- \eta_{65}^1 =1,
\label{eq528}
\end{align}
we obtain the RWE with the spin blocks of the matrices $\Gamma_4$ and $\eta$, being equal
\begin{align}
C^0 = \left(
\begin{tabular}{cccc}
$0$ & $0$ & $1$ & $0$ \\
$0$ & $0$ & $0$ & $1$ \\
$1$ & $0$ & $0$ & $0$ \\
$0$ & $1$ & $0$ & $0$
\end{tabular} \right), \quad
C^1 = \frac{1}{\sqrt{2}} \left(
\begin{tabular}{cccc}
$0$ & $0$ & $1$ & $1$ \\
$0$ & $0$ & $1$ & $-1$ \\
$1$ & $1$ & $0$ & $0$ \\
$1$ & $-1$ & $0$ & $0$
\end{tabular} \right),
\label{eq529}
\end{align}
\begin{align}
\eta^0 &=  \left(
\begin{tabular}{cccc}
$1$ & $0$ & $0$ & $0$ \\
$0$ & $-1$ & $0$ & $0$ \\
$0$ & $0$ & $1$ & $0$ \\
$0$ & $0$ & $0$ & $-1$
\end{tabular} \right), \quad
\eta^1 =  \left(
\begin{tabular}{cccc}
$-1$ & $0$ & $0$ & $0$ \\
$0$ & $1$ & $0$ & $0$ \\
$0$ & $0$ & $0$ & $-1$ \\
$0$ & $0$ & $-1$ & $0$
\end{tabular} \right).
\label{eq530}
\end{align}

The RWE constructed in this way with using the Gel'fand--Yaglom basis satisfies the conditions of an invariance with respect to the full Lorentz group and of a possibility of its derivation from the invariant Lagrangian function. From the formal point of view, this RWE describes a microobject with nonzero mass and the spins $0,1$. The minimal equations for the spin blocks $C^{0}$, $C^{1}$  and the whole matrix $\Gamma_4$ have the equal form
\begin{align}
(C^0)^2  -1 =0, \quad (C^1)^2 -1 = 0, \quad \Gamma_4^2 - 1 =0,
\label{eq531}
\end{align}
from where it follows that the present RWE belongs to the Dirac type with the algebra \eqref{eq54}. The presence of the multiple roots $\pm 1$ in the blocks $C^{0}$, $C^{1}$ implies the presence of an additional (besides spin) internal degree of freedom.

We note that the choice of elements \eqref{eq527} of the matrix $\Gamma_4$ is not unique as long as the derivation of a Dirac-like equation is concerned. In general, to satisfy the characteristic equations \eqref{eq531} it is sufficient to demand the condition
\begin{align}
\Gamma_{\mu} \Gamma_{\nu} + \Gamma_{\nu} \Gamma_{\mu} =2 \delta_{\mu \nu} .
\label{eq532}
\end{align}

It is obvious that only by changing the signs of the numbers $c_{13}^0$, $c_{24}^0$, $c_{35}^1$, $c_{45}^1$ one can define the spin $C^{0}$, $C^{1}$ in  $16$ different ways. However, all these (and other possible) variants are related to each other by similarity transformations, and therefore all of them are physically equivalent.

Thus, on the basis of the linking scheme \eqref{eq523} one ca construct the only (up to a similarity transformation) Dirac-like RWE, which is nondisintegrating in the sense of the full Lorentz group and admitting the Lagrangian formulation.

The above formulated algebraic and group-theoretical substantiation of the dynamical equivalence of the classical Dirac equation and the $SU(2,2)$-invariant Dirac theory is not yet sufficient for a geometrized description of internal degrees of freedom of the Dirac particles in terms of tensor fields. To consistently realize a possibility of such a description one has to demonstrate that this correspondence persists on the quantum level, which is equivalent to proving a possibility of quantization of the DK field in the Fermi--Dirac statistics.

It might seem that such an assumption contradicts to the famous Pauli's theorem about the relation between spin and statistics [34]. However, this is not quite true. Already in the papers [35, 36] it was shown on examples of the simplest equations for particles with integer and half-integer spins that by using an indefinite metric in the Hilbert space of states it is in principle possible to quantize fields by the anomalous statistics   (half-integer spins by the Bose--Einstein statistics and integer spins by the Fermi--Dirac statistics). However, in doing so one obtains unremovable negative probabilities.

An essentially different situation arises in the case of field systems with additional degrees of freedom corresponding to noncompact internal symmetry groups. In such theories, there exist additional conservation laws (ban rules), forbidding transitions which are described by negative probabilities emerging in the quantization with an indefinite metric. Let us consider this question in more detail with regard to the DK equation [37, 38].

With help of the substitution
\begin{align}
\Psi (x) = \Psi (p) e^{i p_{\mu} x_{\mu}}
\label{eq533}
\end{align}
we proceed from the matrix form \eqref{eq:nonzero_mass} of the DK equation in the position representation to the momentum representation
\begin{align}
(\hat{p} + m) \Psi (p) =0,
\label{eq534}
\end{align}
where
\begin{align}
\hat{p} = i p_{\mu} \Gamma_{\mu}
\label{eq535}
\end{align}
is an operator of the $4$-momentum.

As it follows from \eqref{eq531}, the spin blocks $C^{0}$, $C^{1}$ contain the only -- up to a sign -- nonzero root $\pm 1$. A presence of the internal degree of freedom is expressed  in this case in the twofold multiplicity of this nonzero root in the characteristic equations of these blocks. Thus, along with the usual $4$-momentum operators \eqref{eq535}, the spin square operator
\begin{align}
\hat{S}^2 = - [(J^{[12]})^2 + (J^{[23]})^2 + (J^{[31]})^2]
\label{eq536}
\end{align}
and the spin projection operator
\begin{align}
\hat{S}_n =- i \varepsilon_{ijk} n_i J^{[jk]},
\label{eq537}
\end{align}
where $J^{[\mu \nu]} = \frac14 (\Gamma_{[\mu} \Gamma_{\nu]}+\Gamma'_{[\mu} \Gamma'_{\nu]})$, we can assign to this degree of freedom (let us call it $\Pi$-parity for concreteness) an operator $\hat{\Pi}$, which commutes with the above quoted operators and forms together with them the full set of variables for the DK field. Additionally, we complement this assignment by the natural requirements of the diagonalizability of this operator and the real-valuedness of its eigenvalues, and -- in analogy with the operators $\hat{S}^2$, $\hat{S}_n$ -- by the property
\begin{align}
\hat{\Pi} \eta = \eta \hat{\Pi}^{+}.
\label{eq538}
\end{align}

It is not difficult to see that the relativistically invariant definition of the $\Pi$-parity operator, obeying the formulated conditions, has the form
\begin{align}
\hat{\Pi} = \frac{p_{\mu} \Gamma'_{\mu}}{i m};
\label{eq539}
\end{align}
and, in particular, in the rest frame
\begin{align}
\hat{\Pi}_0 = \Gamma'_4 .
\label{eq540}
\end{align}
The eigenvalues of $\hat{\Pi}$ we will denote by $\lambda_i$, $i= 1, 2$. (In the rest frame it holds $\lambda_1= 1$, $\lambda_2 = - 1$).

In the second quantization, the sign factors of the energy and the charge densities of a classical field system acquire an important role. The presence of the spin spectrum and the $\Pi$-parity infers that the signs of these quantities can depend not only on the mass sign (which is meant to be the sign of the matrix $\Gamma_4$ eigenvalues distinguishing between the positive and negative frequency solutions of the equation \eqref{eq534}), but also on the quantum numbers $i$ and $s$. In the other words, both the energy and the charge in such theories appear to be, generally speaking, indefinite. This circumstance is conveniently reflected in the variable  $g_{is}^{(\pm)}$, whose values, evaluated in the rest frame, specify the sign of the energy density in the state  $\psi_{is}^{(\pm)}$. Computing $g_{is}^{(\pm)}$ for the DK equation yields [39]:
\begin{align}
g_{1s}^{(+)} =g_{2s}^{(-)}=1, \quad g_{1s}^{(-)} =g_{2s}^{(+)}=-1 .
\label{eq541}
\end{align}

Now we can directly implement the quantization. Representing the operator wavefuctions $\Psi$, $\overline{\Psi}$ by the series
\begin{align}
\Psi (x) &= \frac{1}{(2 \pi)^{3/2}} \sum_i \sum_s \left[ a_{is} (p) \psi_{is}^{(+)} (p) e^{i p x} + \stackrel{+}{b}_{is} (p) \psi_{is}^{(-)} (p) e^{-i p x} \right] d^3 p,
\label{eq542} \\
\overline{\Psi} (x) &= \frac{1}{(2 \pi)^{3/2}} \sum_i \sum_s \left[  \stackrel{+}{a}_{is} (p) \bar{\psi}_{is}^{(+)} (p) e^{-i p x} + b_{is} (p) \bar{\psi}_{is}^{(-)} (p) e^{i p x} \right] d^3 p ,
\label{eq543}
\end{align}
we postulate the commutation relations for the annihilation and creation operators
\begin{align}
& [a_{is} (p), \stackrel{+}{a}_{i's'} (p') ]_+ = g^{(+)}_{is} \delta_{ii'} \delta_{ss'} \delta (p-p'),
\label{eq544} \\
& [b_{is} (p), \stackrel{+}{b}_{i's'} (p') ]_+ = - g^{(-)}_{is} \delta_{ii'} \delta_{ss'} \delta (p-p')
\label{eq545}
\end{align}
(no summation over the indices $i$ and $s$ here; all the other anticommutators identically vanish), which correspond to the quantization of the DK field in the Fermi--Dirac statistics. The particle and antiparticle number operators leading to the correct eigenvalues are defined in the following way:
\begin{align}
N_{is}^{(+)}  = g^{(+)}_{is} \stackrel{+}{a}_{is} a_{is} , \quad N_{is}^{(-)}  =  - g^{(-)}_{is} \stackrel{+}{b}_{is} b_{is} .
\label{eq546}
\end{align}
Inserting the series  \eqref{eq542},  \eqref{eq543} into the expressions for the energy and the charge operators yields
\begin{align}
E &= \int \{ ( \partial_4 \overline{\Psi})  \Gamma_4 \Psi -\overline{\Psi}  \Gamma_4 \partial_4  \Psi  \}d^3 x ,
\label{eq547} \\
Q &= e \int \overline{\Psi} \Gamma_4 \Psi d^3 x .
\label{eq548}
\end{align}
Taking into account the relations  \eqref{eq544}--\eqref{eq546} and the normalization by charge
\begin{align}
\overline{\Psi} \Gamma_4  \Psi = \pm 1,
\label{eq549}
\end{align}
we obtain the final expressions for the operators  $E$ and $Q$
\begin{align}
E &= \sum_i \sum_s \left( N_{is}^{(+)} \varepsilon_{is}^{(+)} + N_{is}^{(-)} \varepsilon_{is}^{(-)}  \right),
\label{eq550} \\
Q &= e \sum_i \sum_s \left( N_{is}^{(+)} - N_{is}^{(-)}  \right),
\label{eq551}
\end{align}
where $\varepsilon_{is}^{(\pm)} = |p_0 |$, and the indices of $\varepsilon_{is}^{(\pm)}$ indicate the relation to the corresponding state.

The expressions   \eqref{eq550},  \eqref{eq551} mean that the anticommutation relations  \eqref{eq544},  \eqref{eq545} ensure a correct corpuscular picture of the field. Moreover, it is easy to check that they lead to causal commutation relations for the field operators  [40; 41].

Since the right-hand sides of some quantization conditions  \eqref{eq544},  \eqref{eq545} contain the "wrong" minus sign, the corresponding state vectors must have a negative valued norm. In the other words, a quantum description of the DK field in the Fermi--Dirac statistics implies using the space of states  $H$ with the indefinite metrics
\begin{align}
H = H_+ \oplus H_-,
\label{eq552}
\end{align}
where $H_+$ and $H_-$ are the subspaces with positive and negative state vector norms, respectively. In the considered case the subspaces $H_+$ and $H_-$ are spanned by the states
\begin{align}
H_+ : \quad \left( \prod_{N_1}  \stackrel{+}{a}_{1s} \right)  \left( \prod_{N_2}  \stackrel{+}{b}_{2s} \right)  \left( \prod_{N_3}  \stackrel{+}{a}_{2s} \right)  \left( \prod_{N_4}  \stackrel{+}{b}_{1s} \right) | 0 \rangle  ;
\label{eq553} \\
H_- : \quad \left( \prod_{N_5}  \stackrel{+}{a}_{1s} \right)  \left( \prod_{N_6}  \stackrel{+}{b}_{2s} \right)  \left( \prod_{N_7}  \stackrel{+}{a}_{2s} \right)  \left( \prod_{N_8}  \stackrel{+}{b}_{1s} \right) | 0 \rangle .
\label{eq554}
\end{align}
Here $N_1$, $N_2$, $N_5$, and $N_6$ are arbitrary nonnegative integers, $(N_3 + N_4 )$ is an even number, and $( N_7 + N_8 )$ is an odd number. For single particle states the partitions  \eqref{eq553},  \eqref{eq554} acquire the form
\begin{align}
& H_+ : \Psi_{1s}^{(+)}, \Psi_{1s}^{(-)}, \qquad H_- : \Psi_{2s}^{(+)}, \Psi_{2s}^{(-)},
\label{eq555} \\
& H_+ : \Psi_{1s}^{(+)}, \Psi_{2s}^{(-)}, \qquad H_- : \Psi_{2s}^{(+)}, \Psi_{1s}^{(-)} .
\label{eq556}
\end{align}
Additionally, for a correct probabilistic interpretation of the theory it is necessary to ensure the absence of transitions between the states of $H_+$ and $H_-$ by including interactions. Let us show that such transitions are indeed forbidden.

Consider the Lagrangian
\begin{align}
\mathcal{L} = - \overline{\Psi} (x) (\Gamma_{\mu} \partial_{\mu} +m) \Psi (x) + \mathcal{L}_{int},
\label{eq557}
\end{align}
where $ \mathcal{L}_{int}$ is an interaction Lagrangian which does not violate an internal symmetry inherent to a free field. For instance, in the case of electromagnetic interaction it reads
\begin{align}
 \mathcal{L}_{int} =e  \overline{\Psi} \Gamma_{\mu} A_{\mu} \Psi + \overline{\Psi} F_{\mu \nu} \Gamma_{[ \mu} \Gamma_{\nu ]} \Psi.
\label{eq558}
\end{align}
It is evident that the operator $\hat{\Pi}$  \eqref{eq539} belongs to the transformations of the internal symmetry group of the  \eqref{eq557},  \eqref{eq558} (compare  \eqref{eq539} with the generators  \eqref{eq510} of this group). The invariance of the quoted Lagrangian under the transformations
\begin{align}
\Psi \to e^{i \hat{\Pi} \theta} \Psi
\label{eq559}
\end{align}
leads to the conserved "charge"
\begin{align}
G \sim \int \overline{\Psi} (x) \Gamma_4 \hat{\Pi} \Psi (x) d^3 x .
\label{eq560}
\end{align}
The charge $G$ can be transformed to the form
\begin{align}
G \sim \sum_i \sum_s  \lambda_i \left( N_{is}^{(+)}  - N_{is}^{(-)} \right) = \sum_s \left( N_{1s}^{(+)} - N_{2s}^{(+)} - N_{1s}^{(-)} + N_{2s}^{(-)} \right),
\label{eq561}
\end{align}
where it is accounted that $\lambda_1= 1$, $\lambda_2 = - 1$. For the sake of convenience we also rewrite the formula  \eqref{eq551} in the expanded form
\begin{align}
Q \sim \sum_s \left( N_{1s}^{(+)} + N_{2s}^{(+)} - N_{1s}^{(-)} - N_{2s}^{(-)}\right).
\label{eq562}
\end{align}

Comparing the partitions  \eqref{eq555},  \eqref{eq556} with the expressions  \eqref{eq561},  \eqref{eq562}, we draw the conclusion that to the single-particle states belonging to the subspaces $H_+$ and $H_-$ correspond the following signs of the charges $Q$ and $G$:
\begin{align}
H_+ : \quad (1,1), (-1,1), \qquad H_- : \quad (1,-1), (-1,-1)
\label{eq563}
\end{align}
(the first number in the parenthesis corresponds to the electric charge $Q$, while the second number corresponds to the additional charge $G$).

From it follows \eqref{eq563} that a simultaneous fulfilment of the conservation laws for the charges  $Q$ and $G$ leads to a prohibition of physically inappropriate transitions between states from the subspaces with the positive and the negative norms of state vectors.

We remark that if instead of the continuous transformations  \eqref{eq559} we consider the discrete transformations
\begin{align}
\Psi_+ \to \Psi_+, \quad \Psi_- \to - \Psi_- ,
\label{eq564}
\end{align}
they will be reduced to these:
\begin{align}
\begin{split}
& a_{1s}, \stackrel{+}{a}_{1s}\, \,  \to a_{1s}, \stackrel{+}{a}_{1s}, \quad  b_{1s}, \stackrel{+}{b}_{1s} \, \, \to  b_{1s}, \stackrel{+}{b}_{1s}, \\
& a_{2s}, \stackrel{+}{a}_{2s} \, \,  \to - a_{2s},  - \stackrel{+}{a}_{2s}, \quad b_{2l}, \stackrel{+}{b}_{2s} \, \,  \to -  b_{2s}, -  \stackrel{+}{b}_{2s}.
\end{split}
\label{eq565}
\end{align}
In mathematical literature this operation bears the name of a canonical, or $J$-symmetry. It underpins the theory of linear operators in spaces with indefinite metrics, which are also called Hilbert spaces with $J$-metrics, or the Krein spaces [42]. As shown in [43], the $J$-symmetry corresponds to the superselection operator forbidding transitions from $H_+$ to $H_-$, which is in agreement with our result established above. Upon localizing the internal symmetry group, which has a spatiotemporal origin, and considering the corresponding gauge theory the discrete $J$-symmetry also helps to exclude transitions featuring negative probabilities [44].

Thus, the considered quantization procedure of the DK equation by the Fermi--Dirac statistics appears to be correct also from the point of view of a probabilistic interpretation of the theory. This fact in combination with the other algebraic and group-theoretical properties discussed above substantiates a
 possibility in principle to describe internal degrees of freedom of the Dirac particles in the geometrized approach.

We also point out that in the most general formulation the question of a possible physically consistent quantization of RWEs with noncompact groups of internal symmetry either by the normal or by the anomalous statistics using indefinite metrics has been considered in detail in the works  [39--41].

\section{Algebraic generalizations of the Dirac-K\"ahler equation}
\hspace{0.5cm}
Despite of a number of attractive features which the DK equation possesses, a question about its ability to serve, e.g., for a spatiotemporal description of internal quantum numbers of the known fundamental particles still remains open. On one hand, if we mean the only quantum number -- the number of quark (or lepton) generations, then the number of the wavefunction components in the DK equation (equal to 16) is excessive, since up to date only three generations of these particles have been discovered. On the other hand, for a unifying geometrized description of all known degrees of freedom of fundamental particles the sixteen wavefunction components are apparently insufficient. Therefore, a consequent realization of this approach in the four-dimensional spacetime infers the usage of RWEs with a larger number of the wavefunction components and properties analogous to those of the DK equation.

One can obtain a set of such RWEs when passing to a maximal tensorial generalization of the DK equation in space of the dimension $d = 4$. It consists in treating the fields $\psi_0$, $\tilde{\psi}_0$, $\psi_\mu$, $\tilde{\psi}_\mu$, $\psi_{[\mu \nu]}$ as general elements of the Clifford algebra $C_4$. This transition is equivalent to introducing -- along with the Dirac-like equation  \eqref{eq52} for the function $\Psi$ -- analogous equations for the functions $\tilde{\Psi}$ and $\Psi_A$, where $A$ is a free Lorenz index, taking subsequently one of the values $A=\mu , \tilde{\mu}, [\mu \nu]$ (here $\tilde{\mu}$ is a pseudovector index;   $\tilde{\Psi}$ characterizes the same set of antisymmetric tensor fields as $\Psi$ does, differing from the latter by modified transformation properties under spatial inversion).

Before going over to the announced algebraic generalizations of the DK equation, let us first discuss in more detail a question of its closedness (nondisintegration) in the sense of the full Lorentz group. To this end we make substitutions
\begin{align}
\begin{split}
& \varphi_0 = \psi_0 - i \tilde{\psi}_0, \quad \varphi_{\mu} = \psi_\mu - i \tilde{\psi}_\mu , \quad \varphi_{[\mu \nu]} = \psi_{[\mu \nu]} - i \tilde{\psi}_{[\mu  \nu ]} , \\
& \dot{\varphi}_0 =  \psi_0 + i \tilde{\psi}_0, \quad \dot{\varphi}_{\mu} = \psi_\mu + i \tilde{\psi}_\mu , \quad \dot{\varphi}_{[\mu \nu]} = \psi_{[\mu \nu]} + i \tilde{\psi}_{[\mu  \nu ]}
\end{split}
\label{eq61}
\end{align}
and bring the system \eqref{eq51} to the form
\begin{align}
\begin{split}
& \partial_\mu \varphi_\mu + m \varphi_0 =0 ,\\
& \partial_\nu \varphi_{[\mu \nu ]} + \partial_{\mu} \varphi_0 + m \varphi_\mu =0, \\
-& \partial_\mu \varphi_\nu + \partial_\nu \varphi_\mu + i \varepsilon_{\mu \nu \alpha \beta} \partial_\alpha \varphi_\beta + m \varphi_{[\mu \nu ]}  =0;
\end{split}
\label{eq62}
\end{align}
\begin{align}
\begin{split}
& \partial_\mu \dot{\varphi}_\mu + m \dot{\varphi}_0 =0 ,\\
& \partial_\nu  \dot{\varphi}_{[\mu \nu ]} + \partial_{\mu}  \dot{\varphi}_0 + m  \dot{\varphi}_\mu =0, \\
-& \partial_\mu  \dot{\varphi}_\nu + \partial_\nu  \dot{\varphi}_\mu - i \varepsilon_{\mu \nu \alpha \beta} \partial_\alpha  \dot{\varphi}_\beta + m  \dot{\varphi}_{[\mu \nu ]}  =0.
\end{split}
\label{eq63}
\end{align}
It is obvious that the system \eqref{eq62}, \eqref{eq63} falls apart into two ones \eqref{eq62} and \eqref{eq63} which are invariant in the sense of the proper Lorenz group. However, it is alike obvious that under spatial inversion these two subsystems are transformed one into the other, which means that each of them treated alone is not  $P$-invariant. Thus, we conclude that under the full Lorentz group transformations the DK system in question is nondisintegrating.

Let us now consider the matrix equation
\begin{align}
(\Gamma_{\alpha} \partial_{\alpha} + m ) \Psi_{[\eta \xi]}  =0,
\label{eq64}
\end{align}
which is obtained by adding to the wavefunction $\Psi$ \eqref{eq52}, obeying the DK equation, a free bivector index $[\eta \xi]$. This equation cn be represented in the standard form \eqref{eq:nonzero_mass}, where $96$-component wavefunction $\psi$ is being transformed according to the Lorentz group representation
\begin{align}
\left[ (0,0) \oplus 2 (\frac12 , \frac12) \oplus (0,1) \oplus (1,0) \right] \otimes [(0,1) \oplus (1,0)].
\label{eq65}
\end{align}
Irreducible representations, which are contained in the tensor product \eqref{eq65}, form in general the linking scheme
\begin{align}
\begin{tabular}{ccccc}
& & $2 (0,0)$ & & \\
& & $|$ & & \\
$3 (0,1)$ & --- & $4 (\frac12, \frac12)$ & --- & 3 $(1,0)$.
\end{tabular}
\label{eq66}
\end{align}

As it has been already shown, the DK equation can be represented by a direct sum of two Dirac-like $8$-component systems  \eqref{eq62} and \eqref{eq63}, which, in turn, can be written in the form \eqref{eq:nonzero_mass} with the wavefunctions
\begin{align}
\Psi^{(8)}  = (\varphi_0 , \varphi_\mu , \varphi_{[\mu \nu]} )^T , \quad \dot{\Psi}^{(8)}  = (\dot{\varphi}_0 , \dot{\varphi}_\mu , \dot{\varphi}_{[\mu \nu]} )^T
\label{eq67}
\end{align}
(note that the tensors $\varphi_{[\mu \nu]}$, $\dot{\varphi}_{[\mu \nu]}$ are self-dual in the present context, and hence each of them contains three  independent components) and the linking schemes
\begin{align}
(0,0) \quad  \textup{---} \quad  (\frac12 , \frac12) \quad \textup{---}  \quad (0,1),
\label{eq68} \\
(0,0) \quad  \textup{---} \quad (\frac12 , \frac12) \quad  \textup{---} \quad (1,0).
\label{eq69}
\end{align}
Therefore the equation \eqref{eq64} splits into the subequations
\begin{align}
& (\Gamma^{(8)}_{\mu} \partial_{\mu} +m) \Psi^{(8)}_{(0,1)} =0 ,
\label{eq610} \\
& (\Gamma^{(8)}_{\mu} \partial_{\mu} +m) \dot{\Psi}^{(8)}_{(0,1)} =0 ,
\label{eq611} \\
& (\Gamma^{(8)}_{\mu} \partial_{\mu} +m) \Psi^{(8)}_{(1,0)} =0 ,
\label{eq612} \\
& (\Gamma^{(8)}_{\mu} \partial_{\mu} +m) \dot{\Psi}^{(8)}_{(1,0)} =0 ,
\label{eq613}
\end{align}

The equations \eqref{eq610} and \eqref{eq612} are $P$-conjugated to each other. The equations \eqref{eq611} and \eqref{eq613} are conjugated alike. Therefore, considering \eqref{eq610}, \eqref{eq612}, and \eqref{eq611}, \eqref{eq613} jointly, we obtain two invariant -- in the sense of the full Lorentz group --  systems, being transformed by the representations
\begin{align}
& \left\{ [(0,0) \oplus (\frac12 , \frac12) \oplus (0,1)] \otimes (0,1)\right\} \oplus \left\{ [(0,0) \oplus (\frac12 , \frac12) \oplus (1,0)] \otimes (1,0)\right\},
\label{eq614} \\
& \left\{ [(0,0) \oplus (\frac12 , \frac12) \oplus (0,1)] \otimes (1,0)\right\} \oplus \left\{ [(0,0) \oplus (\frac12 , \frac12) \oplus (1,0)] \otimes (0,1)\right\}.
\label{eq615}
\end{align}
In other words, when the linking scheme \eqref{eq66} is used as a basis for constructing the Dirac-like RWE, it falls apart into fragments
\begin{align}
\begin{tabular}{ccccccc}
& & & $2 (0,0)$ & & & \\
& & & $|$ & & &  \\
& $2 (0,1)$ & --- & $2 (\frac12, \frac12)$ & --- & 2 $(1,0)$  & \\
& $|$ & &  & &$|$  & \\
$(0,2) \quad $ --- & $(\frac12, \frac32)$ &  & &  & $(\frac32,\frac12)$ & --- $\quad (2,0)$
\end{tabular}
\label{eq616}
\end{align}
and
\begin{align}
\begin{tabular}{ccccc}
$(0,1)$ &--- & $2 (\frac12,\frac12)$ &--- & $(1,0)$ \\
$|$ & & $|$ & & $|$\\
$(\frac12, \frac32)$ & --- & $2 (1,1)$ & --- & $(\frac32, \frac12)$.
\end{tabular}
\label{eq617}
\end{align}

In turn, out of irreducible components contained in \eqref{eq616} one can compose the two independent linking schemes: the scheme  \eqref{eq515}, corresponding to the DK equation, and the scheme
\begin{align}
\begin{tabular}{ccccccc}
& $(0,1)$ &  & &  & $(1,0)$  & \\
& $|$ & & $\oplus$ & &$|$  & \\
$(0,2) \quad $ --- & $(\frac12, \frac32)$ &  & &  & $(\frac32,\frac12)$ & --- $\quad (2,0)$.
\end{tabular}
\label{eq618}
\end{align}

Let us show that the linking scheme  \eqref{eq618} does provide a possibility to construct a Dirac-type RWE satisfying all necessary physical requirements. We label the irreducible components contained in  \eqref{eq618} like
\begin{align}
\begin{split}
& (0,1) \sim 1, \quad (\frac12 , \frac32) \sim 2, \quad (0,2) \sim 3, \\
& (1,0) \sim 4, \quad (\frac32 , \frac12) \sim 5, \quad (2,0) \sim 6.
\end{split}
\label{eq619}
\end{align}
Then, we obtain an expression for the matrix $\Gamma_4$ in the Gel'fand--Yaglom basis
\begin{align}
& \Gamma_4 = \left(
\begin{tabular}{cc}
$C^1 \otimes I_3$ & $0$  \\
$0$ & $C^2 \otimes I_5$
\end{tabular} \right),
\label{eq620} \\
& C^1 = \left(
\begin{tabular}{cccc}
$0$ & $c^1_{12}$ & $0$ & $0$ \\
$c^1_{21}$ & $0$ & $0$ & $0$ \\
$0$ & $0$ & $0$ & $c^1_{45}$ \\
$0$ & $0$ & $c^1_{54}$ & $0$
\end{tabular} \right), \quad
C^2 = \left(
\begin{tabular}{cccc}
$0$ & $c^2_{23}$ & $0$ & $0$ \\
$c^2_{32}$ & $0$ & $0$ & $0$ \\
$0$ & $0$ & $0$ & $c^2_{56}$ \\
$0$ & $0$ & $c^2_{65}$ & $0$
\end{tabular} \right).
\label{eq621}
\end{align}

The requirement of the $P$-invariance of the theory imposes according to the conditions \eqref{c2_cond}, \eqref{Pc1_cond} the following constraints upon elements of the spin blocks  $C^{1}$, $C^{2}$
\begin{align}
c^1_{12} = c^1_{45}, \quad c^1_{21} = c^1_{54}, \quad c^2_{23} = c^2_{56}, \quad c^2_{32} = c^2_{65}.
\label{eq622}
\end{align}
A possibility of the Lagrangian formulation  ( see equation \eqref{c_eta_cond}) leads to the relations
\begin{align}
c_{12}^1 \eta_{25}^1 = (c_{54}^1)^* \eta_{14}^1, \quad c_{23}^2 \eta_{36}^2 = (c_{65}^2)^* \eta_{25}^2 .
\label{eq623}
\end{align}
Combining \eqref{eq622} and \eqref{eq623}, we obtain
\begin{align}
c^1_{21}=c^1_{54}= f a^* , \quad c^2_{32}=c^2_{65}=g b^* ,
\label{eq624}
\end{align}
where the following notations are introduced
\begin{align}
f = \frac{\eta^1_{25}}{\eta^1_{14}}, \quad g= \frac{\eta^2_{36}}{\eta^2_{25}}, \quad a= c^1_{12}, \quad b= c^2_{23}.
\label{eq625}
\end{align}

A minimal equation for the matrix $\Gamma_4$ would have the form \eqref{eq531} under the fulfilment of the equalities
\begin{align}
f |a|^2 = g |b|^2 =1,
\label{eq626}
\end{align}
which can be satisfied, e.g., by setting
\begin{align}
f=g=a=b=1.
\label{eq627}
\end{align}
At the same time, the spin blocks $C^{1}$, $C^{2}$ acquire the same form
\begin{align}
C^{1} = C^{2} =I_2 \otimes \sigma_1 .
\label{eq628}
\end{align}

Nonzero elements $\eta^s_{ij}$ of the matrix of the bilinear invariant form $\eta$, having in the present case the following structure in the Gel'fand--Yaglom basis
\begin{align}
\eta = \left(
\begin{tabular}{cc}
$\eta^1 \otimes I_3$ & $0$  \\
$0$ & $\eta^2 \otimes I_5$
\end{tabular} \right),
\label{eq629}
\end{align}
can be chosen in accordance with the conditions \eqref{eq625}, \eqref{eq627} as
\begin{align}
\eta^1_{14} = \eta^1_{25} = - \eta^2_{25} = -\eta^2_{36} = 1
\label{eq630}
\end{align}
(recall that $\eta^s_{ij} = -\eta^{s-1}_{ij}$).

The $32$-component RWE obtained in this way is by construction of the Dirac type. It is invariant under transformations of the full Lorentz group, allows for a Lagrangian formulation, and from  the point of view of the standard formulation of the RWE theory it describes a microobject with spins $1,2$, a single mass value and a double degenerate set of states, labeled by some additional quantum number. Like the DK equation, this equation does not fall apart in the sense of the full Lorentz group.

Let us now consider the linking scheme \eqref{eq617}. If we construct on its basis a RWE of the Dirac type, then it falls apart into a direct sum of fragments [45]
\begin{align}
\begin{tabular}{ccccccccc}
$(0,1)$ & --- & $(\frac12, \frac12)$ &  & &  & $(\frac12, \frac12)'$  & --- & $(1,0)$ \\
$|$ & & $|$ & & $\oplus$ & &$|$  & & $|$\\
$(\frac12, \frac32)$ & --- & $(1,1)$ &  & &  & $(1,1)'$ & --- & $(\frac32,\frac12)$.
\end{tabular}
\label{eq631}
\end{align}
where multiple representations $(\frac12, \frac12)$ and $(\frac12, \frac12)'$, as well as $(1, 1)$ and $(1, 1)'$,  are $P$-conjugated  to each other.

As usual, we introduce the labelling of irreducible components contained in the scheme  \eqref{eq631}
\begin{align}
& (\frac12, \frac12) \sim 1, \quad (\frac12, \frac32) \sim 2,  \quad (1,1) \sim 3, \quad (0,1) \sim 4, \notag \\
& (\frac12, \frac12)' \sim 5, \quad (\frac32, \frac12) \sim 6,  \quad (1,1)' \sim 7, \quad (1,0) \sim 8. \notag
\end{align}
Then for the spin blocks $C^{s}$ ($s = 0, 1, 2$) of the matrix $\Gamma_4$
\begin{align}
\Gamma_4 = \left(
\begin{tabular}{ccc}
$C^0$ & $0$ & $0$ \\
$0$ & $C^1 \otimes I_3$ & $0$  \\
$0$ & $0$ & $C^2 \otimes I_5$
\end{tabular} \right)
\label{eq632}
\end{align}
we obtain general expressions
\begin{align}
\begin{split}
C^0 &= \left(
\begin{tabular}{cccc}
$0$ & $0$ & $c^0_{13}$ & $0$ \\
$0$ & $0$ & $0$ & $c^0_{57}$ \\
$c^0_{31}$ & $0$ & $0$ & $0$ \\
$0$ & $c^0_{75}$ & $0$ & $0$
\end{tabular} \right), \quad C^1 = \left(
\begin{tabular}{cc}
$(C^1)'$ & $0$  \\
$0$ & $(C^1)''$
\end{tabular} \right), \\
(C^1)' & = \left(
\begin{tabular}{cccc}
$0$ & $0$ & $c^1_{13}$ & $c^1_{14}$ \\
$0$ & $0$ & $c^1_{23}$ & $c^2_{24}$ \\
$c^1_{31}$ & $c^1_{32}$ & $0$ & $0$ \\
$c^1_{41}$ & $c^1_{42}$ & $0$ & $0$
\end{tabular} \right), \quad
(C^1)'' = \left(
\begin{tabular}{cccc}
$0$ & $0$ & $c^1_{57}$ & $c^1_{58}$ \\
$0$ & $0$ & $c^1_{67}$ & $c^2_{68}$ \\
$c^1_{75}$ & $c^1_{76}$ & $0$ & $0$ \\
$c^1_{85}$ & $c^1_{86}$ & $0$ & $0$
\end{tabular} \right), \\
C^2 &= \left(
\begin{tabular}{cccc}
$0$ & $c^2_{23}$ & $0$ & $0$ \\
$c^2_{32}$ & $0$ & $0$ & $0$ \\
$0$ & $0$ & $0$ & $c^2_{67}$ \\
$0$ & $0$ & $c^2_{76}$ & $0$
\end{tabular} \right).
\end{split}
\label{eq633}
\end{align}

The conditions of the relativistic \eqref{c_cond} and $P$-invariance \eqref{c1_cond}, \eqref{c2_cond} impose on elements of the blocks  \eqref{eq633} the corresponding constarints
\begin{align}
\begin{split}
c^1_{13} &= \sqrt{\frac23} c^0_{13}, \quad c^1_{31} = \sqrt{\frac23} c^0_{31}, \quad c^1_{57} = \sqrt{\frac23} c^0_{57}, \quad c^1_{75} = \sqrt{\frac23} c^0_{75},  \\
c^1_{23} &= \sqrt{\frac13} c^2_{23}, \quad c^1_{32} = \sqrt{\frac13} c^2_{32}, \quad c^1_{67} = \sqrt{\frac13} c^2_{67}, \quad c^1_{76} = \sqrt{\frac13} c^2_{76};
\end{split}
\label{eq634}
\end{align}
\begin{align}
\begin{split}
c^1_{14} & = \pm c^1_{58}, \quad c^1_{23}  = \pm c^1_{67} , \quad c^2_{23} = \pm c^2_{67} ,\quad c^1_{24} =  c^1_{68} , \\
c^1_{41} & = \pm c^1_{85}, \quad c^1_{32}  = \pm c^1_{76} , \quad c^2_{32} = \pm c^2_{76} ,\quad c^1_{42} =  c^1_{86} .
\end{split}
\label{eq635}
\end{align}
The condition \eqref{c_eta_cond} of a possibility to obtain RWE with the linking scheme  \eqref{eq617} from an invariant Lagrangian leads to the constraints
\begin{align}
\begin{split}
& c^s_{31} = \frac{\eta^s_{33}}{\eta^s_{11}} (c^s_{13})^*, \quad  c^s_{75} = \frac{\eta^s_{77}}{\eta^s_{55}} (c^s_{57})^*
 \quad (s=0,1), \\
 & c^s_{76} = \frac{\eta^s_{33}}{\eta^s_{26}} (c^s_{23})^*, \quad  c^s_{67} = \frac{\eta^s_{26}}{\eta^s_{77}} (c^s_{32})^*
 \quad (s=1,2), \\
 & c^1_{85} = \frac{\eta^1_{48}}{\eta^1_{55}} (c^1_{14})^*, \quad  c^1_{58} = \frac{\eta^1_{11}}{\eta^1_{48}} (c^1_{41})^*, \\
 & c^1_{86} = \frac{\eta^1_{48}}{\eta^1_{26}} (c^s_{24})^*, \quad  c^1_{68} = \frac{\eta^1_{26}}{\eta^1_{48}} (c^1_{42})^*.
\end{split}
\label{eq636}
\end{align}
Minimal equations
\begin{align}
(C^0)^2 - 1=0, \quad (C^1)^2 - 1=0, \quad(C^2)^2 - 1=0,
\label{eq637}
\end{align}
leading to the Dirac algebra of the matrices $\Gamma_\mu$, would take place under the fulfilment of equalities
\begin{align}
\begin{split}
& c^0_{13} c^0_{31}=c^2_{23} c^2_{32} =1, \\
& c^1_{13} c^1_{31}+c^1_{14} c^1_{41} =1, \quad c^1_{23} c^1_{32}+c^1_{24} c^1_{42} =1, \\
& c^1_{13} c^1_{32}+c^1_{14} c^1_{42} =0, \quad c^1_{23} c^1_{31}+c^1_{24} c^1_{41} =0, \\
& c^1_{13} c^1_{31}+c^1_{23} c^1_{32} =1, \quad c^1_{14} c^1_{41}+c^1_{24} c^1_{42} =1, \\
& c^1_{14} c^1_{31}+c^1_{24} c^1_{32} =0, \quad c^1_{13} c^1_{41}+c^1_{23} c^1_{42} =0,
\end{split}
\label{eq638}
\end{align}
as well as of equalities following from \eqref{eq638} under the index replacements $1 \to 5$, $2 \to 6$, $3 \to 7$, $4 \to  8$.

The conditions \eqref{eq634}--\eqref{eq638} are simultaneously satisfied, if we set
\begin{align}
\begin{split}
& c^0_{13} =c^0_{31} =c^0_{57} =c^0_{75} =1, \quad - c^2_{23} = - c^2_{32} =c^2_{67} =c^2_{76} =1, \\
& c^1_{13} =c^1_{31} =c^1_{57} =c^1_{75} =\sqrt{\frac23}, \quad -c^1_{23} =-c^1_{32} =c^1_{67} =c^1_{76} =\sqrt{\frac13}, \\
& c^1_{14} =c^1_{41} =-c^1_{58} =-c^1_{85} =\sqrt{\frac13}, \quad c^1_{24} =c^1_{42} =c^1_{68} =c^1_{86} =\sqrt{\frac23}
\label{eq639}
\end{split}
\end{align}
and
\begin{align}
\begin{split}
\eta_{11}^0 &=\eta_{33}^0=\eta_{55}^0=\eta_{77}^0=-\eta_{11}^1=-\eta_{33}^1=-\eta_{55}^1=-\eta_{77}^1 \\
&= \eta_{26}^1=\eta_{62}^1=\eta_{48}^1=\eta_{84}^1=-\eta_{26}^2=-\eta_{62}^2=\eta_{33}^2=\eta_{77}^2 =1.
\label{eq640}
\end{split}
\end{align}
This choice results in the following expressions for the blocks   $C^{s}$, $\eta^{s}$ of the matrices $\Gamma_4$ and $\eta$:
\begin{align}
\begin{split}
& C^0 = \sigma_1 \otimes I_2 ,  \quad C^1 = (C^1)' \oplus (C^1)'', \quad C^2 = - \sigma_3 \otimes \sigma_1 , \\
& (C^1)' = \frac{1}{\sqrt{3}}  \left(
\begin{tabular}{cccc}
$0$ & $0$ & $\sqrt{2}$ & $1$ \\
$0$ & $0$ & $-1$ & $\sqrt{2}$ \\
$\sqrt{2}$ & $-1$ & $0$ & $0$ \\
$1$ & $\sqrt{2}$ & $0$ & $0$
\end{tabular} \right), \quad
(C^1)'' = \frac{1}{\sqrt{3}} \left(
\begin{tabular}{cccc}
$0$ & $0$ & $\sqrt{2}$ & $-1$ \\
$0$ & $0$ & $1$ & $\sqrt{2}$ \\
$\sqrt{2}$ & $1$ & $0$ & $0$ \\
$-1$ & $\sqrt{2}$ & $0$ & $0$
\end{tabular} \right);
\end{split}
\label{eq641}
\end{align}
\begin{align}
\begin{split}
& \eta^0 = \left(
\begin{tabular}{cccc}
$1$ & $0$ & $0$ & $0$ \\
$0$ & $1$ & $0$ & $0$ \\
$0$ & $0$ & $1$ & $0$ \\
$0$ & $0$ & $0$ & $1$ \\
\end{tabular} \right), \quad
\eta^2 = \left(
\begin{tabular}{cccc}
$0$ & $0$ & $-1$ & $0$ \\
$0$ & $1$ & $0$ & $0$ \\
$-1$ & $0$ & $0$ & $0$ \\
$0$ & $0$ & $0$ & $1$ \\
\end{tabular} \right), \\
& \eta^1 =  \left(
\begin{tabular}{cccccccc}
$-1$ & $0$ & $0$ & $0$  & $0$ & $0$ & $0$ & $0$ \\
$0$ & $0$ & $0$ & $0$  & $0$ & $1$ & $0$ & $0$ \\
$0$ & $0$ & $-1$ & $0$ & $0$ & $0$ & $0$ & $0$ \\
$0$ & $0$ & $0$ & $0$ & $0$ & $0$ & $0$ & $1$ \\
$0$ & $0$ & $0$ & $0$  & $-1$ & $0$ & $0$ & $0$ \\
$0$ & $1$ & $0$ & $0$  & $0$ & $0$ & $0$ & $0$ \\
$0$ & $0$ & $0$ & $0$ & $0$ & $0$ & $-1$ & $0$ \\
$0$ & $0$ & $0$ & $1$ & $0$ & $0$ & $0$ & $0$
\end{tabular} \right) .
\end{split}
\label{eq642}
\end{align}

The RWE with matrices $\Gamma_4$ \eqref{eq632}, \eqref{eq641} and $\eta$ \eqref{eq642}, based on the linking scheme \eqref{eq631}, satisfies all the requirements formulated in the Chapter \ref{intro} (besides the energy definiteness) and formally describes a microobject with spins $s= 0, 1, 2$ and a single nonzero mass value. In addition, the states with $s = 0, 2$ of this microobject are two-fold degenerate, and the states with $s = 1$ are four-fold degenerate.

Finally, when considering the equation
\begin{align}
(\Gamma_{\alpha} \partial_{\alpha} +m ) \Psi_\mu =0,
\label{eq643}
\end{align}
which is obtained by adding a free vector index $\mu$ to the function $\Psi$ of the DK equation, we observe the following situation. The wavefunction $\Psi_\mu$ is transformed  by the representation
\begin{align}
\left[ 2 (0,0) \oplus 2 (\frac12, \frac12) \oplus (0,1) \oplus (1,0) \right] \otimes (\frac12, \frac12),
\label{eq644}
\end{align}
which generally corresponds to the linking scheme
\begin{align}
\begin{tabular}{ccccc}
 & & $2 (0,0)$ & &  \\
 & & $|$ & &   \\
 $2 (0,1)$ & --- & $4 (\frac12, \frac12)$ & --- & 2 $(1,0)$  \\
 $|$ & & $|$   & &$|$  \\
 $(\frac12, \frac32)$ & --- & 2 (1,1) & ---  & $(\frac32,\frac12)$ .
\end{tabular}
\label{eq645}
\end{align}

As shown in [45], while constructing the Dirac-type RWE on the basis of the scheme \eqref{eq645}, it falls apart into two schemes  \eqref{eq515} and \eqref{eq617}  which are not linked with each other. Hence, from the viewpoint of our interest the equation \eqref{eq644} does not bring along any new information on top of the above considered RWEs.

Thus, the discussed algebraic generalizations of the DK equation lead to the two new matrix RWEs of the Dirac type: the $32$-component equation with the linking scheme \eqref{eq618} and the  $48$-component equation with the linking scheme \eqref{eq631}.

A tensor formulation of the first of them reads [46]
\begin{align}
\begin{split}
& \partial_{\nu} \varphi_{\nu [\alpha \beta]} + m \varphi_{[\alpha \beta]} =0, \\
& \frac12 \left( - \partial_{\mu} \varphi_{\nu [\alpha \beta]} + \partial_{\nu} \varphi_{\mu [\alpha \beta]} - \partial_{\alpha} \varphi_{\beta [\mu \nu]} + \partial_{\beta} \varphi_{\alpha [\mu \nu]} + \right. \\
& \left. + i \varepsilon_{\mu \nu \eta \xi} \partial_{\eta} \varphi_{\xi [\alpha \beta]} + i \varepsilon_{\alpha \beta \eta \xi} \partial_{\eta} \varphi_{\xi [\mu \nu]}  \right) + m \varphi_{([\mu \nu] [\alpha \beta])} =0 , \\
& \partial_{\nu} \varphi_{([\mu \nu] [\alpha \beta])} + \frac12 \left( \partial_{\beta} \varphi_{[\alpha \mu]}  -\partial_{\alpha} \varphi_{[\beta \mu]} + \delta_{\mu \alpha} \partial_{\nu} \varphi_{[\nu \beta]}  - \delta_{\mu \beta} \partial_{\nu} \varphi_{ [\nu \alpha]}  \right) + \\
& + i \varepsilon_{\alpha \beta \eta \nu} \partial_{\eta} \varphi_{[\nu \mu]} + m \varphi_{\mu  [\alpha \beta]} =0 , \\
& \partial_{\nu} \chi_{\nu [\alpha \beta]} + m \chi_{[\alpha \beta]} =0, \\
& \frac12 \left( - \partial_{\mu} \chi_{\nu [\alpha \beta]} + \partial_{\nu} \chi_{\mu [\alpha \beta]} - \partial_{\alpha} \chi_{\beta [\mu \nu]} + \partial_{\beta} \chi_{\alpha [\mu \nu]} - \right. \\
& \left. - i \varepsilon_{\mu \nu \eta \xi} \partial_{\eta} \chi_{\xi [\alpha \beta]} - i \varepsilon_{\alpha \beta \eta \xi} \partial_{\eta} \chi_{\xi [\mu \nu]}  \right) + m \chi_{([\mu \nu] [\alpha \beta])} =0 , \\
& \partial_{\nu} \chi_{([\mu \nu] [\alpha \beta])} + \frac12 \left( \partial_{\beta} \chi_{[\alpha \mu]}  -\partial_{\alpha} \chi_{[\beta \mu]} + \delta_{\mu \alpha} \partial_{\nu} \chi_{[\nu \beta]}  - \delta_{\mu \beta} \partial_{\nu} \chi_{ [\nu \alpha]}  \right) - \\
& - i \varepsilon_{\alpha \beta \eta \nu} \partial_{\eta} \chi_{[\nu \mu]} + m \chi_{\mu  [\alpha \beta]} =0 .
\end{split}
\label{eq646}
\end{align}
The tensors entering this expression fulfil the self-duality conditions
\begin{align}
& \frac12 \varepsilon_{\mu \nu \alpha \beta} \varphi_{[\alpha \beta ]} = i \varphi_{[\mu \nu]} , \quad \frac12 \varepsilon_{\eta \xi \alpha \beta} \varphi_{\mu [\alpha \beta ]} = i \varphi_{\mu [\eta \xi]} , \quad \frac12 \varepsilon_{\eta \xi \alpha \beta} \varphi_{([\mu \nu] [\alpha \beta ] )} = i \varphi_{([\mu \nu] [\eta \xi ] )} , \label{eq647}  \\
& \frac12 \varepsilon_{\mu \nu \alpha \beta} \chi_{[\alpha \beta ]} = - i \varphi_{[\mu \nu]} , \quad \frac12 \varepsilon_{\eta \xi \alpha \beta} \chi_{\mu [\alpha \beta ]} =- i \chi_{\mu [\eta \xi]} , \quad \frac12 \varepsilon_{\eta \xi \alpha \beta} \chi_{([\mu \nu] [\alpha \beta ] )} = - i \chi_{([\mu \nu] [\eta \xi ] )} . \notag
\end{align}
Moreover, the quantities $\varphi_{\mu [\alpha \beta]} $, $\varphi_{([\mu \nu] [\alpha \beta])} $, $\chi_{\mu [\alpha \beta]}$, and $\chi_{([\mu \nu] [\alpha \beta])}$ obey the conditions
\begin{align}
\begin{split}
& \varphi_{\alpha [\alpha \beta]} =0, \quad \varphi_{([\alpha \beta] [\alpha \nu])} =0 , \\
& \chi_{\alpha [\alpha \beta]} =0, \quad \chi_{([\alpha \beta] [\alpha \nu])} =0  .
\end{split}
\label{eq648}
\end{align}
In other words, the tensor quantities appearing in the system \eqref{eq646} are assigned to the following Lorentz group irreducible representations
\begin{align}
\begin{split}
& \varphi_{[\alpha \beta]} \sim (0,1), \quad \varphi_{\mu [\alpha \beta]} \sim (\frac12,\frac32), \quad \varphi_{([\mu \nu] [\alpha \beta])} \sim (0,2), \\
& \chi_{[\alpha \beta]} \sim (1,0), \quad \chi_{\mu [\alpha \beta]} \sim (\frac32,\frac12), \quad \chi_{([\mu \nu] [\alpha \beta])} \sim (2,0).
\end{split}
\label{eq649}
\end{align}

The $48$-component tensor system, corresponding to the RWE with the linking scheme  \eqref{eq631}, has the form [47]
\begin{align}
\begin{split}
& \partial_{\nu} \varphi_{[\alpha \nu][\alpha \beta]} + \partial_{\alpha} \varphi_{[\alpha \beta]} + m \varphi_{\beta} =0 , \\
& \partial_{\lambda} \varphi_{\lambda [\alpha \beta]} + \frac13 (\partial_{\alpha} \varphi_{\beta} - \partial_{\beta}  \varphi_{\alpha} - i \varepsilon_{\alpha \beta \lambda \rho} \partial_{\lambda} \varphi_{\rho} ) +m \varphi_{[\alpha \beta] } =0, \\
& \partial_{\nu} \varphi_{[\mu \nu] [\alpha \beta]} + \partial_{\mu} \varphi_{[\alpha \beta]} - \frac13 (\delta_{\mu \alpha} \partial_{\eta} \varphi_{[\eta \beta] } - \delta_{\mu \beta} \partial_{\eta} \varphi_{[\eta \alpha] } + \delta_{\mu \alpha} \partial_{\nu} \varphi_{[\eta \nu] [\eta \beta] } - \delta_{\mu \beta} \partial_{\nu} \varphi_{[\eta \nu ] [\eta \alpha] }  +  \\
& + i \varepsilon_{\mu \alpha \beta \rho} \partial_{\eta} \varphi_{[\eta \rho]} + i \varepsilon_{\mu \alpha \beta \rho}  \partial_{\nu} \varphi_{[\eta \nu ] [\eta \rho ]}) + m \varphi_{\mu [\alpha \beta]} = 0, \\
-& \partial_{\eta} \varphi_{\nu [\eta \beta]} + \frac13 (2 \partial_{\nu} \varphi_{\beta} +2 \partial_{\beta}  \varphi_{\nu} -  \delta_{\nu \beta} \partial_{\rho} \varphi_{\rho} )
- i \varepsilon_{\rho \nu \eta \xi} \partial_{\eta} \varphi_{\xi [\rho \beta]}  +m \varphi_{[\eta \nu] [\eta \beta] } =0, \\
& \partial_{\nu} \chi_{[\alpha \nu][\alpha \beta]} + \partial_{\alpha} \chi_{[\alpha \beta]} + m \chi_{\beta} =0 , \\
& \partial_{\lambda} \chi_{\lambda [\alpha \beta]} + \frac13 (\partial_{\alpha} \chi_{\beta} - \partial_{\beta}  \chi_{\alpha} + i \varepsilon_{\alpha \beta \lambda \rho} \partial_{\lambda} \chi_{\rho} ) +m \chi_{[\alpha \beta] } =0, \\
& \partial_{\nu} \chi_{[\mu \nu] [\alpha \beta]} + \partial_{\mu} \chi_{[\alpha \beta]} - \frac13 (\delta_{\mu \alpha} \partial_{\eta} \chi_{[\eta \beta] } - \delta_{\mu \beta} \partial_{\eta} \chi_{[\eta \alpha] } + \delta_{\mu \alpha} \partial_{\nu} \chi_{[\eta \nu] [\eta \beta] } - \delta_{\mu \beta} \partial_{\nu} \chi_{[\eta \nu ] [\eta \alpha] }  + \\
& - i \varepsilon_{\mu \alpha \beta \rho} \partial_{\eta} \chi_{[\eta \rho]} - i \varepsilon_{\mu \alpha \beta \rho}  \partial_{\nu} \chi_{[\eta \nu ] [\eta \rho ]}) + m \chi_{\mu [\alpha \beta]} = 0, \\
-& \partial_{\eta} \chi_{\nu [\eta \beta]} + \frac13 (2 \partial_{\nu} \chi_{\beta} +2 \partial_{\beta}  \chi_{\nu} -  \delta_{\nu \beta} \partial_{\rho} \chi_{\rho} )
+ i \varepsilon_{\rho \nu \eta \xi} \partial_{\eta} \chi_{\xi [\rho \beta]}  +m \chi_{[\eta \nu] [\eta \beta] } =0.
\end{split}
\label{eq650}
\end{align}
Here the tensor quantities are assigned to the representations
\begin{align}
\begin{split}
\varphi_{\beta} & \sim (\frac12 , \frac12), \quad  \varphi_{[ \alpha \beta]} \sim  (0,1), \quad  \varphi_{\mu [ \alpha \beta]} \sim  (\frac12 ,\frac32), \quad  \varphi_{[\mu \nu] [ \alpha \beta]} \sim  (1,1), \\
\chi_{\beta} & \sim (\frac12 , \frac12)', \quad \chi_{[ \alpha \beta]} \sim  (1,0), \quad  \chi_{\mu [ \alpha \beta]} \sim  (\frac32 ,\frac12), \quad  \chi_{[\mu \nu] [ \alpha \beta]} \sim  (1,1)'.
\end{split}
\label{eq651}
\end{align}
Moreover, the tensors $\varphi_{([\mu \nu] [\alpha \beta])}$ and $\chi_{([\mu \nu] [\alpha \beta])}$ satisfy the self-duality conditions
\begin{align}
\begin{split}
& \frac12 \varepsilon_{\eta \xi \alpha \beta} \varphi_{[\mu \nu] [\alpha \beta ]} = i \varphi_{[\mu \nu] [\eta \xi]} , \quad \frac12 \varepsilon_{\eta \xi \mu \nu} \varphi_{[\mu \nu] [\alpha \beta ]} = - i \varphi_{[\eta \xi] [\alpha \beta]}  ,  \\
& \frac12 \varepsilon_{\eta \xi \alpha \beta} \chi_{[\mu \nu] [\alpha \beta ]} = - i \chi_{[\mu \nu] [\eta \xi]} , \quad \frac12 \varepsilon_{\eta \xi \mu \nu} \chi_{[\mu \nu] [\alpha \beta ]} = i \chi_{[\eta \xi] [\alpha \beta]} .
\end{split}
 \label{eq652}
\end{align}

The internal symmetry groups $SU(4,4)$  and  $SU(6,6)$ are inherent in the considered algebraic $32$-component and $48$-component generalizations of the DK equation, respectively. Like in the DK equation case, transformations of the internal symmetry $Q$ do not commute here with the Lorentz transformations $\Lambda$. But at the same time the full invariance group  $G$ can be represented as a direct product $Q = \Lambda' \otimes G$, where $\Lambda'$ corresponds to the Lorentz transformations characterizing the sets of $8$ and $12$ Dirac fields, respectively. Moreover, both systems admit a physically consistent quantization by the Fermi--Dirac statistics. Here we omit their quantization procedure, since it is largely analogous to the quantization of the DK equation considered in the Chapter \ref{TensorRWEDirac}. A comprehensive discussion of this question can be found in the papers [48, 49].

The enumerated properties of the $32$- and $48$-component tensor systems imply -- by the same reasons as in the DK equation case -- a possibility of their usage for a spatiotemporal description of the internal degrees of fredom of fermions. The first of them can serve, e.g., as a quark model with eight flavours, while the second one can serve for a geometrized introducing of the $SU(3)$ gauge interaction in the lattice space [50].

A further generalization of the discussed geometrized way of introducing internal quantum numbers is possible when lifting the constraint associated with dimensionality of the state space, which was discussed in the beginning of the current Chapter. Let us clarify this statement.

The DK equation can be represented in the form
\begin{align}
(\gamma_{\mu} \partial_{\mu} +m) \Psi_{\alpha}^D=0,
\label{eq653}
\end{align}
where $\Psi^D$ is a Dirac bispinor, $\alpha$ is a free index, corresponding to the charge-conjugated bispinor $\bar{\Psi}^c = C (\Psi^D)^*$, and $C$ is a matrix of the charge conjugation. The form of writing \eqref{eq653} means a transition to the basis (we have called it fermionic), in which the representation  \eqref{eq515} is interpreted as a direct product
\begin{align}
[(0, \frac12) \oplus (\frac12, 0)] \otimes [(0, \frac12) \oplus (\frac12, 0)] .
\label{eq654}
\end{align}
By lifting the above mentioned constraint the DK equation allows for generalizations, which consist in considering all possible products
\begin{align}
[(0, \frac12) \oplus (\frac12, 0)] \otimes [(j_1 , j_2) \oplus (j_2 , j_1)]
\label{eq655}
\end{align}
instead of  \eqref{eq654}, under the condition that the sum $(j_1 + j_2)$ takes half-integer values.

Let us dwell on two classes of RWEs, which are most promising from our interest's viewpoint and including the considered  $32$- and $48$-component as particular cases.

Choosing in \eqref{eq655} $j_1 = 0$ (or $j_2 = 0$, which is an equivalent choice), we obtain the representation
\begin{align}
[(0, \frac12) \oplus (\frac12, 0)] \otimes [(0,j) \oplus (j,0)]  \quad (j=j_2).
\label{eq656}
\end{align}
For $j = \frac12$ \eqref{eq656} coincides with \eqref{eq654}, and for $j = \frac32$ it leads to the linking scheme \eqref{eq618} and hence to the $32$-component RWE with the tensor formulation \eqref{eq646}. The cases $j =\frac52, \frac72, \ldots$ yield the linking schemes
\begin{align}
\begin{tabular}{ccccccc}
& $(0,2)$ &  & &  & $(2,0)$  & \\
& $|$ & & $\oplus$ & &$|$  & \\
$(0,3) \quad $ --- & $(\frac12, \frac52)$ &  & &  & $(\frac52,\frac12)$ & --- $\quad (3,0)$,
\end{tabular}
\label{eq657}
\end{align}
\begin{align}
\begin{tabular}{ccccccc}
& $(0,2)$ &  & &  & $(2,0)$  & \\
& $|$ & & $\oplus$ & &$|$  & \\
$(0,4) \quad $ --- & $(\frac12, \frac72)$ &  & &  & $(\frac72,\frac12)$ & --- $\quad (4,0)$,
\end{tabular}
\label{eq658}
\end{align}
and so on. A methodology of constructing Dirac-like RWEs on the basis of  \eqref{eq657}, \eqref{eq658} is analogous to the one applied while analyzing the linking scheme \eqref{eq618}. There are some differences associated with dissimilar spin structures of the corresponding equations~-- they are, however, inessential for the principles of the construction procedure.  While the scheme \eqref{eq618} leads to the matrix $\Gamma_4$ with the spin blocks $C^{1}$, $C^{2}$, the scheme \eqref{eq657} yields the blocks $C^{2}$, $C^{3}$, and the scheme \eqref{eq658} yields the blocks $C^{3}$, $C^{4}$, and so on. When assigning Dirac particles with internal degrees of freedom to these RWEs, the corresponding combined quantum number takes $8, 12, 16, \ldots$ values.

Our second choice the parameters is $ | j_1 - j_2 | = \frac12$. The class of RWEs emerging in this case is based on reducible representations which are direct products of the form
\begin{align}
& [(0, \frac12) \oplus (\frac12, 0)] \otimes [(\frac12,1) \oplus (1,\frac12)] ,
\label{eq659} \\
& [(0, \frac12) \oplus (\frac12, 0)] \otimes [(1,\frac32) \oplus (\frac32,1)] ,
\label{eq660} \\
& [(0, \frac12) \oplus (\frac12, 0)] \otimes [(\frac32,2) \oplus (2,\frac32)] ,
\label{eq661}
\end{align}
and so on. To the product \eqref{eq659} correspond the linking scheme \eqref{eq617} (or \eqref{eq631}) and the $48$-component RWE of the Dirac type, whose matrix and tensor formulations were given above.  To the representations \eqref{eq660}, \eqref{eq661} correspond the linking schemes
\begin{align}
\begin{tabular}{ccccc}
$(\frac12,\frac32)$ &--- & $2 (1,1)$ &--- & $(\frac32, \frac12)$ \\
$|$ & & $|$ & & $|$\\
$(1,2)$ & --- & $2 (\frac32,\frac32)$ & --- & $(2,1)$,
\end{tabular}
\label{eq662}
\end{align}
\begin{align}
\begin{tabular}{ccccc}
$(1,2)$ &--- & $2 (\frac32,\frac32)$ &--- & $(2,1)$ \\
$|$ & & $|$ & & $|$\\
$(\frac32, \frac52)$ & --- & $2 (2,2)$ & --- & $(\frac52, \frac32)$,
\end{tabular}
\label{eq663}
\end{align}
having the structure similar to that of \eqref{eq617}.

A possibility of constructing $P$-invariant RWEs of the Dirac type on the basis of the linking schemes belonging to this class follows from the fact that to each of them we can assign the Dirac equation for a bispinor with a free index, corresponding to the representation $[(j_1, j_2) \oplus (j_2, j_1)]$. Along with that, the scheme \eqref{eq662} describes the spins $s= 0, 1, 2, 3$, and the scheme \eqref{eq663} describes the spins $s=0, 1, 2, 3, 4$, and so on. By assigning (in the above mentioned sense) particles with spin $s=\frac12$ and internal degrees of freedom to these equations, the combined  internal quantum number takes $12, 24, 40, \ldots$ values.

Obviously, the considered generalizations of the DK equation  provide rather broad possibilities for a geometrized description of internal (besides spin) degrees of freedom of Dirac particles.

\section{Joint description of massless fields with different spi\-ra\-li\-ti\-es}
\hspace{0.5cm}
\label{mass_spir}

Let us now discuss which possibilities opens the usage of multiple Lorentz group repre\-sen\-tations in the RWE theory from the viewpoint of a description of massless fields.

Let us first analyze the simplest scheme  \eqref{ls3} concerning a construction of massless RWEs on its basis. To this scheme corresponds the following most general form of a relativistically invariant tensorial system of linear first-order differential equations
\begin{align}
\partial_{\nu} \psi_{[\mu \nu]} + a \psi_{\mu} =0,
\label{eq71} \\
- \partial_{\mu} \psi_{\nu}+  \partial_{\nu} \psi_{\mu}+  b  \psi_{[\mu \nu]} =0,
\label{eq72}
\end{align}
where $a$, $b$ are arbitrary constant coefficient. There exist four essentially different possibilities to choose these coefficients.

The first one implies $a = b = m$ and leads to the Duffin--Kemmer system (see equation \eqref{proca1}) for a microparticle with nonzero mass and spin $s = 1$; this case is currently of no interest for us. The second possible choice $a = b = 0$ leads to physically meaningless independent equations
\begin{align}
\partial_{\nu} \psi_{[\mu \nu]} =0, \quad - \partial_{\mu} \psi_{\nu}+  \partial_{\nu} \psi_{\mu} =0.
\label{eq73}
\end{align}
Making the third choice $a = 0$, $b = 1$ in  \eqref{eq71}, \eqref{eq72}, we obtain the system
\begin{align}
\partial_{\nu} \psi_{[\mu \nu]} =0,
\label{eq74} \\
- \partial_{\mu} \psi_{\nu}+  \partial_{\nu} \psi_{\mu}+   \psi_{[\mu \nu]} =0.
\label{eq75}
\end{align}
If we interpret here the components of $\psi_{\mu}$ as potentials, while interpreting $ \psi_{[\mu \nu]}$ as field's strengths, then the equations \eqref{eq74}, \eqref{eq75} represent themselves the Maxwell's system of equations (in the so called ten-dimensional formulation), describing a photon -- a massless particle with the spirality  $\pm 1$. Moreover, the equation \eqref{eq74} plays the role of an equation of motion, while the equation \eqref{eq75} appears to be a definition of a field's strength in terms of potentials.

Finally, it is also possible to make the forth choice $a = 1$, $b = 0$, which leads to the system
\begin{align}
\partial_{\nu} \psi_{[\mu \nu]} +  \psi_{\mu} =0,
\label{eq76} \\
- \partial_{\mu} \psi_{\nu}+  \partial_{\nu} \psi_{\mu} =0.
\label{eq77}
\end{align}
If we again interpret $\psi_{\mu}$ as potentials and $\psi_{[\mu \nu]}$ as field's strengths, then the system \eqref{eq76}, \eqref{eq77} becomes ill-defined in the following sense: it is impossible to express field's strengths in terms of potentialsHowever, this situation essentially changes, if we stick to another interpretation of variables entering this system, namely if we treat the tensor  $\psi_{[\mu \nu]}$ as a potential and the vector $\psi_{\mu}$ as a field's strength. Then the system \eqref{eq76}, \eqref{eq77} becomes well-defined: the equation \eqref{eq76} serves to define the field's strength via the potential, while the equation \eqref{eq77} serves as the equation of motion.

A physical meaning of the ystem \eqref{eq76}, \eqref{eq77} follows from subsequent considerations. From the equation \eqref{eq76} we have
\begin{align}
\partial_{\mu} \psi_{\mu}= 0 .
\label{eq78}
\end{align}
Taking into account \eqref{eq78}, we straightforwardly derive from the equation of motion \eqref{eq77} the second-order equation
\begin{align}
\square \psi_{\mu}= 0,
\label{eq79}
\end{align}
which points to the lack of mass in a microobject described by the system \eqref{eq76}, \eqref{eq77}.

As known from the theory of a massless vector field based on the equations  \eqref{eq74}, \eqref{eq75}, one can introduce potential's transformations
\begin{align}
\psi_{\mu} \to \psi'_{\mu} =  \psi_{\mu}+ \partial_{\mu} \Lambda (x),
\label{eq710}
\end{align}
which are called gradient transformations, or gauge transformations of the second kind. An arbitrariness in the choice of a gauge function $\Lambda (x) $ allows us to exclude "redundant" states, leaving only two (out of four) transverse components. In turn, the equations \eqref{eq76}--\eqref{eq79} are invariant under the potential's transformations
\begin{align}
\psi_{[\mu \nu]} \to \psi'_{[\mu \nu]} =  \psi_{[\mu \nu]}+ \partial_{\mu} \Lambda_{\nu} -   \partial_{\nu} \Lambda_{\mu},
\label{eq711}
\end{align}
where the gauge functions  $\Lambda_{\mu} (x)$ are constrained by the condition
\begin{align}
\square \Lambda_{\mu} - \partial_{\mu} \partial_{\nu} \Lambda_{\nu} =0.
\label{eq712}
\end{align}
In the paper of Ogievetski and Polubarinov [51] it has been shown that the gauge invariance of such type leaves a single independent component in a tensor-potential, which corresponds to a state with zero spirality.

Let us dwell on the details of this paper. To this end we turn back to the linking scheme \eqref{ls3}, in which we treat the representation $(\frac12, \frac12)$ as a pseudovector one. At first, in this case it is possible to construct a theory of a pseudovector particle with zero mass
\begin{align}
 \frac12 \varepsilon_{\mu \nu \alpha \beta} \partial_{\nu} \psi_{[\alpha \beta]} =0,
\label{eq713} \\
\varepsilon_{\mu \nu \alpha \beta}  \partial_{\alpha} \tilde{\psi}_{\beta} + \psi_{[\mu \nu]} =0
\label{eq714}
\end{align}
(the so called electrodynamics with a pseudovector potential). Introducing  an antisymmetric third-rank tensor $ \psi_{[\mu \nu \alpha]}$ which is conjugate to the pseudovector $\tilde{\psi}_{\mu}$, we obtain instead of \eqref{eq713}, \eqref{eq714} the following system
\begin{align}
\partial_{\mu} \psi_{[\nu \alpha]}+\partial_{\alpha} \psi_{[\mu \nu]}+\partial_{\nu} \psi_{[\alpha \mu]} =0,
\label{eq715} \\
\partial_{\mu}  \psi_{[ \mu\nu \alpha]}+ \psi_{[\nu \alpha]} =0,
\label{eq716}
\end{align}
in which $ \psi_{[\mu \nu \alpha]}$ plays the role of a potential.

At second, it is possible to obtain a system of equations
\begin{align}
 \frac12 \varepsilon_{\mu \nu \alpha \beta} \partial_{\nu} \psi_{[\alpha \beta]} + \tilde{\psi}_{\mu} =0,
\label{eq717} \\
\varepsilon_{\mu \nu \alpha \beta} \partial_{\alpha} \tilde{\psi}_{\beta} =0,
\label{eq718}
\end{align}
or its equivalent
\begin{align}
\partial_{\mu} \psi_{[\nu \alpha]}+\partial_{\alpha} \psi_{[\mu \nu]}+\partial_{\nu} \psi_{[\alpha \mu]}+ \psi_{[\mu \nu \alpha]} =0,
\label{eq719} \\
\partial_{\mu}  \psi_{[\mu \nu \alpha]} =0.
\label{eq720}
\end{align}

Treating here $\psi_{[\mu \nu]}$ as a tensor-potential and $\psi_{[\mu \nu \alpha]}$ as a field's strength, we arrive at the Ogievetski-Polubarinov theory for a massless particle with zero spirality. Indeed, in the work [51] the following second-order equation for the tensor-potential $\psi_{[\mu \nu]}$ is postulated from the outset
\begin{align}
\square \psi_{[\mu \nu]} + \partial_{\mu} \partial_{\alpha}\psi_{[\nu \alpha]} -  \partial_{\nu} \partial_{\alpha}\psi_{[\mu \alpha]} =0.
\label{eq721}
\end{align}
It is easy to check that it agrees with the first-order system \eqref{eq719}, \eqref{eq720}. Moreover, the equation \eqref{eq721} is invariant under the gauge transformations \eqref{eq711}, \eqref{eq712}. This allows us to impose on the potentials $\psi_{[\mu \nu]}$ an additional condition
\begin{align}
\partial_{\nu} \psi_{[\mu \nu]} =0 ,
\label{eq722}
\end{align}
which is equivalent to the condition
\begin{align}
\partial_{\mu} \partial_{\alpha} \psi_{[\nu \alpha]} -\partial_{\nu} \partial_{\alpha} \psi_{[\mu \alpha]} =0 .
\notag
\end{align}
As a result, the equation \eqref{eq721} falls apart into the equations
\begin{align}
\square \psi_{[\mu \nu]} =0
\label{eq723}
\end{align}
and \eqref{eq722}.

What concerns the system \eqref{eq76}, \eqref{eq77}, the equation \eqref{eq722} can be directly derived from it. Thus, in the both variants -- \eqref{eq76}, \eqref{eq77} and \eqref{eq719}, \eqref{eq720} -- of the theory of a massless particle with zero spirality one obtains the same second-order equations for potentials. The difference between these theories consists in the interpretation of the variables: while in the system \eqref{eq76}, \eqref{eq77} the field's strength is a true vector, in the system \eqref{eq719}, \eqref{eq720} it is given by an antisymmetric third-rank tensor (or a pseudovector). Moreover, while for the system \eqref{eq76}, \eqref{eq77} the second-order equation \eqref{eq722} is the main one and the equation \eqref{eq723} is the additional condition, for the system \eqref{eq719}, \eqref{eq720} they exchange their roles, and \eqref{eq723} appears to be the main one, and \eqref{eq722} becomes the additional condition. However, these differences do not influence the number of degrees of freedom corresponding to the both theories.

In the paper [51] a massless particle described by the system \eqref{eq719}, \eqref{eq720} was named a notoph. This name reflects a complementary character of properties of a photon and a notoph, both in the sense of spirality and in the respect of the Lorentz transformation properties of potentials and field's strengths. A notoph, described by the system  \eqref{eq76}, \eqref{eq77} can be naturally called a dual notoph.

It is now worth discussing a question about the so called spin jumping. In several papers (see, e.g., [52, 53]) a particle described by the system \eqref{eq76}, \eqref{eq77} is interpreted as a scalar massless meson. From this follows a conclusion about a change (jumping) of a spin by proceeding from the system  \eqref{proca1} to the system \eqref{eq76}, \eqref{eq77}. However, an analysis of the matrix formulation \eqref{eq:zero_mass} (or \eqref{eq:nonzero_mass} in the case of nonzero mass) of the above discussed tensor systems, which are based on the linking scheme \eqref{ls3}, shows that their only essential difference consists in the form of the matrix  $\Gamma_0$. For a particle with nonzero mass we have $\Gamma_0 = m I$. For a photon and a notoph in the tensor basis we have
\begin{align}
\Gamma_0 = \left( \begin{tabular}{cc}
$0_4$ & \\
& $I_6$
\end{tabular} \right),
\label{eq724} \\
\Gamma_0 = \left( \begin{tabular}{cc}
$I_4$ & \\
& $0_6$
\end{tabular} \right),
\label{eq725}
\end{align}
respectively. Along with this the matrix $\Gamma_0$ \eqref{eq724} cuts out one of the spin-$1$ $S_Z$-projections from the wavefunctions, leaving for a photon the two projections $s_z = \pm1$, while the matrix $\Gamma_0$ \eqref{eq725}, projecting out two projections, leaves for notoph the only one $s_z = 0$. And since the spin block $C^{0}$ in all cases remains equal to zero (see equation \eqref{eq212}), it is obvious that nonzero degrees of freedom of a photon and a notoph are associated with the block  $C^{1}$. Therefore, in fact we have a conversion of degrees of freedom (states) of a massive vector particles with spin projections  $s_z = \pm 1$ into degrees of freedom (states) of a photon with spirality  $\pm 1$, as well as a conversion of the state of massive vector particle with spin projection   $s_z = 0$ into the corresponding notoph's state with spirality $0$. By a reverse transition, e.g., from a real photon and notoph to their virtual analogs possessing finite mass, a virtual photon acquires an additional state with zero spin projection, while a virtual notoph acquires additional states with spin projections $\pm 1$. In other words, a notoph like a photon transmits in interactions the spin $1$. Therefore it is more precise, in our opinion, to consider a notoph as a massless {\it vector} particle with zero spirality. This also agrees with the point of view of the authors in the paper [51].

Summarizing the analysis carried out above, we make a conclusion that the first-order RWE theory of the form \eqref{eq:zero_mass} allows us to describe massless particles (fields) not only with a maximal spirality $\pm s$ for a given set of Lorentz group representations, but  also with intermediate spirality values including zero.

The notoph discoverers [51] did not propose any physical applications for this field. In 1974 Kalb and Ramond essentially re-discovered [52] the notoph whiel considering the question about a phenomenological description of interactions between strings. Later on, for the field system corresponding to the equations  \eqref{eq719}, \eqref{eq720} it was adopted in the literature the name of the Kalb-Ramond field (see, e.g., [53, 54]).

In the paper [51] the tensor $\psi_{[\mu \nu]}$ is proposed as a potential for a field which is a carrier of interactions between closed strings in the space of dimension $d = 4$. For a description of interactions between open strings the Kalb-Ramond field (the notoph of Ogievetski and Polubarinov) is insuffiecient. Modelling endpoints of a string by point-like electric charges, it appears necessary to introduce a vector-potential corresponding to the electromagnetic field.
A since a string is a single  physical entity, it is natural to pose a question about a joint description of the photon and the notoph on a basis of the same first-order system of equations which does not fall apart in the sense of the Lorentz group transformations.

To this end we consider the linking scheme \eqref{eq41}, in which the representation $(\frac12, \frac12)$ is assigned to a true vector, while the representation $(\frac12, \frac12)'$ is assigned to a pseudovector (or a third-rank antisymmetric tensor). The most general first-order system of equations, corresponding to the scheme \eqref{eq41}and satisfying the standard physical requirements has the form
\begin{align}
\begin{split}
& \alpha \partial_{\nu} \psi_{[\mu \nu]} + a \psi_{\mu} =0,  \\
& \beta \partial_{\nu} \tilde{\psi}_{[\mu \nu]} + b \tilde{\psi}_{\mu} =0, \\
& \alpha^* (- \partial_{\mu} \psi_{\nu} +  \partial_{\nu} \psi_{\mu} ) + \beta^* \varepsilon_{\mu \nu \alpha \beta} \partial_{\alpha} \tilde{\psi}_{\beta} + c \psi_{[\mu \nu]} =0,
\end{split}
\label{eq726}
\end{align}
where $\alpha$, $\beta$, $a$, $b$, $c$ are arbitrary parameters. The system \eqref{eq726} can be rewritten as follows
\begin{align}
\begin{split}
& \alpha \partial_{\nu} \psi_{[\mu \nu]} + a \psi_{\mu} =0,  \\
& \beta (\partial_{\mu} \psi_{[\nu \alpha]} + \partial_{\alpha} \psi_{[\mu \nu ]}+ \partial_{\nu} \psi_{[\alpha \mu]} ) + b \psi_{[\mu \nu \alpha]} =0, \\
& \alpha^* (-\partial_{\nu} \psi_{\alpha} + \partial_{\alpha} \psi_{\nu} )  + \beta^* \partial_{\mu} \psi_{[\mu \nu \alpha]} + c \psi_{[\nu \alpha]} =0.
\end{split}
\label{eq727}
\end{align}

Choosing in \eqref{eq727}
\begin{align}
\alpha = \beta =1, \quad a=c=0, \quad b=1,
\label{eq728}
\end{align}
we obtain the system
\begin{align}
\partial_{\nu} \psi_{[\mu \nu]} =0,
\label{eq729} \\
\partial_{\mu} \psi_{[\nu \alpha]} + \partial_{\alpha} \psi_{[\mu \nu ]} + \partial_{\nu} \psi_{[\alpha \mu]} + \psi_{[\mu \nu \alpha]} =0,
\label{eq730} \\
- \partial_{\nu} \psi_{\alpha} + \partial_{\alpha} \psi_{\nu} + \partial_{\mu} \psi_{[\mu \nu \alpha]} =0.
\label{eq731}
\end{align}

Let us adopt the following interpretation of quantities entering \eqref{eq729}--\eqref{eq731}: $ \psi_\mu$ and $\psi_{[\mu \nu ]}$ are interpreted as potentials, while $\psi_{[\mu \nu \alpha]}$ as a field's strength. Then the equation \eqref{eq730} essentially appears to be a definition of the field's strengths via the potentials. The equation \eqref{eq729} plays the role of the additional condition on the potentials $\psi_{[\mu \nu ]}$, which is contained in the system from the outset. This condition leaves in the ptential only two independent components. Moreover, the system \eqref{eq729}--\eqref{eq731} is invariant under the gauge transformations \eqref{eq711}, \eqref{eq712}. The existing ambiguity in the choice of a gauge function allows us to impose a condition excluding one more independent component associated with the tensor-potential  $\psi_{[\mu \nu ]}$. Along with this, for $\psi_{[\mu \nu ]}$ we have the second-order equation
\begin{align}
\square \psi_{[\mu \nu]} + \partial_{\mu} \psi_{\nu} - \partial_{\nu} \psi_{\mu} =0,
\label{eq732}
\end{align}
which describes some massless field with spirality $0$.

Let us now turn to $\psi_\mu$. The system \eqref{eq729}--\eqref{eq731} is also invariant under the gauge transformations
\begin{align}
\psi_{\mu} \to \psi'_{\mu} = \psi_{\mu} + \partial_{\mu} \Lambda,
\label{eq733}
\end{align}
where $\Lambda$ is an arbitrary function. From \eqref{eq731} follows the second-order equation
\begin{align}
\square \psi_{\mu} - \partial_{\mu}  \partial_{\nu}  \psi_{\nu} =0,
\label{eq734}
\end{align}
which along with the gauge invariance \eqref{eq733} means that the vector-potential $\psi_\mu$ characterizes the transverse component  with spirality $\pm1$ of the massless field in question. Then the tensor
\begin{align}
 \partial_{\mu}  \psi_{\nu} - \partial_{\nu}  \psi_{\mu} \equiv F_{[\mu \nu]}
\label{eq735}
\end{align}
can be naturally considered as the field's strength which is directly related to this transverse component. In turn, the equation \eqref{eq731} rewritten with account of the notation \eqref{eq735} in the form
\begin{align}
\partial_{\mu} \psi_{[\mu \nu \alpha]} - F_{[\nu \alpha]} =0,
\label{eq736}
\end{align}
appears to be the equation of motion in the system \eqref{eq729}--\eqref{eq731}.

Thus, the choice \eqref{eq728} of the parameters in the system \eqref{eq727} leads to the theory nondisintegrating in the Lorentz group sense, which provides a joint description of massless fields with spiralities  $0$ and $\pm 1$, that is the Kalb-Ramond field (notoph) and the electromagnetic field. The equation of motion \eqref{eq736} points to their inseparable link with each other which is similar to the link between electric and magnetic components in the Maxwell's theory. It is even more precise to speak not about a joint description of the quoted fields, but  rather about a single massless field with the three possible spirality values $0, \pm 1$.

Interpreting this field as a carrier of interactions between open strings in the space of dimension $d = 4$, we can introduce sources into the system \eqref{eq729}--\eqref{eq731}. In doing so, we take into account that in this case there exist two types of sources: the tensor current  $j_{[\mu \nu]}$, which is generated by the sting's body (body string), and the vector current $j_\mu$, generated by the string's endpoints. The latter are meanwhile considered as point-like electric charges of opposite signs. There exists the relation between the currents $j_\mu$ and $j_{[\mu \nu]}$
\begin{align}
j_{\nu} = \partial_{\mu} j_{[\mu \nu]},
\label{eq737}
\end{align}
from which it follows that the current $j_\mu$ is conserved ($ \partial_{\mu} j_{\mu} = 0$), while $j_{[\mu \nu ]}$ is in general not conserved ($ \partial_{\mu} j_{[\mu \nu ]}\neq 0$). Introducing the current $j_{[\mu \nu ]}$ into the equations of motion \eqref{eq731}, we obtain the system
\begin{align}
\partial_{\nu} \psi_{[\mu \nu]} =0,
\label{eq738} \\
\partial_{\mu} \psi_{[\nu \alpha]} + \partial_{\alpha} \psi_{[\mu \nu ]} + \partial_{\nu} \psi_{[\alpha \mu]} + \psi_{[\mu \nu \alpha]} =0,
\label{eq739} \\
- \partial_{\nu} \psi_{\alpha} + \partial_{\alpha} \psi_{\nu} + \partial_{\mu} \psi_{[\mu \nu \alpha]} =j_{[\nu \alpha]},
\label{eq740}
\end{align}
describing the unified field of an open string in the presence of sources.

In the particular cases, either of closed strings' interactions or of electrically charged particles, the components of the unified field can exist and be described separately. So, setting $j_\mu = 0$ we obtain according to \eqref{eq737}
\begin{align}
\partial_{\mu} j_{[\mu \nu]} =0.
\label{eq741}
\end{align}
As well the system \eqref{eq738}--\eqref{eq740} is transformed into the equations \eqref{eq719}, \eqref{eq720} (with the term $j_{[\nu \alpha]}$ in the right-hand side) and the additional condition \eqref{eq722}, describing the Kalb-Ramond field with the source term. In turn, taking the $\partial_{\alpha}$ derivative of the equation \eqref{eq740}  and accounting the definitions  \eqref{eq735}, \eqref{eq737}, we come to the equation
\begin{align}
\partial_{\nu} F_{[\mu \nu]} = j_{\mu}.
\label{eq742}
\end{align}
Combining the equation \eqref{eq742} with \eqref{eq735} and eliminating the quantities $\psi_{[\mu \nu]}$, $j_{[\mu \nu]}$ related to the string's body, we obtain the Maxwell's system for the electromagnetic field with the source term.

In the matrix formalism the system \eqref{eq729}--\eqref{eq731} corresponds to the RWE of the type \eqref{eq:zero_mass} with the singular matrix $\Gamma_0$, which has in the tensor basis the form
\begin{align}
\Gamma_0 = \left( \begin{tabular}{ccc}
$0_4$ & & \\
& $I_4$ & \\
 & & $0_6$
\end{tabular} \right).
\label{eq743}
\end{align}
Expressions for the spin blocks $C^{0}$, $C^{1}$ of the matrix $\Gamma_4$ in the Gel'fand--Yaglom basis are given by the formulas \eqref{eq43}, \eqref{eq417}, respectively. It follows that a massless field described by this RWE indeed carries the spin $1$, the eigenvalue  $\lambda = \pm 1$ of the spin block $C^{1}$ being double degenerate. In the conext of the above considerations such a degeneracy corresponds to a joint description of the electromagnetic field (photon) and the Kalb-Ramond field (notoph) as constituents of the single massless vector field with the three possible spirality values  $s = 0, +1, -1$. The zero mass value is thereby provided by the projective matrix  $\Gamma_0$, eliminating "redundant"  states which are inherent to a massive analog of this field.

As one can notice, such an analog is given by the particle which was discussed in the Chapter \ref{chirRWE}. Indeed, making in the massless RWE under consideration the replacement  $\Gamma_0 \to m I$,  we arrive at the RWE for a chiral particle with spin $s = 1$ and nonzero mass which was derived in that Chapter. The tensor form \eqref{eq419} of this RWE can be obtained from the system \eqref{eq726} by the choice of parameters $\alpha = \beta = 1$, $a = b = c = m$. This state of affairs sheds some light on a physical meaning of the quantum number "chirality" for particles with nonzero mass, namely: in the very same sense, in which one assigns a vector particle described by the standard Duffin--Kemmer equation to either the virtual photon or the virtual notoph, to the virtual unified field of the photon and the notoph is assigned a vector particle with nonzero mass and an additional internal degree of freedom~-- chirality. By the reverse transition $m I \to \Gamma_0$ the projective matrix  $\Gamma_0$ cuts out the redundant states, leaving for the photon and the notoph three degrees of freedom in total.

Summarizing the discussion of this point, we can say that the RWE theory of the form \eqref{eq:zero_mass} allows us not only to describe massless fields with the maximal (for a given set of the Lorentz group representations) spirality $\pm s$, but also to describe fields with intermediate values of spirality as well as to carry out a joint description of fields with various sprirality values from $+s$ to $-s$, everything being done in the framework of the RWE nondisintegrating under the Lorentz group transformations. For the purposes of the joint description it is necessary to consider an extended set of the Lorentz group representations  in the space of the wavefunction $\Psi$, going beyond the minimal set of representations required for a description of the spirality $\pm s$.

An essentially different way of the unification of massless fields with different spiralities is represented by a mechanism of  the gauge-invariant mixing, or the  $\hat{B} \wedge \hat{F}$-theory [52, 53]. This mechanism leads to an appearance of the mass in the unified field and, in principle, it can lay claim to the role of a mass generation mechanism which is alternative to the Higgs mechanism.

Let us briefly consider the essence of the $\hat{B} \wedge \hat{F}$-theory and give its matrix interpretation. As initial massless fields we choose those described by the systems of equations \eqref{eq74}, \eqref{eq75} and \eqref{eq717}, \eqref{eq718}. The first system describes the photon -- the massless vector field of spirality $\pm 1$, while the second one describes the notoph -- the massless field of spirality $0$. We rewrite the system \eqref{eq717}, \eqref{eq718} in the form
\begin{align}
- \partial_{\mu} \tilde{\varphi}_{\nu} + \partial_{\nu} \tilde{\varphi}_{\mu} =0,
\label{eq744} \\
\partial_{\nu} \tilde{\varphi}_{[\mu \nu]} + \tilde{\varphi}_{\mu} =0,
\label{eq745}
\end{align}
where we used for convenience the noation
\begin{align}
\frac12 \varepsilon_{\mu \nu \alpha \beta} \varphi_{[\alpha \beta]} = \tilde{\varphi}_{[\mu \nu]}.
\label{eq746}
\end{align}
We recall that the quantities $\psi_{\mu}$ and $\tilde{\varphi}_{[\mu \nu]}$ appear in these systems as potentials, while $\psi_{[\mu \nu]}$ and $\tilde{\varphi}_{\mu}$ as field's strengths of these fields.

Into the Lagrangian $\mathcal{L}_0$ of the system \eqref{eq74}, \eqref{eq75}, \eqref{eq744}, \eqref{eq745} (its explicit form is unimportant for the present consideration) we introduce an additional term
\begin{align}
 \mathcal{L}_{int} = m \psi_{\mu} \partial_{\nu} \tilde{\varphi}_{[\mu \nu]} ,
\label{eq747}
\end{align}
which does not violate the invariance of this system under the gauge transformations \eqref{eq733} and the transformations of the type \eqref{eq711}, \eqref{eq712} for the potential $\tilde{\varphi}_{[\mu \nu]}$. This procedure is called a gauge-invariant mixing, or a topological interaction of initial massless fields.

Varying the net Lagrangian $\mathcal{L} = \mathcal{L}_0 + \mathcal{L}_{int}$ yield the system of equations
\begin{align}
\partial_{\nu} \psi_{[\mu \nu]} + m \tilde{\varphi}_{\mu} =0,
\label{eq748} \\
- \partial_{\mu} \tilde{\varphi}_{\nu} + \partial_{\nu} \tilde{\varphi}_{\mu} + m \psi_{[\mu \nu]} =0,
\label{eq749} \\
\partial_{\nu} \tilde{\varphi}_{[\mu \nu]} +  \tilde{\varphi}_{\mu} =0,
\label{eq750} \\
- \partial_{\mu} \psi_{\nu} + \partial_{\nu} \psi_{\mu} + \psi_{[\mu \nu]} =0.
\label{eq751}
\end{align}
Bringing into consideration the quantities
\begin{align}
\tilde{G}_{\mu} = \psi_{\mu} - \frac{1}{m} \tilde{\varphi}_{\mu}, \quad G_{[\mu \nu]} =  \tilde{\varphi}_{[\mu \nu ]} - \frac{1}{m} \psi_{[\mu \nu]},
\label{eq752}
\end{align}
we eventually transform the system \eqref{eq748}--\eqref{eq751} to the form
\begin{align}
\partial_{\nu} \psi_{[\mu \nu]} + m \tilde{\varphi}_{\mu} =0,
\label{eq753} \\
- \partial_{\mu} \tilde{\varphi}_{\nu} + \partial_{\nu} \tilde{\varphi}_{\mu} + m \psi_{[\mu \nu]} =0,
\label{eq754} \\
\partial_{\nu} G_{[\mu \nu]} =0,
\label{eq755} \\
- \partial_{\mu} \tilde{G}_{\nu} +  \partial_{\nu} \tilde{G}_{\mu} =0.
\label{eq756}
\end{align}

The system \eqref{eq753}--\eqref{eq756} falls apart into Lorentz-invariant subsystems \eqref{eq753}, \eqref{eq754} and \eqref{eq755}, \eqref{eq756}. The first of them describes a vector particle with nonzero mass. The subsystem \eqref{eq755}, \eqref{eq756} does not describe any physical field, since it features the zero energy density. Its presence in the system \eqref{eq753}--\eqref{eq756} is dictated by the formal argument of preserving the gauge invariance on every stage of the theory consideration.

In the language of the matrix formalism of the RWE theory the $\hat{B} \wedge \hat{F}$-theory is interpreted in the following way. The initial tensor system \eqref{eq74}, \eqref{eq75}, \eqref{eq744}, \eqref{eq745} can be represented in the form \eqref{eq:zero_mass}, where by using of the basis
\begin{align}
\Psi = (\psi_{\mu},  \psi_{[\mu \nu ]}, \tilde{\varphi}_{[\mu \nu]}, \tilde{\varphi}_{\mu})^T
\label{eq757}
\end{align}
the matrices $\Gamma_\mu$, $\Gamma_0$ have the form
\begin{align}
\Gamma_{\mu}= \left( \begin{tabular}{cc}
$\Gamma_{\mu}^{DK}$ & \\
& $\Gamma_{\mu}^{DK}$
\end{tabular} \right), \quad \Gamma_0 =
 \left( \begin{tabular}{cccc}
$0_4$ & &   \\
& $I_6$ & & \\
& & $0_6$ & \\
& & & $I_4$
\end{tabular} \right),
\label{eq758}
\end{align}
with $\Gamma_{\mu}$ being expressed in terms of the ten-dimensional Duffin--Kemmer matrices $\Gamma_{\mu}^{DK}$.

Introdicing into the Lagrangian the topological term \eqref{eq747} leads to a modification of the matrices $\Gamma_\mu$. The substitutions \eqref{eq752} are equivalent to some unitary transformation of the basis \eqref{eq757}, after which the matrices $\Gamma_\mu$ acquires the initial form \eqref{eq758}. In turn, the matrix $\Gamma_0$ is modified at the same time as follows:
\begin{align}
\Gamma_0 \to
 \left( \begin{tabular}{cccc}
$m I_4$ & &  & \\
& $m I_6$ & & \\
& & $0_6$ & \\
& & & $0_4$
\end{tabular} \right) =
\left( \begin{tabular}{cc}
$m I_{10}$ & \\
& $0_{10}$
\end{tabular} \right).
\label{eq759}
\end{align}

Thus, we obtain the RWE \eqref{eq:zero_mass} with the matrices $\Gamma_\mu$ \eqref{eq758} and $\Gamma_0$ \eqref{eq759}, which represents itself a direct sum of the Duffin--Kemmer equation for the spin   $1$ and the massless limit  ($m \to 0$) of this equation. In its essentials, this way of a mass generation from the point of the RWE theory is reduced to a permutation of zero and unit blocks of the matrix  $\Gamma_0$. While the number (equal to three) of the degrees of freedom of the field system remains the same, a certain redistribution of them takes place: it looks as if the notoph passes its degree of freedom to the photon, which automatically leads to an appearance of a particle with nonzero mass and spin $1$. One can say that happens a sort of "annihilation" between the photon and the notoph, which is accompanied by the creation of a vector particle with nonzero mass.

\section{Massive gauge-invariant fields in the RWE theory}
\hspace{0.5cm}
As it was remarked in the previous Chapter, one of the distinct features in describing bosons with nonzero and zero masses consists in the fact that in the massless case some number of components of the wavefunction $\Psi$ are non-observable (potentials), while the rest of them are observable (field's strengths). For potentials one can define gauge transformations (here we again mean gauge transformations of the second kind) and impose additional conditons upon them, eliminating "redundant" components of the function $\Psi$. In turn, while describing particles with nonzero mass such a separation of the wavefunction components does not take place. Therefore, a notion of the gauge invariance in the above mentioned sense is usually applied to theories of massless particles (fields).

Nevertheless, there are known papers (see, e.g., [55]), which discuss in various approaches the so called massive gauge-invariant theories. In this respect arise the following questions: What is the status of these theories in the approach based on using a matrix form of RWEs? What are the distinctive features of gauge-invariant RWEs for particles with nonzero and zero masses?

Let us consider a set of irreducible Lorentz group representations
\begin{align}
(0,0) \oplus (\frac12 , \frac12) \oplus (0,1) \oplus (1,0),
\label{eq81}
\end{align}
forming the linking scheme
\begin{align}
\begin{tabular}{ccccc}
& & $(0,0)$ & & \\
& & $|$ & & \\
$(0,1)$ & --- & $(\frac12, \frac12)$ & --- & $(1,0)$.
\end{tabular}
\label{eq82}
\end{align}
To the scheme \eqref{eq82} in general corresponds a tensor system of first-order equations
\begin{align}
\alpha \partial_{\mu} \psi_{\mu} + a \psi_0 =0,
\label{eq83} \\
\beta^* \partial_{\nu} \psi_{[\mu \nu]} + \alpha^* \partial_{\mu} \psi_0 + b \psi_{\mu} =0,
\label{eq84} \\
\beta (- \partial_{\mu} \psi_{\nu}+\partial_{\nu} \psi_{\mu}) + c \psi_{[\mu \nu]} =0.
\label{eq85}
\end{align}

In the case when neither of parameters in this system is zero, it describes a microobject with spins $0,1$ and two mass values
\begin{align}
m_1 = \frac{\sqrt{ab}}{|\alpha |}, \quad m_2 = \frac{\sqrt{bc}}{|\beta |},
\label{eq86}
\end{align}
the mass $m_1$ corresponding to the spin $0$, and the mass $m_2$ to the spin $1$. If we impose on the parameters of the system \eqref{eq83}--\eqref{eq85} the condition
\begin{align}
\frac{\sqrt{a}}{|\alpha |} = \frac{\sqrt{b}}{|\beta |} ,
\label{eq87}
\end{align}
then we obtain a RWE for a microobject with spns $0, 1$ and a single mass value $m = m_1 = m_2$. At $\alpha = 0$ the system in question turns into an equation of the Duffin--Kemmer type for a particle with spin $1$ and mass $m = m_2$
\begin{align}
\beta^* \partial_{\nu} \psi_{[\mu \nu]} + b \psi_{\mu} =0,
\label{eq88} \\
\beta (- \partial_{\mu} \psi_{\nu}+\partial_{\nu} \psi_{\mu}) + c \psi_{[\mu \nu]} =0.
\label{eq89}
\end{align}
At $\beta = 0$ the system  \eqref{eq83}--\eqref{eq85} turns into an equation of the Duffin--Kemmer type for a particle with spin $0$ and mass $m = m_1$
\begin{align}
\alpha \partial_{\mu} \psi_{\mu} + a \psi_0 =0 ,
\label{eq810} \\
\alpha^* \partial_{\mu} \psi_0 + b \psi_{\mu} =0.
\label{eq811}
\end{align}

Let us now go over to the cases of our main interest. Setting $a=0$ in  \eqref{eq83}, we obtain a system
\begin{align}
\partial_{\mu} \psi_{\mu} =0,
\label{eq812} \\
\beta^* \partial_{\nu} \psi_{[\mu \nu]} + \alpha^* \partial_{\mu} \psi_0 + b \psi_{\mu} =0,
\label{eq813} \\
\beta (- \partial_{\mu} \psi_{\nu} +  \partial_{\nu} \psi_{\mu}) + c \psi_{[\mu \nu]} =0.
\label{eq814}
\end{align}
Introducing the notation
\begin{align}
\varphi_{\mu} = \psi_{\mu} + \frac{\alpha^*}{b} \partial_{\mu} \psi_0 ,
\label{eq815}
\end{align}
one can cast the system \eqref{eq812}--\eqref{eq814} to the form
\begin{align}
\beta^* \partial_{\nu} \psi_{[\mu \nu]} + b \varphi_\mu =0,
\label{eq816} \\
\beta (- \partial_{\mu} \varphi_{\nu} +  \partial_{\nu} \varphi_{\mu})+ c \psi_{[\mu \nu]} =0,
\label{eq817}
\end{align}
coinciding -- up to the notations -- with the system \eqref{eq88}, \eqref{eq89}.

The same result can be alternatively obtained using the arguments based on the gauge invariance. The system \eqref{eq812}--\eqref{eq814} is invariant under the transformations
\begin{align}
 \psi_0 \to \psi'_0 = \psi_0 - \frac{1}{\alpha^*} \Lambda, \quad  \psi_\mu \to \psi'_\mu = \psi_\mu + \frac{1}{b}  \partial_{\mu} \Lambda,
\label{eq818}
\end{align}
where a freedom in choosing the gauge function $\Lambda$ is limited by the condition
\begin{align}
\square \Lambda =0.
\label{eq819}
\end{align}
But exactly the same equation can be easily derived from the system \eqref{eq812}--\eqref{eq814} also for the function $\psi_0$. This means that the function  $\psi_0$ in the system \eqref{eq812}--\eqref{eq814} can be treated as a gauge function. Imposing an additional condition $\psi_0 = 0$ on it, we cast the system \eqref{eq812}--\eqref{eq814} to the form of the Duffin--Kemmer system \eqref{eq88}, \eqref{eq89}.

Such a formulation of the gauge-invariant  theory  was first proposed in the paper [56], and it is known in the literature as the St\"uckelberg approach to a description of a particle with spin  $1$ and nonzero mass. A detailed analysis of this approach and its advantages in comparison with the usual Duffin--Kemmer and Proca approaches given in the paper [57].

Let us now set $c=0$ in \eqref{eq83}--\eqref{eq85}, which leads to the system of equations
\begin{align}
\alpha \partial_{\mu} \psi_{\mu} + a \psi_0 =0,
\label{eq820} \\
\beta^* \partial_{\nu} \psi_{[\mu \nu]} + \alpha^* \partial_{\mu} \psi_0 + b \psi_{\mu} =0,
\label{eq821} \\
- \partial_{\mu} \psi_{\nu}+\partial_{\nu} \psi_{\mu} =0.
\label{eq822}
\end{align}
From \eqref{eq820}--\eqref{eq822} follows the second-order equation
\begin{align}
\square \psi_0 - \frac{a b}{|\alpha |^2} \psi_0 =0,
\label{eq823}
\end{align}
which means that this system describes a particle with nonzero mass and spin $s=0$.

The system \eqref{eq820}--\eqref{eq822} is invariant under the transformations
\begin{align}
\psi_{[\mu \nu]} \to \psi'_{[\mu \nu]} = \psi_{[\mu \nu]}- \frac{1}{\beta^*} \Lambda_{[\mu \nu]}, \quad \psi_{\mu} \to \psi'_{\mu} = \psi_{\mu}+ \frac{1}{b} \partial_{\nu} \Lambda_{[\mu \nu]},
\label{eq824}
\end{align}
where a freedom in choosing gauge functions $\Lambda_{[\mu \nu]} (x)$ is limited by the condition
\begin{align}
\partial_{\alpha} \partial_{\nu} \Lambda_{[\mu \nu]} -\partial_{\mu} \partial_{\nu}  \Lambda_{[\alpha \nu]} =0.
\label{eq825}
\end{align}
On the other hand, as it follows from the equations \eqref{eq821}, \eqref{eq822}, the tensor $\psi_{[\mu \nu]}$ satisfies an analogous condition:
\begin{align}
\partial_{\alpha} \partial_{\nu} \psi_{[\mu \nu]} -\partial_{\mu} \partial_{\nu}  \psi_{[\alpha \nu]} =0.
\label{eq826}
\end{align}
Speaking in other words, a freedom in choosing gauge functions $\Lambda_{[\mu \nu]}$ is sufficient to impose on  $\psi_{[\mu \nu]}$ an additional condition
\begin{align}
\partial_{\nu} \psi_{[\mu \nu]} =0 .
\label{eq827}
\end{align}
It excludes the states with spin $1$, and the system \eqref{eq820}--\eqref{eq822} is thus brought to the form  \eqref{eq810}, \eqref{eq811}.

Such a variant of the gauge-invariant theory is an analog of the St\"uckelberg approach, which is adapted for a description of a particle with nonzero mass and spin $s=0$. It was proposed for the first time in the paper [58]. We again note that   with help of the substitution
\begin{align}
\varphi_\mu = \psi_\mu + \frac{\beta^*}{b} \partial_{\nu} \psi_{[\mu \nu]}
\label{eq828}
\end{align}
the system \eqref{eq820}--\eqref{eq822} can be transformed to the form
\begin{align}
\alpha \partial_{\mu} \varphi_{\mu} + a \psi_0 = 0 , \quad \alpha^* \partial_{\mu} \psi_0 + b \varphi_{\mu} =0,
\label{eq829}
\end{align}
coinciding with \eqref{eq810}, \eqref{eq811} up to the notations.

Let us finally consider the case when $b=0$ in \eqref{eq83}--\eqref{eq85}. We obtain the system
\begin{align}
\alpha \partial_{\mu} \psi_{\mu} + a \psi_0 =0,
\label{eq830} \\
\beta^* \partial_{\nu} \psi_{[\mu \nu]} + \alpha^* \partial_{\mu} \psi_0 =0,
\label{eq831} \\
\beta (- \partial_{\mu} \psi_{\nu} + \partial_{\nu} \psi_{\mu} ) +c \psi_{[\mu \nu]} =0,
\label{eq832}
\end{align}
which is invariant under the transformation
\begin{align}
\psi_\mu \to \psi'_\mu = \psi_{\mu} + \partial_{\mu} \Lambda,
\label{eq833}
\end{align}
where the gauge function $\Lambda$ satisfies the condition \eqref{eq819}.The same condition (equation) follows from the system \eqref{eq830}, \eqref{eq831} for the scalar function $\psi_0$. This means that the function $\psi_0$ appears to be the gauge function, and without loss of generality it can be chosen equal to zero. As a result, the system   \eqref{eq830}--\eqref{eq832}  is transformed to the form
\begin{align}
\partial_{\mu} \psi_{\mu} =0,
\label{eq834} \\
\partial_{\nu} \psi_{[\mu \nu]} =0,
\label{eq835} \\
- \partial_{\mu} \psi_{\nu} + \partial_{\nu} \psi_{\mu} + \psi_{[\mu \nu]} =0.
\label{eq836}
\end{align}
The content of the system \eqref{eq834}--\eqref{eq836} coincides with that of the Maxwell's equations up to  the only inessential difference that in the Maxwell's theory the equation  \eqref{eq834} appears as an additional condition, while in the system \eqref{eq834}--\eqref{eq836} it is an independent equation. In the both cases we speak about a massless particle with spirality $\pm 1$. It is obvious that the system \eqref{eq830}--\eqref{eq832},  which is equivalent to the system \eqref{eq834}--\eqref{eq836},  has the same physical meaning.

When being written in the matrix form, the tensor systems considered above are characterized by the same form of the matrices $\Gamma_{\mu}$. The difference between these systems consists only in the form of the matrix $\Gamma_0$. In the basis, in which the wavefunction $\Psi$ has a form of the column
\begin{align}
\Psi = (\psi_0 , \psi_{\mu}, \psi_{[\mu \nu]})^T,
\label{eq837}
\end{align}
we find the following expressions for $\Gamma_0$:

1) $a=0$, the system \eqref{eq812}--\eqref{eq814}
\begin{align}
\Gamma_0 =
 \left( \begin{tabular}{ccc}
$0$ & &  \\
& $I_4$ & \\
& & $I_6$
\end{tabular} \right);
\label{eq838}
\end{align}

2) $c=0$, the system \eqref{eq820}--\eqref{eq822}
\begin{align}
\Gamma_0 =
 \left( \begin{tabular}{ccc}
$1$ & &  \\
& $I_4$ & \\
& & $0_6$
\end{tabular} \right);
\label{eq839}
\end{align}

3) $b=0$, the system \eqref{eq830}--\eqref{eq832}
\begin{align}
\Gamma_0 =
 \left( \begin{tabular}{ccc}
$1$ & &  \\
& $0_4$ & \\
& & $I_6$
\end{tabular} \right).
\label{eq840}
\end{align}

Comparing the expression \eqref{eq837}--\eqref{eq840}, one can notice that, all other things being equal,  either a massive or a massless type of the field described by the RWE  \eqref{eq:zero_mass} depends on the Lorentz structure of the term $\Gamma_0 \Psi$. In the case when this term contains a set of the Lorentz covariants which is minimally necessary for constructing an equation of the form  \eqref{eq:nonzero_mass} for a particle (field) with nonzero mass, a construction of the gauge-invariant theory for a massive field on the basis of a RWE of the form  \eqref{eq:zero_mass} is possible. In the opposite case, one can only speak about a theory of a massless field. So, the system \eqref{eq812}--\eqref{eq814}, represented in the form  \eqref{eq:zero_mass}, contains the term $\Gamma_0 \Psi$, in which the covariants $\psi_{\mu}$, $\psi_{[\mu \nu]}$ are present. On their basis one can construct a RWE of the form \eqref{eq:nonzero_mass} for a particle with nonzero mass --  the $10$-component Duffin--Kemmer equation for a spin-$1$ particle. Accordingly the  $11$-component gauge-invariant  RWE of the form \eqref{eq:zero_mass} with the wavefunction \eqref{eq837} and the matrix $\Gamma_0$ \eqref{eq838} also describes a vector particle with nonzero mass. A gauge-invariant analog of the Duffin--Kemmer equation for a spin-$0$ particle is given by the system \eqref{eq820}--\eqref{eq822}, or by the $11$-component matrix RWE  \eqref{eq:zero_mass} with the matrix $\Gamma_0$ \eqref{eq839}, in which the term $\Gamma_0 \Psi$ contains the covariants $\psi_0$, $\psi_\mu$. In the gauge-invariant RWE with the wavefunction \eqref{eq837}, which corresponds to the tensor system \eqref{eq830}--\eqref{eq832}, the term $\Gamma_0 \Psi$ contains the covariant $\psi_0$, $\psi_{[\mu \nu]}$. On the basis of the latter it is impossible to construct a first-order equation for a particle with a nonzero mass, as it follows from the basic principles of the RWE theory. Therefore, the RWE with the matrix $\Gamma_0$ \eqref{eq840} can only describe a massless field -- the fact that we explicitly established above.

\section{RWE theory and the electroweak field}
\hspace{0.5cm}

The standard RWE theory assumes only a possibility of a separate description of microobjects with nonzero and zero masses. However, in the modern high energy physics there are studied the fields, whose quanta can have either zero or nonzero mass (for the same field). A remarkable example illustrating this situation is given by the electroweak field.

If we wish to treat the free electroweak field as a single physical object (and at sufficiently high energies this is indeed the case), then it is natural to ask the question: can one jointly describe a field with nonzero and zero masses (the so called massive-massless field) in the RWE framework? We remind that by a joint description we understand  -- just as before -- a description in terms of the same system of equations which does not fall apart in the relativistically invariant sense.

Let us consider this question in detail, and turn back to the linking scheme \eqref{eq41}, where the representation $(\frac12, \frac12)$ is the vector one, while $(\frac12, \frac12)'$ is a pseudovector one. If we set in the most general tensor formulation of the relativistically invariant first-order system  \eqref{eq727}, corresponding to the scheme \eqref{eq41}, the parameters to the values
\begin{align}
a = 0, \quad b = c = m ,
\label{eq91}
\end{align}
then we obtain a system
\begin{align}
& \partial_{\nu} \psi_{[\mu \nu]} =0,
\label{eq92} \\
\beta & ( \partial_{\mu} \psi_{[\nu \alpha]}+ \partial_{\alpha} \psi_{[\mu \nu]}+ \partial_{\nu} \psi_{[\alpha \mu]}) + m \psi_{[\mu \nu \alpha]} =0,
\label{eq93} \\
-& \partial_{\mu} \psi_{\nu }+ \partial_{\nu} \psi_{\mu }+ \beta^*  \partial_{\alpha} \psi_{[\mu \nu \alpha]} + m \psi_{[\mu \nu]} =0.
\label{eq94}
\end{align}
From \eqref{eq92}--\eqref{eq94} it is easy to get second-order equations
\begin{align}
& (\square - \frac{m^2}{|\beta |^2}) \psi_{[\mu \alpha \nu]} =0 ,
\label{eq95} \\
& \varepsilon_{\mu \nu \alpha \beta} \partial_{\beta} \psi_{[\mu \nu \alpha]} =0,
\label{eq96} \\
& \square \psi_{\mu}  - \partial_{\mu}  \partial_{\nu} \psi_{\nu} =0 .
\label{eq97}
\end{align}

The equations \eqref{eq95}, \eqref{eq96} mean that the system  \eqref{eq92}--\eqref{eq94} contains a description of a pseudovector particle with nonzero mass. In turn , the equation \eqref{eq97} points to the fact that this system also describes a massless vector field with the potential $\psi_\mu$. The latter circumstance allows us to use a gauge transformation of the type \eqref{eq710}, with respect to which both the system \eqref{eq92}--\eqref{eq94} and the equation \eqref{eq97} are invariant. This invariance means that the quoted massless field is a field of the Maxwell's type with spirality $\pm 1$.

Thus, the tensor system  \eqref{eq92}--\eqref{eq94} nondisintegrating in the Lorentz sense gives a joint description of a pseudovector particle with nonzero mass and a massless vector field of the electromagnetic type.

Let us also remark that choosing in  \eqref{eq727}
\begin{align}
b = 0, \quad a = c = m,
\label{eq98}
\end{align}
leads to a theory which is dually symmetric to \eqref{eq92}--\eqref{eq94} in the sense that by the choice  \eqref{eq98} the system \eqref{eq727} leads to a joint description of a massive vector particle and a massless field of the Maxwell's type with a pseudovector potential.

Moreover, as it was remarked in the Chapter \ref{chirRWE}, on the basis of the scheme \eqref{eq41} it is possible to carry out a joint description of a microobject with two different nonzero masses. To this end,  in \eqref{eq727} it is necessary to set
\begin{align}
a = b = c = m.
\label{eq99}
\end{align}
As a result, we get the first-order system
\begin{align}
& \alpha \partial_{\nu} \psi_{[\mu \nu]} + m \psi_{\mu} =0 ,
\label{eq910} \\
& \beta (\partial_{\mu} \psi_{[ \nu \alpha ]} + \partial_{\alpha} \psi_{[ \mu \nu  ]} + \partial_{\nu} \psi_{[ \alpha \mu]}) + m \psi_{[\mu \nu \alpha]} =0 ,
\label{eq911} \\
& \alpha^* (-\partial_{\nu} \psi_{\alpha}+ \partial_{\alpha} \psi_{\nu} ) + \beta^* \partial_{\mu} \psi_{[\mu \nu \alpha]} + m \psi_{[\nu \alpha]} =0 .
\label{eq912}
\end{align}
From \eqref{eq910}--\eqref{eq912} follow the second-order equations \eqref{eq95}, \eqref{eq96} as well as the equations
\begin{align}
(\square - \frac{m^2}{|\alpha |^2}) \psi_{\mu } =0 , \quad \partial_{\mu} \psi_{\mu} =0 ,
\label{eq913}
\end{align}
and this justifies the statement made above.

Let us now consider a possibiliy of a joint desription of vector particles (fields), which are assigned to the systems  \eqref{eq92}--\eqref{eq94} and \eqref{eq910}--\eqref{eq912}. It is obvious that a simple mechanical conjunction of these systems would not lead to a desired result, since the so obtained system of equations falls apart in the relativistically invariant sense, and therefore it can not describe a single physical object from the RWE theory point of view. The corresponding linking scheme also falls apart into two independent fragments of the form \eqref{eq41}.

The most simple and natural way of constructing a nondisintegrating system of equations which would provide a joint description of vector fields quoted above consists in introducing into our treatment of an additional scalar representations $(0, 0)$.
As a result, one can obtain, e.g., the linking scheme
\begin{align}
& \hspace{7.45mm} \line(0,-1){10} \line(1,0){185} \line(0,-1){10} \notag  \\[-0.5em]
& \begin{tabular}{ccccccccccccc}
\vline & & $(0,1)$ & & & & \vline  & & &  & $(0,1)$ &  &     \\
\vline & $\diagup$  &  & $\diagdown$ &   & & \vline & & &  $\diagup$  &         & $\diagdown$  &  \\
$(\frac12,\frac12)'$ & & & & $(\frac12,\frac12)$  & --- & $(0,0)$ &  --- & $(\frac12,\frac12)'$   &  &   &  & $(\frac12,\frac12) $ . \\
 & $\diagdown$  & & $\diagup$ & & &  &   & &   $\diagdown$ &  & $\diagup$  &   \\
 & & $(1,0)$ & & & &   &   & &  & $ (1,0)$ &  &
\end{tabular}
\label{eq914}
\end{align}

A first-order system of equations corresponding to the scheme \eqref{eq914} reads in the most general form as follows:
\begin{align}
& \partial_{\nu} \psi_{[\mu \nu]} =0 ,
\label{eq915} \\
& \alpha (\partial_{\mu} \psi_{[\nu \alpha]}+ \partial_{\alpha} \psi_{[\mu \nu]}+ \partial_{\nu} \psi_{[\alpha \mu]}) + \beta \varepsilon_{\mu \nu \alpha \beta} \partial_{\beta} \psi_0 + m \psi_{[\mu \nu \alpha]} =0 ,
\label{eq916} \\
- & \partial_{\mu} \psi_{\nu} + \partial_{\nu} \psi_{\mu} + \alpha^* \partial_{\alpha} \psi_{[\mu \nu \alpha]}
+ m \psi_{[\mu \nu]} =0,
 \label{eq917} \\
 & \rho \partial_{\nu} \varphi_{[\mu \nu]} + \gamma \partial_{\mu} \psi_0 + m \varphi_{\mu} =0 ,
\label{eq918} \\
 & \delta (\partial_{\mu} \varphi_{[\nu \alpha ]} + \partial_{\alpha} \varphi_{[\mu \nu ]} +\partial_{\nu} \varphi_{[\alpha \mu]} ) + \sigma \varepsilon_{\mu \nu \alpha \beta} \partial_{\beta} \psi_0 +m \varphi_{[\mu \nu \alpha]} =0,
\label{eq919} \\
 & \rho^* (- \partial_{\mu} \varphi_{\nu} + \partial_{\nu} \varphi_{\mu}) + \delta^* \partial_{\alpha}  \varphi_{[\mu \nu \alpha]} + m  \varphi_{[\mu \nu ]} =0 ,
\label{eq920} \\
 & \frac{1}{3!} \beta^* \varepsilon_{\beta \mu \nu \alpha} \partial_{\beta} \psi_{[\mu \nu \alpha]} + \gamma^*
 \partial_{\mu} \psi_{\mu} + \frac{1}{3!} \sigma^* \varepsilon_{\beta \mu \nu \alpha} \partial_{\beta} \varphi_{[\mu \nu \alpha]} + m \psi_0 =0 .
\label{eq921}
\end{align}

Let us now establish second-order equation which follow from the system \eqref{eq915}--\eqref{eq921}.

First, we act on \eqref{eq917} with the operator $\partial_{\nu}$. Taking into account \eqref{eq915} we obtain
\begin{align}
\square \psi_{\mu} + \partial_{\mu} \partial_{\nu} \psi_{\nu} =0.
\label{eq922}
\end{align}
Using the gauge transformation \eqref{eq710}, the equation \eqref{eq922} is transformed in the standard way to the form
\begin{align}
\square \psi_{\mu} =0, \quad \partial_{\mu} \psi_{\mu} =0 .
\label{eq923}
\end{align}

Applying the operator $\partial_{\nu}$ to the equation \eqref{eq920} yields
\begin{align}
\rho^* (\square \varphi_{\mu} - \partial_{\mu} \partial_{\nu} \varphi_{\nu}) + m \partial_{\nu} \varphi_{[\mu \nu]}= 0.
\label{eq924}
\end{align}
From \eqref{eq918} we express the term $\partial_{\nu} \varphi_{[\mu \nu]}$
\begin{align}
\partial_{\nu} \varphi_{[\mu \nu]} = - \frac{\gamma}{\rho} \partial_{\mu} \psi_0  - \frac{m}{\rho}  \varphi_{\mu}.
\label{eq925}
\end{align}
It follows that
\begin{align}
 \partial_{\mu} \varphi_{\mu} = - \frac{\gamma}{m} \square \psi_0  .
\label{eq926}
\end{align}
Substituting \eqref{eq925}, \eqref{eq926}  in \eqref{eq924}, we obtain
\begin{align}
\square \varphi_{\mu} +  \frac{\gamma}{m} \partial_{\mu} \square \psi_0  - \frac{m \gamma}{|\rho |^2}  \partial_{\mu} \psi_0
- \frac{m^2}{|\rho |^2} \varphi_{\mu} =0.
\label{eq927}
\end{align}

If we bring into consideration the vector
\begin{align}
\Phi_{\mu} = \varphi_{\mu} +  \frac{\gamma}{m} \partial_{\mu} \psi_0 ,
\label{eq928}
\end{align}
then the equation \eqref{eq927} acquires the form
\begin{align}
\square \Phi_{\mu} - \frac{m^2}{|\rho |^2} \Phi_{\mu} =0.
\label{eq929}
\end{align}
In addition, with account of \eqref{eq926} the vector $\Phi_{\mu}$ satisfies the condition
\begin{align}
\partial_{\mu} \Phi_{\mu} =0.
\label{eq930}
\end{align}

Let us now apply the operator $\varepsilon_{\rho \mu \nu \sigma} \partial_{\sigma}$ to the equation \eqref{eq917}. This yields
\begin{align}
\alpha^* \varepsilon_{\rho \mu \nu \sigma} \partial_{\sigma} \partial_{\alpha} \psi_{[ \mu \nu \alpha ]} + m \varepsilon_{\rho \mu \nu \sigma} \partial_{\sigma} \psi_{[ \mu \nu ]} =0.
\label{eq931}
\end{align}
With help of the identity
\begin{align}
\varepsilon_{\rho \mu \nu \sigma} \partial_{\sigma} \partial_{\alpha} \psi_{[\mu \nu \alpha] } = \varepsilon_{\rho \mu \nu \alpha} \square \psi_{[\mu \nu \alpha] }- \partial_{\rho} \partial_{\beta} \varepsilon_{\beta \mu \nu \alpha}  \psi_{[\mu \nu \alpha] } \notag
\end{align}
which can be directly checked, the equation \eqref{eq931} is brought to the form
\begin{align}
\alpha^* ( \varepsilon_{\rho \mu \nu \alpha} \square \psi_{[\mu \nu \alpha] }- \partial_{\rho} \partial_{\beta} \varepsilon_{\beta \mu \nu \alpha}  \psi_{[\mu \nu \alpha] } ) + m \varepsilon_{\rho \mu \nu \alpha} \partial_{\alpha} \psi_{[ \mu \nu ]} =0.
\label{eq932}
\end{align}
From the equation \eqref{eq916} it follows
\begin{align}
\partial_{\mu} \psi_{[\nu \alpha]}+ \partial_{\alpha} \psi_{[\mu \nu]} + \partial_{\nu} \psi_{[ \alpha \mu]} = - \frac{\beta}{\alpha} \varepsilon_{\mu \nu \alpha  \beta} \partial_{\beta} \psi_0 - \frac{m}{\alpha} \psi_{[\mu \nu \alpha]} .
\label{eq933}
\end{align}
Applying the operator $\frac{1}{3!} \varepsilon_{\mu \nu \alpha \rho} \partial_{\rho}$ to   \eqref{eq916}, we obtain
\begin{align}
\frac{1}{3!} \varepsilon_{\mu \nu \alpha \rho} \partial_{\rho}  \psi_{[\mu \nu \alpha] } = \frac{\beta}{m} \square \psi_0 .
\label{eq934}
\end{align}
Combining the equations \eqref{eq932}--\eqref{eq934} yields
\begin{align}
\square \psi_{[\mu \nu \alpha] } - \frac{m^2}{|\alpha |^2} \psi_{[\mu \nu \alpha] }+ \frac{\beta}{m}  \varepsilon_{\mu \nu \alpha \beta} \partial_{\beta} \square \psi_0 - \frac{\beta m}{|\alpha |^2}  \varepsilon_{\mu \nu \alpha \beta} \partial_{\beta} \psi_0 =0.
\label{eq935}
\end{align}
By means of the notation
\begin{align}
 \Psi_{[\mu \nu \alpha]}  =  \psi_{[\mu \nu \alpha]} + \frac{\beta}{m} \varepsilon_{\mu \nu \alpha \beta} \partial_{\beta} \psi_0
\label{eq936}
\end{align}
we cast the equation \eqref{eq935} to the form
\begin{align}
\square \Psi_{[\mu \nu \alpha]} - \frac{m^2}{| \alpha |^2} \Psi_{[\mu \nu \alpha]} =0.
\label{eq937}
\end{align}
Along with \eqref{eq937} we have in accordance with  \eqref{eq934} the condition
\begin{align}
\varepsilon_{\mu \nu \alpha \beta} \partial_{\beta} \Psi_{[\mu \nu \alpha]} =0 .
\label{eq938}
\end{align}

Analogously one can obtain from the system \eqref{eq915}--\eqref{eq921} the second-order equation
\begin{align}
\square \Phi_{[\mu \nu \alpha]} - \frac{m^2}{| \delta |^2} \Phi_{[\mu \nu \alpha]} =0
\label{eq939}
\end{align}
with the additional condition
\begin{align}
\varepsilon_{\mu \nu \alpha \beta} \partial_{\beta} \Phi_{[\mu \nu \alpha]} =0,
\label{eq940}
\end{align}
where
\begin{align}
 \Phi_{[\mu \nu \alpha]}  =  \varphi_{[\mu \nu \alpha]} + \frac{\sigma}{m} \varepsilon_{\mu \nu \alpha \beta} \partial_{\beta} \psi_0.
\label{eq941}
\end{align}

Let us finally establish the second-order equation for the scalar $\psi_0$. To this end we apply the operator $\frac{1}{3!} \varepsilon_{\mu \nu \alpha \rho} \partial_{\rho}$ to the equation \eqref{eq919} and find
\begin{align}
\frac{1}{3!} \varepsilon_{\mu \nu \alpha \rho} \partial_{\rho} \varphi_{[\mu \nu \alpha]}  = \frac{\sigma}{m} \square \psi_0 .
\label{eq942}
\end{align}
Substituting now \eqref{eq926}, \eqref{eq934} and \eqref{eq942} into \eqref{eq921}, we obtain
\begin{align}
\square \psi_0- \frac{m^2}{|\beta |^2 + |\gamma |^2 + |\sigma |^2} \psi_0  = 0.
\label{eq943}
\end{align}

The equations \eqref{eq923}, \eqref{eq929}, \eqref{eq930}, \eqref{eq937}--\eqref{eq940} and \eqref{eq943}  show that the first-order system   \eqref{eq915}--\eqref{eq921} nondisintegrating in the Lorentz sense does contain a description of both four particles (fields) with spin $1$, one of them having zero mass, and a scalar particle with nonzero mass.

The tensor system \eqref{eq915}--\eqref{eq921} can be represented in the standard RWE matrix form  \eqref{eq:zero_mass}  with a singular matrix $\Gamma_0$ of the form
\begin{align}
\Gamma_0 = \left(\begin{tabular}{cc}
$0_4$ & \\
  & $m I_{25}$
\end{tabular} \right).
\label{eq944}
\end{align}
Let us give a formulation of this RWE in the Gel'fand--Yaglom basis.

As usual, we introduce a labelling of the irreducible representation contained in the linking scheme  \eqref{eq914}, e.g., in the following form
\begin{align}
\begin{split}
& (0,0) \sim 1 \,\, (\psi_0), \quad (\frac12, \frac12) \sim 2 \,\, (\psi_\mu), \quad (\frac12, \frac12)' \sim 3 \,\, (\psi_{[\mu \nu \alpha]}), \\
& (0,1),(1,0) \sim 4,5 \,\, (\psi_{[\mu \nu]}), \quad (\frac12, \frac12) \sim 6 \,\, (\varphi_\mu), \\
& (\frac12, \frac12)' \sim 7\,\, (\varphi_{[\mu \nu \alpha]}), \quad (0,1), (1,0) \sim 8,9 \,\, (\varphi_{[\mu \nu ]}).
\end{split}
\label{eq945}
\end{align}
Then, for the spin blocks $C^0$, $C^1$ of the matrix $\Gamma_4$ we obtain the following general expressions
\begin{align}
C^0 &= \left(\begin{tabular}{ccccc}
$0$ & $0$ & $c^0_{13}$ & $c^0_{16}$ & $c^0_{17}$  \\
$0$ & $0$ & $0$ & $0$ & $0$  \\
$c^0_{31}$ & $0$ & $0$ & $0$ & $0$  \\
$c^0_{61}$ & $0$ & $0$ & $0$ & $0$  \\
$c^0_{71}$ & $0$ & $0$ & $0$ & $0$
\end{tabular} \right), \quad
C^1 = \left(\begin{tabular}{cc}
$(C^1)'$ & \\
  & $(C^1)''$
\end{tabular} \right),
\label{eq946} \\
(C^1)' &= \left(\begin{tabular}{cccc}
$0$ & $0$ & $c^1_{24}$ & $c^1_{25}$  \\
$0$ & $0$ & $c^1_{34}$ & $c^1_{35}$  \\
$c^1_{42}$ & $c^1_{43}$ & $0$ & $0$   \\
$c^1_{52}$ & $c^1_{53}$ & $0$ & $0$
\end{tabular} \right), \quad
(C^1)'' = \left(\begin{tabular}{cccc}
$0$ & $0$ & $c^1_{68}$ & $c^1_{69}$  \\
$0$ & $0$ & $c^1_{78}$ & $c^1_{79}$  \\
$c^1_{86}$ & $c^1_{87}$ & $0$ & $0$   \\
$c^1_{96}$ & $c^1_{97}$ & $0$ & $0$
\end{tabular} \right).
\label{eq947}
\end{align}

The invariance conditions \eqref{c_cond} of the RWE under consideration with respect to transformations of the proper Lorentz group do not impose any constraints on the elements $c^0_{ij}$, $c^1_{ij}$. The condition of the  $P$-invariance of the theory is not applicable to the electroweak field. The condition \eqref{c_eta_cond} of a possible Lagrangian formulation of the theory leads to the relations
\begin{align}
\begin{split}
c^0_{31} &= f (c^0_{13})^* , \quad c^0_{61} = g (c^0_{16})^* , \quad c^0_{71} = h (c^0_{17})^*, \\
c^1_{42} &= p (c^1_{25})^* , \quad c^1_{52} = p (c^1_{24})^* , \quad c^1_{43} = q (c^1_{35})^*, \quad c^1_{53} = q (c^1_{34})^*, \\
c^1_{86} &= r (c^1_{69})^* , \quad c^1_{96} = r (c^1_{68})^* , \quad c^1_{87} = s (c^1_{79})^*, \quad c^1_{97} = s (c^1_{78})^*, \\
\end{split}
\label{eq948}
\end{align}
where
\begin{align}
\begin{split}
f &=\frac{\eta^0_{33}}{\eta^0_{11}}, \quad g = \frac{\eta^0_{66}}{\eta^0_{11}}, \quad h=\frac{\eta^0_{77}}{\eta^0_{11}} ,\\
p &= \frac{\eta^1_{45}}{\eta^1_{22}} , \quad q = \frac{\eta^1_{45}}{\eta^1_{33}} , \quad r= \frac{\eta^1_{89}}{\eta^1_{66}} , \quad s= \frac{\eta^1_{89}}{\eta^1_{77}}.
\end{split}
\label{eq949}
\end{align}

Introducing for convenience the notations
\begin{align}
\begin{split}
c^0_{13} &= \lambda_1 , \quad  c^0_{16} = \lambda_2 ,\quad c^0_{17} = \lambda_3,  \quad c^1_{24} = \lambda_4,  \quad c^1_{25} = \lambda_5, \\
c^1_{34} &= \lambda_6 , \quad  c^1_{35} = \lambda_7 , \quad c^1_{68} = \lambda_8,  \quad c^1_{69} = \lambda_9,  \quad c^1_{78} = \lambda_{10}, \quad c^1_{79} = \lambda_{11},
\end{split}
\label{eq950}
\end{align}
and taking into account \eqref{eq948}, \eqref{eq949}, we obtain for the blocks $C^0$ \eqref{eq946}, $(C^1)'$, $(C^1)''$ \eqref{eq947} the following expressions
\begin{align}
C^0 &= \left(\begin{tabular}{ccccc}
$0$ & $0$ & $\lambda_1$ & $\lambda_2$ & $\lambda_3$  \\
$0$ & $0$ & $0$ & $0$ & $0$  \\
$f \lambda_1^*$ & $0$ & $0$ & $0$ & $0$  \\
$g \lambda_2^*$ & $0$ & $0$ & $0$ & $0$  \\
$h \lambda_3^*$ & $0$ & $0$ & $0$ & $0$
\end{tabular} \right), \quad
\label{eq951} \\
(C^1)' &= \left(\begin{tabular}{cccc}
$0$ & $0$ & $\lambda_4$ & $\lambda_5$  \\
$0$ & $0$ & $\lambda_6$ & $\lambda_7$  \\
$p \lambda_5^*$ & $q \lambda_7^*$ & $0$ & $0$   \\
$p \lambda_4^*$ & $q \lambda_6^*$ & $0$ & $0$
\end{tabular} \right), \quad
(C^1)'' = \left(\begin{tabular}{cccc}
$0$ & $0$ & $\lambda_8$ & $\lambda_9$  \\
$0$ & $0$ & $\lambda_{10}$ & $\lambda_{11}$  \\
$r \lambda_9^*$ & $s \lambda_{11}^*$ & $0$ & $0$   \\
$r \lambda_8^*$ & $s \lambda_{10}^*$ & $0$ & $0$
\end{tabular} \right).
\label{eq952}
\end{align}

Characteristic equations for the blocks \eqref{eq951}, \eqref{eq952} read
\begin{align}
& \lambda^3 (\lambda^2 -  f |\lambda_1 |^2- g |\lambda_2 |^2- h |\lambda_3 |^2) =0,
\label{eq953} \\
\begin{split}
& \lambda^4 - \lambda^2 (p \lambda_4^* \lambda_5 + p  \lambda_4 \lambda_5^*  + q  \lambda_6^* \lambda_7  + q \lambda_6 \lambda_7^* ) + \\
& + pq (\lambda_4 \lambda_5^* \lambda_6^* \lambda_7  + \lambda_4^* \lambda_5 \lambda_6 \lambda_7^* - |\lambda_4 |^2 |\lambda_7 |^2 - |\lambda_5 |^2 |\lambda_6 |^2)=0 ,
\end{split}
\label{eq954} \\
\begin{split}
& \lambda^4 - \lambda^2 (r \lambda_8^* \lambda_9 + r  \lambda_8 \lambda_9^* + s  \lambda_{10}^* \lambda_{11}  + s \lambda_{10} \lambda_{11}^*) + \\
& + rs (\lambda_8 \lambda_9^* \lambda_{10}^* \lambda_{11}  + \lambda_8^* \lambda_9 \lambda_{10} \lambda_{11}^* - |\lambda_8 |^2 |\lambda_{11} |^2 - |\lambda_9 |^2 |\lambda_{10} |^2)=0 ,
\end{split}
\label{eq955}
\end{align}
respectively. A RWE equivalent to the tensor system \eqref{eq915}--\eqref{eq921} is obtained, when setting
\begin{align}
\begin{split}
& f |\lambda_1 |^2+ g |\lambda_2 |^2+ h |\lambda_3 |^2= |\beta |^2 + |\gamma |^2 + |\sigma |^2 , \\
& p \lambda_4^* \lambda_5 + p  \lambda_4 \lambda_5^*  + q  \lambda_6^* \lambda_7  + q \lambda_6 \lambda_7^* = |\alpha |^2 +1, \\
& p q ( \lambda_4 \lambda_5^* \lambda_6^* \lambda_7  + \lambda_4^* \lambda_5 \lambda_6 \lambda_7^* - |\lambda_4 |^2 |\lambda_7 |^2 - |\lambda_5 |^2 |\lambda_6 |^2) = |\alpha |^2, \\
& r \lambda_8^* \lambda_9 + r  \lambda_8 \lambda_9^* + s  \lambda_{10}^* \lambda_{11}  + s \lambda_{10} \lambda_{11}^* = |\rho |^2 +|\delta |^2 , \\
& r s (\lambda_8 \lambda_9^* \lambda_{10}^* \lambda_{11}  + \lambda_8^* \lambda_9 \lambda_{10} \lambda_{11}^* - |\lambda_8 |^2 |\lambda_{11} |^2 - |\lambda_9 |^2 |\lambda_{10} |^2)=  |\rho |^2 |\delta |^2 .
\end{split}
\label{eq956}
\end{align}
The relations  \eqref{eq956} are satisfied, e.g., by choosing
\begin{align}
\begin{split}
\lambda_1 &= \beta, \quad \lambda_2 = \sqrt{2 (|\gamma |^2 + |\sigma |^2)} , \quad \lambda_3 = \sqrt{|\gamma |^2 + |\sigma |^2}, \\
\lambda_4 &= - \lambda_5 = \frac{i |\alpha|}{\sqrt{2}}, \quad \lambda_6 =  \lambda_7 = \frac{1}{\sqrt{2}}, \quad  \lambda_8 =  -\lambda_9 = \frac{i |\rho|}{\sqrt{2}}, \quad  \lambda_{10} =  \lambda_{11} = \frac{|\delta|}{\sqrt{2}},
\end{split}
\label{eq957} \\
f &= g = -h = -p = q =-r=s=1.
\label{eq958}
\end{align}
In turn, the equalities \eqref{eq958} lead to the following values of the matrix elements $\eta_{\tau \tau'}^s$ of the Lorentz-invariant bilinear form $\eta$:
\begin{align}
\eta^0_{11} = \eta^0_{33} =\eta^0_{66} =-\eta^0_{77} =\eta^1_{22} = -\eta^1_{33} = -\eta^1_{45} = - \eta^1_{66} = \eta^1_{77} = \eta^1_{89} = 1.
\label{eq959}
\end{align}

The spin blocks $C^0$ \eqref{eq951}, $(C^1)'$, $(C^1)''$ \eqref{eq952} in accordance with \eqref{eq957}, \eqref{eq958} acquire the form
\begin{align}
C^0 &= \left(\begin{tabular}{ccccc}
$0$ & $0$ & $\beta$ & $\sqrt{2 (|\gamma |^2 + |\sigma |^2)}$ & $\sqrt{|\gamma |^2 + |\sigma |^2}$  \\
$0$ & $0$ & $0$ & $0$ & $0$  \\
$\beta^*$ & $0$ & $0$ & $0$ & $0$  \\
$\sqrt{2 (|\gamma |^2 + |\sigma |^2)}$ & $0$ & $0$ & $0$ & $0$  \\
$-\sqrt{|\gamma |^2 + |\sigma |^2}$ & $0$ & $0$ & $0$ & $0$
\end{tabular} \right), \quad
\label{eq960} \\
\begin{split}
& (C^1)' = \frac{1}{\sqrt{2}} \left(\begin{tabular}{cccc}
$0$ & $0$ & $i |\alpha |$ & $-i |\alpha |$  \\
$0$ & $0$ & $1$ & $1$  \\
$-i |\alpha |$ & $1$ & $0$ & $0$   \\
$i |\alpha |$ & $1$ & $0$ & $0$
\end{tabular} \right), \\
& (C^1)'' =  \frac{1}{\sqrt{2}}  \left(\begin{tabular}{cccc}
$0$ & $0$ & $i | \rho | $ & $- i | \rho |$  \\
$0$ & $0$ & $ | \delta | $ & $ |\delta |$  \\
$- i |\rho |$ & $|\delta |$ & $0$ & $0$   \\
$i |\rho |$  & $| \delta |$ & $0$ & $0$
\end{tabular} \right).
\end{split}
\label{eq961}
\end{align}
The form of the blocks $\eta^0$, $\eta^1$ of the matrix $\eta$ follows from \eqref{eq959}.

The spin block $C^0$ \eqref{eq960} has a single (up to a sign) nonzero root
\begin{align}
\pm \sqrt{|\beta |^2 +|\gamma |^2 +|\sigma |^2  },
\label{eq962}
\end{align}
which corresponds to the mass of a scalar boson
\begin{align}
m^{(0)} = \frac{m}{ \sqrt{|\beta |^2 +|\gamma |^2 +|\sigma |^2  }}.
\label{eq963}
\end{align}
The block $(C^1)'$ has the roots $\pm 1$, $\pm |\alpha|$. By virtue of the projective property of the matrix $\Gamma_0$ \eqref{eq944} the first of them corresponds to a massless vector field of the Maxwell's type, while the second one corresponds to a vector particle with the mass
\begin{align}
m_1^{(1)} = \frac{m}{ |\alpha|}.
\label{eq964}
\end{align}
The roots $\pm |\delta|$, $\pm |\rho|$ of block $(C^1)''$ correspond to the masses
\begin{align}
m_2^{(1)} = \frac{m}{ |\delta|}, \quad m_3^{(1)} = \frac{m}{ |\rho|}
\label{eq965}
\end{align}
of two other vector particles, whose description is contained in the discussed RWE and in the tensor system equivalent to it.

So, the linking scheme \eqref{eq914} allows us to construct a relativistic wave equation describing a vector field with four types of quanta -- one massless  and three massive. At the same time, a scalar particle with nonzero mass, which provides the unity of the components of the vector field in question, necessarily appears in the theory. One can interpret the vector part of the field as the "electroweak" field, while the scalar part is interpreted as a linear analog of the Higgs boson. The linear character of the equation  \eqref{eq943} describing the scalar particle is caused by the fact that the question of the mass generation is not touched within the RWE theory, whose bounds we do not exceed in our considerations. Either a presence or an absence of a mass appears for us as a given fact, and therefore there is no need for us to include nonlinear terms into \eqref{eq943}. Thus, while avoiding contradictions with the commonly accepted Higgs mechanism of the mass generation, our model proposed above is capable of pointing to yet another possible reason of the emergence of a scalar component in the theory of the vector electroweak field. However, if one prefers to consider the other -- Kalb-Ramond -- mechanism of the mass generation which is described in the Chapter \ref{mass_spir} and which is alternative to the Higgs one, then one arrives at the alternative theory of the electroweak interaction, in which the scalar boson plays exclusively the role of a link between the vector constituents of the electroweak field.
\vspace{3mm}

\section*{Conclusion}
\addcontentsline{toc}{chapter}{ \bf{Conclusion}     }

\vspace{3mm}
\hspace{0.5cm}
Let us list once again in a concise form the presented results. On
the basis of the use of extended sets of irreducible
representations of the Lorentz group, it is given:
\begin{itemize}
  \item semiphenomenological description of the internal structure of
microobjects with lower spins;
  \item a description of the
isospin degrees of freedom, in particular the chirality of massive
microobjects, by means of the RWE that does not decay in the
Lorentz group and which has internal symmetry of geometric origin;
   \item a joint description of massless fields with helicities as a single
physical object, on this basis the possibility of a semiphenomenological
description of the interaction of strings and membranes in Minkowski space
is shown;
  \item  matrix interpretation of the mechanism of mass generation of vector
fields, which differs from the well-known Higgs mechanism and does not lead
to the appearance of additional scalar or any other massive particles;
 \item matrix interpretation of massive gauge-invariant fields in the approach of
the RWE theory;
 \item  finally, a non-disintegrating RWE is described that describes a
massively massless vector field with three types of massive and one massless
quanta. This field may well be interpreted as an electroweak field. The need
for the appearance of a scalar massive field is justified in a completely
new way. It turns out that in the approach of the theory of RWE, the
indicated vector field can exist only in a "bundle" with a massive scalar
field, forming together with it a single unified physical object. Otherwise,
the free massive and massless vector fields appear as independent ones, i.e.
are unconnected in a relativistically invariant sense by equations.
 \end{itemize}

The novelty and the possibility of applying the results obtained is as
follows. Global unitary symmetries, which are used in modern gauge models of
fundamental particles and their interactions, are based on non-geometric
origin. In other words, in these models the simplest Dirac equation is taken
as the initial one, the free function of which is "hung" by the free
non-Lorentz index. Thus, the relationship between the properties of
space-time and the material world is manifested only after the localization
of these symmetries. Our proposal is to rely on internal symmetries in
local-calibration models, which already in the original global version have
a geometric origin, i.e. are inherent in equations that do not decay over
the full Lorentz group. The Dirac--K\"{a}hler equation and its algebraic
generalizations considered in Chapters 5 and 6 can serve as a possible
candidate for the role of such RWE.

This approach, in our opinion, provides a closer relationship between
space-time and the material world. In addition, the expansion of the class
of basic RWE should lead to new physical effects and, possibly, eliminate
some of the difficulties that occur in the Standard Model and the theory of
superstrings.

\end{document}

\begin{center}
{{\bf
 РЕЛЯТИВИСТСКИЕ ВОЛНОВЫЕ УРАВНЕНИЯ С РАСШИРЕННЫМ \\
НАБОРОМ ПРЕДСТАВЛЕНИЙ ГРУППЫ ЛОРЕНЦА \\[5mm]
}}
\end{center}
\begin{center}
В.А. Плетюхов \\[3mm]
д-р физ.-мат. наук, проф. каф. общей и теоретической физики \\
Брестского государственного университета имени А.С. Пушкина\\
\end{center}

\begin{abstract}
Показано, что использование расширенных наборов неприводимых представлений группы Лоренца открывает новые возможности теории релятивистских волновых уравнений с точки зрения пространственно-временного описания как внутренней структуры, так и изоспиновых степеней свободы элементарных частиц. Развиваемый в работе подход позволяет также применять методы теории релятивистских волновых уравнений в суперструнных и калибровочных моделях фундаментальных взаимодействий.
\end{abstract}

\section*{Введение}
\addcontentsline{toc}{chapter}{ \bf{Введение} }

\hspace{1.25cm}
Уравнения Ньютона, Максвелла, Эйнштейна, Шрёдингера, Дирака, Янга--Миллса... .

Каждое из них являлось эпохальным событием в физике.
Уравнения Ньютона положили начало теоретической физике. С уравнениями Максвелла связано  введение принципиально нового физического понятия поля, объединение электрических  и магнитных явлений, предсказание существования электромагнитных волн. Уравнения Эйнштейна  соединили воедино свойства материи и пространства-времени, создали основу для описания Вселенной как единого физического объекта. Уравнение Шрёдингера привело к пониманию вероятностного характера физических процессов в микромире. Уравнение Дирака~-- вершина квантово-механического описания физических явлений~-- создало основу для квантовой теории  поля, предсказало существование нового вида материи(античастиц). Уравнения Янга--Миллса лежат  в фундаменте теории калибровочных  полей, единой теории электромагнитного и слабого взаимодействий, Стандартной Модели.

Уравнение Дирака послужило одновременно в качестве исходной модели для создания общей теории релятивистских волновых уравнений (РВУ)~-- релятивистских квантово-механических уравнений первого порядка, записанных в матрично-дифференциальной форме. Принципиальной основой данной теории является взаимосвязь РВУ с соответствующим, в зависимости от спина микрообъекта, набором неприводимых представлений группы геометрической (пространственно-ременной) симметрии пространства Минковского. Помимо \, Дирака [1] \, наиболее \, существен\-ный
вклад в разработку теории РВУ внесли Паули, Фирц [2; 3], Баба [4; 5], Хариш--Чандра [6; 7], Гельфанд, Яглом [8; 9], белорусский физик-теоретик \-Ф.И.~Фёдоров [10--13].

Если собрать воедино те исходные положения, по которым у создателей теории РВУ сложился консенсус, то можно сформулировать следующий постулативный базис данной теории:

    1) любое РВУ должно удовлетворять требованиям инвариантности относительно преобразований собственной группы Лоренца и операции пространственного отражения, а также обеспечивать возможность лагранжевой формулировки теории;

    2) РВУ, описывающее единый физический микрообъект, должно быть не распадающимся в смысле полной группы Лоренца;

    3) среди состояний микрообъекта не могут присутствовать такие, которым соответствует нулевая энергия;

    4) «правильное» РВУ обязано приводить к положительно определённой плотности энергии (заряда) в случае целого (полуцелого) спина;

    5) микрообъект  с целым (полуцелым) спином описывается на основе тензорных (спинорных) представлений группы Лоренца;

    6) если под  микрообъектом понимается элементарная частица со спином $s$ (что обычно подразумевается), то для построения соответствующего  РВУ достаточно ограничится минимально необходимым набором неприводимых представлений группы Лоренца, содержащих данное значение спина.

Перечисленные положения теории РВУ были сформулированы в 20-х--50-х годах прошлого столетия. Они соответствуют представлениям того времени об элементарных частицах как бесструктурных точечных микрообъектах с единственной внутренней степенью свободы (спином), имеющей пространственно-временную трактовку. Однако по мере установления новых экспериментальных фактов (существование структуры у некоторых из частиц, наличие у них дополнительных помимо спина внутренних степеней свободы и др.) вышеуказанные представления претерпели существенные изменения. Изменилось и само понятие «элементарная частица». В отношении действительно (на сегодняшний день) бесструктурных микрообъектов прочно вошёл в обиход термин «фундаментальная частица». Возникло представление о существовании принципиально новых физических объектов, объединяющих качества микрочастиц (полей) с ненулевой и нулевой массой (например, электрослабое поле), а также свойства безмассовых микрообъектов с различными значениями спиральности (поля, осуществляющие взаимодействие незамкнутых струн). Оставались нерешёнными и «старые» проблемы теории РВУ. С одной стороны, её постулативный базис представляется недостаточно полным, поскольку не ограничивает
жёстко спектр возможных элементарных микрообъектов. С другой, использование в теории РВУ только симметрий пространственно-временного происхождения затрудняет описание дополнительных помимо спина (изоспиновых) внутренних степеней свободы. Вне поля зрения теории РВУ оставался и вопрос о происхождении массы.

Таким образом, теория РВУ столкнулась с вызовами, на которые она, казалось бы, не способна дать адекватные ответы. Поэтому интерес к ней в последние два-три десятилетия существенно снизился.

Вместе с тем, ещё в 1955--1957 гг. чешскими физиками Петрашем и Улеглой было построено и исследовано РВУ для микрообъекта со спином $1/2$ на основе использования расширенного по сравнению с уравнением Дирака набора неприводимых представлений группы Лоренца. Было показано, что во внешнем магнитном поле у данного микрообъекта проявляется внутренняя электромагнитная структура в виде аномального магнитного момента. Ещё ранее, в 1928 году английским физиком Дарвиным  в качестве своего рода альтернативы уравнению  Дирака было  предложено не распадающееся по полной группе Лоренца РВУ, которое впоследствии не совсем справедливо, на наш взгляд, получило  в литературе название уравнения Дирака--Кэлера. Уравнение Дирака--Кэлера содержит двукратные (дважды повторяющиеся) скалярное и векторное представления собственной группы Лоренца. Вследствие этого оно обладает внутренней симметрией, описываемой некомпактной группой $SU(2,2)$, которая образует полупрямое произведение, т.е. не коммутирует, с группой лоренцевской симметрии. Последнее обстоятельство позволяет интерпретировать уравнение Дирака--Кэлера, несмотря на тензорный характер компонент его волновой функции, как РВУ для микрообъекта со спином  $1/2$ и некоторой изоспиновой степенью свободы пространственно-временного происхождения.

Уравнения Дирака--Кэлера и Петраша--Улеглы можно считать  первыми успешными попытками выйти за рамки постулатов 1) -- 6)  и тем самым существенно
расширить возможности теории  РВУ. Они показали, в частности, что при отказе от постулата 6) появляется возможность пространственно-временного описания  как внутренней структуры, так и изоспиновых  степеней свободы микрообъектов.

С середины 1960-х годов данное направление начало активно развиваться в ряде научных центров Республики Беларусь  по инициативе и под руководством академика АНБ Ф.И.~Фёдорова. За прошедшее десятилетие был  накоплен богатый материал по теории РВУ с расширенным набором представлений группы Лоренца. В настоящей работе изложены некоторые значимые результаты, которые, на наш  взгляд, могут  не только быть адаптированы к современным экспериментальным достижениям и теоретическим тенденциям в области физики высоких энергий, но и в определённой мере помочь в решении остающихся здесь проблем.
В Главе 1 подробно рассматриваются те ограничения, которые накладывают на алгебраическую и групповую структуру матрично-дифференциальных РВУ условия 1)~-- 5). Анализ  проводится в так называемом каноническом базисе, или базисе Гельфанда--Яглома, наиболее удобном для установления спиновых  и массовых  характеристик микрообъекта. В Главе 2 показано, как конкретно работает подход Гельфанда--Яглома применительно к простейшим известным уравнениям для частиц с низшим спинами. Помимо уравнения Дирака это уравнения Даффина--Кеммера для скалярной и векторной частиц, уравнение Фирца--Паули для спина $\frac{3}{2}$.

В Главе 3  представлены РВУ для микрообъектов со спинами $s=0, \frac{1}{2}, 1, \frac{3}{2}$, основанные на расширенных наборах представлений группы Лоренца, включая кратные. Показано, что такое расширение позволяет отразить  внутреннюю электромагнитную структуру микрообъекта в рамках обычного подхода теории РВУ. Все рассмотренные в данном параграфе «расширенные»  РВУ являются перенормируемыми и не содержат непричинных решений.

В Главе 4 построено РВУ для векторной частицы с ненулевой массой и дополнительным внутренним квантовым числом -- киральностью. Как показано далее (Глава 7), такая частица в определённом смысле может рассматриваться как массивный аналог объединённого безмассового поля Максвелла--Кальба--Рамонда.

В Главах 5, 6 дана матричная формулировка уравнения Дирака--Кэлера и его максимальных тензорных обобщений, ограниченных размерностью пространства-времени $d=4$. Показана возможность непротиворечивого квантования этих уравнений по статистике Ферми--Дирака как с точки зрения правильной корпускулярной картины вторично-квантованного поля, так и в отношении вероятностной интерпретации теории.

В Главе 7 показана возможность совместного описания полей максвелловского типа и типа Кальба--Рамонда (нотоф Огиевецкого-Полубаринова) в подходе теории РВУ. Это  позволяет говорить о едином безмассовом векторном поле с тремя значениями спиральности $0, \pm 1$, которое может претендовать на роль  переносчика взаимодействия открытых  струн в теории  струн, редуцированной на четырёхмерное пространство Минковского. Дана также матричная интерпретация механизма генерации массы путём калибровочно-инвариантного смешивания безмассовых полей с различными значениями спиральности. Указанный механизм не приводит к появлению дополнительного массивного скалярного поля и может выступать  в качестве альтернативы общепринятому в настоящее время механизму Хиггса.

В Главе 8 детально исследован статус так называемых массивных калибровоч\-но-инвариантых полей.
Установлено, в чём состоит сходство и различие массивных и безмассовых калибровочно-инвариантных полей с точки  зрения формализма теории РВУ.

Наконец, в Главе 9 предложены тензорная и матричная формулировки массив\-но-безмассового векторного поля с тремя массивными и одним безмассовым квантами, которое может трактоваться как электрослабое поле. Требование нераспадения соответствующего РВУ приводит к необходимости введения массивного скалярного поля как неотъемлемой составляющей электрослабого поля. Отсюда вытекает принципиально иная трактовка происхождения бозона Хиггса, не связанная с механизмом спонтанного нарушения симметрии и генерации массы.

\section{Релятивистские волновые уравнения. \\
Подход Гельфанда--Яглома}
\hspace{1.25cm}
%Теория релятивистских волновых уравнений (РВУ) является одним из самых устоявшихся и разработанных в математическом отношении направлений в классической и квантовой теории поля. Как уже отмечалось во Введении, она берет начало с уравнения Дирака. Постулативный базис данной теории разрабатывался Дираком [1], Фирцем и Паули [2; 3], Баба [4; 5], Хариш–Чандра [6; 7] и другими авторами. В систематизированном изложении основные положения теории РВУ можно найти, например, в работах [8; 9]. Приведем их в сжатом виде.
Релятивистское квантово-механическое описание свободных элементарных микрообъектов всегда может быть сведено к системе линейных дифференциальных уравнений первого порядка с постоянными коэффициентами. Как уже отмечалось во Введении, в случае микрообъектов с ненулевой массой такая система представима в матрично-дифференциальной форме\footnote{Здесь и везде далее мы используем метрику $g_{\mu \nu} = \mathrm{diag} (1,1,1,1)$, поэтому нет необходимости различать ковариантные и контравариантные лоренцевские индексы. По повторяющимся индексам в соответствии с известным правилом Эйнштейна подразумевается суммирование.}
\begin{align}
(\Gamma_{\mu} \partial_{\mu} + m) \Psi (x) =0,
\label{eq:nonzero_mass}
\end{align}
где $\Psi$ – многокомпонентная волновая функция, $\Gamma_{\mu}$ – квадратные матрицы, $m$ – скалярный параметр, связанный с массой. Для микрообъектов с нулевой массой имеем
\begin{align}
(\Gamma_{\mu} \partial_{\mu} + \Gamma_0) \Psi (x) =0,
\label{eq:zero_mass}
\end{align}
где $\Gamma_0$ – особенная матрица ($\det \Gamma_0 =0$), которая может быть и нулевой. Именно эти формы записи ассоциируются с термином «релятивистские волновые уравнения». Отметим еще, что если в форме РВУ  \eqref{eq:zero_mass} матрица $\Gamma_0$ – неособенная, то форма \eqref{eq:zero_mass} всегда может быть приведена к виду \eqref{eq:nonzero_mass}. Поэтому в дальнейшем в записи \eqref{eq:zero_mass} мы будем подразумевать только особенную матрицу $\Gamma_0$, если специально не будет оговорено иное.

Основным и безусловным требованием, предъявляемым к уравнениям \eqref{eq:nonzero_mass}, \eqref{eq:zero_mass}, является их инвариантность относительно преобразований собственной группы Лоренца. Отсюда следует, что, во-первых, функция $\Psi$ преобразуется по некоторому представлению $T$ собственной группы Лоренца (далее будет показано, что представление $T$ должно быть приводимым). Во-вторых, матрицы $\Gamma_\mu$ и $\Gamma_0$ должны удовлетворять условиям
\begin{align}
T^{-1} \Gamma_{\mu} T &= L_{\mu \nu} \Gamma_{\nu}, \label{nonsing_trans} \\
T^{-1} \Gamma_{0} T &= \Gamma_{0},  \label{sing_trans}
\end{align}
где $L_{\mu \nu}$ – матрица Лоренца. Применяя \eqref{nonsing_trans} к бесконечно малым преобразованиям Лоренца
\begin{align}
T = 1 + \delta \omega_{[\mu \nu]} J^{[\mu \nu]},
\label{InfLoTra}
\end{align}
придем к соотношению
\begin{align}
[J^{[\mu \nu]}, \Gamma_{\alpha}]_- = \delta_{\nu \alpha} \Gamma_{\mu} - \delta_{\mu \alpha} \Gamma_{\nu}.
\label{Jcond}
\end{align}
Полагая в \eqref{Jcond} $\mu = i$ и $\nu = \alpha =4$, матрицы $\Gamma_i \, (i=1,2,3)$ можно выразить через $\Gamma_4$ и «бусты» $J^{[i 4]}$ лоренцевских преобразований:
\begin{align}
\Gamma_i = [J^{[i 4]}, \Gamma_4 ]_- .
\label{gam_i_gen}
\end{align}
Таким образом, среди матриц $\Gamma_{\mu}$ матрица $\Gamma_4$ играет выделенную роль.

Теперь напомним, что всякое неприводимое конечномерное представление $\tau$ группы Лоренца может быть задано парой чисел $l_1$, $l_2$, которые одновременно или порознь принимают целые (включая ноль) либо полуцелые положительные значения [14]. Представление $\tau$, действующее в пространстве $R^{\tau}$, порождает, вообще говоря, приводимое представление своей подгруппы – группы вращений. Другими словами, пространство $R^{\tau}$ можно разложить в прямую сумму инвариантных подпространств $R^{\tau}_s$, в каждом из которых представление группы вращений, индуцированное представлением $\tau \sim (l_1, l_2)$ группы Лоренца, неприводимо и задается целочисленным либо полуцелочисленным весом $s$. При этом в представлении $\tau$ присутствуют все веса от $| l_1 - l_2 |$ до $l_1 + l_2$. Размерность пространства $R^{\tau}$ равна $(2 l_1 + 1) (2 l_2 + 1)$.

В зависимости от значений (целочисленность или полуцелочисленность) чисел $l_1$,
$l_2$ все неприводимые конечномерные представления группы Лоренца удобно разбить на четыре класса:
\begin{align}
\begin{split}
\textup{класс} \,\, +1: & \quad  l_1, l_2 \,\, \textup{оба  целые}; \\
 \textup{класс} \,\, -1: & \quad  l_1, l_2 \,\, \textup{оба полуцелые}; \\
 \textup{класс} \,\, +\varepsilon : & \quad  l_1 \,\, \textup{целое}, l_2 \,\, \textup{полуцелое}; \\
 \textup{класс} \,\, -\varepsilon : & \quad  l_1 \,\, \textup{полуцелое}, l_2 \,\, \textup{целое}.
\end{split}
\end{align}
Неприводимые представления $\tau \sim (l_1 , l_2)$ и $\tau' \sim (l'_1, l'_2)$ называются зацепляющимися, если одновременно выполняются условия
\begin{align}
l'_ 1 = l_1 \pm \frac12 , \quad l'_2 = l_2 \pm \frac12 ,
\end{align}
причем знаки «+» и «–» могут не коррелировать между собой.

Нагляд\-ное графическое изображение зацепля\-ющих\-ся неприводи\-мых пред\-став\-ле\-ний группы
Лоренца удобно осуществлять посредством так называемой схемы зацепле\-ний,
в которой зацепляющиеся компоненты соединяются чертой. Очевидно,
что зацепляться друг с другом могут только представ\-ле\-ния
классов «+1» с «–1» и «$+\varepsilon$» с «$- \varepsilon$».
Таким образом, существуют две несмешивающиеся между собой раз\-но\-вид\-нос\-ти схем зацеплений. Наиболее полная их реализация имеет вид
\begin{align}
\begin{tabular}{ccccccccc}
& & &   & $(0,0)$ &  & & & \\
& & &   &   $|$       &  & & & \\
 & & $(0,1)$   & --- & $( \frac12, \frac12)$  & --- & $(1,0)$ & &  \\
 &  & $|$ &  & $|$ &  & $|$ &  &  \\
$(0,2)$ &  --- & $ (\frac12, \frac32) $ & --- & $ (1,1) $ & --- & $(\frac32, \frac12)$ & --- & $ (2,0)$ \\
$|$ &  & $|$ &  & $|$ &  & $|$ &  & $|$
\end{tabular}
\label{type_1}
\end{align}
в первом случае и
\begin{align}
\begin{tabular}{ccccccccccc}
& & &   & $(0,\frac12)$ &  --- & $(\frac12 , 0)$ & & & & \\
& & &   &   $|$       &  & $|$ & & \\
 & & $(0,\frac32)$   & --- & $( \frac12, 1)$  & --- & $(1,\frac12)$ & --- &  $(\frac32,0)$ & & \\
 &  & $|$ &  & $|$ &  & $|$ &  & $|$ & &  \\
$(0,\frac52)$ &  --- & $ (\frac12, 2) $ & --- & $ (1,\frac32) $ & --- & $(\frac32, 1)$ & --- & $ (2,\frac12)$ & --- & $(\frac52,0)$ \\
$|$ &  & $|$ &  & $|$ &  & $|$ &  & $|$ & & $|$
\end{tabular}
\label{type_eps}
\end{align}
во втором.

Важно подчеркнуть, что в обозначенных схемах зацеплений каждая из неприводимых компонент может встречаться более одного раза. Тогда говорят об РВУ с кратными (повторяющимися) представлениями группы Лоренца. Очевидно, что любая схема зацеплений, которая относится к типу \eqref{type_1}, содержит только целочисленные веса s неприводимых представлений группы вращений и, следовательно, соответствующее ей РВУ описывает микрообъекты с целым спином. Аналогично схемы зацеплений типа \eqref{type_eps} служат для описания микрообъектов с полуцелым спином.

Рассмотрим сначала, какие ограничения на представление $T$, действующее в пространстве $R$ волновой функции $\Psi$, накладывает требование инвариантности РВУ \eqref{eq:nonzero_mass} относительно преобразований собственной группы Лоренца.

Предположим, что представление $T$ состоит из одного неприводимого представления $\tau \sim (l_1 , l_2)$ собственной группы Лоренца. Тогда слагаемое $m \Psi$ в \eqref{eq:nonzero_mass} преобразуется по представлению $\tau$. Слагаемое же $\Gamma_{\mu} \partial_{\mu} \Psi$ преобразуется по представлению, которое полностью или частично состоит из неприводимых компонент, содержащихся в прямом произведении представлений
\begin{align}
(l_1 , l_2) & \otimes (\frac12 , \frac12) = \nonumber \\
&= (l_1 + \frac12 , l_2 + \frac12) \oplus (l_1 + \frac12 , l_2 - \frac12) \oplus (l_1 - \frac12 , l_2 + \frac12) \oplus (l_1 - \frac12 , l_2 - \frac12).
\label{dp_decomp}
\end{align}
Но поскольку ни одно из неприводимых представлений, фигурирующих в разложении \eqref{dp_decomp}, не совпадает с представлением $( l_1 , l_2 )$, члены $\Gamma_{\mu} \partial_{\mu} \Psi$ и $ m \Psi$
в уравнении \eqref{eq:nonzero_mass} не могут преобразовываться одинаково при преобразованиях Лоренца.
Это означает, что РВУ \eqref{eq:nonzero_mass} не может базироваться на единственном неприводимом представлении группы Лоренца. Подобным образом рассуждая, приходим к выводу, что представление $T$ волновой функции $\Psi$ не может состоять из незацепляющихся неприводимых компонент. Таким образом, представление $T$ должно быть приводимым и состоять из зацепляющихся неприводимых представлений.

Например, наиболее известные простейшие РВУ – уравнение Дирака (спин $\frac12$), уравнения Даффина–Кеммера (спины $0$ и $1$), уравнение Фирца–Паули (спин $\frac32$) – базируются соответственно на схемах зацеплений
\begin{align}
(0, \frac12) - (\frac12 , 0),
\label{ls1}
\end{align}
\begin{align}
\begin{tabular}{c}
$(0,0)$ \\
$|$ \\
$(\frac12,\frac12)$
\end{tabular},
\label{ls2}
\end{align}
\begin{align}
(0, 1) - (\frac12 , \frac12) - (1,0),
\label{ls3}
\end{align}
и
\begin{align}
\begin{tabular}{ccc}
$(0,\frac12)$ & --- & $(\frac12,0 )$ \\
$|$ & & $|$ \\
$(\frac12,1)$ & --- & $(1, \frac12)$
\end{tabular}.
\label{ls4}
\end{align}
Возможен также вариант, когда рассматриваемая схема зацеплений состоит из отдельных фрагментов, каждый из которых удовлетворяет выше сформулированным условиям, но которые не зацепляются друг с другом. В этом случае соответствующее РВУ будет распадающимся в смысле собственной группы Лоренца.

На РВУ \eqref{eq:nonzero_mass} еще обычно накладывается требование инвариантности относительно операции пространственного отражения, совместно с которой чисто лоренцевские преобразования образуют полную группу Лоренца. Указанное требование приводит к тому, что в используемой схеме зацеплений наряду с каждым неприводимым представлением $\tau \sim (l_1 , l_2)$ при $l_1 \neq l_2$  должно присутствовать также представление $\dot{\tau} \sim (l_2 , l_1)$ (его называют сопряженным к $\tau$).

Принципиальным моментом в теории РВУ является положение, согласно которому единый физический микрообъект должен описываться уравнением, не распадающимся в смысле полной группы Лоренца.

Все сказанное выше в отношении РВУ \eqref{eq:nonzero_mass} остается в силе и для РВУ \eqref{eq:zero_mass}, за исключением случая, когда $\Gamma_0 = 0$. РВУ с $\Gamma_0 = 0$ в нашей работе рассматриваться не будут. Поэтому, чтобы не загромождать изложение, на этом случае не останавливаемся. Интересующегося читателя можем отослать, например, к работам [8; 10].

Физический интерес представляют те РВУ, которые могут быть получены на основе вариационного принципа из лоренц-инвариантной функции Лагранжа (плотности лагранжиана). Для ее построения необходимо иметь инвариантные квадратичные комбинации (квадратичные формы), составленные из функций поля, их первых производных и матриц $\Gamma_\mu$, $\Gamma_0$. Проблема здесь заключается в том, что матрицы $T$ конечномерных представлений группы Лоренца не являются унитарными. Вследствие этого обычная квадратичная форма
\begin{align}
\Psi^{\dagger} \Psi =( \Psi^T)^* \Psi
\label{QuFo}
\end{align}
не является лоренцевским инвариантом. Вместо квадратичной формы \eqref{QuFo} приходится вводить так называемую билинейную форму
\begin{align}
\overline{\Psi} \Psi = \Psi^{\dagger}\eta  \Psi, \quad \overline{\Psi} =  \Psi^{\dagger}\eta ,
\label{InBiFo}
\end{align}
где $\eta$ – некоторая числовая матрица. Для релятивистской инвариантности формы \eqref{InBiFo} матрица $\eta$ должна удовлетворять условию
\begin{align}
T^{\dagger} \eta T = \eta .
\label{InvCond}
\end{align}
Условие  \eqref{InvCond} можно представить в ином виде, если в качестве преобразования T использовать бесконечно малое преобразование \eqref{InfLoTra}. Подставляя \eqref{InfLoTra} в  \eqref{InvCond} и ограничиваясь членами первого порядка малости по параметрам $\delta \omega_{[\mu \nu]}$, с учетом независимости этих параметров получим эквивалентные  \eqref{InvCond} условия
\begin{align}
\eta J^{[ij]} & =  J^{[ij]} \eta , \label{inv1} \\
\eta J^{[i4]} & = - J^{[i4]} \eta . \label{inv2}
\end{align}
Матрица $\eta$, фигурирующая в выражении \eqref{InBiFo} и удовлетворяющая условиям  \eqref{inv1},  \eqref{inv2}, называется матрицей билинейной лоренц-инвариантной формы.

При выполнении условий \eqref{nonsing_trans}, \eqref{sing_trans}, \eqref{inv1}, \eqref{inv2} лоренц-инвариантными являются также квадратичные комбинации
\begin{align}
\partial_{\mu} (\overline{\Psi} \Gamma_{\mu} \Psi), \quad (\partial_{\mu} \overline{\Psi}) \Gamma_{\mu} \Psi, \quad \overline{\Psi} \Gamma_{\mu} (\partial_{\mu} \Psi).
\end{align}

Лагранжиан, приводящий к уравнению \eqref{eq:nonzero_mass}, может быть выбран, например, в виде
\begin{align}
\mathcal{L} (x) = - \frac12 \overline{\Psi} (\Gamma_{\mu} \partial_{\mu} + m) \Psi + \frac12( (\partial_{\mu} \overline{\Psi}) \Gamma_{\mu} -m \overline{\Psi} )\Psi.
\label{lagr_dens}
\end{align}
Варьируя в соответствии с принципом наименьшего действия функцию действия (функции $\Psi$ и $\overline{\Psi}$ при этом считаются независимыми), придем к уравнению \eqref{eq:nonzero_mass} для функции $\Psi$ и уравнению
\begin{align}
- (\partial_{\mu} \overline{\Psi}) \Gamma_{\mu} + m \overline{\Psi} = 0
\label{eqOPsi}
\end{align}
для функции $\overline{\Psi}$. Динамические переменные получаются из лагранжиана \eqref{lagr_dens} по общим формулам
\begin{align}
E = - \int \overline{\Psi}\,  \Gamma_4 \, \partial_4 \Psi\,  d^3 x , \quad P=  - \frac{i}{c} \int \overline{\Psi}\,  \Gamma_4 \, \partial_i \Psi \, d^3 x ,
\end{align}
вытекающим из теоремы Нетер.

Часто используется упрощенный вид лагранжиана
\begin{align}
\mathcal{L} (x) = - \overline{\Psi} (\Gamma_{\mu} \partial_{\mu}+ m) \Psi ,
\label{eq126}
\end{align}
также позволяющий получить РВУ \eqref{eq:nonzero_mass} и правильные выражения для динамических переменных. Уравнение \eqref{eq:zero_mass} может быть получено из лагранжиана
\begin{align}
\mathcal {L} (x) =  - \overline{\Psi} (\Gamma_{\mu} \partial_{\mu} +\Gamma_0 )\Psi .
\end{align}

Из физических соображений очевидно, что билинейная форма \eqref{InBiFo} должна быть вещественной, то есть
\begin{align}
( \Psi^{\dagger} \eta \Psi)^{\dagger} = \Psi^{\dagger} \eta^{\dagger} \Psi = \Psi^{\dagger} \eta \Psi,
\nonumber
\end{align}
откуда следует
\begin{align}
\eta^{\dagger}= \eta .
\label{etaHerm}
\end{align}
Кроме того, в пространстве волновой функции $\Psi$ базис можно выбрать так, чтобы выполнялось условие
\begin{align}
\eta^2 =1, \quad \textup{или} \quad \eta^{-1} = \eta .
\nonumber
\end{align}
В сочетании с \eqref{etaHerm} это приводит к соотношениям
\begin{align}
\eta = \eta^T = \eta^* = \eta^{\dagger} = \eta^{-1} .
\label{etaCond}
\end{align}

Условия \eqref{etaHerm}, \eqref{etaCond} еще не определяют полностью матрицу $\eta$, поскольку они не учитывают того, что функция $\overline{\Psi} (x)$ является решением уравнения \eqref{eqOPsi}. Чтобы найти дополнительные ограничения, накладываемые благодаря этому на матрицу $\eta$, проведем над уравнением \eqref{eq:nonzero_mass} операции эрмитовского сопряжения и умножения на $\eta$. В результате получим
\begin{align}
(\partial_i \Psi^{\dagger} \Gamma_i^{\dagger} - \partial_4 \Psi^{\dagger} \Gamma_4^{\dagger} +\Psi^{\dagger} m) =0 .
\label{eq:Psi_spec}
\end{align}
Потребуем, чтобы от уравнения \eqref{eq:Psi_spec} можно было перейти к уравнению \eqref{eqOPsi}. Очевидно, что для этого должны выполняться перестановочные соотношения
\begin{align}
\Gamma_i^{\dagger} \eta = - \eta \Gamma_i, \quad \Gamma_4^{\dagger} \eta = \eta \Gamma_4 .
\label{cond1}
\end{align}

В случае РВУ \eqref{eq:zero_mass} к соотношениям \eqref{cond1} надо добавить условие
\begin{align}
 \Gamma_0^{\dagger} \eta = \eta \Gamma_0 ,
\label{cond2}
\end{align}
которое совместно с \eqref{cond1} приводит к уравнению
\begin{align}
- (\partial_{\mu} \overline{\Psi}) \Gamma_{\mu} + \overline{\Psi} \Gamma_0 =0 .
\end{align}
для функции $\overline{\Psi}$.

На выбор всевозможных РВУ накладывается также условие: среди состояний микрообъекта должны отсутствовать такие, которым соответствует нулевая энергия. Это условие существенно ограничивает возможный вид минимальных полиномов матриц $\Gamma_{\mu}$.

Как показано в работе [11], минимальный полином матриц $\Gamma_4$ (а также $\Gamma_\mu$) должен иметь структуру
\begin{align}
\Gamma_4^n (\Gamma_4^2 -\lambda_1^2) (\Gamma_4^2 - \lambda_2^2) \ldots =0,
\label{min_pol}
\end{align}
где все $\lambda_i$ – вещественные и различные числа, $n$ – целое положительное число либо ноль.

Условия дефинитности энергии и заряда для микрообъектов с одной массой могут быть представлены соответственно в виде неравенств [13]
\begin{align}
(-1)^{n+1} \left[ (\mathrm{Sp} (\Gamma_4^{n+1} \eta))^2 - (\mathrm{Sp} (\Gamma_4^{n} \eta))^2 \right] & >0, \label{ineq1} \\
(-1)^n \left[[ (\mathrm{Sp} (\Gamma_4^{n+1} \eta))^2 - (\mathrm{Sp} (\Gamma_4^{n} \eta))^2 \right] & >0 . \label{ineq2}
\end{align}
При этом в соответствии с теоремой Паули полагается, что условие \eqref{ineq1} должно иметь место в случае целого, а \eqref{ineq2} – в случае полуцелого спина.

Наконец, что касается матрицы $\Gamma_0$ в уравнении \eqref{eq:zero_mass}, то при выборе соответствующего базиса в пространстве волновой функции она всегда может быть приведена к диагональному виду. При этом матрица $\Gamma_0$ состоит из независимых скалярных блоков $a_\tau$, сопоставляемых неприводимым представлениям $\tau \in T$. Часть указанных блоков являются нулевыми. Из требования инвариантности РВУ \eqref{eq:zero_mass} относительно преобразований полной группы Лоренца следует, что ненулевые блоки $a_\tau$ удовлетворяют равенству
\begin{align}
a_{\tau} = a_{\dot{\tau}}.
\end{align}
В случае конечномерных РВУ к такому же равенству приводит и условие возможности лагранжевой формулировки РВУ \eqref{eq:zero_mass}.

При построении РВУ с заданным спектром массовых и спиновых состояний удобно использовать так называемый канонический базис, или базис Гельфанда– Яглома [8]. В этом базисе компоненты волновой функции $\psi^{\tau}_{sk}$ описывают «чистые» состояния,
то есть состояния с определенным значением спина $s$ и его проекции $k$; верхний индекс указывает на принадлежность данного состояния к неприводимому представлению $\tau$. В базисе Гельфанда–Яглома матрица $\Gamma_4$ имеет квазидиагональный вид
\begin{align}
\Gamma_4 = \bigoplus_s C^s \otimes I_{2s+1},
\label{osumG4}
\end{align}
где $I_{2s+1}$ – единичная матрица размерности $2s + 1$; $C^{s}$ – матричный блок, отвечающий спину $s$, в том смысле, что если матрица $C^{s}$ имеет ненулевые корни (собственные значения), то частица обладает спином $s$. Возможные значения массы микрообъекта в случае РВУ \eqref{eq:nonzero_mass} выражаются через ненулевые корни $\lambda_i$ блока
$C^{s}$ по формуле
\begin{align}
m_i^{(s)} = \frac{m}{|\lambda_i^{(s)}|}.
\label{eq_ms}
\end{align}

Спиновый блок $C^{s}$ строится следующим образом. Из схемы зацеплений, на которой базируется РВУ, выбираются все неприводимые компоненты $\tau \sim (l_1 , l_2)$, удовлетворяющие условию
\begin{align}
|l_1 - l_2 | \leq s\leq l_1 + l_2 .
\label{s_range}
\end{align}
(Про такие представления говорят, что они формируют блок  $C^s$). Затем эти компоненты для удобства нумеруются. Матрица $C^{s}$ состоит из элементов $c^s_{\tau \tau'}$, в обозначении которых номера представлений $\tau, \tau'$ играют роль матричных индексов. При этом незацепляющимся компонентам $\tau, \tau'$ соответствуют нулевые элементы $c^s_{\tau \tau'}$. Отсюда следует, что в схеме зацеплений, предназначенной для описания спина $s$, должно содержаться, как минимум, два зацепляющихся неприводимых представления, которые удовлетворяют условию \eqref{s_range}.

Требование релятивистской инвариантности РВУ \eqref{eq:nonzero_mass}, \eqref{eq:zero_mass} накладывает следующие ограничения на элементы $c_{\tau \tau'}^s$:
\begin{align}
c^s_{\tau \tau'} &=c_{\tau \tau'} \sqrt{(s+l_+ +2)(s - l_+ - 1)}  \quad \textup{если} \quad l'_+ = l_+ + 1, \quad l'_- = l_-  , \nonumber \\
c^s_{\tau \tau'} &=c_{\tau \tau'} \sqrt{(s+l_- +1)(s - l_-)} \quad \textup{если} \quad l'_+ = l_+ , \quad l'_- = l_- + 1 ,\label{c_cond} \\
c^s_{\tau \tau'} &=c_{\tau \tau'} (s+ \frac12)\quad \textup{если} \quad l'_+ = l_+ , \quad l'_- = l_-  , \nonumber
\end{align}
где $l_+ = l_1 + l_2$, $l_- = |l_1 - l_2|$, $l'_+ = l'_1 + l'_2$, $l'_- = |l'_1 - l'_2|$; $c_{\tau \tau'}$ – произвольные комплексные числа в указанных случаях, а в остальных случаях – равные нулю.

Инвариантность РВУ относительно операции пространственного отражения, задаваемой матрицей $P$, налагает на числа $c_{\tau \tau'}$ ограничения
\begin{align}
c_{\tau \tau'} &= c_{\dot{\tau} \dot{\tau}'} \quad \textup{если} \quad \dot{\tau} \neq \tau, \,\, \dot{\tau}' \neq \tau' ; \label{c1_cond}  \\
c_{\tau \tau'} &=\pm  c_{\dot{\tau} \dot{\tau}'} \quad \textup{если} \quad \dot{\tau} = \tau, \,\, \dot{\tau}' \neq \tau' \quad  \textup{или} \quad  \dot{\tau} \neq \tau, \,\, \dot{\tau}' = \tau'  . \label{c2_cond}
\end{align}
Знак «+» в условии \eqref{c2_cond} (для определенности берем первый вариант) соответствует случаю, при котором оператор $P$ действует в подпространствах $R^{\tau}$ и $R^{\tau'}$ одинаково, то есть
\begin{align}
P \psi_{sk}^{\tau} = (-1)^{[s]} \psi_{sk}^{\tau}, \quad P \psi_{sk}^{\tau'}= (-1)^{[s]} \psi_{sk}^{\dot{\tau}'}.
\label{Pc1_cond}
\end{align}
Если же оператор $P$ действует в $R^{\tau}$ и $R^{\tau'}$ по-разному
\begin{align}
P \psi_{sk}^{\tau} = (-1)^{[s]+1} \psi_{sk}^{\tau}, \quad P \psi_{sk}^{\tau'}= (-1)^{[s]} \psi_{sk}^{\dot{\tau}'},
\label{Pc2_cond}
\end{align}
то в условии \eqref{c2_cond} выбирается знак «–».

При $\tau = \dot{\tau}$, $\tau' = \dot{\tau}'$ имеем $c_{\tau \tau'} \neq 0$ тогда, когда оператор $P$ действует одинаково в пространствах $R^{\tau}$, $R^{\tau'}$; каких-либо других ограничений на числа $c_{\tau \tau'}$ в этом случае не накладывается.

Матрица $\eta$ билинейной формы \eqref{QuFo} в базисе Гельфанда–Яглома имеет структуру, аналогичную \eqref{osumG4}:
\begin{align}
\eta =\bigoplus_s \eta^s \otimes I_{2 s+1}.
\label{eq146}
\end{align}
Условия \eqref{inv1}, \eqref{inv2}, \eqref{etaCond} приводят к тому, что отличными от нуля являются лишь элементы $\eta^s_{\tau \dot{\tau}}$, причем
\begin{align}
\eta_{\tau  \dot{\tau}}^s = \eta_{\dot{\tau} \tau}^s = - \eta_{\tau \dot{\tau}}^{s+1}.
\label{eta_cond}
\end{align}

 Условие \eqref{cond1}  приводит к соотношению
\begin{align}
c_{\tau \tau'}^s \eta^s_{\tau' \dot{\tau}'} = (c^s_{\dot{\tau}' \dot{\tau}} )^* \eta^s_{\tau \dot{\tau}}.
\label{c_eta_cond}
\end{align}

После наложения на элементы матриц $\Gamma_4$, $\Gamma_0$, $\eta$ связей \eqref{c_cond}--\eqref{c2_cond}, \eqref{eta_cond}, \eqref{c_eta_cond} в выборе этих элементов остается, как правило, произвол, который может быть использован для удовлетворения условиям \eqref{min_pol} и \eqref{ineq1} (или \eqref{ineq2}). Если это оказывается невозможным, значит на основе рассматриваемой схемы зацеплений РВУ для описания микрообъекта с заданным спектром массовых и спиновых состояний построить нельзя.

\section{Простейшие релятивистские волновые уравнения для частиц с низшими спинами}

\hspace{1.25cm}
В дальнейшем мы будем часто применять подход Гельфанда–Яглома, суть которого в общих чертах была изложена в предыдущем разделе. Покажем, как им конкретно пользоваться, на примерах известных простейших РВУ для частиц с низшими спинами.

Рассмотрим сначала схему зацеплений \eqref{ls2}, в которой неприводимое представление $(0, 0)$ соответствует скалярной функции $\psi_0$, а представление $(\frac12, \frac12)$ – векторной функции $\psi_\mu$. Матрица $\Gamma_4$ РВУ \eqref{eq:nonzero_mass}, базирующегося на данном наборе неприводимых представлений группы Лоренца, в базисе Гельфанда–Яглома будет иметь вид
\begin{align}
\Gamma_4 = \left( \begin{tabular}{cc}
$C^0$ &  $0$ \\
$0$ & $C^1 \otimes I_3$
\end{tabular} \right),
\label{gam4_s2}
\end{align}
где $C^{0}$ и $C^{1}$ – блоки, сопоставляемые спинам $0$ и $1$. Снабдив рассматриваемые представления номерами
\begin{align}
(0,0) \sim 1 , \quad (\frac12 , \frac12) \sim 2,
\end{align}
играющими роль матричных индексов, получим для матриц $C^{0}$, $C^{1}$ выражения
\begin{align}
C^0 = \left( \begin{tabular}{cc}
$0$ &  $c_{12}^0$ \\
 $c_{21}^0$ & $0$
\end{tabular} \right), \qquad C^1 =0 .
\end{align}
При этом, как нетрудно видеть, соотношения \eqref{c_cond}--\eqref{c2_cond} никаких ограничений на числа
$c^{0}_{12}$, $c^{0}_{21}$ не накладывают.

Элементы матрицы $\eta$ лоренц-инвариантной билинейной формы
\begin{align}
\eta =  \left( \begin{tabular}{cc}
$\eta^0$ &  $0$  \\
$0$  & $\eta^1 \otimes I_3$
\end{tabular} \right), \quad \eta^0 =   \left( \begin{tabular}{cc}
$\eta^0_{11}$ &   $0$ \\
$0$  & $\eta^0_{22}$
\end{tabular} \right), \quad \eta^1 = \eta_{22}^1
\label{eq24}
\end{align}
выберем следующим образом:
\begin{align}
\eta^0_{11} = \eta^0_{22} = - \eta_{21}^1 =1.
\end{align}
При таком выборе условие  \eqref{c_eta_cond} приводит к равенству
\begin{align}
c_{21}^0 = (c_{12}^0)^* .
\end{align}
Полагая $c^{0}_{12} = 1$, получим для матриц $C^0$ и $\Gamma_4$ окончательно выражения
\begin{align}
C^0 = \left(  \begin{tabular}{cc} $0$ & $1$ \\ $1$ & $0$  \end{tabular} \right) , \quad
\Gamma_4 = \left( \begin{tabular}{ccccc}
$ 0$ & $1$ & $ 0$ & $0$ & $0$ \\
$ 1$ & $0$ & $ 0$ & $0$ & $0$ \\
$ 0$ & $0$ & $ 0$ & $0$ & $0$ \\
$ 0$ & $0$ & $ 0$ & $0$ & $0$ \\
$ 0$ & $0$ & $ 0$ & $0$ & $0$
\end{tabular}\right).
\label{gam4_2}
\end{align}
Легко убедиться, что минимальное уравнение для матрицы $\Gamma_4$ имеет вид
\begin{align}
\Gamma_4 (\Gamma_4^2 -1) =0.
\label{gam4_2min}
\end{align}
Вид остальных матриц $\Gamma_i$ определяется по формуле \eqref{gam_i_gen}.

Для дефинитности энергии должно в данном случае $(n = 1)$ выполняться неравенство
\begin{align}
(-1)^2 \left[ \left(\mathrm{Sp} (\Gamma_4^2 \eta ) \right)^2 - \left(\mathrm{Sp} (\Gamma_4 \eta ) \right)^2 \right] >0 ,
\label{def_en_s2}
\end{align}
справедливость которого легко проверяется.

Из \eqref{gam4_2}, \eqref{gam4_2min} следует, что состоянию со спином $s = 0$ в силу формулы \eqref{eq_ms} соответствует одно значение массы, а состояние со спином $s =1$ отсутствует. В литературе такое уравнение называется обычно уравнением Даффина–Кеммера для скалярной частицы.

Тензорная формулировка этого уравнения имеет вид
\begin{align}
\partial_{\mu} \psi_{\mu} + m\psi_0 = 0 , \quad \partial_{\mu} \psi_0 + m \psi_{\mu} = 0.
\end{align}
Отсюда нетрудно получить уравнение второго порядка для скалярной функции $\psi_0$
\begin{align}
(\square -m^2) \psi_0 = 0,
\end{align}
означающее, что речь действительно идет об описании частицы с ненулевой массой и спином $s = 0$.

Для построения простейшего РВУ для частицы с ненулевой массой и спином $s = 1$ служит схема зацеплений \eqref{ls3}, где представления $(0,1)$, $(1,0)$ в совокупности соответствуют антисимметричному тензору второго ранга $\psi_{[ \mu \nu]}$. Матрица $\Gamma_4$ в базисе Гельфанда–Яглома в данном случае по-прежнему имеет блочную структуру \eqref{gam4_s2}, где с учетом нумерации $(\frac12, \frac12) \sim 1$, $(0,1) \sim 2$, $(1,0) \sim 3$ для спиновых блоков $C^{0}$, $C^{1}$ имеют место выражения
\begin{align}
C^1  =  \left(
\begin{tabular}{ccc}
$0$ & $c_{12}^1$ & $c_{13}^1$ \\
$c_{21}^1$ & $0$ & $0$ \\
$c_{31}^1$ & $0$ & $0$
\end{tabular} \right), \quad C^0 = 0.
\label{eq212}
\end{align}
Соотношения \eqref{c_cond} здесь, как и в случае спина $s = 0$, никаких ограничений на числа $c^{1}_{ij}$ не накладывают. Инвариантность относительно преобразований пространственных отражений (смотри условия \eqref{c2_cond}) дает
\begin{align}
c_{12}^1 = \pm c_{13}^1 , \quad c_{21}^1  = \pm c_{31}^1 .
\label{Pcond_s2}
\end{align}

Возможность лагранжевой формулировки теории приводит к условию
\begin{align}
c_{12}^1 = \frac{\eta_{11}^1}{\eta_{23}^1} (c_{31}^1)^* .
\label{c_eta_s2}
\end{align}
Ненулевые элементы матрицы билинейной инвариантной формы
\begin{align}
 \eta &= \left(\begin{tabular}{cc} $\eta^0$ & $0$  \\ $0$ & $\eta^1 \otimes I_3$ \end{tabular}  \right), \quad \eta^0 = \eta_{11}^0 , \nonumber  \\
\eta^1 &= \left(\begin{tabular}{ccc}
$\eta^1$ & $0$ & $0$ \\
$0$ & $0$ & $\eta_{23}^1$ \\
$0$ & $\eta_{32}^1$ & $0$
\end{tabular} \right)
\end{align}
выберем следующим способом:
\begin{align}
-\eta_{11}^0 = \eta_{11}^1 = \pm \eta_{23}^1 = \pm \eta_{32}^1 = 1 .
\end{align}
Тогда из соотношения \eqref{c_eta_s2} с учетом \eqref{Pcond_s2} приходим к равенствам
\begin{align}
c_{21}^1 = (c_{12}^1)^* , \quad c_{31}^1 = (c_{13}^1)^* .
\end{align}

Выбирая $c_{12}^1 =\frac{1}{\sqrt{2}} $, для блока $C^{1}$ получаем выражение
\begin{align}
C^1  = \frac{1}{\sqrt{2}} \left(
\begin{tabular}{ccc}
$0$ & $1$ & $\pm 1$ \\
$1$ & $0$ & $0$ \\
$ \pm 1$ & $0$ & $0$
\end{tabular} \right),
\label{C1_1}
\end{align}
где знаки «+» и «–» коррелируют между собой. Блок $C^{1}$ \eqref{C1_1} имеет единственное с точностью до знака ненулевое собственное значение $1$, то есть построенное РВУ описывает частицу с ненулевой массой и спином $s = 1$. Соотношения \eqref{gam4_2min}, \eqref{def_en_s2} здесь также выполняются. В литературе данное РВУ называют уравнением Даффина–Кеммера для векторной частицы.

Приведем еще его тензорную формулировку:
\begin{align}
\partial_{\nu} \psi_{[\mu \nu]} + m \psi_{\mu} = 0 , \quad - \partial_{\mu} \psi_{\nu} + \partial_{\nu} \psi_{\mu} + m \psi_{[\mu \nu]} = 0 .
\label{proca1}
\end{align}
Из системы первого порядка \eqref{proca1} получается уравнение второго порядка (уравнение Прока)
\begin{align}
(\square -m^2) \psi_{\mu} =0 , \quad \partial_{\mu} \psi_{\mu} =0 .
\end{align}

Простейшее уравнение для частицы со спином $s =\frac12$ (уравнение Дирака) основывается на схеме зацеплений \eqref{ls1}, соответствующей биспинору первого ранга. Матрица $\Gamma_4$ этого уравнения, представленного в стандартной форме  \eqref{eq:nonzero_mass}, в базисе Гельфанда–Яглома запишется следующим образом:
\begin{align}
\Gamma_4 = C^{1/2} \otimes I_2 = \left( \begin{tabular}{cccc}
$0$ & $0$ & $c_{12}^{1/2}$ & $0$ \\
$0$ & $0$ & $0$ & $c_{12}^{1/2}$ \\
$c_{21}^{1/2}$ & $0$ & $0$ & $0$ \\
$0$ & $c_{21}^{1/2}$ & $0$ & $0$ \\
\end{tabular} \right),
\end{align}
(смысл нижних индексов здесь очевиден: $(0, \frac12) \sim 1$, $(\frac12 ,0) \sim 2$). Применяя  к элементам этой матрицы условия \eqref{c_cond}, \eqref{c1_cond}, \eqref{c_eta_cond}, будем иметь
\begin{align}
\Gamma_4 = \left( \begin{tabular}{cc} $0$ &  $I_2$ \\  $I_2$ &  $0$  \end{tabular} \right) = \sigma_1 \otimes I_2 .
\end{align}

Остальные матрицы $\Gamma_i$ имеют при этом вид
\begin{align}
\Gamma_1 = \sigma_2 \otimes \sigma_1 , \quad \Gamma_2 = \sigma_2 \otimes \sigma_2 , \quad \Gamma_3 = \sigma_2 \otimes \sigma_3 ,
\end{align}
где  $\sigma_i$ – матрицы Паули. Матрица билинейной формы совпадает с матрицей $\Gamma_4$:
\begin{align}
\eta = \Gamma_4 .
\end{align}
Матрицы Дирака удовлетворяют алгебре
\begin{align}
\Gamma_{\mu} \Gamma_{\nu} + \Gamma_{\nu} \Gamma_{\mu} = 2 \delta_{\mu \nu}
\end{align}
и минимальному уравнению
\begin{align}
\Gamma_{\mu}^2 - 1 = 0 \quad (\textup{где нет суммирования по \,\,} \mu).
\label{min_pol_dir}
\end{align}
В дальнейшем, следуя сложившейся традиции, матрицы Дирака будем обозначать через $\gamma_{\mu}$.

Заметим, что с помощью унитарного преобразования
\begin{align}
U = \frac{1}{\sqrt{2}} \left( \begin{tabular}{cc} $I_2$ &  $I_2$ \\  $I_2$ &  $-I_2$  \end{tabular} \right)
\end{align}
матрицу $\Gamma_4 = \gamma_4$ можно привести к виду
\begin{align}
\gamma_4 = \left( \begin{tabular}{cc} $I_2$ &  $0$ \\  $0$ &  $-I_2$  \end{tabular}  \right) = \sigma_3 \otimes I_2,
\end{align}
не изменяя при этом вида выражений для матриц $\Gamma_i = \gamma_i$.

Условие положительной дефинитности заряда \eqref{ineq2} имеет в рассматриваемом случае ($n=0$) вид
\begin{align}
\left(\mathrm{Sp} (\gamma_4 \eta) \right)^2 -\left (\mathrm{Sp} \, \eta \right)^2 >0.
\label{pd_ch1}
\end{align}
Поскольку
\begin{align}
\mathrm{Sp} \, \eta = \mathrm{Sp} \,  \gamma_4= 0 , \quad \mathrm{Sp} (\gamma_4 \eta) = 4,
\nonumber
\end{align}
условие \eqref{pd_ch1} выполняется.

Необходимо указать на важное свойство, которое объединяет рассмотренные выше РВУ. Нетрудно видеть, что матрицы  $\Gamma_4$ в уравнениях Даффина–Кеммера для скалярной и векторной частиц, также как и в случае уравнения Дирака, приводимы к диагональному виду.

В работах [8; 9] показано, что из всех конечномерных уравнений с приводимой к диагональному виду матрицей $\Gamma_4$ только уравнения Даффина–Кеммера имеют положительно определенную энергию и только уравнение Дирака имеет положительно определенный заряд. Возможны, однако, уравнения с положительной энергией или зарядом, у которых $\Gamma_4$ неприводима к диагональному виду. Ниже мы рассмотрим примеры таких уравнений.

Самым простым и известным среди РВУ такого типа является уравнение Фирца–Паули для спина $s = \frac32$. Дадим его формулировку в подходе Гельфанда–Яглома. С этой целью возьмем схему зацеплений  \eqref{ls4} неприводимых представлений группы Лоренца и проанализируем ее с точки зрения возможности построения всевозможных теорий
спина $s=\frac32$.

Матрица $\Gamma_4$ уравнения \eqref{eq:nonzero_mass}, соответствующего набору представлений, содержащихся в \eqref{ls4},
будет иметь в базисе Гельфанда–Яглома вид
\begin{align}
\Gamma_4 &= \left(\begin{tabular}{cc} $C^{1/2} \otimes I_2$ & $0$ \\
$0$ & $C^{3/2} \otimes I_4$
\end{tabular} \right), \label{gam4_s2init} \\
C^{3/2} &=  \left( \begin{tabular}{cc}
$0$ & $c_{34}^{3/2}$  \\
 $c_{43}^{3/2}$  &  $0$
\end{tabular} \right), \quad
C^{1/2} = \left( \begin{tabular}{cccc}
$0$ & $c_{12}^{1/2}$ & $c_{13}^{1/2}$ & $0$ \\
$c_{21}^{1/2}$ & $0$ & $0$ & $c_{24}^{1/2}$ \\
$c_{31}^{1/2}$ & $0$ & $0$ & $c_{34}^{1/2}$ \\
$0$ & $c_{42}^{1/2} $ & $c_{43}^{1/2} $ & $0$
 \end{tabular} \right),
\end{align}
где использована нумерация $(0, \frac12) \sim 1$, $(\frac12, 0 ) \sim 2$, $(\frac12 , 1) \sim 3$, $(1, \frac12) \sim 4$. Матрица $\eta$ имеет аналогичную структуру
\begin{align}
\eta = (\eta^{1/2} \otimes I_2 ) \oplus (\eta^{3/2} \otimes I_4 ),
\nonumber
\end{align}
причем отличными от нуля элементами ее блоков $\eta^{1/2}$, $\eta^{3/2}$ являются
\begin{align}
\eta_{12}^{1/2} = \eta_{21}^{1/2}, \quad \eta_{34}^{1/2} = \eta_{43}^{1/2} =- \eta_{34}^{3/2} =- \eta_{43}^{3/2}.
\nonumber
\end{align}

Требование инвариантности теории относительно преобразований собственной группы Лоренца приводит в данном случае к соотношениям
\begin{align}
c_{34}^{3/2} = 2 c_{34}^{1/2}, \quad c_{43}^{3/2} = 2 c_{43}^{1/2}.
\label{invLor}
\end{align}

Для $P$-инвариантности строящегося РВУ необходимо выполнение равенств (см. условия \eqref{c1_cond}-\eqref{Pc2_cond})
\begin{align}
\begin{split}
c_{12}^{1/2} & = c_{21}^{1/2}, \quad c_{34}^{1/2} = c_{43}^{1/2}, \\
c_{13}^{1/2} & = c_{24}^{1/2}, \quad c_{31}^{1/2} = c_{42}^{1/2}, \quad c_{34}^{3/2} = c_{43}^{3/2}.
\end{split}
\label{invP}
\end{align}

Наконец, возможность получения искомого РВУ из инвариантной функции Лагранжа в соответствии с  \eqref{c_eta_cond}
накладывает на элементы $c_{ij}^s$ матрицы $\Gamma_4$ ограничения
\begin{align}
c_{12}^{1/2},c_{21}^{1/2}, c_{34}^{1/2},c_{43}^{1/2} \in \mathbb{R}
\label{invLag1}
\end{align}
и
\begin{align}
\begin{split}
c_{42}^{1/2} \eta_{21}^{1/2} &= (c_{13}^{1/2})^* \eta_{43}^{1/2} , \\
c_{31}^{1/2} \eta_{12}^{1/2} &= (c_{24}^{1/2})^* \eta_{34}^{1/2} .
\end{split}
\label{c_eta_cond2}
\end{align}

Для удобства дальнейших рассуждений перепишем условия  \eqref{c_eta_cond2} следующим образом:
\begin{align}
c_{42}^{1/2} =( c_{13}^{1/2})^* f , \quad c_{31}^{1/2} =( c_{24}^{1/2})^* f , \quad f = \frac{\eta_{34}^{1/2}}{\eta_{12}^{1/2}} =\frac{\eta_{43}^{1/2}}{\eta_{21}^{1/2}}.
\end{align}

Не уменьшая общности, параметр $f$ можно выбрать равным $+1$ или $-1$, что равносильно двум существенно различным способам задания матрицы $\eta$:
\begin{align}
- \eta_{12}^{1/2} &= - \eta_{34}^{1/2} = \eta_{34}^{3/2} = 1, \label{def_eta1} \\
\eta_{12}^{1/2} &= - \eta_{34}^{1/2} = \eta_{34}^{3/2} = 1 . \label{def_eta2}
\end{align}

На первом этапе рассмотрим всевозможные РВУ для спина $s=\frac32$, накладывая лишь ограничения \eqref{invLor}, \eqref{invLag1}, \eqref{c_eta_cond2}, то есть включая в множество полученных таким образом РВУ как  $P$-инвариантные, так и  $P$-неинвариантные уравнения. Для спинового блока $C^{3/2}$ существует, с точностью до эквивалентности, единственный вариант выбора его элементов: $c_{34}^{3/2}= c_{43}^{3/2}=1$. Отсюда согласно \eqref{invLor} имеем $c_{34}^{1/2}=c_{43}^{1/2} = \frac12$, и блоки $C^{1/2}$, $C^{3/2}$ принимают вид
\begin{align}
C^{1/2} = \left(  \begin{tabular}{cccc}
$0$ & $a$ & $c$ & $0$ \\
$b$ & $0$ & $0$ & $d$ \\
$f d^*$ & $0$ & $0$ & $1/2$ \\
$0$ & $ f c^* $ & $1/2 $ & $0$
 \end{tabular}\right) , \quad C^{3/2} = \left(  \begin{tabular}{cc} $0$ & $1$ \\ $1$ & $0$ \end{tabular} \right),
\label{c12c32}
\end{align}
где для краткости введены обозначения
\begin{align}
c_{12}^{1/2} = a, \quad c_{21}^{1/2} = b , \quad c_{13}^{1/2} = c , \quad c_{24}^{1/2} =d .
\nonumber
\end{align}

Чтобы на основе матриц \eqref{c12c32} получить РВУ, описывающее чистый спин $s = \frac32$, необходимо потребовать обращения в нуль всех корней характеристического уравнения для спинового блока $C^{1/2}$
\begin{align}
\lambda^4 - \left( f c^* d + f c d^* + a b+ \frac14\right) \lambda^2+ \left( \frac14 a b - \frac12 fa |d|^2  - \frac12 f b |c|^2 + |c|^2 |d|^2 \right) = 0.
\end{align}
Корни этого уравнения будут равными нулю при выполнении условий
\begin{align}
\begin{split}
&  \frac14 a b - \frac12 fa |d|^2  - \frac12 f b |c|^2 + |c|^2 |d|^2  = 0 , \\
&   f c^* d + f c d^* + a b+ \frac14  = 0 .
\end{split}
\label{zero_cond}
\end{align}

Структура \eqref{c12c32} спиновых блоков $C^{1/2}$, $C^{3/2}$ матрицы $\Gamma_4$ соответствует некоторому обобщенному РВУ, из которого при определенных способах выбора параметров  $a$, $b$, $c$, $d$ получаются формулировки тех или иных
конкретных уравнений для частицы со спином $\frac32$. Учитывая возможности \eqref{def_eta1} и \eqref{def_eta2} задания матрицы билинейной формы $\eta$, целесообразно ввести классы РВУ первого типа с матрицами
\begin{align}
C^{1/2} =  \left( \begin{tabular}{cccc}
$0$ & $a$ & $c$ & $0$ \\
$b$ & $0$ & $0$ & $d$ \\
$d^*$ & $0$ & $0$ & $1/2$ \\
$0$ & $c^* $ & $1/2 $ & $0$
 \end{tabular} \right), \quad C^{3/2} = \left(  \begin{tabular}{cc} $0$ & $1$ \\ $1$ & $0$ \end{tabular} \right) \quad \textup{для} \quad f=+1,
\label{C_type1}
\end{align}
и РВУ второго типа с матрицами
\begin{align}
C^{1/2} =  \left(  \begin{tabular}{cccc}
$0$ & $a$ & $c$ & $0$ \\
$b$ & $0$ & $0$ & $d$ \\
$-d^*$ & $0$ & $0$ & $1/2$ \\
$0$ & $-c^* $ & $1/2 $ & $0$
 \end{tabular}\right), \quad C^{3/2} = \left(  \begin{tabular}{cc} $0$ & $1$ \\ $1$ & $0$ \end{tabular} \right)  \quad \textup{для} \quad f=-1.
\label{C_type2}
\end{align}

В случае РВУ второго типа система \eqref{zero_cond} допускает решение
\begin{align}
a=b= - \frac12 , \quad c=d= \frac12,
\nonumber
\end{align}
которое приводит к матрице $\Gamma_4$ с блоками
\begin{align}
C^{1/2} = \frac12 \left(  \begin{tabular}{cccc}
$0$ & $-1$ & $1$ & $0$ \\
$-1$ & $0$ & $0$ & $1$ \\
$-1$ & $0$ & $0$ & $1$ \\
$0$ & $-1$ & $1$ & $0$
 \end{tabular}\right), \quad C^{3/2} = \left( \begin{tabular}{cc} $0$ & $1$ \\ $1$ & $0$ \end{tabular} \right).
\end{align}
Нетрудно убедиться, что при этом условия \eqref{invP} также выполняются. Таким образом, получается $P$-инвариантное РВУ, которое представляет собой уравнение Фирца–Паули в формализме Гельфанда–Яглома.

Матрица $\Gamma_4$ \eqref{gam4_s2init}, \eqref{C_type2} уравнения Фирца–Паули удовлетворяет минимальному уравнению
\begin{align}
\Gamma_4^2 (\Gamma_4^2 - 1) = 0
\end{align}
и, следовательно, неприводима к диагональному виду. Условие дефинитности заряда \eqref{ineq2} принимает в данном случае вид
\begin{align}
\left(\mathrm{Sp} (\Gamma_4^3 \eta) \right)^2 - \left(\mathrm{Sp} (\Gamma_4^2 \eta) \right)^2 >0.
\label{ch_def}
\end{align}

Учитывая явные выражения для блоков $\eta^{1/2}$, $\eta^{3/2}$ матрицы $\eta$
\begin{align}
\eta^{1/2} = \left(  \begin{tabular}{cccc}
$0$ & $1$ & $0$ & $0$ \\
$1$ & $0$ & $0$ & $0$ \\
$0$ & $0$ & $0$ & $-1$ \\
$0$ & $0$ & $-1$ & $0$
 \end{tabular}\right), \quad \eta^{3/2} = \left( \begin{tabular}{cc} $0$ & $1$ \\ $1$ & $0$ \end{tabular} \right),
\end{align}
получаем
\begin{align}
\mathrm{Sp} (\Gamma_4^2 \eta) =0, \quad \mathrm{Sp} (\Gamma_4^3 \eta)  \in \mathbb{R}\backslash \{  0 \}.
\end{align}
Таким образом, условие \eqref{ch_def} выполняется.

В рамках РВУ первого типа построить $P$-инвариантное РВУ, допускающее лагранжеву формулировку, нельзя. Действительно, условия  \eqref{invP} означают (в случае $f=+1$), что $a=b$, $c=d$, и тогда из \eqref{zero_cond} имеем $2 |c|^2 + a^2 + \frac14 =0$. Но это соотношение не может быть выполнено, поскольку параметр $a=c_{12}^{1/2}$, согласно \eqref{invLag1}, является вещественным.

Если же требование \eqref{invP} не накладывать, то у системы \eqref{zero_cond} при $f=-1$ имеются решения
\begin{align}
a=b=c=-d = \frac12 \quad \mathrm{and} \quad a=b=-c=d = \frac12,
\nonumber
\end{align}
которые приводят к спиновым блокам
\begin{align}
C^{1/2} = \frac12 \left(  \begin{tabular}{cccc}
$0$ & $1$ & $\pm 1$ & $0$ \\
$1$ & $0$ & $0$ & $\mp 1$ \\
$\mp 1$ & $0$ & $0$ & $1$ \\
$0$ & $\pm 1 $ & $1$ & $0$
 \end{tabular}\right), \quad C^{3/2} = \left(  \begin{tabular}{cc} $0$ & $1$ \\ $1$ & $0$ \end{tabular} \right),
\end{align}
где знаки «+» и «–» соответствуют первому и второму решениям соответственно. Полученные указанным способом РВУ описывают частицу со спином $s=\frac32$, допускают лагранжеву формулировку, но не являются инвариантными по отношению к операции пространственного отражения.

Анализ схемы зацеплений \eqref{ls4} в подходе Гельфанда–Яглома позволяет сделать одно важное заключение. В рамках РВУ первого типа, полагая
\begin{align}
a=b= \frac12 , \quad c=d= \frac{\sqrt{3}}{2},
\end{align}
приходим к матрице $\Gamma_4$, минимальный полином которой имеет вид \eqref{min_pol_dir}. Такой выбор согласуется с условиями \eqref{invLor}-\eqref{c_eta_cond2} и приводит к РВУ дираковского типа, описывающему {\it переменный спин} $\frac12, \frac32$. В спин-тензорной форме оно имеет хорошо известный вид
\begin{align}
(\gamma_{\mu} \partial_{\mu} + m)_{\alpha \beta} \psi_{\nu}^{\beta} =0,
\end{align}
являясь исходным для получения уравнения Рариты–Швингера путем наложения условий
\begin{align}
\gamma_{\nu} \psi_{\nu}^{\beta} = 0 , \quad \partial_{\nu} \psi_{\nu}^{\beta} = 0.
\end{align}
Поскольку эти условия, вырезающие лишний спин $\frac12$, $P$-инвариантны и не изменяют структуры \eqref{def_eta1} матрицы билинейной формы, то очевидно, что уравнение Рариты–Швингера относится к РВУ первого типа. Уравнение же Фирца–Паули относится ко второму типу. Таким образом, обычно встречающееся в литературе [15; 16] отождествление этих уравнений нуждается в критическом переосмыслении.

\section{РВУ с расширенным набором представлений \\ группы Лоренца и внутренняя структура микрообъектов}

\hspace{1.25cm}
Характерной особенностью всех рассмотренных выше РВУ является то, что в них используется набор неприводимых представлений группы Лоренца, минимально необходимый для построения теории данного спина. При этом, в соответствии с идеологией релятивистской квантовой механики, трактующей элементарные частицы как точечные бесструктурные объекты, такие РВУ учитывают только спиновые свойства частиц. Возможность описания иных внутренних свойств частиц в ортодоксальном варианте теории РВУ не предусматривается.

Отказ от требования минимальности используемых наборов представлений группы Лоренца открывает новые возможности метода теории РВУ с точки зрения пространственно-временного (геометризованного) описания внутренних свойств частиц. Получение нераспадающихся по группе Лоренца уравнений, способных отображать внутреннюю структуру частицы с заданным спином $s$, возможно либо за счет включения в схему зацеплений представлений с более высокими весами, либо за счет использования кратных представлений группы Лоренца. В настоящей главе мы покажем, как в рамках подхода теории РВУ с расширенными наборами представлений группы Лоренца может осуществляться описание внутренней электромагнитной структуры частиц с низшими спинами.

Впервые РВУ для частицы со спином $s = \frac12$, которое возникает за счет привлечения дополнительных по отношению к биспинору неприводимых компонент в пространстве представления волновой функции, было предложено Петрашем [17]. Дадим сжатое изложение теории уравнения Петраша в подходе Гельфанда–Яглома. Для этого рассмотрим схему зацеплений
\begin{align}
\begin{tabular}{ccc}
$ (0,\frac12)'$  & --- &  $(\frac12 ,0)'$ \\
$|$ & & $|$ \\
$(\frac12 ,1)$ & --- &  $(1,\frac12 )$ \\
$|$ & & $|$ \\
$ (0,\frac12 )$  & --- &  $(\frac12 ,0)$ .
\end{tabular}
\label{scheme31}
\end{align}

Пронумеруем неприводимые представления,  содержащиеся в \eqref{scheme31}:
\begin{align}
\begin{split}
& (0, \frac12) \sim 1, \quad  (0, \frac12)' \sim 2 , \quad (1, \frac12) \sim 3 , \\
& (\frac12 ,0) \sim 4 , \quad (\frac12 , 0)' \sim 5 , \quad (\frac12 ,1) \sim 6 .
\end{split}
\end{align}
Тогда для спиновых блоков $C^{1/2}$, $C^{3/2}$ матрицы $\Gamma_4$
\begin{align}
\Gamma_4 = (C^{1/2} \otimes I_2) \oplus (C^{3/2} \otimes I_4)
\end{align}
получим выражения
\begin{align}
C^{1/2} = \left( \begin{tabular}{cccccc}
$0$ & $0$ & $0$ & $c_{14}^{1/2}$ & $0$ & $c_{16}^{1/2}$ \\
$0$ & $0$ & $0$ & $0$ & $c_{25}^{1/2}$ & $c_{26}^{1/2}$ \\
$0$ & $0$ & $0$ & $c_{34}^{1/2}$ & $c_{35}^{1/2}$ & $c_{36}^{1/2}$ \\
$c_{41}^{1/2}$ & $0$ & $c_{43}^{1/2}$ & $0$ & $0$ & $0$ \\
$0$ & $c_{52}^{1/2}$ & $c_{53}^{1/2}$ & $0$ & $0$ & $0$ \\
$c_{61}^{1/2}$ & $c_{62}^{1/2}$ & $c_{63}^{1/2}$ & $0$ & $0$ & $0$
\end{tabular} \right), \quad
C^{3/2} = \left( \begin{tabular}{cc}
$0$ & $c_{36}^{3/2}$ \\
$c_{63}^{3/2} $ & $0$ \end{tabular} \right).
\label{eq34}
\end{align}
Для исключения спина $\frac32$ надо положить
\begin{align}
c_{36}^{3/2} = c_{63}^{3/2} =0, \quad \textup{или} \quad C^{3/2} =0.
\label{eq35}
\end{align}
Отсюда в силу \eqref{c_cond} следует
\begin{align}
c_{36}^{1/2} = c_{63}^{1/2} =0.
\label{eq36}
\end{align}
Условие \eqref{c1_cond} $P$-инвариатности теории приводит к соотношениям
\begin{align}
\begin{split}
& c_{14}^{1/2} =c_{14}^{1/2} , \quad  c_{25}^{1/2} =c_{52}^{1/2} , \quad  c_{16}^{1/2} =c_{43}^{1/2} , \\
& c_{26}^{1/2} =c_{53}^{1/2} , \quad  c_{34}^{1/2} =c_{61}^{1/2} , \quad  c_{35}^{1/2} =c_{62}^{1/2} .
\end{split}
\label{eq37}
\end{align}
Возможность лагранжевой формулировки (формула \eqref{c_eta_cond}) дает
\begin{align}
c_{14}^{1/2}, \,\, c_{25}^{1/2} \in \mathbb{R}; \quad
c_{34}^{1/2} = \frac{\eta_{63}^{1/2}}{\eta_{14}^{1/2}} (c_{16}^{1/2})^*,  \quad
c_{35}^{1/2} = \frac{\eta_{63}^{1/2}}{\eta_{25}^{1/2}} (c_{26}^{1/2})^*.
\label{eq38}
\end{align}

С учетом ограничений \eqref{eq35}--\eqref{eq38} для спинового блока $C^{1/2}$ получится выражение
\begin{align}
C^{1/2} = \left( \begin{tabular}{cc} $0$ & $C$ \\ $C$ & $0$ \end{tabular} \right), \quad
C = \left( \begin{tabular}{ccc}
$c_1$  & $0$ & $c_3$ \\
$0$  & $c_2$ & $c_4$ \\
$f_1 c_3^*$  & $f_2 c_4^*$ & $0$
 \end{tabular} \right),
\label{eq39}
\end{align}
где для упрощения записи использованы обозначения
\begin{align}
\begin{split}
c_1 = c_{14}^{1/2}, \quad c_2 & = c_{25}^{1/2}, \quad c_3 =c_{16}^{1/2} , \quad c_4 =c_{26}^{1/2} , \\
f_1 &= \frac{\eta_{63}^{1/2}}{\eta_{14}^{1/2}}, \quad f_2 = \frac{\eta_{63}^{1/2}}{\eta_{25}^{1/2}}.
\end{split}
\label{eq310}
\end{align}

Характеристическое уравнение для блока $C$ имеет вид
\begin{align}
\lambda^3 - (c_1 + c_2) \lambda^2 + (c_1 c_2 - f_1 |c_3|^2 - f_2 |c_4|^2) \lambda + f_1 c_2 |c_3|^2 + f_2 c_1 |c_4|^2 =0 .
\end{align}
Чтобы получить одно значение массы, надо наложить условия
\begin{align}
\begin{split}
c_1 c_2 - f_1 |c_3|^2 - f_2 |c_4|^2 &=0 , \\
f_1 c_2 |c_3|^2 + f_2 c_1 |c_4|^2 &=0.
\end{split}
\label{eq312}
\end{align}
При этом ненулевое собственное значение блока $C$, не уменьшая общности, можно выбрать равным единице:
\begin{align}
\lambda = c_1 + c_2 =1.
\label{eq313}
\end{align}
Такой выбор приводит к следующим минимальным полиномам для спинового блока $C^{1/2}$ и матрицы $\Gamma_4$
\begin{align}
(C^{1/2})^2 [(C^{1/2})^2-1]=0, \quad \Gamma_4^2 (\Gamma_4^2 -1) =0.
\end{align}

Остается наложить условие дефинитности заряда \eqref{ineq2}, которое в рассматриваемом случае ($C^{3/2} = 0$, $n = 2$) принимает вид
\begin{align}
\mathrm{Sp} \, \, \left( (C^{1/2} )^3 \eta^{1/2} \right) \neq 0 ,
\end{align}
где
\begin{align}
\eta^{1/2} = \left( \begin{tabular}{cc} $0$ & $\eta'$ \\ $\eta'$ & $0$ \end{tabular} \right), \quad
\eta' = \left( \begin{tabular}{ccc}
$\eta_{14}^{1/2}$  & $0$ & $0$ \\
$0$  & $\eta_{25}^{1/2}$ & $0$ \\
$0$  & $0$ & $\eta_{36}^{1/2}$
 \end{tabular} \right).
\end{align}
Отсюда с учетом выражения \eqref{eq39} получаем неравенство
\begin{align}
\eta_{14}^{1/2} c_1^3 + \eta_{25}^{1/2} c_2^3 + (2 \eta_{63}^{1/2} + \eta_{14}^{1/2}) c_1 |c_3|^2 +(2 \eta_{63}^{1/2} + \eta_{25}^{1/2})  c_2 |c_4|^2 \neq 0 .
\label{eq317}
\end{align}

Совместное выполнение условий \eqref{eq312}, \eqref{eq313}, \eqref{eq317} можно обеспечить, выбирая, например
\begin{align}
& c_1 = \frac13, \quad c_2 = \frac23 , \quad  c_3 = \frac{\sqrt{2}}{3}, \quad  c_4 =\frac23 , \\
& \eta_{14}^{1/2} =-1, \quad \eta_{25}^{1/2} = 1, \quad \eta_{36}^{1/2} = 1.
\end{align}
При этом матрицы $C$ и $\eta'$ принимают вид
\begin{align}
C = \left( \begin{tabular}{ccc}
$1/3$  & $0$ & $\sqrt{2}/3$ \\
$0$  & $2/3$ & $2/3$ \\
$-\sqrt{2}/3$  & $2/3$ & $0$
 \end{tabular} \right), \quad
\eta' = \left( \begin{tabular}{ccc}
$-1$  & $0$ & $0$ \\
$0$  & $1$ & $0$ \\
$0$  & $0$ & $1$
 \end{tabular} \right).
\end{align}

Таким образом, мы получили 20–компонентное РВУ со схемой зацеплений \eqref{scheme31}, с недиагонализируемой матрицей $\Gamma_4$, которое описывает частицу со спином $s = \frac12$ и удовлетворяет необходимым физическим требованиям.

Теперь покажем, что на основе схемы зацеплений \eqref{scheme31} можно построить также РВУ для частицы со спином $s = \frac32$ [18]. Сохраняя прежнюю нумерацию неприводимых представлений, содержащихся в \eqref{scheme31}, опять приходим к общему виду \eqref{eq34} спиновых блоков $C^{1/2}$, $C^{3/2}$ матрицы $\Gamma_4$ в базисе Гельфанда–Яглома.

Условия \eqref{c_cond} релятивистской инвариантности приводят к ограничениям
\begin{align}
c_{36}^{3/2} = 2 c_{36}^{1/2} , \quad c_{63}^{3/2}= 2 c_{63}^{1/2}.
\label{eq321}
\end{align}
Инвариантность РВУ относительно пространственных отражений помимо соотношений \eqref{eq37} дает
\begin{align}
c_{36}^{1/2} = c_{63}^{1/2}, \quad
c_{36}^{3/2} = c_{63}^{3/2}.
\label{eq322}
\end{align}
Возможность лагранжевой формулировки теории дополняет условия \eqref{eq38} ограничением
\begin{align}
c_{36}^{1/2} \in \mathbb{R} .
\label{eq323}
\end{align}
Поскольку нас в данном случае интересует спин $\frac32$, не уменьшая общности, можно положить
\begin{align}
c_{36}^{3/2} = c_{63}^{3/2} = 1.
\end{align}
При таком выборе условия \eqref{eq321}--\eqref{eq323} приводят к спиновым блокам $C^{1/2}$, $C^{3/2}$ вида
\begin{align}
C^{1/2} &= \left(\begin{tabular}{cc} $0$ & $C$ \\ $C$ & $0$ \end{tabular} \right) , \quad
C = \left(\begin{tabular}{ccc}
$c_1$ & $0$ & $c_3$ \\
$0$ & $c_2$ & $c_4$  \\
$f_1 c_3^*$ & $f_2 c_4^*$ & $1/2$
\end{tabular} \right), \label{eq325} \\
C^{3/2} &= \left(\begin{tabular}{cc} $0$ & $1$ \\ $1$ & $0$ \end{tabular} \right),
\label{eq326}
\end{align}
где использованы обозначения \eqref{eq310}.

Характеристическое уравнение для матрицы $C$ имеет вид
\begin{align}
\lambda^3 - \left(c_1 +c_2 +\frac12 \right) \lambda^2 + \left(\frac{c_1 +c_2}{2}+ c_1 c_2 - f_1 |c_3|^2 - f_2 |c_4|^2 \right) \lambda & \nonumber \\
-\frac{c_1 c_2}{2} +f_1 c_2 |c_3|^2 + f_2 c_1 |c_4|^2 &=0  .
\label{eq327}
\end{align}
Для исключения состояний со спином $\frac12$ надо потребовать, чтобы все собственные значения блока $C^{1/2}$ были равными нулю. Это требование приводит к условиям
\begin{align}
\begin{split}
c_1 +c_2 +\frac12  &=0, \\
\frac{c_1 +c_2}{2}+ c_1 c_2 - f_1 |c_3|^2 - f_2 |c_4|^2 &=0, \\
-\frac{c_1 c_2}{2} +f_1 c_2 |c_3|^2 + f_2 c_1 |c_4|^2 &=0,
\end{split}
\label{eq328}
\end{align}
в которых числа $f_1$, $f_2$ могут независимо друг от друга принимать значения $+1$ или $-1$.

Нетрудно видеть, что условия \eqref{eq328} совместны, причем здесь (как и для спина $\frac12$) в выборе параметров $c_1$, $c_2$, $c_3$, $c_4$, $f_1$, $f_2$ существует достаточно широкий произвол. Но при этом во всех случаях минимальные уравнения для спиновых блоков $C^{1/2}$ \eqref{eq325}, $C^{3/2}$ \eqref{eq326} и матрицы $\Gamma_4$ имеют один и тот же вид
\begin{align}
& (C^{1/2})^3 =0, \quad (C^{3/2})^2-1 =0 , \label{eq329} \\
& \qquad \qquad \Gamma_4^3 (\Gamma_4^2 -1) =0. \label{eq330}
\end{align}

Поскольку в данной работе мы не ставим цель рассмотреть все имеющиеся возможности построения РВУ для спина $\frac32$ на основе схемы зацеплений \eqref{scheme31}, то положим, например:
\begin{align}
f_1 &= -1, \quad f_2 = -1 , \label{eq331} \\
c_1 &= \frac12, \quad c_2 = -1, \quad |c_3| = \frac{1}{2\sqrt{3}}, \quad |c_4| = \sqrt{\frac23} . \label{eq332}
\end{align}
Такой выбор приводит к матрице билинейной формы с блоками
\begin{align}
\eta^{1/2} =\left(\begin{tabular}{cccccc}
$0$ & $0$ & $0$ & $1$ & $0$ & $0$ \\
$0$ & $0$ & $0$ & $0$ & $1$ & $0$ \\
$0$ & $0$ & $0$ & $0$ & $0$ & $-1$ \\
$1$ & $0$ & $0$ & $0$ & $0$ & $0$ \\
$0$ & $1$ & $0$ & $0$ & $0$ & $0$ \\
$0$ & $0$ & $-1$ & $0$ & $0$ & $0$
\end{tabular} \right) , \quad
\eta^{3/2} = \left(\begin{tabular}{cc} $0$ & $1$ \\ $1$ & $0$ \end{tabular} \right),
\label{eq333}
\end{align}
и спиновому блоку $C^{1/2}$ матрицы $\Gamma_4$
\begin{align}
C^{1/2}  =\left(\begin{tabular}{cccccc}
$0$ & $0$ & $0$ & $\frac12$ & $0$ & $\frac{1}{2 \sqrt{3}}$ \\
$0$ & $0$ & $0$ & $0$ & $-1$ & $\sqrt{\frac23}$ \\
$0$ & $0$ & $0$ & $-\frac{1}{2 \sqrt{3}}$ & $-\sqrt{\frac23}$ & $\frac12$ \\
$\frac12$ & $0$ & $\frac{1}{2 \sqrt{3}}$ & $0$ & $0$ & $0$ \\
$0$ & $-1$ & $\sqrt{\frac23}$ & $0$ & $0$ & $0$ \\
$-\frac{1}{2 \sqrt{3}}$ & $-\sqrt{\frac23}$ & $\frac12$ & $0$ & $0$ & $0$
\end{tabular} \right).
\label{eq334}
\end{align}

Условие дефинитности заряда в соответствии с \eqref{eq330} ($n =3$) принимает вид
\begin{align}
\left( \mathrm{Sp} \, (\Gamma_4^4 \eta )\right)^2 -\left(  \mathrm{Sp} \, (\Gamma_4^3 \eta ) \right)^2 <0 .
\label{ineq43}
\end{align}
Используя полученные явные выражения для матриц $\Gamma_4$ и $\eta$, находим
\begin{align}
  \mathrm{Sp} \, (\Gamma_4^4 \eta ) =0, \quad  \mathrm{Sp} \, (\Gamma_4^3 \eta ) =8,
  \label{eq336}
\end{align}
откуда следует, что неравенство \eqref{ineq43} выполняется.

Для построения РВУ, описывающих микрообъекты со спинами $s = 0, 1$ и отличающихся от известных уравнений Даффина–Кеммера, рассмотрим набор неприводимых представлений группы Лоренца [19; 20]
\begin{align}
(0,0) \oplus 2 (\frac12, \frac12) \oplus (0,1) \oplus (1,0),
\label{eq337}
\end{align}
образующих схему зацеплений
\begin{align}
\begin{tabular}{ccccc}
&  & $(0,0)$ &  & \\
& & $|$ & & \\
$(0,1)$ & --- & $2 (\frac12, \frac12)$ & --- & $(1,0)$,
\end{tabular}
\label{eq338}
\end{align}
где векторное представление $(\frac12, \frac12)$ имеет кратность, равную двум (в дальнейшем для различения этих представлений одно из них будем помечать штрихом).

Блочная структура матрицы $\Gamma_4$ РВУ, соответствующего схеме \eqref{eq337}, в базисе Гельфанда–Яглома имеет вид \eqref{gam4_s2}.
Пронумеровав содержащиеся в  \eqref{eq337} неприводимые компоненты способом
\begin{align}
(0,0) \sim 1, \quad  (\frac12 , \frac12)' \sim 2 , \quad  (\frac12 , \frac12) \sim 3  ,\quad (0,1) \sim 4, \quad (1,0) \sim 5 ,
\label{eq339}
\end{align}
после применения условий релятивистской и $P$-инвариантности теории получим для блоков $C^{0}$, $C^{1}$ выражения
\begin{align}
C^0 = \left(\begin{tabular}{ccc}
$0$ & $c_{12}^0$ & $c_{13}^0$ \\
$c_{21}^0$ & $0$ & $0$ \\
$c_{31}^0$ & $0$ & $0$
\end{tabular}\right) , \quad
C^1 = \left( \begin{tabular}{cccc}
$0$ & $0$ & $c_{24}^1$ & $c_{24}^1$ \\
$0$ & $0$ & $c_{34}^1$ & $c_{34}^1$ \\
$c_{42}^1$ & $c_{43}^1$ & $0$ & $0$ \\
$c_{42}^1$ & $c_{43}^1$ & $0$ & $0$
\end{tabular} \right).
\label{eq340}
\end{align}

Ненулевые элементы матрицы билинейной формы $\eta$, которая в базисе Гельфанда–Яглома имеет вид
\begin{align}
\eta &= \left( \begin{tabular}{cc} $\eta^0$ & $0$ \\ $0$ & $\eta^1 \otimes I_3$ \end{tabular} \right)  ,\label{eq341}  \\
\eta^0 &= \left( \begin{tabular}{ccc}
$\eta_{11}^0$ & $0$ & $0$ \\
$0$ & $\eta_{22}^0$ & $0$ \\
$0$ & $0$ & $\eta_{33}^0$
\end{tabular} \right)  , \quad
\eta^1 = \left( \begin{tabular}{cccc}
$\eta_{22}^1$ & $0$ & $0$ & $0$ \\
$0$ & $\eta_{33}^1$ & $0$ & $0$ \\
$0$ & $0$ & $0$ & $\eta_{45}^1$ \\
$0$ & $0$ & $\eta_{54}^1$ & $0$
\end{tabular} \right), \nonumber
\end{align}
выберем следующим образом:
\begin{align}
\eta_{11}^0 = - \eta_{22}^0 = \eta_{33}^0 = \eta_{22}^1 = -\eta_{33}^1 =- \eta_{45}^1 = - \eta_{54}^1 = 1 .
\label{eq342}
\end{align}
Тогда условие \eqref{c_eta_cond} возможности лагранжевой формулировки строящегося РВУ приводит к соотношениям
\begin{align}
c_{21}^0 = -(c_{12}^0)^* , \quad c_{31}^0 = (c_{13}^0)^* , \quad c_{42}^1 = -(c_{24}^1)^* , \quad c_{43}^1 = (c_{34}^1)^* .
\label{eq343}
\end{align}

Остающийся произвол в выборе элементов матрицы $\Gamma_4$ можно использовать для получения РВУ с требуемым значением спина.

Так, полагая
\begin{align}
c_{12}^0 = 0, \quad c_{13}^0 = c_{24}^1 = c_{34}^1 = 1,
\label{eq344}
\end{align}
придем к спиновым блокам
\begin{align}
C^0 = \left(\begin{tabular}{ccc}
$0$ & $0$ & $1$ \\
$0$ & $0$ & $0$ \\
$1$ & $0$ & $0$
\end{tabular}\right) , \quad
C^1 = \left( \begin{tabular}{cccc}
$0$ & $0$ & $1$ & $1$ \\
$0$ & $0$ & $1$ & $1$ \\
$-1$ & $1$ & $0$ & $0$ \\
$-1$ & $1$ & $0$ & $0$
\end{tabular} \right).
\label{eq345}
\end{align}
Нетрудно убедиться в справедливости минимальных уравнений
\begin{align}
& (C^{1})^3 =0, \quad C^0 [ (C^{0})^2-1] =0 , \label{eq346} \\
& \qquad \qquad \Gamma_4^3 (\Gamma_4^2 -1) =0. \label{eq347}
\end{align}

Из  \eqref{eq346} вытекает, что состоянию со спином $s = 0$ соответствует одно значение массы, а все собственные значения блока $C^{1}$ равны нулю, то есть состояния со спином $s = 1$ отсутствуют. Таким образом, получаем РВУ для микрообъекта со спином $s = 0$ и одним значением массы.

Используя выражения \eqref{eq341}, \eqref{eq342}, \eqref{eq345} для матриц $\Gamma_4$ и $\eta$, нетрудно убедиться, что
\begin{align}
  \mathrm{Sp} \, (\Gamma_4^3 \eta ) =0, \quad  \mathrm{Sp} \, (\Gamma_4^4 \eta ) =2.
  \label{eq348}
\end{align}
Отсюда вытекает справедливость условия дефинитности энергии, имеющего для данного РВУ вид неравенства (смотри \eqref{ineq1}, $n = 3$)
\begin{align}
(-1)^4 \left[ \left( \mathrm{Sp} \, (\Gamma_4^4 \eta )\right)^2 -\left(  \mathrm{Sp} \, (\Gamma_4^3 \eta ) \right)^2  \right] > 0 .
\label{eq349}
\end{align}

Тензорная формулировка построенного РВУ имеет вид
\begin{align}
\begin{split}
\partial_{\mu} \psi_{\mu} + m \psi_0 = 0, \\
\partial_{\nu} \psi_{[\mu \nu ]} + \partial_{\mu} \psi_0 + m \psi_{\mu} =0, \\
-\partial_{\nu} \psi_{[\mu \nu ]} + m \psi'_{\mu} =0, \\
-\partial_{\mu} \psi_{\nu}+ \partial_{\nu} \psi_{\mu} -\partial_{\mu} \psi'_{\nu}+ \partial_{\nu} \psi'_{\mu} + m \psi_{[\mu \nu]} =0 ,
\end{split}
\label{eq350}
\end{align}
где $\psi_0$ – скаляр, $\psi_{\mu}$ и $\psi'_{\mu}$ – 4-векторы, $\psi_{[\mu \nu]}$ – антисимметричный тензор второго ранга. Из \eqref{eq350} нетрудно получить уравнение второго порядка
\begin{align}
(\square- m^2) \psi_0 = 0 ,
\label{eq351}
\end{align}
означающее, что система \eqref{eq350} действительно описывает микрочастицу с ненулевой массой и спином $s = 0$.

Для построения на основе схемы зацеплений \eqref{eq337} РВУ для микрочастицы со спином $s = 1$ выберем элементы матрицы \eqref{eq341} следующим образом:
\begin{align}
\eta_{11}^0 =\eta_{22}^0 =-\eta_{33}^0 = -\eta_{22}^1 =\eta_{33}^1=\eta_{45}^1 =\eta_{54}^1 =  1 .
\label{eq352}
\end{align}
Тогда в соответствии с условиями \eqref{c_eta_cond} будем иметь
\begin{align}
c_{21}^0 = (c_{12}^0)^* , \quad c_{31}^0 =-  (c_{13}^0)^* , \quad c_{42}^1 =- (c_{24}^1)^* , \quad c_{43}^1 =  (c_{34}^1)^* .
\label{eq353}
\end{align}

В рамках остающегося произвола в выборе элементов спиновых блоков $C^{0}$, $C^{1}$ \eqref{eq340} возьмем
\begin{align}
\begin{split}
c_{12}^0 &= c_{13}^0 = c_{21}^0 = - c_{31}^0 = 1,\\
c_{24}^1 &= c_{42}^1 =0 ,\quad c_{34}^1 = c_{43}^1 = \frac{1}{\sqrt{2}}.
\end{split}
\label{eq354}
\end{align}
В результате получаем для матриц $\eta^0$, $\eta^1$, $C^{0}$, $C^{1}$ окончательные выражения
\begin{align}
\eta^0 &= \left( \begin{tabular}{ccc}
$1$ & $0$ & $0$ \\
$0$ & $1$ & $0$ \\
$0$ & $0$ & $-1$
\end{tabular} \right)  , \quad
\eta^1 = \left( \begin{tabular}{cccc}
$-1$ & $0$ & $0$ & $0$ \\
$0$ & $1$ & $0$ & $0$ \\
$0$ & $0$ & $0$ & $1$ \\
$0$ & $0$ & $1$ & $0$
\end{tabular} \right) ,
\label{eq355}
\end{align}
\begin{align}
C^0 = \left(\begin{tabular}{ccc}
$0$ & $1$ & $1$ \\
$1$ & $0$ & $0$ \\
$-1$ & $0$ & $0$
\end{tabular}\right) , \quad
C^1 = \frac{1}{\sqrt{2}} \left( \begin{tabular}{cccc}
$0$ & $0$ & $0$ & $0$ \\
$0$ & $0$ & $1$ & $\pm 1$ \\
$0$ & $1$ & $0$ & $0$ \\
$0$ & $\pm 1$ & $0$ & $0$
\end{tabular} \right).
\label{eq356}
\end{align}

Легко убедиться, что минимальные уравнения для спиновых блоков \eqref{eq356} матрицы
$\Gamma_4$ имеют вид
\begin{align}
(C^0)^3 = 0 , \quad C^1 [(C^1)^2 - 1] =0,
\label{eq357}
\end{align}
то есть данное РВУ действительно описывает микрочастицу со спином $s = 1$.

Из \eqref{eq357} следует, что минимальное уравнение для матрицы $\Gamma_4$ совпадает с аналогичным по смыслу уравнением \eqref{eq347} для скалярной частицы. Поэтому и условие дефинитности энергии совпадает здесь с \eqref{eq349}. Используя определения \eqref{eq355}, \eqref{eq356}, убеждаемся в справедливости условия \eqref{eq349} и в случае обсуждаемого РВУ для векторной частицы.

Тензорная формулировка полученного РВУ с расширенным набором представлений для частицы со спином $1$ такова:
\begin{align}
\begin{split}
\partial_{\mu} \psi_{\mu} +\partial_{\mu} \psi'_{\mu} +m \psi_0 = 0, \\
\partial_{\lambda} \psi_{[\mu \lambda ]} -\partial_{\mu} \psi_0 +m \psi_{\mu} = 0, \\
\partial_{\mu} \psi_0 +m \psi'_{\mu} = 0, \\
-\partial_{\mu} \psi_{\nu} +\partial_{\nu} \psi'_{\mu} +m \psi_{[\mu \nu]} = 0.
\end{split}
\label{eq358}
\end{align}
Из системы \eqref{eq358} вытекают уравнения
\begin{align}
(\square - m^2) (\psi_{\mu} + \psi'_{\mu}) =0 , \quad \partial_{\mu} (\psi_{\mu} + \psi'_{\mu}) = 0 ,
\label{eq359}
\end{align}
которые однозначно указывают на то, что данная система действительно описывает векторную частицу с ненулевой массой.

Иные варианты расширенных РВУ для частиц с низшими спинами предлагаются в работах [21] (спин $\frac12$), [22] (спин $0$), [23] (спин $1$).

Вопрос о физической неэквивалентности РВУ с минимальным и расширенным наборами представлений группы Лоренца для конкретных уравнений впервые был
рассмотрен в работах [24; 25] (спин $\frac12$), [19; 20] (спины $0, 1$), [26] (спин $\frac32$). Общее исследование этого вопроса для частиц с произвольным спином $s$ и одной массой, взаимодействующих с электромагнитным полем, проведено в работах [27; 28].

Суть и основные результаты проведенного в [27; 28] исследования заключаются в следующем.

Сначала рассматриваются минимальное и расширенное уравнения для свободных частиц
\begin{align}
(\Gamma_{\mu}^{(0)} \partial_{\mu} + m) \Psi_0 (x) = 0, \label{eq360} \\
(\Gamma_{\mu}^{(1)} \partial_{\mu} + m) \Psi_1 (x) = 0, \label{eq361}
\end{align}
заданные в пространствах неприводимых представлений группы Лоренца $T_0$ и $T_1= T_0 + T'$ соответственно. Далее находится вид операторов $R$ и $K$, переводящих $\psi_0$ в $\psi_1$ и обратно:
\begin{align}
R & = (A, 0  ) , \quad    K= \left(\begin{tabular}{c} $F$ \\ $G$ \end{tabular} \right) , \label{eq362}  \\
R \Psi_1 &= (A, 0  )\left(\begin{tabular}{c} $\Psi_1^0$ \\ $\Psi_1^1$ \end{tabular} \right) = A \Psi_1^0 = \Psi_0 , \label{eq363}  \\
K \Psi_0 & =\left(\begin{tabular}{c} $F$ \\ $G$ \end{tabular} \right) \Psi_0 =\left(\begin{tabular}{c} $F \Psi_0$ \\ $G \Psi_0$ \end{tabular} \right)   =\left(\begin{tabular}{c} $\Psi_1^0$ \\ $\Psi_1^1$ \end{tabular} \right)  = \Psi_1 . \label{eq364}
\end{align}
Здесь $A$, $F$ – прямоугольные числовые матрицы, удовлетворяющие условию
\begin{align}
A F = I ,
\label{eq365}
\end{align}
матрица $G$ в общем случае содержит операторы дифференцирования. При этом для операторов $R$ и $K$ будем иметь
\begin{align}
R \Gamma_{\mu}^{(1)} K = \Gamma_{\mu}^{(0)} + B_{\mu},
\label{eq366}
\end{align}
где матрицы $B_{\mu}$ удовлетворяют уравнению
\begin{align}
B_{\mu} \partial_{\mu} \Psi_0 (x) = 0 .
\label{eq367}
\end{align}

Затем рассматриваются уравнения
\begin{align}
(\Gamma_{\mu}^{(0)}  \mathcal{ D}_{\mu}   + m) \Phi_0 (x) =0, \label{eq368} \\
(\Gamma_{\mu}^{(1)}  \mathcal{ D}_{\mu}   + m) \Phi_1 (x) =0, \label{eq369}
\end{align}
которые описывают частицы, взаимодействующие с электромагнитным полем $A_{\mu}(x)$, введенным минимальным образом:
\begin{align}
\partial_{\mu} \to \mathcal{D}_{\mu} = \partial_{\mu} - i e A_{\mu} .
\label{eq370}
\end{align}
Оператор $R$, осуществляющий переход от функции $\Phi_1(x)$ к $\Phi_0 (x)$, имеет тот же вид,
что и в случае свободной частицы. Для оператора $K'$, осуществляющего обратный переход в случае взаимодействующей частицы, получим
\begin{align}
K' = \left( \begin{tabular}{c} $F$ \\ $G+G'$ \end{tabular} \right),
\label{eq371}
\end{align}
где добавка $G'$ обусловлена подстановкой \eqref{eq370}. Уравнение \eqref{eq368} при этом приводится к виду
\begin{align}
(R \Gamma_{\mu}^{(1)}  \mathcal{ D}_{\mu}  K' + m) \Phi_0 (x) =0 ,
\label{eq372}
\end{align}
или
\begin{align}
(\Gamma_{\mu}^{(0)}  \mathcal{ D}_{\mu}  + Q + m) \Phi_0 (x) =0 ,
\label{eq373}
\end{align}
где
\begin{align}
Q \sim R \Gamma_{\mu}^{(1)} \mathcal{D}_{\mu} G' + B_{\mu} \mathcal{ D}_{\mu} .
\label{eq374}
\end{align}
Таким образом, после приведения уравнения \eqref{eq369} к уравнению типа \eqref{eq368} относительно волновой функции с минимально необходимым числом компонент в последнем появляется дополнительное слагаемое $Q$ \eqref{eq374}. Для частиц со спинами $s = \frac12$ и $\frac32$, описываемых рассмотренными выше РВУ с расширенными наборами представлений, данное слагаемое принимает вид
\begin{align}
Q  \sim \frac{i e}{m} (\partial_{\mu} A_{\nu} - \partial_{\nu} A_{\mu}) J^{[\mu \nu]} =   \frac{i e}{m} F_{[\mu \nu]}  J^{[\mu \nu]} ,
\label{eq375}
\end{align}
где $J^{[\mu \nu]}$ – генераторы представлений группы Лоренца в пространствах представлений \eqref{ls1} и \eqref{ls4}, соответственно. В нерелятивистском приближении оно описывает дополнительный (аномальный) магнитный момент и приводит в лагранжиане к члену взаимодействия типа Паули.

В случае рассмотренных выше расширенных РВУ для частиц с целыми низшими спинами дополнительное слагаемое $Q$ имеет вид
\begin{align}
Q & \sim \frac{e^2}{m} F_{[ \mu \nu]} F_{[ \mu \nu]} e^{00}, \label{eq376} \\
Q & \sim \frac{e^2}{m} F_{[ \mu \nu]} F_{[ \rho \sigma]} e^{[\mu \nu], [\rho \sigma]} , \label{eq377}
\end{align}
где $e^{AB}$ – обобщенные символы Кронекера, определяемые по формулам [29]
\begin{align}
(e^{AB})_{CD} = \delta_{AC} \delta_{BD} .
\label{eq378}
\end{align}
В случае частицы со спином $s = 0$ член $Q$ \eqref{eq376} в нерелятивистском приближении описывает наведенные во внешнем электромагнитном поле дипольные электрическую и магнитную поляризуемости этой частицы. Для частицы со спином $s = 1$ аналогичный член \eqref{eq377} описывает ее статическую тензорную электрическую поляризуемость.

Очевидно, что дополнительное взаимодействие с внешним электромагнитным полем должно влиять на вид матричных элементов конкретных процессов рассеяния. Подробные расчеты некоторых таких процессов проделаны в работах [19–28]. Показано, что в первом порядке теории возмущений указанное взаимодействие не проявляется. Например, рассеяние на кулоновском центре происходит одинаковым образом как в случае РВУ с минимальным, так и расширенным набором представлений группы Лоренца. Расчет сечений типичного процесса второго порядка – комптоновского рассеяния света на
частицах, описываемых РВУ с расширенными наборами представлений – приводит во всех случаях к матричным элементам вида
\begin{align}
M_1 = M_0 + M' .
\label{eq379}
\end{align}
Здесь $M_0$ – матричный элемент, отвечающий частице, которая описывается РВУ с минимальным набором представлений; $M'$ – добавка, обусловленная наличием у частицы внутренней электромагнитной структуры. Явные выражения для этих добавок можно найти в вышеуказанных работах.

Таким образом, простое расширение используемого набора представлений, в том числе за счет включения в него повторяющихся неприводимых компонент, позволяет отразить внутреннюю структуру частиц в рамках обычного пространственно-временного описания методами теории РВУ. Очевидно, что в принципиальном отношении такой подход обладает преимуществом по сравнению с распространенным феноменологическим подходом, при котором дополнительные члены, описывающие специфические структурные эффекты, вводятся в лагранжиан вручную.

В заключение отметим то немаловажное обстоятельство, что все рассмотренные в данном параграфе «расширенные» РВУ для низших спинов свободны от трудностей, имеющих место при введении электромагнитного взаимодействия минимальным образом в теориях высших спинов. Они являются перенормируемыми и не содержат непричинных решений, несмотря на недиагонализируемый характер матрицы $\Gamma_4$ (смотри в этой связи, например, [22; 30]).

\section{РВУ для киральной частицы со спином $1$}

\hspace{1.25cm}
В предыдущем параграфе было показано, что РВУ с расширенным наборами неприводимых представлений группы Лоренца (включая кратные) позволяют учитывать внутреннюю структуру элементарных частиц в рамках не распадающихся в релятивистски-инвариантном смысле уравнений. Вместе с тем, использование кратных представлений в теории РВУ позволяет осуществлять также пространственно-временное описание\footnote{Во избежание недопонимания отметим, что под пространственно-временным описанием в данном контексте понимается использование РВУ, не распадающихся в смысле полной группы Лоренца. Иными словами, речь идет об описании внутренних свойств элементарных частиц без включения в рассмотрение полевых индексов нелоренцевского происхождения.} дополнительных (помимо спина) внутренних степеней свободы частиц.

Продемонстрируем сказанное на примере киральности – степени свободы, которая связана с двукратным вырождением состояний, сопряженных относительно операции пространственной инверсии. Понятие киральности в настоящее время широко используется в адронной физике, в которой оно появляется при пренебрежении массами легких $u$- и $d$-кварков. Уже существует ряд экспериментов при низких энергиях, результаты которых свидетельствуют в пользу существования взаимодействий, переносимых киральными частицами со спином $1$. Подробный обзор по всем этим вопросам можно найти в [31].

Рассмотрим схему зацеплений неприводимых представлений группы Лоренца
\begin{align}
\begin{tabular}{ccccccc}
& &   & $(\frac12,\frac12)'$ &  & & \\
& &  $\diagup$  &         & $\diagdown$  & & \\
 & $(0,1)$   &  &   &  & $(1,0)$, &  \\
   &   &  $\diagdown$ &  & $\diagup$  &  &    \\
   &   &  & $ (\frac12,\frac12) $ &  & &
\end{tabular}
\label{eq41}
\end{align}
которая содержит двукратное представление $(\frac12, \frac12)$ (как и в \eqref{eq338}, штрих здесь введен для различения одинаковых компонент). Матрица $\Gamma_4$ РВУ, отвечающего схеме \eqref{eq41}, имеет в базисе Гельфанда–Яглома квазидиагональную форму \eqref{gam4_s2}. Введем, как обычно, нумерацию представлений
\begin{align}
(\frac12,\frac12) \sim 1, \quad (\frac12,\frac12)' \sim 2, \quad (0,1) \sim 3, \quad (1,0) \sim 4 .
\label{eq42}
\end{align}
Поскольку компоненты $1$ и $2$, формирующие спиновый блок $C^{0}$, не зацепляются, получаем
\begin{align}
C^0 = \left(
\begin{tabular}{cc}
$0$ & $0$ \\
$0$ & $0$
\end{tabular} \right),
\label{eq43}
\end{align}
то есть спин $s = 0$ здесь отсутствует.

Для блока $C^{1}$ имеем следующее общее выражение
\begin{align}
C^1 = \left(
\begin{tabular}{cccc}
$0$ & $0$ & $c_{13}^1$ & $c_{14}^1$ \\
$0$ & $0$ & $c_{23}^1$ & $c_{24}^1$ \\
$c_{31}^1$ & $c_{32}^1$ & $0$ & $0$ \\
$c_{41}^1$ & $c_{42}^1$ & $0$ & $0$
\end{tabular} \right).
\label{eq44}
\end{align}

Требование инвариантности строящегося РВУ относительно преобразований собственной группы Лоренца в данном случае на элементы $c_{ij}^1$ никаких ограничений не накладывает. Инвариантность относительно операции $P$-инверсии приводит к условиям
\begin{align}
c_{14}^1 = \pm c_{13}^1, \quad c_{41}^1 = \pm c_{31}^1, \quad c_{24}^1 = \pm c_{23}^1, \quad c_{42}^1 = \pm c_{32}^1.
\label{eq45}
\end{align}
При этом в соответствии с формулами \eqref{c2_cond}--\eqref{Pc2_cond} знак «+» («–») здесь берется тогда, когда оба представления $(\frac12, \frac12)$ и $(\frac12, \frac12)'$ в схеме \eqref{eq41} являются истинно векторными (псевдовекторными). В указанных случаях получаем для спинового блока $C^{1}$ выражение
\begin{align}
C^1 = \left(
\begin{tabular}{cccc}
$0$ & $0$ & $c_{13}^1$ & $\pm c_{13}^1$ \\
$0$ & $0$ & $c_{23}^1$ & $\pm c_{23}^1$ \\
$c_{31}^1$ & $c_{32}^1$ & $0$ & $0$ \\
$\pm c_{31}^1$ & $\pm c_{32}^1$ & $0$ & $0$
\end{tabular} \right),
\label{eq46}
\end{align}
где одновременно берутся верхние либо нижние знаки.

Характеристическое уравнение блока $C^{1}$ имеет вид
\begin{align}
\lambda^4 -2 \lambda^2 (c_{13}^1 c_{31}^1 + c_{23}^1 c_{32}^1) =0 .
\label{eq47}
\end{align}
Отсюда следует, что этот блок имеет единственное с точностью до знака ненулевое собственное значение
\begin{align}
\lambda = \pm \sqrt{2 (c_{13}^1 c_{31}^1 + c_{23}^1 c_{32}^1)} .
\label{eq48}
\end{align}
Другими словами, состояние со спином $1$ не содержит дополнительных внутренних степеней свободы. Очевидно, что полученное таким образом РВУ сводится к уравнению Дирака–Кэлера для векторной частицы.

Иначе обстоит дело, когда одно из векторных представлений в схеме \eqref{eq41} – векторное,
а второе – псевдовекторное. В этом случае требование $P$-инвариантности теории приводит к условиям
\begin{align}
c_{14}^1 =  c_{13}^1, \quad c_{41}^1 = c_{31}^1, \quad c_{24}^1 = - c_{23}^1, \quad c_{42}^1 = - c_{32}^1
\label{eq49}
\end{align}
и соответственно к спиновому блоку $C^{1}$
\begin{align}
C^1 = \left(
\begin{tabular}{cccc}
$0$ & $0$ & $c_{13}^1$ & $c_{13}^1$ \\
$0$ & $0$ & $c_{23}^1$ & $- c_{23}^1$ \\
$c_{31}^1$ & $c_{32}^1$ & $0$ & $0$ \\
$c_{31}^1$ & $- c_{32}^1$ & $0$ & $0$
\end{tabular} \right).
\label{eq410}
\end{align}

Его характеристическое уравнение имеет вид
\begin{align}
\lambda^4 -2 \lambda^2 (c_{13}^1 c_{31}^1 + c_{23}^1 c_{32}^1) + 4 c_{13}^1 c_{31}^1 c_{23}^1 c_{32}^1=0 .
\label{eq411}
\end{align}
Решения уравнения \eqref{eq411} $\lambda_1 = \pm \sqrt{2 c_{13}^1 c_{31}^1}$ и $\lambda_2 = \pm \sqrt{2 c_{23}^1 c_{32}^1}$ означают, что речь может идти о частице со спином $s = 1$ и, вообще говоря, двумя значениями массы
\begin{align}
 m_1 = \frac{m}{\sqrt{2 c_{13}^1 c_{31}^1}}, \quad m_2 = \frac{m}{\sqrt{2 c_{23}^1 c_{32}^1}} .
\label{eq412}
\end{align}
Однако, если положить
\begin{align}
c_{13}^1 c_{31}^1 = c_{23}^1 c_{32}^1 ,
\label{eq413}
\end{align}
мы приходим к РВУ для частицы с одной массой, но двукратным вырождением состояний по некоторому дополнительному квантовому числу.

Произвол, остающийся в выборе элементов $c^{1}_{ij}$ блока $C^{1}$ \eqref{eq410} после наложения на них условия \eqref{eq413}, используем для получения лагранжевой формулировки интересующего нас РВУ. Для этого элементы $\eta^0_{ij}$, $\eta^1_{ij}$ матрицы билинейной формы $\eta$ зададим следующим образом:
\begin{align}
-\eta^0_{11} = \eta^0_{22} = \eta^1_{11} = -\eta^1_{22} = \eta^1_{34} = \eta^1_{43} =1 .
\label{eq414}
\end{align}

Тогда требование \eqref{c_eta_cond} релятивистской инвариантности лагранжиана теории приводит к равенствам
\begin{align}
c_{31}^1 = (c_{13}^1 )^* , \quad c_{32}^1 = -  (c_{23}^1 )^* .
\label{eq415}
\end{align}
Не уменьшая общности, все еще остающиеся произвольными элементы $c^1_{13}$, $c^1_{23}$
можно выбрать, например, так:
\begin{align}
c_{13}^1 = c_{23}^1 = \frac{1}{\sqrt{2}}
\label{eq416}
\end{align}
(при этом собственные значения блока $C^{1}$ будут равными $\pm 1$).

Таким образом, для ненулевого блока $C^{1}$ матрицы $\Gamma_4$ и блоков $\eta^0$, $\eta^1$ матрицы $\eta$ находим окончательно вид
\begin{align}
C^1 &= \frac{1}{\sqrt{2}}  \left(
\begin{tabular}{cccc}
$0$ & $0$ & $1$ & $1$ \\
$0$ & $0$ & $1$ & $- 1$ \\
$1$ & $1$ & $0$ & $0$ \\
$1$ & $- 1$ & $0$ & $0$
\end{tabular} \right),
\label{eq417} \\
\eta^0 &= \left(
\begin{tabular}{cc}
$-1$ & $0$ \\
$0$ & $1$
\end{tabular} \right),  \quad \eta^1 = \left(
\begin{tabular}{cccc}
$1$ & $0$ & $0$ & $0$ \\
$0$ & $-1$ & $0$ & $0$ \\
$0$ & $0$ & $0$ & $1$ \\
$0$ & $0$ & $1$ & $0$
\end{tabular} \right). \label{eq418}
\end{align}

Для выяснения смысла двукратного вырождения состояний микрообъекта, описываемого полученным РВУ, удобно использовать его тензорную формулировку
\begin{align}
\begin{split}
& \partial_{\nu} \psi_{[\mu \nu]} + m \psi_{\mu} =0, \\
& \partial_{\nu}  \tilde{\psi}_{[\mu \nu]} + m \tilde{\psi}_{\mu}  =0, \\
-& \partial_{\mu}  \psi_{\nu} +  \partial_{\nu}  \psi_{\mu}+ \varepsilon_{\mu \nu \alpha \beta} \partial_{\alpha} \tilde{\psi}_{\beta} + m \psi_{[\mu \nu]}=0,
\end{split}
\label{eq419}
\end{align}
где величины $\psi_{\mu}$, $\tilde{\psi}_{\mu}$, $\psi_{[\mu \nu]} $ сопоставляются представлениям $(\frac12, \frac12)$, $(\frac12, \frac12)'$, $(0, 1) \oplus (1, 0)$ соответственно,
\begin{align}
\tilde{\psi}_{[\mu \nu]} = \frac12 \varepsilon_{\mu \nu \alpha \beta} \psi_{[\alpha \beta]} ,
\label{eq420}
\end{align}
и $\varepsilon_{\mu \nu \alpha \beta}$ – тензор Леви–Чивита ($ \varepsilon_{1234}= - i$).

Осуществляя в системе \eqref{eq419} подстановки
\begin{align}
\begin{split}
\varphi_{\mu} &= \frac{1}{\sqrt{2}} (\psi_{\mu} - i \tilde{\psi}_{\mu}), \quad \varphi_{[\mu \nu]} = \frac{1}{\sqrt{2}} (\psi_{[\mu \nu]}  -i \tilde{\psi}_{[\mu \nu]} ), \\
\chi_{\mu} &=  \frac{1}{\sqrt{2}} (\psi_{\mu} + i \tilde{\psi}_{\mu}), \quad \chi_{[\mu \nu]} = \frac{1}{\sqrt{2}} (\psi_{[\mu \nu]}  + i \tilde{\psi}_{[\mu \nu]} ),
\end{split}
\label{eq421}
\end{align}
преобразуем ее к прямой сумме двух семикомпонентных подсистем:
\begin{align}
\begin{split}
& \partial_{\nu} \varphi_{[\mu  \nu]} + m \varphi_{\mu} =0 , \\
- &\partial_{\mu} \varphi_{\nu} + \partial_{\nu} \varphi_{\mu} + i \varepsilon_{\mu \nu \alpha \beta} \partial_{\alpha} \varphi_{\beta} + m \varphi_{[\mu  \nu]}  =0
\end{split}
\label{eq422}
\end{align}
и
\begin{align}
\begin{split}
& \partial_{\nu} \chi_{[\mu  \nu]} + m \chi_{\mu} =0 , \\
-&  \partial_{\mu} \chi_{\nu} + \partial_{\nu} \chi_{\mu} - i \varepsilon_{\mu \nu \alpha \beta} \partial_{\alpha} \chi_{\beta} + m \chi_{[\mu  \nu]}  =0 .
\end{split}
\label{eq423}
\end{align}

Подсистемы \eqref{eq422}, \eqref{eq423} инвариантны в смысле преобразований собственной группы Лоренца. Однако при операции пространственного отражения они переходят друг в друга. Так что по отношению к преобразованиям полной группы Лоренца система \eqref{eq422}, \eqref{eq423}, а значит и система \eqref{eq419}, является нераспадающейся. Кроме того, заметим, что подсистемам \eqref{eq422}, \eqref{eq423} по отдельности нельзя сопоставить удовлетворяющий стандартным требованиям лагранжиан. Корректная лагранжева формулировка возможна только для системы \eqref{eq422}, \eqref{eq423}, рассматриваемой в целом.

Итак, двукратно вырожденные состояния векторной частицы, описываемой построенным РВУ, связаны между собой операцией $P$-инверсии. Следовательно, дополнительное квантовое число, о котором идет речь, различает указанные $P$- сопряженные состояния и может трактоваться как киральность по аналогии с понятием киральности для безмассовых частиц.

Здесь важно отметить следующее. Как известно, для построения теории безмассовых частиц с киральностью $S$ достаточно использования только представлений $(0, S)$ и $(S, 0)$, причем понятия киральности и спиральности (проекции спина на направление импульса) в этом случае, по существу, совпадают. В случае частиц с ненулевой массой это уже не так: во-первых, понятия киральности и спиральности для них не совпадают; во-вторых, теория таких частиц, как ясно из вышеизложенного, с необходимостью базируется на наборе {\it зацепляющихся} неприводимых представлений группы Лоренца, включая кратные компоненты. Возможный физический смысл киральности применительно к обладающим массой виртуальным частицам будет обсужден в главе 7.

\section{Тензорные РВУ дираковского типа и геометризованное описание
внутренних степеней свободы \\ фундаментальных частиц}

\hspace{1.25cm}
Поскольку существование дополнительных внутренних квантовых чисел у фундаментальных частиц является в настоящее время твердо установленным фактом, возникает вопрос о возможности применения теории РВУ к описанию степеней свободы, связанных с внутренними, в том числе калибровочными, симметриями. Традиционные калибровочные теории фундаментальных частиц и их взаимодействий базируются, как правило, на уравнении Дирака, волновая функция которого снабжается свободным нелоренцевским индексом, играющим роль внутренней переменной. С точки зрения теории РВУ такой подход означает фактически использование распадающихся по группе Лоренца уравнений. На этой основе строятся известные модели электрослабых и сильных взаимодействий, стандартная $SU(3)\times  SU(2) \times U(1)$ модель.

Однако данный подход не в состоянии решить ряд проблем, в частности, он оказывается малоэффективным при включении в общую схему гравитационного взаимодействия. Решение этой и многих других проблем в настоящее время связывается, главным образом, с использованием групп симметрии, преобразования которых содержали бы на равных пространственно-временные и внутренние переменные. Иначе говоря, речь идет о возможности геометризованного введения внутренних степеней свободы.

Перечислим вкратце наиболее известные подходы в этом направлении:
\begin{itemize}
  \item
  теории типа Калуцы–Клейна, в которых пространство–время имеет размерность, большую четырех, причем дополнительные измерения рассматриваются как равноправные с четырьмя наблюдаемыми. Компактификация «лишних» измерений приводит к выделению внутренних степеней свободы, сохраняя их геометрический статус;
  \item
  суперсимметрия–супергравитация, объединяющая частицы с разными спинами и статистикой в единые супермультиплеты. Одна из исходных посылок здесь заключается в том, что существует новая математическая структура – преобразования суперсимметрии, которые перемешивают бозонные и фермионные поля. В результате точно так же, как преобразования Лоренца обнаруживают связь между собственно пространством и временем, преобразования суперсимметрии связывают в одно целое пространство–время и внутренние степени свободы частиц;
  \item
  струнные и суперструнные модели, включающие в себя идеи Калуцы–Клейна и суперсимметрии, калибровочного подхода и теории относительности.
\end{itemize}

Однако можно предложить и иной способ геометризованного описания внутренних степеней свободы, основанный на применении расширенного (включая кратные) набора неприводимых представлений группы Лоренца в подходе теории РВУ. Естественной возможностью в этом плане является использование не распадающихся по полной группе Лоренца уравнений, волновая функция которых обладает трансформационными свойствами прямого произведения (полного или усеченного) дираковских биспиноров, а матрицы $\Gamma_\mu$ удовлетворяют перестановочным соотношениям алгебры матриц Дирака. В дальнейшем такие РВУ будем называть диракоподобными, или уравнениями дираковского типа.

Наиболее известным РВУ указанного типа является уравнение Дирака–Кэлера (ДК), которое представляет собой максимально общее дифференциальное уравнение (систему уравнений) первого порядка над полем комплексных чисел для полного набора антисимметричных тензорных полей в пространстве Минковского. С другой стороны, в
соответствующем базисе (будем называть его фермионным) волновая функция уравнения ДК обладает лоренцевскими трансформационными свойствами прямого произведения дираковского биспинора на зарядово-сопряженный биспинор.
В тензорной формулировке уравнение ДК может быть представлено в виде системы
\begin{align}
\begin{split}
& \partial_{\mu} \psi_{\mu} + m \psi_0 =0 , \\
&  \partial_{\nu} \psi_{[\mu \nu]} +  \partial_{\mu}  \psi_0 + m \psi_{\mu}  =0 , \\
-&  \partial_{\mu}\psi_{\nu} + \partial_{\nu}\psi_{\mu}  + i \varepsilon_{\mu \nu \alpha \beta} \partial_{\alpha} \tilde{\psi}_{\beta}  + m \psi_{[\mu \nu]} =0 , \\
& \partial_{\mu} \tilde{\psi}_{\mu} + m \tilde{\psi}_0 =0 , \\
&  \partial_{\nu} \tilde{\psi}_{[\mu \nu]} +  \partial_{\mu}  \tilde{\psi}_0 + m \tilde{\psi}_{\mu}  =0 , \\
& \frac12 \varepsilon_{\mu \nu \alpha \beta} \partial_{\nu} \psi_{[ \alpha \beta ]}  + \partial_{\mu}  \tilde{\psi}_0  +  m \tilde{\psi}_{\mu} =0 .
\end{split}
\label{eq51}
\end{align}
Здесь $\psi_0$ – скаляр, $\psi_{\mu}$ – вектор, $\psi_{[\mu \nu]}$ – антисимметричный тензор второго ранга,
$\tilde{\psi}_{\mu} = \frac{1}{3!} \varepsilon_{\mu \nu \alpha \beta} \psi_{[\nu \alpha \beta]} $ – псевдовектор, дуально сопряженный антисимметричному тензору третьего ранга $\psi_{[\nu \alpha \beta]} $, и $\tilde{\psi}_{0} = \frac{1}{4!} \varepsilon_{\mu \nu \alpha \beta} \psi_{[ \mu \nu \alpha \beta]} $ – псевдоскаляр, дуально сопряженный антисимметричному тензору четвертого ранга $\psi_{[ \mu \nu \alpha \beta]}$.

Система \eqref{eq51} является не распадающейся в смысле полной группы Лоренца. Она может быть записана в стандартной для теории РВУ матрично- дифференциальной форме \eqref{eq:nonzero_mass}, где волновая функция $\Psi$ представляет собой столбец с тензорными компонентами
\begin{align}
\Psi = (\psi_0 , \tilde{\psi}_0 , \psi_{\mu}, \tilde{\psi}_{\mu}, \psi_{[\mu \nu]})^T ,
\label{eq52}
\end{align}
а матрицы $\Gamma_\mu$ размерности $16 \times 16$ имеют вид
\begin{align}
\begin{split}
& \Gamma_{\mu} = \Gamma_{\mu}^{(+)}+  \Gamma_{\mu}^{(-)} , \\
& \Gamma_{\mu}^{(+)} = e^{\tilde{0} \tilde{\mu}} + e^{\tilde{\mu} \tilde{0}} + e^{\lambda, [ \lambda \mu ]} + e^{[\lambda \mu ],\lambda}, \\
& \Gamma_{\mu}^{(-)} = e^{0 \mu} + e^{\mu 0} + \frac{i}{2} \varepsilon_{\lambda \mu \alpha \beta} ( e^{\tilde{\lambda}, [ \alpha \beta ]} + e^{[\alpha \beta ], \tilde{\lambda}}).
\end{split}
\label{eq53}
\end{align}

Используя известные правила перемножения обобщенных символов Кронекера [29], можно убедиться, что матрицы $\Gamma_\mu$ \eqref{eq53} удовлетворяют алгебре матриц Дирака
\begin{align}
\Gamma_{\mu} \Gamma_{\nu} + \Gamma_{\nu} \Gamma_{\mu} = 2 \delta_{\mu \nu} .
\label{eq54}
\end{align}

Для установления группы внутренней симметрии поля ДК удобно от тензорного базиса \eqref{eq52} перейти к фермионному базису, в котором матрицы $\Gamma_\mu$ и матрица $\eta$ лоренц-инвариантной билинейной формы имеют вид
\begin{align}
\Gamma_{\mu} &= I_4 \otimes \gamma_4 ,
\label{eq55} \\
\eta &= \gamma_4 \otimes \gamma_4 .
\label{eq56}
\end{align}

Напомним, что под преобразованием внутренней симметрии РВУ \eqref{eq:nonzero_mass} понимаются линейные преобразования волновой функции
\begin{align}
\Psi' (x) = Q \Psi (x),
\label{eq57}
\end{align}
не затрагивающие пространственно-временных координат и оставляющие инвариантным уравнение \eqref{eq:nonzero_mass} и его лагранжиан \eqref{eq126}. Для этого матрицы $Q$ должны удовлетворять условиям
\begin{align}
 \, [ \Gamma_\mu , Q]_- =0 , \label{eq58} \\
 Q^+ \eta Q = \eta . \label{eq59}
\end{align}

Применение условий \eqref{eq58}, \eqref{eq59} к матрицам $\Gamma_\mu$ и $\eta$ приводит к некомпактной $15$-параметрической группе $SU (2, 2)$, генераторами которой могут служить эрмитовые матрицы
\begin{align}
\Gamma'_{\mu}, \quad \Gamma'_5 , \quad i \Gamma'_{\mu} \Gamma'_5 , \quad i \Gamma'_{[\mu} \Gamma'_{\nu ]} = \frac{i}{2} (\Gamma'_{\mu} \Gamma'_{\nu} - \Gamma'_{\nu} \Gamma'_{\mu}).
\label{eq510}
\end{align}
Здесь
\begin{align}
\Gamma'_5 = \Gamma'_1 \Gamma'_2 \Gamma'_3 \Gamma'_4
\label{eq511}
\end{align}
и
\begin{align}
\Gamma'_{\mu} = \Gamma^{(+)}_{\mu} -\Gamma^{(-)}_{\mu}
\label{eq512}
\end{align}
второй набор матриц размерности $16 \times 16$, удовлетворяющих, как и $\Gamma_\mu$, алгебре матриц Дирака и коммутирующих с матрицами $\Gamma_\mu$. В фермионном базисе эти матрицы имеют вид
\begin{align}
\Gamma'_{\mu} = \gamma_{\mu} \otimes I_4 .
\label{eq513}
\end{align}
Отличительной особенностью группы внутренней симметрии уравнения ДК является то, что ее генераторы \eqref{eq510} не коммутируют с лоренцевскими генераторами
\begin{align}
J_{[\mu \nu]} = \frac14 (\Gamma_{[ \mu} \Gamma_{\nu ]} + \Gamma'_{[ \mu} \Gamma'_{\nu ]})
\label{eq514}
\end{align}
представления волновой функции $\Psi$. При этом группа $G$ алгебры полной инвариантности уравнения ДК является полупрямым произведением группы лоренцевских преобразований $\Lambda$ и группы внутренней симметрии $Q$: $G = \Lambda \oslash Q$. С другой стороны, группу $G$ можно представить в виде прямого произведения $G = \Lambda' \otimes Q$, где $\Lambda'$  – переопределенная группа Лоренца, по отношению к которой волновая функция $\Psi$ характеризует уже не совокупность тензорных величин, а набор четырех дираковских полей с обычной, то есть коммутирующей с преобразованиями группы Лоренца, внутренней симметрией.

Приведенные соображения сохраняют силу для всех взаимодействий (в том числе калибровочных), не нарушающих внутреннюю симметрию свободного лагранжиана. Они означают принципиальную применимость уравнения ДК для описания частиц со спином $s = \frac12$ и внутренними степенями свободы, имеющими, таким образом, геометрическое происхождение (более подробно по этому поводу смотри, например, [32]). Так, идея о том, что уравнение ДК может выступать, например, в качестве геометрической модели поколений кварков (или лептонов), впервые была выдвинута в работах [33; 34].

Теперь дадим матричную формулировку уравнения ДК в базисе Гельфанда– Яглома, которая нам понадобится в дальнейшем.

Возьмем в качестве исходного набор неприводимых представлений собственной группы Лоренца
\begin{align}
\begin{tabular}{ccccc}
& & $2 (0,0)$ & & \\
& & $|$ & & \\
$(0,1)$ & --- & $2 (\frac12, \frac12)$ & --- & $(1,0)$,
\end{tabular}
\label{eq515}
\end{align}
содержащий двукратные компоненты компоненты $(0, 0)$ и $(\frac12, \frac12)$. Матрица $\Gamma_4$ соответствующего РВУ в базисе Гельфанда–Яглома будет иметь вид \eqref{gam4_s2}, где
\begin{align}
C^0 = \left(
\begin{tabular}{cccc}
$0$ & $0$ & $c_{13}^0$ & $c_{14}^0$ \\
$0$ & $0$ & $c_{23}^0$ & $c_{24}^0$ \\
$c_{31}^0$ & $c_{32}^0$ & $0$ & $0$ \\
$c_{41}^0$ & $c_{42}^0$ & $0$ & $0$
\end{tabular} \right), \quad
C^1 = \left(
\begin{tabular}{cccc}
$0$ & $0$ & $c_{35}^1$ & $c_{36}^1$ \\
$0$ & $0$ & $c_{45}^1$ & $c_{46}^1$ \\
$c_{53}^1$ & $c_{54}^1$ & $0$ & $0$ \\
$c_{63}^1$ & $c_{64}^1$ & $0$ & $0$
\end{tabular} \right),
\label{eq516}
\end{align}
и принята следующая нумерация содержащихся в наборе \eqref{eq515} неприводимых представлений:
\begin{align}
(0,0) \sim 1, \quad (0,0)' \sim 2, \quad (\frac12 , \frac12) \sim 3 , \quad (\frac12 , \frac12)' \sim 4, \quad (0,1) \sim 5, \quad (1,0) \sim 6.
\label{eq517}
\end{align}
Здесь, как и ранее, штрих используется для различения кратных представлений.

Рассмотрим сначала спиновый блок $C^{1}$. Условия релятивистской и $P$- инвариантности теории накладывают на элементы $c^{1}_{ij}$ в общем случае ограничения
\begin{align}
c^1_{35} = \pm c^1_{36}, \quad c^1_{45} = \pm c^1_{46}, \quad c^1_{53} = \pm c^1_{63}, \quad c^1_{54} = \pm c^1_{64}.
\label{eq518}
\end{align}
При этом выбор знаков «+» или «–» в \eqref{eq518} зависит от определения оператора пространственного отражения согласно формулам \eqref{Pc1_cond} или \eqref{Pc2_cond}. Применительно к рассматриваемому случаю сказанное означает, что знак «+» («–») в \eqref{eq518} имеет место при истинно векторном (псевдовекторном) характере кратных представлений $(\frac12,\frac12)$. Удовлетворяющее всем необходимым физическим требованиям РВУ можно построить, если выбрать одно из представлений $(\frac12,\frac12)$ истинно векторным, а второе~-- псевдовекторным (в дальнейшем будем обозначать его $(\frac12,\frac12)'$). Тогда соотношения \eqref{eq518} принимают вид
\begin{align}
c^1_{35} =  c^1_{36} , \quad c^1_{45} = - c^1_{46}, \quad  c^1_{53} =  c^1_{63} , \quad c^1_{54} = - c^1_{64},
\label{eq519}
\end{align}
и для блока $C^{1}$ получается выражение
\begin{align}
C^1 = \left(
\begin{tabular}{cccc}
$0$ & $0$ & $c_{35}^1$ & $c_{35}^1$ \\
$0$ & $0$ & $c_{45}^1$ & $-c_{45}^1$ \\
$c_{53}^1$ & $c_{54}^1$ & $0$ & $0$ \\
$c_{53}^1$ & $-c_{54}^1$ & $0$ & $0$
\end{tabular} \right).
\label{eq520}
\end{align}

Аналогичным образом, одно из представлений $(0,0)$ в \eqref{eq515} выберем скалярным, а второе – псевдоскалярным (будем также помечать его штрихом). И поскольку в $P$-инвариантном РВУ векторное (псевдовекторное) представление не может зацепляться с псевдоскалярным (скалярным), то будут иметь место равенства
\begin{align}
c^0_{14}=c^0_{23}=c^0_{41}=c^0_{32}=0.
\label{eq521}
\end{align}
Блок $C^{0}$ \eqref{eq518} при этом преобразуется к виду
\begin{align}
C^0 = \left(
\begin{tabular}{cccc}
$0$ & $0$ & $c_{13}^0$ & $0$ \\
$0$ & $0$ & $0$ & $c_{24}^0$ \\
$c_{31}^0$ & $0$ & $0$ & $0$ \\
$0$ & $c_{42}^0$ & $0$ & $0$
\end{tabular} \right),
\label{eq522}
\end{align}
а представления \eqref{eq515} образуют схему зацеплений
\begin{align}
\begin{tabular}{ccccccccc}
& & &   & $(0,1)$ &  & & & \\
& & &  $\diagup$  &         & $\diagdown$  & & & \\
$(0,0)'$ &  --- & $(\frac12,\frac12)'$   &  &   &  & $(\frac12,\frac12) $ &  --- & $(0,0)$ . \\
 &  &   &  $\diagdown$ &  & $\diagup$  &  &   & \\
&   &   &  & $ (1,0)$ &  & &  &
\end{tabular}
\label{eq523}
\end{align}

Блоки $\eta^{0}$, $\eta^{1}$ матрицы билинейной инвариантной формы $\eta$ \eqref{eq24}, имеют в данном случае вид
\begin{align}
\begin{split}
\eta &= \eta^0 \oplus (\eta^1 \otimes I_3), \\
\eta^0 &=  \left(
\begin{tabular}{cccc}
$\eta_{11}^0$ & $0$ & $0$ & $0$ \\
$0$ & $\eta_{22}^0$ & $0$ & $0$ \\
$0$ & $0$ & $\eta_{33}^0$ & $0$ \\
$0$ & $0$ & $0$ & $\eta_{44}^0$
\end{tabular} \right), \quad
\eta^1 =  \left(
\begin{tabular}{cccc}
$\eta_{33}^1$ & $0$ & $0$ & $0$ \\
$0$ & $\eta_{44}^1$ & $0$ & $0$ \\
$0$ & $0$ & $0$ & $\eta_{56}^1$ \\
$0$ & $0$ & $\eta_{65}^1$ & $0$
\end{tabular} \right),
\end{split}
\label{eq524}
\end{align}
причем согласно \eqref{eta_cond}
\begin{align}
\eta_{33}^1 = - \eta_{33}^0, \quad \eta_{44}^1 = - \eta_{44}^0, \quad \eta_{65}^1 = \pm \eta_{56}^1 .
\label{eq525}
\end{align}

Условия \eqref{c_eta_cond} совместно с равенствами \eqref{eq519} приводят к соотношениям
\begin{align}
c_{31}^0 = \frac{\eta_{33}^0}{\eta_{11}^0}(c_{13}^0)^*, \quad c_{42}^0 = \frac{\eta_{44}^0}{\eta_{22}^0}(c_{24}^0)^*, \quad
c_{53}^1 = \frac{\eta_{56}^1}{\eta_{33}^1}(c_{35}^1)^*, \quad c_{54}^1 = \frac{\eta_{56}^1}{\eta_{44}^1}(c_{45}^1)^*.
\label{eq526}
\end{align}
Выбирая теперь для остающихся произвольными элементов $c^s_{\tau \tau'}$ и $\eta^s_{\tau \dot{\tau}}$, например, значения
\begin{align}
c_{13}^0 &= c_{24}^0=1 , \quad c_{35}^1 = c_{45}^1 = \frac{1}{\sqrt{2}}, \label{eq527} \\
\eta_{11}^0 &=- \eta_{22}^0 =\eta_{33}^0 =- \eta_{44}^0 =- \eta_{56}^1 =- \eta_{65}^1 =1,
\label{eq528}
\end{align}
получим РВУ со спиновыми блоками матриц $\Gamma_4$ и $\eta$ вида
\begin{align}
C^0 = \left(
\begin{tabular}{cccc}
$0$ & $0$ & $1$ & $0$ \\
$0$ & $0$ & $0$ & $1$ \\
$1$ & $0$ & $0$ & $0$ \\
$0$ & $1$ & $0$ & $0$
\end{tabular} \right), \quad
C^1 = \frac{1}{\sqrt{2}} \left(
\begin{tabular}{cccc}
$0$ & $0$ & $1$ & $1$ \\
$0$ & $0$ & $1$ & $-1$ \\
$1$ & $1$ & $0$ & $0$ \\
$1$ & $-1$ & $0$ & $0$
\end{tabular} \right),
\label{eq529}
\end{align}
\begin{align}
\eta^0 &=  \left(
\begin{tabular}{cccc}
$1$ & $0$ & $0$ & $0$ \\
$0$ & $-1$ & $0$ & $0$ \\
$0$ & $0$ & $1$ & $0$ \\
$0$ & $0$ & $0$ & $-1$
\end{tabular} \right), \quad
\eta^1 =  \left(
\begin{tabular}{cccc}
$-1$ & $0$ & $0$ & $0$ \\
$0$ & $1$ & $0$ & $0$ \\
$0$ & $0$ & $0$ & $-1$ \\
$0$ & $0$ & $-1$ & $0$
\end{tabular} \right).
\label{eq530}
\end{align}

Построенное таким образом с использованием базиса Гельфанда–Яглома РВУ удовлетворяет условиям инвариантности относительно преобразований полной группы Лоренца и возможности его получения из инвариантной функции Лагранжа. С формальной точки зрения оно описывает микрообъект с ненулевой массой и набором спинов $0, 1$. Минимальные уравнения для спиновых блоков $C^{0}$, $C^{1}$ и матрицы $\Gamma_4$ в целом имеют одинаковый вид
\begin{align}
(C^0)^2  -1 =0, \quad (C^1)^2 -1 = 0, \quad \Gamma_4^2 - 1 =0,
\label{eq531}
\end{align}
из которого вытекает, что данное РВУ относится к уравнениям дираковского типа с алгеброй \eqref{eq54}. Наличие повторяющихся корней $\pm 1$ у блоков $C^{0}$, $C^{1}$ означает наличие дополнительной (помимо спина) внутренней степени свободы.

Отметим, что выбор \eqref{eq527} элементов матрицы $\Gamma_4$ не является единственно возможным с точки зрения получения диракоподобного РВУ. В общем случае для удовлетворения характеристическим уравнениям \eqref{eq531} достаточно выполнения условий
\begin{align}
\Gamma_{\mu} \Gamma_{\nu} + \Gamma_{\nu} \Gamma_{\mu} =2 \delta_{\mu \nu} .
\label{eq532}
\end{align}

Очевидно, что только за счет изменения знаков у чисел $c_{13}^0$, $c_{24}^0$, $c_{35}^1$, $c_{45}^1$ можно реализовать $16$ способов задания спиновых блоков $C^{0}$, $C^{1}$. Однако очевидно, что все такие (и иные) модификации связаны преобразованием подобия и, следовательно, физически эквивалентны.

Итак, на основе схемы зацеплений \eqref{eq523} может быть построено единственное (с точностью до преобразования подобия) РВУ дираковского типа, не распадающееся в смысле полной группы Лоренца и допускающее лагранжеву формулировку.

Приведенное выше алгебраическое и теоретико-групповое обоснование динамического соответствия (эквивалентности) классического уравнения ДК и $SU(2,2)$-инвариантной теории Дирака является еще недостаточным для геометризованного описания внутренних степеней свободы дираковских частиц посредством тензорных полей. Последовательная реализация возможности такого описания предполагает существование такого соответствия и на квантовом уровне, что равносильно возможности квантования поля ДК по статистике Ферми–Дирака.

Казалось бы, такое допущение противоречит известной теореме Паули о связи спина и статистики [34]. Однако, это не совсем так. Еще в работах [35; 36] на примере простейших уравнений для частиц с целым и полуцелым спином было показано, что при использовании индефинитной метрики в гильбертовом пространстве состояний, в принципе, допускается квантование по аномальной статистике (полуцелый спин – по статистике Бозе–Эйнштейна, целый – по статистике Ферми–Дирака). Но при этом в теории появляются неустранимые отрицательные вероятности.

Существенно иная ситуация возникает в случае полевых систем с дополнительными степенями свободы, соответствующими некомпактным группам внутренней симметрии. В таких теориях имеют место дополнительные законы сохранения (правила запрета), исключающие переходы, сопровождающиеся отрицательными вероятностями при квантовании с индефинитной метрикой. Рассмотрим подробно этот вопрос применительно к уравнению ДК [37; 38].

Сначала с помощью подстановки
\begin{align}
\Psi (x) = \Psi (p) e^{i p_{\mu} x_{\mu}}
\label{eq533}
\end{align}
перейдем от матричной формы \eqref{eq:nonzero_mass} уравнения ДК в координатном представлении к импульсному представлению
\begin{align}
(\hat{p} + m) \Psi (p) =0,
\label{eq534}
\end{align}
где
\begin{align}
\hat{p} = i p_{\mu} \Gamma_{\mu}
\label{eq535}
\end{align}
оператор $4$-импульса.

Как следует из \eqref{eq531}, спиновые блоки $C^{0}$, $C^{1}$ содержат один с точностью до знака ненулевой корень $\pm 1$. Наличие же внутренней степени свободы выражается здесь в том, в характеристических полиномах этих блоков указанный ненулевой корень имеет кратность, равную двум. Так что, наряду с обычными операторами $4$-импульса \eqref{eq535}, квадрата спина
\begin{align}
\hat{S}^2 = - [(J^{[12]})^2 + (J^{[23]})^2 + (J^{[31]})^2]
\label{eq536}
\end{align}
и проекции спина
\begin{align}
\hat{S}_n =- i \varepsilon_{ijk} n_i J^{[jk]},
\label{eq537}
\end{align}
где $J^{[\mu \nu]} = \frac14 (\Gamma_{[\mu} \Gamma_{\nu]}+\Gamma'_{[\mu} \Gamma'_{\nu]})$, этой степени свободы (назовем ее для определенности $\Pi$-четностью) можно сопоставить некоторый оператор $\hat{\Pi}$, коммутирующий с указанными операторами и образующий вместе с ними полный набор переменных для поля ДК. Дополним это условие естественными требованиями диагонализируемости этого оператора и вещественности его собственных значений, а также по аналогии с операторами $\hat{S}^2$, $\hat{S}_n$ свойством
\begin{align}
\hat{\Pi} \eta = \eta \hat{\Pi}^{+}.
\label{eq538}
\end{align}

Нетрудно убедиться, что релятивистски-инвариантное определение оператора $\Pi$-четности, удовлетворяющее сформулированным условиям, имеет вид
\begin{align}
\hat{\Pi} = \frac{p_{\mu} \Gamma'_{\mu}}{i m};
\label{eq539}
\end{align}
в частности, в системе покоя
\begin{align}
\hat{\Pi}_0 = \Gamma'_4.
\label{eq540}
\end{align}
Собственные значения оператора $\hat{\Pi}$ будем обозначать через $\lambda_i$, $i= 1, 2$. (В системе покоя $\lambda_1= 1$, $\lambda_2 = - 1$).

При вторичном квантовании существенную роль играет знакоповедение плотности энергии (и заряда) классической полевой системы. Наличие спектра спинов и $\Pi$-четности приводит к тому, что знаки указанных величин могут зависеть не только от знака массы (под которым понимается знак собственных значений матрицы $\Gamma_4$, различающий положительно- и отрицательно-частотные решения уравнения \eqref{eq534}), но и от квантовых чисел $i$ и $s$. Другими словами, и энергия и заряд в таких теориях являются, вообще говоря, индефинитными. Данное обстоятельство удобно отразить, вводя в рассмотрение переменную $g_{is}^{(\pm)}$, значения которой, вычисленные в системе покоя, характеризуют знак плотности энергии в состоянии $\psi_{is}^{(\pm)}$. Расчет величин $g_{is}^{(\pm)}$ для уравнения ДК дает [39]:
\begin{align}
g_{1s}^{(+)} =g_{2s}^{(-)}=1, \quad g_{1s}^{(-)} =g_{2s}^{(+)}=-1 .
\label{eq541}
\end{align}

Теперь приступаем непосредственно к квантованию. Операторные волновые функции $\Psi$, $\overline{\Psi}$ представим в виде разложений
\begin{align}
\Psi (x) &= \frac{1}{(2 \pi)^{3/2}} \sum_i \sum_s \left[ a_{is} (p) \psi_{is}^{(+)} (p) e^{i p x} + \stackrel{+}{b}_{is} (p) \psi_{is}^{(-)} (p) e^{-i p x} \right] d^3 p,
\label{eq542} \\
\overline{\Psi} (x) &= \frac{1}{(2 \pi)^{3/2}} \sum_i \sum_s \left[  \stackrel{+}{a}_{is} (p) \bar{\psi}_{is}^{(+)} (p) e^{-i p x} + b_{is} (p) \bar{\psi}_{is}^{(-)} (p) e^{i p x} \right] d^3 p .
\label{eq543}
\end{align}
Для операторов рождения и уничтожения постулируем перестановочные соотношения
\begin{align}
& [a_{is} (p), \stackrel{+}{a}_{i's'} (p') ]_+ = g^{(+)}_{is} \delta_{ii'} \delta_{ss'} \delta (p-p'),
\label{eq544} \\
& [b_{is} (p), \stackrel{+}{b}_{i's'} (p') ]_+ = - g^{(-)}_{is} \delta_{ii'} \delta_{ss'} \delta (p-p')
\label{eq545}
\end{align}
(по индексам $i$ и $s$ суммирования здесь нет; все остальные антикоммутаторы равны нулю), которые соответствуют квантованию поля ДК по статистике Ферми–Дирака. Приводящие к правильным собственным значениям операторы числа частиц и античастиц определим при этом следующим образом:
\begin{align}
N_{is}^{(+)}  = g^{(+)}_{is} \stackrel{+}{a}_{is} a_{is} , \quad N_{is}^{(-)}  =  - g^{(-)}_{is} \stackrel{+}{b}_{is} b_{is} .
\label{eq546}
\end{align}
Подставляем разложения  \eqref{eq542},  \eqref{eq543} в выражения для операторов энергии и заряда
\begin{align}
E &= \int \{ ( \partial_4 \overline{\Psi})  \Gamma_4 \Psi -\overline{\Psi}  \Gamma_4 \partial_4  \Psi  \}d^3 x ,
\label{eq547} \\
Q &= e \int \overline{\Psi} \Gamma_4 \Psi d^3 x .
\label{eq548}
\end{align}
Учитывая соотношения  \eqref{eq544}--\eqref{eq546} и нормировку по заряду
\begin{align}
\overline{\Psi} \Gamma_4  \Psi = \pm 1,
\label{eq549}
\end{align}
получим в конечном счете для операторов $E$ и $Q$ выражения
\begin{align}
E &= \sum_i \sum_s \left( N_{is}^{(+)} \varepsilon_{is}^{(+)} + N_{is}^{(-)} \varepsilon_{is}^{(-)}  \right),
\label{eq550} \\
Q &= e \sum_i \sum_s \left( N_{is}^{(+)} - N_{is}^{(-)}  \right),
\label{eq551}
\end{align}
где $\varepsilon_{is}^{(\pm)} = |p_0 |$, индексы у $\varepsilon_{is}^{(\pm)}$ указывают на принадлежность к соответствующему состоянию.

Формулы  \eqref{eq550},  \eqref{eq551} означают, что антикоммутационные соотношения  \eqref{eq544},  \eqref{eq545} обеспечивают правильную корпускулярную картину поля. Кроме того, нетрудно показать, что они приводят к причинным перестановочным соотношениям для операторных волновых функций [40; 41].

Поскольку правые части некоторых из условий квантования  \eqref{eq544},  \eqref{eq545} содержат в правой части «неправильный» знак (минус), соответствующие векторы состояний должны иметь отрицательно определенную норму. Иными словами, квантовое описание поля ДК по статистике Ферми–Дирака предполагает использование пространства состояний $H$ c индефинитной метрикой
\begin{align}
H = H_+ \oplus H_-,
\label{eq552}
\end{align}
где $H_+$ и $H_-$ – подпространства с положительной и отрицательной нормами векторов состояний соответственно. В рассматриваемом случае в подпространства $H_+$ и $H_-$ попадают состояния
\begin{align}
H_+ : \quad \left( \prod_{N_1}  \stackrel{+}{a}_{1s} \right)  \left( \prod_{N_2}  \stackrel{+}{b}_{2s} \right)  \left( \prod_{N_3}  \stackrel{+}{a}_{2s} \right)  \left( \prod_{N_4}  \stackrel{+}{b}_{1s} \right) | 0 \rangle  ;
\label{eq553} \\
H_- : \quad \left( \prod_{N_5}  \stackrel{+}{a}_{1s} \right)  \left( \prod_{N_6}  \stackrel{+}{b}_{2s} \right)  \left( \prod_{N_7}  \stackrel{+}{a}_{2s} \right)  \left( \prod_{N_8}  \stackrel{+}{b}_{1s} \right) | 0 \rangle .
\label{eq554}
\end{align}
Здесь $N_1$, $N_2$, $N_5$, $N_6$ – произвольные неотрицательные целые числа, $(N_3 + N_4 )$ – четное и $( N_7 + N_8 )$ – нечетное числа. Для одночастичных состояний разбиения  \eqref{eq553},  \eqref{eq554} принимают соответственно вид
\begin{align}
& H_+ : \Psi_{1s}^{(+)}, \Psi_{1s}^{(-)}, \qquad H_- : \Psi_{2s}^{(+)}, \Psi_{2s}^{(-)},
\label{eq555} \\
& H_+ : \Psi_{1s}^{(+)}, \Psi_{2s}^{(-)}, \qquad H_- : \Psi_{2s}^{(+)}, \Psi_{1s}^{(-)} .
\label{eq556}
\end{align}
При этом для корректной вероятностной интерпретации теории необходимо, чтобы в ней при включении взаимодействия отсутствовали переходы между состояниями $H_+$ и $H_-$. Покажем, что здесь такие переходы действительно запрещены.

Рассмотрим лагранжиан
\begin{align}
\mathcal{L} = - \overline{\Psi} (x) (\Gamma_{\mu} \partial_{\mu} +m) \Psi (x) + \mathcal{L}_{int},
\label{eq557}
\end{align}
где $ \mathcal{L}_{int}$ описывает взаимодействие, которое не нарушает внутренней симметрии, присущей свободному полю. Для электромагнитного взаимодействия, например:
\begin{align}
 \mathcal{L}_{int} =e  \overline{\Psi} \Gamma_{\mu} A_{\mu} \Psi + \overline{\Psi} F_{\mu \nu} \Gamma_{[ \mu} \Gamma_{\nu ]} \Psi.
\label{eq558}
\end{align}
Очевидно, что оператор $\hat{\Pi}$  \eqref{eq539} содержится среди преобразований группы внутренней симметрии лагранжиана  \eqref{eq557},  \eqref{eq558} (сравни  \eqref{eq539} с генераторами  \eqref{eq510} этой группы). Инвариантность указанного лагранжиана относительно преобразований
\begin{align}
\Psi \to e^{i \hat{\Pi} \theta} \Psi
\label{eq559}
\end{align}
приводит к сохраняющемуся «заряду»
\begin{align}
G \sim \int \overline{\Psi} (x) \Gamma_4 \hat{\Pi} \Psi (x) d^3 x .
\label{eq560}
\end{align}
Заряд $G$  может быть преобразован к виду
\begin{align}
G \sim \sum_i \sum_s  \lambda_i \left( N_{is}^{(+)}  - N_{is}^{(-)} \right) = \sum_s \left( N_{1s}^{(+)} - N_{2s}^{(+)} - N_{1s}^{(-)} + N_{2s}^{(-)} \right),
\label{eq561}
\end{align}
где учтено, что $\lambda_1= 1$, $\lambda_2 = - 1$. Перепишем для удобства формулу  \eqref{eq551} также в развернутом виде
\begin{align}
Q \sim \sum_s \left( N_{1s}^{(+)} + N_{2s}^{(+)} - N_{1s}^{(-)} - N_{2s}^{(-)}\right).
\label{eq562}
\end{align}

Сравнивая разбиения  \eqref{eq555},  \eqref{eq556} с формулами  \eqref{eq561},  \eqref{eq562}, приходим к заключению, что одночастичным состояниям, относящимся к подпространствам $H_+$ и $H_-$, соответствуют следующие знаки зарядов $Q$ и $G$:
\begin{align}
H_+ : \quad (1,1), (-1,1), \qquad H_- : \quad (1,-1), (-1,-1)
\label{eq563}
\end{align}
(первая цифра в скобках относится к электрическому заряду $Q$, вторая – к дополнительному заряду $G$).

Из \eqref{eq563} очевидно, что совместное выполнение законов сохранения для зарядов $Q$ и $G$ приводит к запрету физически неприемлемых переходов между состояниями из подпространств с положительной и отрицательной нормами векторов состояний.

Отметим, что если вместо непрерывных преобразований \eqref{eq559} рассмотреть дискретные преобразования
\begin{align}
\Psi_+ \to \Psi_+, \quad \Psi_- \to - \Psi_- ,
\label{eq564}
\end{align}
то они сводятся к следующему:
\begin{align}
\begin{split}
& a_{1s}, \stackrel{+}{a}_{1s}\, \,  \to a_{1s}, \stackrel{+}{a}_{1s}, \quad  b_{1s}, \stackrel{+}{b}_{1s} \, \, \to  b_{1s}, \stackrel{+}{b}_{1s}, \\
& a_{2s}, \stackrel{+}{a}_{2s} \, \,  \to - a_{2s},  - \stackrel{+}{a}_{2s}, \quad b_{2l}, \stackrel{+}{b}_{2s} \, \,  \to -  b_{2s}, -  \stackrel{+}{b}_{2s}.
\end{split}
\label{eq565}
\end{align}
Такая операция носит в математической литературе название канонической, или $J$-симметрии. Она лежит в основе теории линейных операторов в пространстве с индефинитной метрикой, которое называется еще гильбертовым пространством с $J$-метрикой, или пространством Крейна [42]. Как показано в [43], $J$-симметрия соответствует оператору суперотбора, запрещающему переходы из $H_+$ в $H_-$, что находится в согласии с установленным выше результатом. При локализации некомпактной группы внутренней симметрии, имеющей пространственно-временное происхождение, и рассмотрении соответствующей калибровочной теории дискретная $J$-симметрия также позволяет исключить переходы, характеризующиеся отрицательными вероятностями [44].

Таким образом, рассмотренная процедура квантования уравнения ДК по статистике Ферми–Дирака является корректной и с точки зрения вероятностной интерпретации теории. Данный факт в совокупности с другими (алгебраическими, групповыми) выше отмеченными свойствами этого уравнения указывает на то, что на его основе действительно существует принципиальная возможность геометризованного описания внутренних степеней свободы дираковских частиц.

Укажем еще, что в самом общем виде вопрос о возможности физически непротиворечивого квантования РВУ с некомпактными группами внутренней симметрии как по нормальной, так и аномальной статистике с использованием индефинитной метрики подробно рассмотрен в работах [39–41].

\section{Алгебраические обобщения уравнения Дирака – Кэлера}

\hspace{1.25cm}
Несмотря на ряд привлекательных черт уравнения ДК, вопрос о его способности служить, например, для пространственно-временного описания внутренних квантовых чисел известных фундаментальных частиц до сих пор остается открытым. С одной стороны, если иметь в виду только одно квантовое число – поколение кварков (или лептонов), то число компонент волновой функции в уравнении ДК (равное $16$) является слишком большим, поскольку в настоящее время известно всего три поколения этих частиц. С другой, для геометризованного описания с единых позиций всех известных степеней свободы фундаментальных частиц $16$-ти компонент волновой функции явно недостаточно. Поэтому последовательная реализация данного подхода в рамках $4$-мерного пространства-времени предполагает использование РВУ с большим числом компонент волновой функции и аналогичными уравнению ДК свойствами.

Набор таких РВУ можно получить, если перейти к максимальному тензорному алгебраическому обобщению уравнения ДК в пространстве размерности $d = 4$, заключающемуся в рассмотрении полей $\psi_0$, $\tilde{\psi}_0$, $\psi_\mu$, $\tilde{\psi}_\mu$, $\psi_{[\mu \nu]}$ как общих элементов алгебры Клиффорда $C_4$. Этот переход эквивалентен введению наряду с диракоподобным РВУ для функции $\Psi$ \eqref{eq52} аналогичных уравнений для функций $\tilde{\Psi}$ и $\Psi_A$, где $A$~-- свободный лоренцевский индекс, принимающий по очереди значения $A=\mu , \tilde{\mu}, [\mu \nu]$ (здесь $\tilde{\mu}$~-- псевдовекторный индекс; $\tilde{\Psi}$ характеризует тот же набор
антисимметричных тензорных полей, что и $\Psi$, но с измененными трансформационными свойствами относительно операции пространственной инверсии).

Прежде, чем переходить к указанным алгебраическим обобщениям уравнения ДК, обсудим более подробно вопрос о его замкнутости (нераспадении) в смысле полной группы Лоренца. С этой целью с помощью подстановок
\begin{align}
\begin{split}
& \varphi_0 = \psi_0 - i \tilde{\psi}_0, \quad \varphi_{\mu} = \psi_\mu - i \tilde{\psi}_\mu , \quad \varphi_{[\mu \nu]} = \psi_{[\mu \nu]} - i \tilde{\psi}_{[\mu  \nu ]} , \\
& \dot{\varphi}_0 =  \psi_0 + i \tilde{\psi}_0, \quad \dot{\varphi}_{\mu} = \psi_\mu + i \tilde{\psi}_\mu , \quad \dot{\varphi}_{[\mu \nu]} = \psi_{[\mu \nu]} + i \tilde{\psi}_{[\mu  \nu ]}
\end{split}
\label{eq61}
\end{align}
приведем систему \eqref{eq51} к виду
\begin{align}
\begin{split}
& \partial_\mu \varphi_\mu + m \varphi_0 =0 ,\\
& \partial_\nu \varphi_{[\mu \nu ]} + \partial_{\mu} \varphi_0 + m \varphi_\mu =0, \\
-& \partial_\mu \varphi_\nu + \partial_\nu \varphi_\mu + i \varepsilon_{\mu \nu \alpha \beta} \partial_\alpha \varphi_\beta + m \varphi_{[\mu \nu ]}  =0;
\end{split}
\label{eq62}
\end{align}
\begin{align}
\begin{split}
& \partial_\mu \dot{\varphi}_\mu + m \dot{\varphi}_0 =0 ,\\
& \partial_\nu  \dot{\varphi}_{[\mu \nu ]} + \partial_{\mu}  \dot{\varphi}_0 + m  \dot{\varphi}_\mu =0, \\
-& \partial_\mu  \dot{\varphi}_\nu + \partial_\nu  \dot{\varphi}_\mu - i \varepsilon_{\mu \nu \alpha \beta} \partial_\alpha  \dot{\varphi}_\beta + m  \dot{\varphi}_{[\mu \nu ]}  =0,
\end{split}
\label{eq63}
\end{align}
Очевидно, что система \eqref{eq62}, \eqref{eq63} распадается на две инвариантные в смысле собственной группы Лоренца подсистемы \eqref{eq62} и \eqref{eq63}. Однако так же очевидно, что при пространственном отражении эти подсистемы переходят друг в друга, то есть по отдельности каждая из них $P$-неинвариантна. Таким образом, действительно, по отношению к преобразованиям полной группы Лоренца система ДК является не распадающейся.

Теперь рассмотрим матричное уравнение
\begin{align}
(\Gamma_{\alpha} \partial_{\alpha} + m ) \Psi_{[\eta \xi]}  =0,
\label{eq64}
\end{align}
которое получается в результате навешивания на волновую функцию $\Psi$ \eqref{eq52} уравнения ДК свободного бивекторного индекса. Данное уравнение можно представить в стандартной форме \eqref{eq:nonzero_mass}, где $96$-компонентная волновая функция $\psi$ преобразуется согласно представлению группы Лоренца
\begin{align}
\left[ (0,0) \oplus 2 (\frac12 , \frac12) \oplus (0,1) \oplus (1,0) \right] \otimes [(0,1) \oplus (1,0)].
\label{eq65}
\end{align}
Неприводимые представления, содержащиеся в прямом произведении \eqref{eq65}, формируют в общем случае схему зацеплений
\begin{align}
\begin{tabular}{ccccc}
& & $2 (0,0)$ & & \\
& & $|$ & & \\
$3 (0,1)$ & --- & $4 (\frac12, \frac12)$ & --- & 3 $(1,0)$.
\end{tabular}
\label{eq66}
\end{align}

Как было показано, уравнение ДК представимо в виде прямой суммы двух диракоподобных $8$-компонентных систем \eqref{eq62} и \eqref{eq63}, которые, в свою очередь, можно записать в форме \eqref{eq:nonzero_mass} с волновыми функциями
\begin{align}
\Psi^{(8)}  = (\varphi_0 , \varphi_\mu , \varphi_{[\mu \nu]} )^T , \quad \dot{\Psi}^{(8)}  = (\dot{\varphi}_0 , \dot{\varphi}_\mu , \dot{\varphi}_{[\mu \nu]} )^T
\label{eq67}
\end{align}
(отметим, что тензоры $\varphi_{[\mu \nu]}$, $\dot{\varphi}_{[\mu \nu]}$ являясь здесь самодуальными, содержат по три независимые компоненты) и схемами зацеплений
\begin{align}
(0,0) \quad  \textup{---} \quad  (\frac12 , \frac12) \quad \textup{---}  \quad (0,1),
\label{eq68} \\
(0,0) \quad  \textup{---} \quad (\frac12 , \frac12) \quad  \textup{---} \quad (1,0).
\label{eq69}
\end{align}
Поэтому уравнение \eqref{eq64} распадается на составляющие
\begin{align}
& (\Gamma^{(8)}_{\mu} \partial_{\mu} +m) \Psi^{(8)}_{(0,1)} =0 ,
\label{eq610} \\
& (\Gamma^{(8)}_{\mu} \partial_{\mu} +m) \dot{\Psi}^{(8)}_{(0,1)} =0 ,
\label{eq611} \\
& (\Gamma^{(8)}_{\mu} \partial_{\mu} +m) \Psi^{(8)}_{(1,0)} =0 ,
\label{eq612} \\
& (\Gamma^{(8)}_{\mu} \partial_{\mu} +m) \dot{\Psi}^{(8)}_{(1,0)} =0 ,
\label{eq613}
\end{align}

Уравнения \eqref{eq610} и \eqref{eq612} $P$-сопряжены по отношению друг к другу. Так же сопряжены уравнения \eqref{eq611} и \eqref{eq613}. Поэтому, рассматривая \eqref{eq610}, \eqref{eq612} и \eqref{eq611}, \eqref{eq613} совместно, получим две инвариантные в смысле полной группы Лоренца системы, преобразующиеся по представлениям
\begin{align}
& \left\{ [(0,0) \oplus (\frac12 , \frac12) \oplus (0,1)] \otimes (0,1)\right\} \oplus \left\{ [(0,0) \oplus (\frac12 , \frac12) \oplus (1,0)] \otimes (1,0)\right\},
\label{eq614} \\
& \left\{ [(0,0) \oplus (\frac12 , \frac12) \oplus (0,1)] \otimes (1,0)\right\} \oplus \left\{ [(0,0) \oplus (\frac12 , \frac12) \oplus (1,0)] \otimes (0,1)\right\}.
\label{eq615}
\end{align}
Другими словами, схема зацеплений \eqref{eq66} при построении на ее основе диракоподобного РВУ распадается на фрагменты
\begin{align}
\begin{tabular}{ccccccc}
& & & $2 (0,0)$ & & & \\
& & & $|$ & & &  \\
& $2 (0,1)$ & --- & $2 (\frac12, \frac12)$ & --- & 2 $(1,0)$  & \\
& $|$ & &  & &$|$  & \\
$(0,2) \quad $ --- & $(\frac12, \frac32)$ &  & &  & $(\frac32,\frac12)$ & --- $\quad (2,0)$
\end{tabular}
\label{eq616}
\end{align}
и
\begin{align}
\begin{tabular}{ccccc}
$(0,1)$ &--- & $2 (\frac12,\frac12)$ &--- & $(1,0)$ \\
$|$ & & $|$ & & $|$\\
$(\frac12, \frac32)$ & --- & $2 (1,1)$ & --- & $(\frac32, \frac12)$.
\end{tabular}
\label{eq617}
\end{align}

В свою очередь, из неприводимых компонент, входящих в \eqref{eq616}, можно сформировать две самостоятельные схемы зацеплений: схему \eqref{eq515}, соответствующую уравнению ДК, и схему
\begin{align}
\begin{tabular}{ccccccc}
& $(0,1)$ &  & &  & $(1,0)$  & \\
& $|$ & & $\oplus$ & &$|$  & \\
$(0,2) \quad $ --- & $(\frac12, \frac32)$ &  & &  & $(\frac32,\frac12)$ & --- $\quad (2,0)$.
\end{tabular}
\label{eq618}
\end{align}

Покажем, что схема зацеплений \eqref{eq618} действительно обеспечивает возможность построения удовлетворяющего всем необходимым физическим требованиям РВУ дираковского типа. Пронумеруем содержащиеся \eqref{eq618} неприводимые компоненты следующим образом:
\begin{align}
\begin{split}
& (0,1) \sim 1, \quad (\frac12 , \frac32) \sim 2, \quad (0,2) \sim 3, \\
& (1,0) \sim 4, \quad (\frac32 , \frac12) \sim 5, \quad (2,0) \sim 6.
\end{split}
\label{eq619}
\end{align}
Тогда в базисе Гельфанда–Яглома для матрицы $\Gamma_4$ получим выражение
\begin{align}
& \Gamma_4 = \left(
\begin{tabular}{cc}
$C^1 \otimes I_3$ & $0$  \\
$0$ & $C^2 \otimes I_5$
\end{tabular} \right),
\label{eq620} \\
& C^1 = \left(
\begin{tabular}{cccc}
$0$ & $c^1_{12}$ & $0$ & $0$ \\
$c^1_{21}$ & $0$ & $0$ & $0$ \\
$0$ & $0$ & $0$ & $c^1_{45}$ \\
$0$ & $0$ & $c^1_{54}$ & $0$
\end{tabular} \right), \quad
C^2 = \left(
\begin{tabular}{cccc}
$0$ & $c^2_{23}$ & $0$ & $0$ \\
$c^2_{32}$ & $0$ & $0$ & $0$ \\
$0$ & $0$ & $0$ & $c^2_{56}$ \\
$0$ & $0$ & $c^2_{65}$ & $0$
\end{tabular} \right).
\label{eq621}
\end{align}

Требование $P$-инвариантности теории в соответствии с условиями \eqref{c2_cond}, \eqref{Pc1_cond} накладывает на элементы спиновых блоков $C^{1}$, $C^{2}$ ограничения
\begin{align}
c^1_{12} = c^1_{45}, \quad c^1_{21} = c^1_{54}, \quad c^2_{23} = c^2_{56}, \quad c^2_{32} = c^2_{65}.
\label{eq622}
\end{align}
Возможность лагранжевой формулировки (смотри (1.48)) приводит к соотношениям
\begin{align}
c_{12}^1 \eta_{25}^1 = (c_{54}^1)^* \eta_{14}^1, \quad c_{23}^2 \eta_{36}^2 = (c_{65}^2)^* \eta_{25}^2 .
\label{eq623}
\end{align}
Объединяя \eqref{eq622} и \eqref{eq623}, получаем
\begin{align}
c^1_{21}=c^1_{54}= f a^* , \quad c^2_{32}=c^2_{65}=g b^* ,
\label{eq624}
\end{align}
где введены обозначения
\begin{align}
f = \frac{\eta^1_{25}}{\eta^1_{14}}, \quad g= \frac{\eta^2_{36}}{\eta^2_{25}}, \quad a= c^1_{12}, \quad b= c^2_{23}.
\label{eq625}
\end{align}

Минимальное уравнение для матрицы $\Gamma_4$ будет иметь вид \eqref{eq531} при выполнении равенств
\begin{align}
f |a|^2 = g |b|^2 =1,
\label{eq626}
\end{align}
которым можно удовлетворить, полагая, например,
\begin{align}
f=g=a=b=1.
\label{eq627}
\end{align}
При этом спиновые блоки $C^{1}$, $C^{2}$ принимают одинаковый вид
\begin{align}
C^{1} = C^{2} =I_2 \otimes \sigma_1 .
\label{eq628}
\end{align}

Ненулевые элементы $\eta^s_{ij}$ матрицы билинейной инвариантной формы $\eta$, имеющей в рассматриваемом случае в базисе Гельфанда–Яглома структуру
\begin{align}
\eta = \left(
\begin{tabular}{cc}
$\eta^1 \otimes I_3$ & $0$  \\
$0$ & $\eta^2 \otimes I_5$
\end{tabular} \right),
\label{eq629}
\end{align}
в соответствии с условиями \eqref{eq625}, \eqref{eq627} могут быть выбраны следующим образом:
\begin{align}
\eta^1_{14} = \eta^1_{25} = - \eta^2_{25} = -\eta^2_{36} = 1
\label{eq630}
\end{align}
(напомним, что $\eta^s_{ij} = -\eta^{s-1}_{ij}$).

Полученное таким образом $32$-компонентное РВУ является по построению уравнением дираковского типа, инвариантно относительно преобразований полной группы Лоренца, допускает лагранжеву формулировку и с точки зрения стандартной трактовки теории РВУ описывает микрообъект с набором спинов $1, 2$, одним значением массы и удвоенным набором состояний, вырожденных по некоторому дополнительному квантовому числу. Это уравнение, как и уравнение ДК, не распадается в смысле полной группы Лоренца.

Теперь рассмотрим схему зацеплений \eqref{eq617}. Если строить на ее основе РВУ дираковского типа, то она распадается в прямую сумму фрагментов [45]
\begin{align}
\begin{tabular}{ccccccccc}
$(0,1)$ & --- & $(\frac12, \frac12)$ &  & &  & $(\frac12, \frac12)'$  & --- & $(1,0)$ \\
$|$ & & $|$ & & $\oplus$ & &$|$  & & $|$\\
$(\frac12, \frac32)$ & --- & $(1,1)$ &  & &  & $(1,1)'$ & --- & $(\frac32,\frac12)$.
\end{tabular}
\label{eq631}
\end{align}
где кратные представления $(\frac12, \frac12)$ и $(\frac12, \frac12)'$, а также $(1, 1)$ и $(1, 1)'$ $P$-сопряжены друг другу.

Введем нумерацию содержащихся в схеме \eqref{eq631} неприводимых компонент
\begin{align}
& (\frac12, \frac12) \sim 1, \quad (\frac12, \frac32) \sim 2,  \quad (1,1) \sim 3, \quad (0,1) \sim 4, \notag \\
& (\frac12, \frac12)' \sim 5, \quad (\frac32, \frac12) \sim 6,  \quad (1,1)' \sim 7, \quad (1,0) \sim 8. \notag
\end{align}
Тогда для спиновых блоков $C^{s}$ ($s = 0, 1, 2$) матрицы $\Gamma_4$
\begin{align}
\Gamma_4 = \left(
\begin{tabular}{ccc}
$C^0$ & $0$ & $0$ \\
$0$ & $C^1 \otimes I_3$ & $0$  \\
$0$ & $0$ & $C^2 \otimes I_5$
\end{tabular} \right)
\label{eq632}
\end{align}
получим общие выражения
\begin{align}
\begin{split}
C^0 &= \left(
\begin{tabular}{cccc}
$0$ & $0$ & $c^0_{13}$ & $0$ \\
$0$ & $0$ & $0$ & $c^0_{57}$ \\
$c^0_{31}$ & $0$ & $0$ & $0$ \\
$0$ & $c^0_{75}$ & $0$ & $0$
\end{tabular} \right), \quad C^1 = \left(
\begin{tabular}{cc}
$(C^1)'$ & $0$  \\
$0$ & $(C^1)''$
\end{tabular} \right), \\
(C^1)' & = \left(
\begin{tabular}{cccc}
$0$ & $0$ & $c^1_{13}$ & $c^1_{14}$ \\
$0$ & $0$ & $c^1_{23}$ & $c^2_{24}$ \\
$c^1_{31}$ & $c^1_{32}$ & $0$ & $0$ \\
$c^1_{41}$ & $c^1_{42}$ & $0$ & $0$
\end{tabular} \right), \quad
(C^1)'' = \left(
\begin{tabular}{cccc}
$0$ & $0$ & $c^1_{57}$ & $c^1_{58}$ \\
$0$ & $0$ & $c^1_{67}$ & $c^2_{68}$ \\
$c^1_{75}$ & $c^1_{76}$ & $0$ & $0$ \\
$c^1_{85}$ & $c^1_{86}$ & $0$ & $0$
\end{tabular} \right), \\
C^2 &= \left(
\begin{tabular}{cccc}
$0$ & $c^2_{23}$ & $0$ & $0$ \\
$c^2_{32}$ & $0$ & $0$ & $0$ \\
$0$ & $0$ & $0$ & $c^2_{67}$ \\
$0$ & $0$ & $c^2_{76}$ & $0$
\end{tabular} \right).
\end{split}
\label{eq633}
\end{align}

Условия релятивистской \eqref{c_cond} и $P$-инвариантности \eqref{c1_cond}, \eqref{c2_cond} накладывают на элементы блоков \eqref{eq633} соответственно ограничения
\begin{align}
\begin{split}
c^1_{13} &= \sqrt{\frac23} c^0_{13}, \quad c^1_{31} = \sqrt{\frac23} c^0_{31}, \quad c^1_{57} = \sqrt{\frac23} c^0_{57}, \quad c^1_{75} = \sqrt{\frac23} c^0_{75},  \\
c^1_{23} &= \sqrt{\frac13} c^2_{23}, \quad c^1_{32} = \sqrt{\frac13} c^2_{32}, \quad c^1_{67} = \sqrt{\frac13} c^2_{67}, \quad c^1_{76} = \sqrt{\frac13} c^2_{76};
\end{split}
\label{eq634}
\end{align}
\begin{align}
\begin{split}
c^1_{14} & = \pm c^1_{58}, \quad c^1_{23}  = \pm c^1_{67} , \quad c^2_{23} = \pm c^2_{67} ,\quad c^1_{24} =  c^1_{68} , \\
c^1_{41} & = \pm c^1_{85}, \quad c^1_{32}  = \pm c^1_{76} , \quad c^2_{32} = \pm c^2_{76} ,\quad c^1_{42} =  c^1_{86} .
\end{split}
\label{eq635}
\end{align}
Условие \eqref{c_eta_cond} возможности получения РВУ со схемой зацеплений \eqref{eq617} из инвариантной функции Лагранжа приводит к соотношениям
\begin{align}
\begin{split}
& c^s_{31} = \frac{\eta^s_{33}}{\eta^s_{11}} (c^s_{13})^*, \quad  c^s_{75} = \frac{\eta^s_{77}}{\eta^s_{55}} (c^s_{57})^*
 \quad (s=0,1), \\
 & c^s_{76} = \frac{\eta^s_{33}}{\eta^s_{26}} (c^s_{23})^*, \quad  c^s_{67} = \frac{\eta^s_{26}}{\eta^s_{77}} (c^s_{32})^*
 \quad (s=1,2), \\
 & c^1_{85} = \frac{\eta^1_{48}}{\eta^1_{55}} (c^1_{14})^*, \quad  c^1_{58} = \frac{\eta^1_{11}}{\eta^1_{48}} (c^1_{41})^*, \\
 & c^1_{86} = \frac{\eta^1_{48}}{\eta^1_{26}} (c^s_{24})^*, \quad  c^1_{68} = \frac{\eta^1_{26}}{\eta^1_{48}} (c^1_{42})^*.
\end{split}
\label{eq636}
\end{align}
Минимальные уравнения
\begin{align}
(C^0)^2 - 1=0, \quad (C^1)^2 - 1=0, \quad(C^2)^2 - 1=0,
\label{eq637}
\end{align}
приводящие к дираковской алгебре матриц $\Gamma_\mu$, будут иметь место при выполнении равенств
\begin{align}
\begin{split}
& c^0_{13} c^0_{31}=c^2_{23} c^2_{32} =1, \\
& c^1_{13} c^1_{31}+c^1_{14} c^1_{41} =1, \quad c^1_{23} c^1_{32}+c^1_{24} c^1_{42} =1, \\
& c^1_{13} c^1_{32}+c^1_{14} c^1_{42} =0, \quad c^1_{23} c^1_{31}+c^1_{24} c^1_{41} =0, \\
& c^1_{13} c^1_{31}+c^1_{23} c^1_{32} =1, \quad c^1_{14} c^1_{41}+c^1_{24} c^1_{42} =1, \\
& c^1_{14} c^1_{31}+c^1_{24} c^1_{32} =0, \quad c^1_{13} c^1_{41}+c^1_{23} c^1_{42} =0,
\end{split}
\label{eq638}
\end{align}
плюс аналогичные равенства, которые получаются из \eqref{eq638} при замене индексов $1 \to 5$, $2 \to 6$, $3 \to 7$, $4 \to  8$.

Условия \eqref{eq634}--\eqref{eq638} одновременно выполняются, если положить
\begin{align}
\begin{split}
& c^0_{13} =c^0_{31} =c^0_{57} =c^0_{75} =1, \quad - c^2_{23} = - c^2_{32} =c^2_{67} =c^2_{76} =1, \\
& c^1_{13} =c^1_{31} =c^1_{57} =c^1_{75} =\sqrt{\frac23}, \quad -c^1_{23} =-c^1_{32} =c^1_{67} =c^1_{76} =\sqrt{\frac13}, \\
& c^1_{14} =c^1_{41} =-c^1_{58} =-c^1_{85} =\sqrt{\frac13}, \quad c^1_{24} =c^1_{42} =c^1_{68} =c^1_{86} =\sqrt{\frac23}
\label{eq639}
\end{split}
\end{align}
и
\begin{align}
\begin{split}
\eta_{11}^0 &=\eta_{33}^0=\eta_{55}^0=\eta_{77}^0=-\eta_{11}^1=-\eta_{33}^1=-\eta_{55}^1=-\eta_{77}^1 \\
&= \eta_{26}^1=\eta_{62}^1=\eta_{48}^1=\eta_{84}^1=-\eta_{26}^2=-\eta_{62}^2=\eta_{33}^2=\eta_{77}^2 =1.
\label{eq640}
\end{split}
\end{align}
В результате приходим к следующим выражениям для блоков $C^{s}$, $\eta^{s}$ матриц $\Gamma_4$ и $\eta$:
\begin{align}
\begin{split}
& C^0 = \sigma_1 \otimes I_2 ,  \quad C^1 = (C^1)' \oplus (C^1)'', \quad C^2 = - \sigma_3 \otimes \sigma_1 , \\
& (C^1)' = \frac{1}{\sqrt{3}}  \left(
\begin{tabular}{cccc}
$0$ & $0$ & $\sqrt{2}$ & $1$ \\
$0$ & $0$ & $-1$ & $\sqrt{2}$ \\
$\sqrt{2}$ & $-1$ & $0$ & $0$ \\
$1$ & $\sqrt{2}$ & $0$ & $0$
\end{tabular} \right), \quad
(C^1)'' = \frac{1}{\sqrt{3}} \left(
\begin{tabular}{cccc}
$0$ & $0$ & $\sqrt{2}$ & $-1$ \\
$0$ & $0$ & $1$ & $\sqrt{2}$ \\
$\sqrt{2}$ & $1$ & $0$ & $0$ \\
$-1$ & $\sqrt{2}$ & $0$ & $0$
\end{tabular} \right);
\end{split}
\label{eq641}
\end{align}
\begin{align}
\begin{split}
& \eta^0 = \left(
\begin{tabular}{cccc}
$1$ & $0$ & $0$ & $0$ \\
$0$ & $1$ & $0$ & $0$ \\
$0$ & $0$ & $1$ & $0$ \\
$0$ & $0$ & $0$ & $1$ \\
\end{tabular} \right), \quad
\eta^2 = \left(
\begin{tabular}{cccc}
$0$ & $0$ & $-1$ & $0$ \\
$0$ & $1$ & $0$ & $0$ \\
$-1$ & $0$ & $0$ & $0$ \\
$0$ & $0$ & $0$ & $1$ \\
\end{tabular} \right), \\
& \eta^1 =  \left(
\begin{tabular}{cccccccc}
$-1$ & $0$ & $0$ & $0$  & $0$ & $0$ & $0$ & $0$ \\
$0$ & $0$ & $0$ & $0$  & $0$ & $1$ & $0$ & $0$ \\
$0$ & $0$ & $-1$ & $0$ & $0$ & $0$ & $0$ & $0$ \\
$0$ & $0$ & $0$ & $0$ & $0$ & $0$ & $0$ & $1$ \\
$0$ & $0$ & $0$ & $0$  & $-1$ & $0$ & $0$ & $0$ \\
$0$ & $1$ & $0$ & $0$  & $0$ & $0$ & $0$ & $0$ \\
$0$ & $0$ & $0$ & $0$ & $0$ & $0$ & $-1$ & $0$ \\
$0$ & $0$ & $0$ & $1$ & $0$ & $0$ & $0$ & $0$
\end{tabular} \right) .
\end{split}
\label{eq642}
\end{align}

РВУ с матрицами $\Gamma_4$ \eqref{eq632}, \eqref{eq641} и $\eta$ \eqref{eq642}, базирующееся на схеме зацеплений \eqref{eq631}, удовлетворяет всем требованиям, сформулированным в главе 1 (кроме дефинитности энергии), и формально описывает микрообъект с набором спинов $0, 1, 2$ и единственной ненулевой массой. При этом состояния с $s = 0, 2$ двукратно вырождены, а состояния с $s = 1$ вырождены четырехкратно.

Если, наконец, рассмотреть уравнение
\begin{align}
(\Gamma_{\alpha} \partial_{\alpha} +m ) \Psi_\mu =0,
\label{eq643}
\end{align}
которое получается в результате навешивания на волновую функцию $\Psi$ уравнения ДК свободного векторного индекса $\mu$, то тут имеет место следующая ситуация. Волновая функция $\Psi_\mu$ преобразуется по представлению
\begin{align}
\left[ 2 (0,0) \oplus 2 (\frac12, \frac12) \oplus (0,1) \oplus (1,0) \right] \otimes (\frac12, \frac12),
\label{eq644}
\end{align}
соответствующему в общем случае схеме зацеплений
\begin{align}
\begin{tabular}{ccccc}
 & & $2 (0,0)$ & &  \\
 & & $|$ & &   \\
 $2 (0,1)$ & --- & $4 (\frac12, \frac12)$ & --- & 2 $(1,0)$  \\
 $|$ & & $|$   & &$|$  \\
 $(\frac12, \frac32)$ & --- & 2 (1,1) & ---  & $(\frac32,\frac12)$ .
\end{tabular}
\label{eq645}
\end{align}

Как показано в [45], при построении на основе схемы \eqref{eq645} РВУ дираковского типа она распадается на две не зацепляющиеся между собой схемы \eqref{eq515} и \eqref{eq617}. Следовательно, с интересующих нас позиций, уравнение \eqref{eq644} не дает ничего нового по сравнению с рассмотренными выше РВУ.

Таким образом, обсуждаемые алгебраические обобщения уравнения ДК приводят к двум новым матричным РВУ дираковского типа: $32$-компонентному – со схемой зацеплений \eqref{eq618} и $48$-компонентному – со схемой \eqref{eq631}.

Тензорная формулировка первого из них имеет вид [46]
\begin{align}
\begin{split}
& \partial_{\nu} \varphi_{\nu [\alpha \beta]} + m \varphi_{[\alpha \beta]} =0, \\
& \frac12 \left( - \partial_{\mu} \varphi_{\nu [\alpha \beta]} + \partial_{\nu} \varphi_{\mu [\alpha \beta]} - \partial_{\alpha} \varphi_{\beta [\mu \nu]} + \partial_{\beta} \varphi_{\alpha [\mu \nu]} + \right. \\
& \left. + i \varepsilon_{\mu \nu \eta \xi} \partial_{\eta} \varphi_{\xi [\alpha \beta]} + i \varepsilon_{\alpha \beta \eta \xi} \partial_{\eta} \varphi_{\xi [\mu \nu]}  \right) + m \varphi_{([\mu \nu] [\alpha \beta])} =0 , \\
& \partial_{\nu} \varphi_{([\mu \nu] [\alpha \beta])} + \frac12 \left( \partial_{\beta} \varphi_{[\alpha \mu]}  -\partial_{\alpha} \varphi_{[\beta \mu]} + \delta_{\mu \alpha} \partial_{\nu} \varphi_{[\nu \beta]}  - \delta_{\mu \beta} \partial_{\nu} \varphi_{ [\nu \alpha]}  \right) + \\
& + i \varepsilon_{\alpha \beta \eta \nu} \partial_{\eta} \varphi_{[\nu \mu]} + m \varphi_{\mu  [\alpha \beta]} =0 , \\
& \partial_{\nu} \chi_{\nu [\alpha \beta]} + m \chi_{[\alpha \beta]} =0, \\
& \frac12 \left( - \partial_{\mu} \chi_{\nu [\alpha \beta]} + \partial_{\nu} \chi_{\mu [\alpha \beta]} - \partial_{\alpha} \chi_{\beta [\mu \nu]} + \partial_{\beta} \chi_{\alpha [\mu \nu]} - \right. \\
& \left. - i \varepsilon_{\mu \nu \eta \xi} \partial_{\eta} \chi_{\xi [\alpha \beta]} - i \varepsilon_{\alpha \beta \eta \xi} \partial_{\eta} \chi_{\xi [\mu \nu]}  \right) + m \chi_{([\mu \nu] [\alpha \beta])} =0 , \\
& \partial_{\nu} \chi_{([\mu \nu] [\alpha \beta])} + \frac12 \left( \partial_{\beta} \chi_{[\alpha \mu]}  -\partial_{\alpha} \chi_{[\beta \mu]} + \delta_{\mu \alpha} \partial_{\nu} \chi_{[\nu \beta]}  - \delta_{\mu \beta} \partial_{\nu} \chi_{ [\nu \alpha]}  \right) - \\
& - i \varepsilon_{\alpha \beta \eta \nu} \partial_{\eta} \chi_{[\nu \mu]} + m \chi_{\mu  [\alpha \beta]} =0 .
\end{split}
\label{eq646}
\end{align}
Входящие сюда тензоры удовлетворяют условиям самодуальности
\begin{align}
& \frac12 \varepsilon_{\mu \nu \alpha \beta} \varphi_{[\alpha \beta ]} = i \varphi_{[\mu \nu]} , \quad \frac12 \varepsilon_{\eta \xi \alpha \beta} \varphi_{\mu [\alpha \beta ]} = i \varphi_{\mu [\eta \xi]} , \quad \frac12 \varepsilon_{\eta \xi \alpha \beta} \varphi_{([\mu \nu] [\alpha \beta ] )} = i \varphi_{([\mu \nu] [\eta \xi ] )} , \label{eq647}  \\
& \frac12 \varepsilon_{\mu \nu \alpha \beta} \chi_{[\alpha \beta ]} = - i \varphi_{[\mu \nu]} , \quad \frac12 \varepsilon_{\eta \xi \alpha \beta} \chi_{\mu [\alpha \beta ]} =- i \chi_{\mu [\eta \xi]} , \quad \frac12 \varepsilon_{\eta \xi \alpha \beta} \chi_{([\mu \nu] [\alpha \beta ] )} = - i \chi_{([\mu \nu] [\eta \xi ] )} . \notag
\end{align}
Кроме того, величины $\varphi_{\mu [\alpha \beta]} $, $\varphi_{([\mu \nu] [\alpha \beta])} $, $\chi_{\mu [\alpha \beta]}$, $\chi_{([\mu \nu] [\alpha \beta])}$ подчиняются условиям
\begin{align}
\begin{split}
& \varphi_{\alpha [\alpha \beta]} =0, \quad \varphi_{([\alpha \beta] [\alpha \nu])} =0 , \\
& \chi_{\alpha [\alpha \beta]} =0, \quad \chi_{([\alpha \beta] [\alpha \nu])} =0  .
\end{split}
\label{eq648}
\end{align}
Иными словами, фигурирующие в системе \eqref{eq646} тензорные величины сопоставляются следующим неприводимым представлениям группы Лоренца:
\begin{align}
\begin{split}
& \varphi_{[\alpha \beta]} \sim (0,1), \quad \varphi_{\mu [\alpha \beta]} \sim (\frac12,\frac32), \quad \varphi_{([\mu \nu] [\alpha \beta])} \sim (0,2), \\
& \chi_{[\alpha \beta]} \sim (1,0), \quad \chi_{\mu [\alpha \beta]} \sim (\frac32,\frac12), \quad \chi_{([\mu \nu] [\alpha \beta])} \sim (2,0).
\end{split}
\label{eq649}
\end{align}

$48$-компонентная тензорная система, соответствующая РВУ со схемой зацеплений \eqref{eq631}, имеет вид [47]
\begin{align}
\begin{split}
& \partial_{\nu} \varphi_{[\alpha \nu][\alpha \beta]} + \partial_{\alpha} \varphi_{[\alpha \beta]} + m \varphi_{\beta} =0 , \\
& \partial_{\lambda} \varphi_{\lambda [\alpha \beta]} + \frac13 (\partial_{\alpha} \varphi_{\beta} - \partial_{\beta}  \varphi_{\alpha} - i \varepsilon_{\alpha \beta \lambda \rho} \partial_{\lambda} \varphi_{\rho} ) +m \varphi_{[\alpha \beta] } =0, \\
& \partial_{\nu} \varphi_{[\mu \nu] [\alpha \beta]} + \partial_{\mu} \varphi_{[\alpha \beta]} - \frac13 (\delta_{\mu \alpha} \partial_{\eta} \varphi_{[\eta \beta] } - \delta_{\mu \beta} \partial_{\eta} \varphi_{[\eta \alpha] } + \delta_{\mu \alpha} \partial_{\nu} \varphi_{[\eta \nu] [\eta \beta] } - \delta_{\mu \beta} \partial_{\nu} \varphi_{[\eta \nu ] [\eta \alpha] }  +  \\
& + i \varepsilon_{\mu \alpha \beta \rho} \partial_{\eta} \varphi_{[\eta \rho]} + i \varepsilon_{\mu \alpha \beta \rho}  \partial_{\nu} \varphi_{[\eta \nu ] [\eta \rho ]}) + m \varphi_{\mu [\alpha \beta]} = 0, \\
-& \partial_{\eta} \varphi_{\nu [\eta \beta]} + \frac13 (2 \partial_{\nu} \varphi_{\beta} +2 \partial_{\beta}  \varphi_{\nu} -  \delta_{\nu \beta} \partial_{\rho} \varphi_{\rho} )
- i \varepsilon_{\rho \nu \eta \xi} \partial_{\eta} \varphi_{\xi [\rho \beta]}  +m \varphi_{[\eta \nu] [\eta \beta] } =0, \\
& \partial_{\nu} \chi_{[\alpha \nu][\alpha \beta]} + \partial_{\alpha} \chi_{[\alpha \beta]} + m \chi_{\beta} =0 , \\
& \partial_{\lambda} \chi_{\lambda [\alpha \beta]} + \frac13 (\partial_{\alpha} \chi_{\beta} - \partial_{\beta}  \chi_{\alpha} + i \varepsilon_{\alpha \beta \lambda \rho} \partial_{\lambda} \chi_{\rho} ) +m \chi_{[\alpha \beta] } =0, \\
& \partial_{\nu} \chi_{[\mu \nu] [\alpha \beta]} + \partial_{\mu} \chi_{[\alpha \beta]} - \frac13 (\delta_{\mu \alpha} \partial_{\eta} \chi_{[\eta \beta] } - \delta_{\mu \beta} \partial_{\eta} \chi_{[\eta \alpha] } + \delta_{\mu \alpha} \partial_{\nu} \chi_{[\eta \nu] [\eta \beta] } - \delta_{\mu \beta} \partial_{\nu} \chi_{[\eta \nu ] [\eta \alpha] }  + \\
& + i \varepsilon_{\mu \alpha \beta \rho} \partial_{\eta} \chi_{[\eta \rho]} + i \varepsilon_{\mu \alpha \beta \rho}  \partial_{\nu} \chi_{[\eta \nu ] [\eta \rho ]}) + m \chi_{\mu [\alpha \beta]} = 0, \\
-& \partial_{\eta} \chi_{\nu [\eta \beta]} + \frac13 (2 \partial_{\nu} \chi_{\beta} +2 \partial_{\beta}  \chi_{\nu} -  \delta_{\nu \beta} \partial_{\rho} \chi_{\rho} )
+ i \varepsilon_{\rho \nu \eta \xi} \partial_{\eta} \chi_{\xi [\rho \beta]}  +m \chi_{[\eta \nu] [\eta \beta] } =0.
\end{split}
\label{eq650}
\end{align}
Здесь тензорные величины сопоставляются представлениям
\begin{align}
\begin{split}
\varphi_{\beta} & \sim (\frac12 , \frac12), \quad  \varphi_{[ \alpha \beta]} \sim  (0,1), \quad  \varphi_{\mu [ \alpha \beta]} \sim  (\frac12 ,\frac32), \quad  \varphi_{[\mu \nu] [ \alpha \beta]} \sim  (1,1), \\
\chi_{\beta} & \sim (\frac12 , \frac12)', \quad \chi_{[ \alpha \beta]} \sim  (1,0), \quad  \chi_{\mu [ \alpha \beta]} \sim  (\frac32 ,\frac12), \quad  \chi_{[\mu \nu] [ \alpha \beta]} \sim  (1,1)'.
\end{split}
\label{eq651}
\end{align}
При этом тензоры $\varphi_{([\mu \nu] [\alpha \beta])}$ и $\chi_{([\mu \nu] [\alpha \beta])}$ удовлетворяют условиям самодуальности
\begin{align}
\begin{split}
& \frac12 \varepsilon_{\eta \xi \alpha \beta} \varphi_{[\mu \nu] [\alpha \beta ]} = i \varphi_{[\mu \nu] [\eta \xi]} , \quad \frac12 \varepsilon_{\eta \xi \mu \nu} \varphi_{[\mu \nu] [\alpha \beta ]} = - i \varphi_{[\eta \xi] [\alpha \beta]}  ,  \\
& \frac12 \varepsilon_{\eta \xi \alpha \beta} \chi_{[\mu \nu] [\alpha \beta ]} = - i \chi_{[\mu \nu] [\eta \xi]} , \quad \frac12 \varepsilon_{\eta \xi \mu \nu} \chi_{[\mu \nu] [\alpha \beta ]} = i \chi_{[\eta \xi] [\alpha \beta]} .
\end{split}
 \label{eq652}
\end{align}

Рассматриваемым алгебраическим обобщениям уравнения ДК присущи группы внутренней симметрии $SU(4,4)$ ($32$-компонентная система) и $SU(6,6)$ ($48$- компонентная система). Как и в случае уравнения ДК, преобразования внутренней симметрии $Q$ здесь не коммутируют с преобразованиями Лоренца $\Lambda$. Но при этом группу полной инвариантности $G$ можно представить в виде прямого произведения $Q = \Lambda' \otimes G$, где $\Lambda'$ – лоренцевские преобразования, которые характеризуют набор из восьми и двенадцати дираковских полей соответственно. Кроме того, обе системы допускают физически непротиворечивое квантование по статистике Ферми–Дирака. Процедуру их квантования мы рассматривать не будем, поскольку она вполне аналогична рассмотренному в главе 5 квантованию уравнения ДК. Полное изложения этого вопроса можно найти в работах [48; 49].

Перечисленные свойства $32$- и $48$-компонентной тензорных систем по тем же соображениям, что и в случае уравнения ДК, предполагают возможность их использования для пространственно-временного описания внутренних степеней фермионов. Первая из них может служить, например, в качестве кварковой модели с восемью ароматами, вторая – для геометризованного введения $SU(3)$-калибровочного взаимодействия в решеточном пространстве [50].

Дальнейшее обобщение обсуждаемого геометризованного способа введения \linebreak
внутренних квантовых чисел возможно при отказе от ограничения, сформулированного в начале данного пункта и связанного с размерностью пространства состояний. Поясним сказанное.

Уравнение ДК может быть представлено в форме
\begin{align}
(\gamma_{\mu} \partial_{\mu} +m) \Psi_{\alpha}^D=0,
\label{eq653}
\end{align}
где $\Psi^D$ – дираковский биспинор, $\alpha$ – свободный индекс, соответствующий зарядово-сопряженному биспинору $\bar{\Psi}^c = C (\Psi^D)^*$, $C$ – матрица зарядового сопряжения. Запись \eqref{eq653} означает переход в базис (мы его назвали фермионным), в котором представление \eqref{eq515} трактуется как прямое произведение
\begin{align}
[(0, \frac12) \oplus (\frac12, 0)] \otimes [(0, \frac12) \oplus (\frac12, 0)] .
\label{eq654}
\end{align}
При отказе от вышеуказанного ограничения уравнение ДК допускает обобщения, которые заключаются в рассмотрении вместо \eqref{eq654} всевозможных произведений вида
\begin{align}
[(0, \frac12) \oplus (\frac12, 0)] \otimes [(j_1 , j_2) \oplus (j_2 , j_1)] ,
\label{eq655}
\end{align}
при условии, что сумма $(j_1 + j_2)$ принимает полуцелые значения.

Остановимся на двух классах РВУ, наиболее перспективных с интересующей нас точки зрения и включающих рассмотренные $32$- и $48$-компонентную системы дираковского типа в качестве частных случаев.

Возьмем в \eqref{eq655} $j_1 = 0$ (либо $j_2 = 0$, что то же самое). Получим представление
\begin{align}
[(0, \frac12) \oplus (\frac12, 0)] \otimes [(0,j) \oplus (j,0)]  \quad (j=j_2).
\label{eq656}
\end{align}
При $j = \frac12$ \eqref{eq656} совпадает с \eqref{eq654}, а при $j = \frac32$ приводит к схеме зацеплений \eqref{eq618} и соответственно к $32$-компонентному РВУ с тензорной формулировкой \eqref{eq646}. Случаи $j =\frac52, \frac72, \ldots$ дают схемы зацеплений
\begin{align}
\begin{tabular}{ccccccc}
& $(0,2)$ &  & &  & $(2,0)$  & \\
& $|$ & & $\oplus$ & &$|$  & \\
$(0,3) \quad $ --- & $(\frac12, \frac52)$ &  & &  & $(\frac52,\frac12)$ & --- $\quad (3,0)$,
\end{tabular}
\label{eq657}
\end{align}
\begin{align}
\begin{tabular}{ccccccc}
& $(0,2)$ &  & &  & $(2,0)$  & \\
& $|$ & & $\oplus$ & &$|$  & \\
$(0,4) \quad $ --- & $(\frac12, \frac72)$ &  & &  & $(\frac72,\frac12)$ & --- $\quad (4,0)$
\end{tabular}
\label{eq658}
\end{align}
и так далее. Методика построения на основе \eqref{eq657}, \eqref{eq658} диракоподобных РВУ аналогична той, что применялась при исследовании схемы зацеплений \eqref{eq618}. Несущественные с точки зрения процедуры различия заключаются в неодинаковости спиновой структуры получаемых уравнений. Если схема \eqref{eq618} приводит к матрице $\Gamma_4$ со спиновыми блоками $C^{1}$, $C^{2}$, то в случае \eqref{eq657} получаем блоки $C^{2}$, $C^{3}$, в случае \eqref{eq658}~-- $C^{3}$, $C^{4}$ и так далее. При сопоставлении этим РВУ дираковских частиц с внутренними степенями свободы соответствующее собирательное квантовое число принимает $8, 12, 16, \ldots$ значений.

Второй случай: $ | j_1 - j_2 | = \frac12$. Возникающий при этом класс РВУ базируется на приводимых представлениях, которые являются прямыми произведениями вида
\begin{align}
& [(0, \frac12) \oplus (\frac12, 0)] \otimes [(\frac12,1) \oplus (1,\frac12)] ,
\label{eq659} \\
& [(0, \frac12) \oplus (\frac12, 0)] \otimes [(1,\frac32) \oplus (\frac32,1)] ,
\label{eq660} \\
& [(0, \frac12) \oplus (\frac12, 0)] \otimes [(\frac32,2) \oplus (2,\frac32)] ,
\label{eq661}
\end{align}
и тому подобное. Произведению \eqref{eq659} соответствует схема зацеплений \eqref{eq617} (или \eqref{eq631}) и $48$-компонентное РВУ дираковского типа, матричная и тензорная формулировки которых даны выше. Представлениям \eqref{eq660}, \eqref{eq661} отвечают схемы зацеплений
\begin{align}
\begin{tabular}{ccccc}
$(\frac12,\frac32)$ &--- & $2 (1,1)$ &--- & $(\frac32, \frac12)$ \\
$|$ & & $|$ & & $|$\\
$(1,2)$ & --- & $2 (\frac32,\frac32)$ & --- & $(2,1)$,
\end{tabular}
\label{eq662}
\end{align}
\begin{align}
\begin{tabular}{ccccc}
$(1,2)$ &--- & $2 (\frac32,\frac32)$ &--- & $(2,1)$ \\
$|$ & & $|$ & & $|$\\
$(\frac32, \frac52)$ & --- & $2 (2,2)$ & --- & $(\frac52, \frac32)$,
\end{tabular}
\label{eq663}
\end{align}
имеющие структуру, сходную с \eqref{eq617}.

Возможность построения на основе схем зацеплений данного класса $P$- инвариантных РВУ дираковского типа вытекает из того, что каждой из них можно сопоставить уравнение Дирака для биспинора со свободным индексом, соответствующим представлению $[(j_1, j_2) \oplus (j_2, j_1)]$. При этом схема \eqref{eq662} описывает спины $0, 1, 2, 3$, схема \eqref{eq663} – спины $0, 1, 2, 3, 4$ и так далее. При сопоставлении (в указанном выше смысле) данным уравнениям частиц со спином $\frac12$ и внутренними степенями свободы собирательное внутреннее квантовое число принимает $12, 24, 40, \ldots$ значений. Очевидно, что рассмотренные обобщения уравнения ДК предоставляют весьма широкие возможности с точки зрения геометризованного описания внутренних (помимо спина) степеней свободы дираковских частиц.

\section{Совместное описание безмассовых полей с различными спиральностями}

\hspace{1.25cm}
Обсудим теперь, какие возможности открывает использование кратных представлений группы Лоренца в теории РВУ с точки зрения описания безмассовых полей.

Для начала проанализируем простейшую схему зацеплений \eqref{ls3} на предмет построения на ее основе различных безмассовых РВУ. Схеме \eqref{ls3} соответствует следующий наиболее общий вид релятивистски-инвариантной тензорной системы линейных дифференциальных уравнений первого порядка:
\begin{align}
\partial_{\nu} \psi_{[\mu \nu]} + a \psi_{\mu} =0,
\label{eq71} \\
- \partial_{\mu} \psi_{\nu}+  \partial_{\nu} \psi_{\mu}+  b  \psi_{[\mu \nu]} =0,
\label{eq72}
\end{align}
где $a$, $b$ – произвольные постоянные коэффициенты. Существуют четыре принципиально различные возможности в выборе этих коэффициентов.

Первая, когда $a = b = m$, приводит к системе Даффина–Кеммера (смотри \eqref{proca1}) для микрочастицы с ненулевой массой и спином $s = 1$. Этот случай нас сейчас не интересует. Вторая $a = b = 0$ приводит к не имеющим физического смысла независимым уравнениям
\begin{align}
\partial_{\nu} \psi_{[\mu \nu]} =0, \quad - \partial_{\mu} \psi_{\nu}+  \partial_{\nu} \psi_{\mu} =0.
\label{eq73}
\end{align}
Выбирая в \eqref{eq71}, \eqref{eq72} $a = 0$, $b = 1$, получим систему
\begin{align}
\partial_{\nu} \psi_{[\mu \nu]} =0,
\label{eq74} \\
- \partial_{\mu} \psi_{\nu}+  \partial_{\nu} \psi_{\mu}+   \psi_{[\mu \nu]} =0.
\label{eq75}
\end{align}
Если трактовать здесь компоненты вектора $\psi_{\mu}$ как потенциалы, а $ \psi_{[\mu \nu]}$ – как напряженности, то уравнения \eqref{eq74}, \eqref{eq75} представляют собой систему уравнений Максвелла (так называемая десятимерная формулировка), описывающую фотон – безмассовую частицу со спиральностью $\pm 1$. При этом первое из них является уравнением движения, а второе выступает как определение напряженности через потенциалы.

Наконец, возможен выбор $a = 1$, $b = 0$, который приводит к системе
\begin{align}
\partial_{\nu} \psi_{[\mu \nu]} +  \psi_{\mu} =0,
\label{eq76} \\
- \partial_{\mu} \psi_{\nu}+  \partial_{\nu} \psi_{\mu} =0.
\label{eq77}
\end{align}
Если в данном случае по-прежнему трактовать $\psi_{\mu}$ как потенциалы, а $\psi_{[\mu \nu]}$ – как напряженности, то система \eqref{eq76}, \eqref{eq77} становится неопределенной в том смысле, что напряженности не могут быть выражены через потенциалы. Ситуация, однако, существенно изменяется, если придерживаться иной интерпретации входящих в эту систему величин, а именно: считать потенциалом тензор $\psi_{[\mu \nu]}$, а напряженностью  – вектор $\psi_{\mu}$. Тогда система \eqref{eq76}, \eqref{eq77} становится вполне определенной: уравнение \eqref{eq76} выступает как определение напряженности через потенциал, уравнение \eqref{eq77} – как уравнение движения.

Физический смысл системы \eqref{eq76}, \eqref{eq77} вытекает из следующих соображений. Из уравнения \eqref{eq76} имеем
\begin{align}
\partial_{\mu} \psi_{\mu}= 0 .
\label{eq78}
\end{align}
С учетом \eqref{eq78} из уравнения движения \eqref{eq77} легко получить уравнение второго порядка
\begin{align}
\square \psi_{\mu}= 0,
\label{eq79}
\end{align}
которое указывает на отсутствие массы у микрообъекта, описываемого системой \eqref{eq76}, \eqref{eq77}.

Как известно, в теории безмассового векторного поля, базирующейся на уравнениях \eqref{eq74}, \eqref{eq75}, на потенциалах можно задать преобразования
\begin{align}
\psi_{\mu} \to \psi'_{\mu} =  \psi_{\mu}+ \partial_{\mu} \Lambda (x),
\label{eq710}
\end{align}
называемые градиентными, или калибровочными преобразованиями второго рода. Произвол в выборе калибровочной функции  $\Lambda (x) $ позволяет исключить «лишние» состояния, оставляя лишь две (из четырех) поперечные составляющие. В свою очередь, уравнения \eqref{eq76}--\eqref{eq79} инвариантны относительно преобразований потенциалов
\begin{align}
\psi_{[\mu \nu]} \to \psi'_{[\mu \nu]} =  \psi_{[\mu \nu]}+ \partial_{\mu} \Lambda_{\nu} -   \partial_{\nu} \Lambda_{\mu},
\label{eq711}
\end{align}
где калибровочные функции $\Lambda_{\mu} (x)$ ограничены условием
\begin{align}
\square \Lambda_{\mu} - \partial_{\mu} \partial_{\nu} \Lambda_{\nu} =0.
\label{eq712}
\end{align}
В работе Огиевецкого и Полубаринова [51] показано, что калибровочная инвариантность такого рода оставляет у тензор-потенциала только одну независимую компоненту, соответствующую состоянию с нулевой спиральностью.

Остановимся подробнее на указанной работе. Для этого вернемся к схеме зацеплений \eqref{ls3}, в которой представление $(\frac12, \frac12)$ будем считать псевдовекторным. В этом случае можно построить, во-первых, теорию псевдовекторной частицы с нулевой массой
\begin{align}
 \frac12 \varepsilon_{\mu \nu \alpha \beta} \partial_{\nu} \psi_{[\alpha \beta]} =0,
\label{eq713} \\
\varepsilon_{\mu \nu \alpha \beta}  \partial_{\alpha} \tilde{\psi}_{\beta} + \psi_{[\mu \nu]} =0
\label{eq714}
\end{align}
(так называемая электродинамика с псевдовекторным потенциалом). Вводя вместо псевдовектора $\tilde{\psi}_{\mu}$ сопряженный ему антисимметричный тензор третьего ранга $ \psi_{[\mu \nu \alpha]}$, вместо \eqref{eq713}, \eqref{eq714} прийдем к системе
\begin{align}
\partial_{\mu} \psi_{[\nu \alpha]}+\partial_{\alpha} \psi_{[\mu \nu]}+\partial_{\nu} \psi_{[\alpha \mu]} =0,
\label{eq715} \\
\partial_{\mu}  \psi_{[ \mu\nu \alpha]}+ \psi_{[\nu \alpha]} =0,
\label{eq716}
\end{align}
в которой $ \psi_{[\mu \nu \alpha]}$ выступает в роли потенциала.

Во-вторых, можно получить систему уравнений
\begin{align}
 \frac12 \varepsilon_{\mu \nu \alpha \beta} \partial_{\nu} \psi_{[\alpha \beta]} + \tilde{\psi}_{\mu} =0,
\label{eq717} \\
\varepsilon_{\mu \nu \alpha \beta} \partial_{\alpha} \tilde{\psi}_{\beta} =0
\label{eq718}
\end{align}
или эквивалентную ей систему
\begin{align}
\partial_{\mu} \psi_{[\nu \alpha]}+\partial_{\alpha} \psi_{[\mu \nu]}+\partial_{\nu} \psi_{[\alpha \mu]}+ \psi_{[\mu \nu \alpha]} =0,
\label{eq719} \\
\partial_{\mu}  \psi_{[\mu \nu \alpha]} =0.
\label{eq720}
\end{align}

Рассматривая здесь $\psi_{[\mu \nu]}$ как тензор-потенциал, а $\psi_{[\mu \nu \alpha]}$ – как напряженность, мы приходим к теории Огиевецкого-Полубаринова для безмассовой частицы со спиральностью $0$. Действительно, в работе [51] для тензор-потенциала $\psi_{[\mu \nu]}$ в качестве исходного постулируется уравнение второго порядка
\begin{align}
\square \psi_{[\mu \nu]} + \partial_{\mu} \partial_{\alpha}\psi_{[\nu \alpha]} -  \partial_{\nu} \partial_{\alpha}\psi_{[\mu \alpha]} =0.
\label{eq721}
\end{align}
Нетрудно убедиться, что оно коррелирует с системой первого порядка \eqref{eq719}, \eqref{eq720}. Кроме того, уравнение \eqref{eq721} инвариантно относительно калибровочных преобразований \eqref{eq711}, \eqref{eq712}, что позволяет наложить на потенциалы $\psi_{[\mu \nu]}$ дополнительное условие
\begin{align}
\partial_{\nu} \psi_{[\mu \nu]} =0 ,
\label{eq722}
\end{align}
равносильное условию
\begin{align}
\partial_{\mu} \partial_{\alpha} \psi_{[\nu \alpha]} -\partial_{\nu} \partial_{\alpha} \psi_{[\mu \alpha]} =0 .
\notag
\end{align}
В результате уравнение \eqref{eq721} распадается на уравнения
\begin{align}
\square \psi_{[\mu \nu]} =0
\label{eq723}
\end{align}
и \eqref{eq722}.

Что касается системы \eqref{eq76}, \eqref{eq77}, то уравнение \eqref{eq722} можно получить из нее непосредственно. Таким образом, в обоих вариантах теории безмассовой частицы со спиральностью $0$ (имеются в виду системы \eqref{eq76}, \eqref{eq77} и \eqref{eq719}, \eqref{eq720}) получаются одинаковые уравнения второго порядка для потенциалов. Различие же этих двух теорий заключается в том, что в системе \eqref{eq76}, \eqref{eq77} напряженность является истинным вектором, а в системе \eqref{eq719}, \eqref{eq720} – антисимметричным тензором третьего ранга, или, иначе говоря, псевдовектором. Кроме того, если для системы \eqref{eq76}, \eqref{eq77} уравнение второго порядка \eqref{eq722} выступает как основное, а уравнение \eqref{eq723} – как дополнительное условие, то по отношению к системе \eqref{eq719}, \eqref{eq720}, наоборот, \eqref{eq723} является основным уравнением, а \eqref{eq722} – дополнительным условием. Однако эти различия не сказываются на числе степеней свободы, соответствующим обеим теориям.

В [51] безмассовая частица, описываемая системой \eqref{eq719}, \eqref{eq720}, была названа нотофом. Это название отражает дополнительность свойств фотона и нотофа как в смысле спиральности, так и в отношении лоренцевских трансформационных свойств потенциалов и напряженностей. Нотоф, описываемый системой \eqref{eq76}, \eqref{eq77}, естественно назвать дуальным нотофом.

Здесь будет кстати обсудить вопрос о так называемом «спиновом скачке» (spin jumping). В работах ряда авторов (смотри, напр., [52; 53]) частица, описываемая системой \eqref{eq76}, \eqref{eq77}, трактуется как скалярный безмассовый мезон. Отсюда делается вывод об изменении (скачке) спина при переходе от системы \eqref{proca1} к системе \eqref{eq76}, \eqref{eq77}. Однако, анализ матричной формулировки \eqref{eq:zero_mass} (или \eqref{eq:nonzero_mass} в случае ненулевой массы) рассмотренных выше тензорных систем, базирующихся на схеме зацеплений \eqref{ls3}, показывает, что их единственное существенное отличие состоит в виде матрицы $\Gamma_0$. Для частицы с ненулевой массой $\Gamma_0 = m I$. Для фотона и нотофа в тензорном базисе будем иметь соответственно
\begin{align}
\Gamma_0 = \left( \begin{tabular}{cc}
$0_4$ & \\
& $I_6$
\end{tabular} \right),
\label{eq724} \\
\Gamma_0 = \left( \begin{tabular}{cc}
$I_4$ & \\
& $0_6$
\end{tabular} \right).
\label{eq725}
\end{align}
При этом матрица $\Gamma_0$ \eqref{eq724} вырезает из волновой функции одну из трех проекций спина $s =1$, оставляя для фотона две проекции ($s_z = \pm1$), а матрица $\Gamma_0$ \eqref{eq725}, вырезая две проекции, оставляет для нотофа лишь одну ($s_z = 0$). И поскольку спиновый блок $C^{0}$ во всех случаях остается равным нулю (смотри \eqref{eq212}), очевидно, что ненулевые степени и фотона и нотофа связаны со спиновым блоком $C^{1}$. Так что, на самом деле, в данном случае имеет место переход степеней свободы (состояний) массивной векторной частицы с проекциями спина $s_z = \pm 1$ в степени свободы (состояния) фотона со спиральностью $\pm 1$, и точно также переход состояния массивной векторной частицы с проекцией спина $s_z = 0$ в соответствующее состояние нотофа со спиральностью $0$. При обратном же переходе, например, от реальных фотона и нотофа к их виртуальным аналогам, обладающим массой, виртуальный фотон приобретает дополнительное состояние с нулевой проекцией спина, а виртуальный – дополнительные состояния с проекцией спина $\pm 1$. Другими словами, нотоф, как и фотон, переносит во взаимодействиях спин $1$. Поэтому более точно, на наш взгляд, рассматривать нотоф как безмассовую {\it векторную} частицу со спиральностью равной $0$, что соответствует и точке зрения авторов работы [51].

Обобщая проделанный выше анализ, можно сделать вывод, что теория РВУ первого порядка вида \eqref{eq:zero_mass} позволяет описывать безмассовые частицы (поля) не только с максимальной для данного набора представлений группы Лоренца спиральностью $\pm s$, но и с промежуточными значениями спиральности, включая нулевую.

Первооткрыватели нотофа не смогли предложить для него каких-либо физических приложений. В 1974 году Кальб и Рамонд [52], по существу, переоткрыли нотоф, рассматривая вопрос о феноменологическом описании взаимодействия струн. Впоследствии за полевой системой, сопоставляемой уравнениям \eqref{eq719}, \eqref{eq720}, в литературе утвердилось название поля Кальба--Рамонда (смотри, напр., [53; 54]).

В [51] тензор $\psi_{[\mu \nu]}$ предлагается в качестве потенциала поля – переносчика взаимодействия замкнутых струн в пространстве размерности $d = 4$. Очевидно, что для описания взаимодействия открытых струн одного лишь поля Кальба--Рамонда (нотофа Огиевецкого-Полубаринова) недостаточно. Моделируя концы струны как точечные электрические заряды, необходимо ввести вектор-потенциал, соответствующий электромагнитному полю. И поскольку струна является единым физическим объектом, естественна постановка задачи о совместном описании фотона и нотофа на основе одной не распадающейся по группе Лоренца системы уравнений первого порядка.

С этой целью рассмотрим схему зацеплений \eqref{eq41}, в которой представление $(\frac12, \frac12)$ сопоставляется истинному вектору, а представление $(\frac12, \frac12)'$ – псевдовектору, или антисимметричному тензору третьего ранга. Наиболее общая тензорная система уравнений первого порядка, соответствующая схеме \eqref{eq41} и удовлетворяющая стандартным физическим требованиям, имеет вид
\begin{align}
\begin{split}
& \alpha \partial_{\nu} \psi_{[\mu \nu]} + a \psi_{\mu} =0,  \\
& \beta \partial_{\nu} \tilde{\psi}_{[\mu \nu]} + b \tilde{\psi}_{\mu} =0, \\
& \alpha^* (- \partial_{\mu} \psi_{\nu} +  \partial_{\nu} \psi_{\mu} ) + \beta^* \varepsilon_{\mu \nu \alpha \beta} \partial_{\alpha} \tilde{\psi}_{\beta} + c \psi_{[\mu \nu]} =0,
\end{split}
\label{eq726}
\end{align}
где $\alpha$, $\beta$, $a$, $b$, $c$ – произвольные параметры. Систему \eqref{eq726} можно переписать следующим образом:
\begin{align}
\begin{split}
& \alpha \partial_{\nu} \psi_{[\mu \nu]} + a \psi_{\mu} =0,  \\
& \beta (\partial_{\mu} \psi_{[\nu \alpha]} + \partial_{\alpha} \psi_{[\mu \nu ]}+ \partial_{\nu} \psi_{[\alpha \mu]} ) + b \psi_{[\mu \nu \alpha]} =0, \\
& \alpha^* (-\partial_{\nu} \psi_{\alpha} + \partial_{\alpha} \psi_{\nu} )  + \beta^* \partial_{\mu} \psi_{[\mu \nu \alpha]} + c \psi_{[\nu \alpha]} =0.
\end{split}
\label{eq727}
\end{align}

Положим в \eqref{eq727}
\begin{align}
\alpha = \beta =1, \quad a=c=0, \quad b=1.
\label{eq728}
\end{align}
Получим систему
\begin{align}
\partial_{\nu} \psi_{[\mu \nu]} =0,
\label{eq729} \\
\partial_{\mu} \psi_{[\nu \alpha]} + \partial_{\alpha} \psi_{[\mu \nu ]} + \partial_{\nu} \psi_{[\alpha \mu]} + \psi_{[\mu \nu \alpha]} =0,
\label{eq730} \\
- \partial_{\nu} \psi_{\alpha} + \partial_{\alpha} \psi_{\nu} + \partial_{\mu} \psi_{[\mu \nu \alpha]} =0.
\label{eq731}
\end{align}

Примем следующую трактовку входящих в \eqref{eq729}--\eqref{eq731} величин: $ \psi_\mu$ и $\psi_{[\mu \nu ]}$ – потенциалы, $\psi_{[\mu \nu \alpha]}$ – напряженность. Тогда уравнение \eqref{eq730} является, по существу, определением напряженности через потенциалы. Уравнение \eqref{eq729} играет роль дополнительного условия на потенциалы $\psi_{[\mu \nu ]}$, которое изначально содержится в самой системе. Данное условие оставляет у потенциала только две независимые компоненты. Кроме того, система \eqref{eq729}--\eqref{eq731} инвариантна относительно калибровочных преобразований \eqref{eq711}, \eqref{eq712}. Имеющийся произвол в выборе калибровочной функции позволяет наложить условие, исключающее еще одну независимую степень свободы, связанную с тензор-потенциалом $\psi_{[\mu \nu ]}$. При этом для $\psi_{[\mu \nu ]}$ имеем уравнение второго порядка
\begin{align}
\square \psi_{[\mu \nu]} + \partial_{\mu} \psi_{\nu} - \partial_{\nu} \psi_{\mu} =0,
\label{eq732}
\end{align}
которое описывает состояние некоторого безмассового поля со спиральностью $0$.

Обратимся к потенциалу $\psi_\mu$. Система \eqref{eq729}--\eqref{eq731} инвариантна также относительно калибровочных преобразований
\begin{align}
\psi_{\mu} \to \psi'_{\mu} = \psi_{\mu} + \partial_{\mu} \Lambda,
\label{eq733}
\end{align}
где $\Lambda$ – произвольная скалярная функция. Из уравнения \eqref{eq731} вытекает уравнение второго порядка
\begin{align}
\square \psi_{\mu} - \partial_{\mu}  \partial_{\nu}  \psi_{\nu} =0,
\label{eq734}
\end{align}
которое совместно с калибровочной инвариантностью \eqref{eq733} означает, что вектор- потенциал $\psi_\mu$ характеризует поперечную составляющую (спиральность $\pm1$) обсуждаемого безмассового поля. Тогда тензор
\begin{align}
 \partial_{\mu}  \psi_{\nu} - \partial_{\nu}  \psi_{\mu} \equiv F_{[\mu \nu]}
\label{eq735}
\end{align}
естественно рассматривать как напряженность, непосредственно связанную с этой поперечной составляющей. Уравнение же \eqref{eq731}, переписанное с учетом обозначения \eqref{eq735} в виде
\begin{align}
\partial_{\mu} \psi_{[\mu \nu \alpha]} - F_{[\nu \alpha]} =0,
\label{eq736}
\end{align}
выступает, очевидно, в качестве уравнения движения в системе \eqref{eq729}--\eqref{eq731}.

Таким образом, выбор \eqref{eq728} параметров в системе \eqref{eq727} приводит к не распадающейся по группе Лоренца теории, которая дает совместное описание безмассовых полей со спиральностями $0$ и $\pm 1$, то есть поля Кальба–Рамонда (нотофа) и электромагнитного поля. Уравнение движения \eqref{eq736} указывает на неразрывную связь этих полей подобно тому, как связаны электрическая и магнитная составляющие в теории Максвелла. Точнее даже говорить не о совместном описании указанных полей, а о едином безмассовом поле с тремя возможными значениями спиральности $0, \pm 1$.

Интерпретируя данное поле в качестве переносчика взаимодействия между открытыми струнами в пространстве размерности $d = 4$, можем ввести в систему \eqref{eq729}--\eqref{eq731} источники. При этом учтем, что в данном случае существует два типа источников: тензорный ток $j_{[\mu \nu]}$, который создается телом струны (body string), и векторный ток $j_\mu$, создаваемый концами струны. Последние при этом рассматриваются как точечные электрические заряды противоположных знаков. Между токами $j_\mu$ и $j_{[\mu \nu]}$ существует связь
\begin{align}
j_{\nu} = \partial_{\mu} j_{[\mu \nu]},
\label{eq737}
\end{align}
из которой следует, что $j_\mu$ сохраняется ($ \partial_{\mu} j_{\mu} = 0$), а $j_{[\mu \nu ]}$, вообще говоря, не сохраняется ($ \partial_{\mu} j_{[\mu \nu ]}\neq 0$). Вводя ток $j_{[\mu \nu ]}$ в уравнение движения \eqref{eq731}, получим систему
\begin{align}
\partial_{\nu} \psi_{[\mu \nu]} =0,
\label{eq738} \\
\partial_{\mu} \psi_{[\nu \alpha]} + \partial_{\alpha} \psi_{[\mu \nu ]} + \partial_{\nu} \psi_{[\alpha \mu]} + \psi_{[\mu \nu \alpha]} =0,
\label{eq739} \\
- \partial_{\nu} \psi_{\alpha} + \partial_{\alpha} \psi_{\nu} + \partial_{\mu} \psi_{[\mu \nu \alpha]} =j_{[\nu \alpha]}.
\label{eq740}
\end{align}
описывающую единое поле открытой струны в присутствии источников.

В частных случаях, когда рассматривается взаимодействие замкнутых струн или электрически заряженных частиц, компоненты единого поля могут существовать и описываться по отдельности. Так, полагая в системе $j_\mu = 0$, получим согласно \eqref{eq737}
\begin{align}
\partial_{\mu} j_{[\mu \nu]} =0.
\label{eq741}
\end{align}
И система \eqref{eq738}--\eqref{eq740} трансформируется в уравнения \eqref{eq719}, \eqref{eq720} (c членом $j_{[\nu \alpha]}$ в правой части) и дополнительное условие \eqref{eq722}, описывающие поле Кальба--Рамонда с источником. В свою очередь, беря от уравнения \eqref{eq740} производную $\partial_{\alpha}$ и учитывая определения \eqref{eq735}, \eqref{eq737}, придем к уравнению
\begin{align}
\partial_{\nu} F_{[\mu \nu]} = j_{\mu}.
\label{eq742}
\end{align}
Объединяя уравнение \eqref{eq742} с \eqref{eq735} и исключая из рассмотрения величины $\psi_{[\mu \nu]}$, $j_{[\mu \nu]}$, относящиеся к телу струны, получим максвелловскую систему для электромагнитного поля с источником.

В матричном формализме система \eqref{eq729}--\eqref{eq731} соответствует РВУ типа \eqref{eq:zero_mass} с особенной матрицей $\Gamma_0$, имеющей в тензорном базисе вид
\begin{align}
\Gamma_0 = \left( \begin{tabular}{ccc}
$0_4$ & & \\
& $I_4$ & \\
 & & $0_6$
\end{tabular} \right).
\label{eq743}
\end{align}
Выражения для спиновых блоков $C^{0}$, $C^{1}$ матрицы $\Gamma_4$ в базисе Гельфанда–Яглома задаются соответственно формулами \eqref{eq43}, \eqref{eq417}. Отсюда следует, что безмассовое поле, описываемое этим РВУ, действительно переносит спин $1$, причем собственное значение  $\lambda = \pm 1$ спинового блока $C^{1}$ двукратно вырождено. В контексте вышесказанного такое вырождение соответствует совместному описанию электромагнитного поля (фотона) и поля Кальба–Рамонда (нотофа) как составляющих единого безмассового векторного поля с тремя возможными значениями спиральности $s = 0, +1, -1$. Нулевое значение массы при этом обеспечивает проективная матрица $\Gamma_0$, которая устраняет «лишние» состояния, присущие массивному аналогу данного поля.

Таковым аналогом, как нетрудно видеть, является частица, которая обсуждалась в главе 4. Действительно, совершая в рассматриваемом безмассовом РВУ замену $\Gamma_0 \to m I$,  мы придем к полученному в указанном пункте РВУ для киральной частицы со спином $s = 1$ и ненулевой массой. Тензорная форма \eqref{eq419} этого РВУ может быть получена из системы \eqref{eq726} при выборе параметров $\alpha = \beta = 1$, $a = b = c = m$. Данное положение вещей проливает определенный свет на физический смысл квантового числа «киральность» для частиц с ненулевой массой, а именно: в том смысле, в каком виртуальному фотону или нотофу сопоставляется векторная частица, описываемая обычным уравнением Даффина–Кеммера, виртуальному объединенному полю фотона и нотофа сопоставляется векторная частица с ненулевой массой и дополнительной внутренней степенью свободы – киральностью. При обратном переходе $m I \to \Gamma_0$ проективная матрица $\Gamma_0$ «вырезает» лишние состояния, оставляя в сумме три степени свободы на фотон и нотоф.

Подводя итог данному пункту, можно сказать, что теория РВУ вида \eqref{eq:zero_mass} позволяет описывать безмассовые поля не только с максимальной (для данного набора представлений группы Лоренца) спиральностью $\pm s$, но и поля с промежуточными значениями спиральности, а также осуществлять совместное описание полей с различными значениями спиральности от $+s$ до $-s$, включая промежуточные, в рамках не распадающегося по группе Лоренца РВУ. В последнем случае требуется расширенный (по сравнению с необходимым для описания спиральности $\pm s$) набор неприводимых представлений группы Лоренца в пространстве волновой функции $\Psi$.

Принципиально иным способом объединения безмассовых полей с различными спиральностями является механизм калибровочно-инвариантного смешивания, или $\hat{B} \wedge \hat{F}$-теория [52; 53]. Этот способ приводит к появлению массы у объединенного поля и, в принципе, может претендовать на роль механизма генерации массы, альтернативного механизму Хиггса.

Рассмотрим вкратце суть $\hat{B} \wedge \hat{F}$-теории и дадим ее матричную интерпретацию. Возьмем в качестве исходных безмассовые поля, описываемые системами уравнений \eqref{eq74}, \eqref{eq75} и \eqref{eq717}, \eqref{eq718}. Первая из них описывает фотон – безмассовое векторное поле со спиральностью $\pm 1$, вторая – нотоф, или безмассовое поле со спиральностью $0$. Перепишем систему \eqref{eq717}, \eqref{eq718} в виде
\begin{align}
- \partial_{\mu} \tilde{\varphi}_{\nu} + \partial_{\nu} \tilde{\varphi}_{\mu} =0,
\label{eq744} \\
\partial_{\nu} \tilde{\varphi}_{[\mu \nu]} + \tilde{\varphi}_{\mu} =0,
\label{eq745}
\end{align}
где для удобства использовано обозначение
\begin{align}
\frac12 \varepsilon_{\mu \nu \alpha \beta} \varphi_{[\alpha \beta]} = \tilde{\varphi}_{[\mu \nu]}.
\label{eq746}
\end{align}
Напомним, что величины $\psi_{\mu}$ и $\tilde{\varphi}_{[\mu \nu]}$ выступают в этих системах в качестве потенциалов, а $\psi_{[\mu \nu]}$ и $\tilde{\varphi}_{\mu}$ – в качестве напряженностей указанных полей.

Далее в лагранжиан $\mathcal{L}_0$ системы \eqref{eq74}, \eqref{eq75}, \eqref{eq744}, \eqref{eq745} (его вид не выписываем за ненадобностью) вводится дополнительный член
\begin{align}
 \mathcal{L}_{int} = m \psi_{\mu} \partial_{\nu} \tilde{\varphi}_{[\mu \nu]} ,
\label{eq747}
\end{align}
который не нарушает инвариантность этой системы относительно калибровочных преобразований \eqref{eq733} и преобразований вида \eqref{eq711}, \eqref{eq712} для потенциала $\tilde{\varphi}_{[\mu \nu]}$. Данную процедуру называют калибровочно-инвариантным смешиванием или топологическим взаимодействием исходных безмассовых полей.

Варьирование суммарного лагранжиана $\mathcal{L} = \mathcal{L}_0 + \mathcal{L}_{int}$ дает систему уравнений
\begin{align}
\partial_{\nu} \psi_{[\mu \nu]} + m \tilde{\varphi}_{\mu} =0,
\label{eq748} \\
- \partial_{\mu} \tilde{\varphi}_{\nu} + \partial_{\nu} \tilde{\varphi}_{\mu} + m \psi_{[\mu \nu]} =0,
\label{eq749} \\
\partial_{\nu} \tilde{\varphi}_{[\mu \nu]} +  \tilde{\varphi}_{\mu} =0,
\label{eq750} \\
- \partial_{\mu} \psi_{\nu} + \partial_{\nu} \psi_{\mu} + \psi_{[\mu \nu]} =0.
\label{eq751}
\end{align}
Вводя в рассмотрение величины
\begin{align}
\tilde{G}_{\mu} = \psi_{\mu} - \frac{1}{m} \tilde{\varphi}_{\mu}, \quad G_{[\mu \nu]} =  \tilde{\varphi}_{[\mu \nu ]} - \frac{1}{m} \psi_{[\mu \nu]},
\label{eq752}
\end{align}
систему \eqref{eq748}--\eqref{eq751} можно в конечном счете привести к виду
\begin{align}
\partial_{\nu} \psi_{[\mu \nu]} + m \tilde{\varphi}_{\mu} =0,
\label{eq753} \\
- \partial_{\mu} \tilde{\varphi}_{\nu} + \partial_{\nu} \tilde{\varphi}_{\mu} + m \psi_{[\mu \nu]} =0,
\label{eq754} \\
\partial_{\nu} G_{[\mu \nu]} =0,
\label{eq755} \\
- \partial_{\mu} \tilde{G}_{\nu} +  \partial_{\nu} \tilde{G}_{\mu} =0.
\label{eq756}
\end{align}

Система \eqref{eq753}--\eqref{eq756} распадается на лоренц-инвариантные подсистемы \eqref{eq753}, \eqref{eq754} и \eqref{eq755}, \eqref{eq756}. Первая из них описывает векторную частицу с ненулевой массой. Подсистема \eqref{eq755}, \eqref{eq756} физического поля не описывает, поскольку ей соответствует нулевая плотность энергии. Ее присутствие в системе \eqref{eq753}--\eqref{eq756} связано с формальными соображениями сохранения калибровочной инвариантности теории на всех этапах.

На языке матричного формализма теории РВУ $\hat{B} \wedge \hat{F}$-теория интерпретируется следующим образом. Исходную тензорную систему \eqref{eq74}, \eqref{eq75}, \eqref{eq744}, \eqref{eq745} можно представить в форме \eqref{eq:zero_mass}, где при использовании базиса
\begin{align}
\Psi = (\psi_{\mu},  \psi_{[\mu \nu ]}, \tilde{\varphi}_{[\mu \nu]}, \tilde{\varphi}_{\mu})^T
\label{eq757}
\end{align}
матрицы $\Gamma_\mu$, $\Gamma_0$ имеют вид
\begin{align}
\Gamma_{\mu}= \left( \begin{tabular}{cc}
$\Gamma_{\mu}^{DK}$ & \\
& $\Gamma_{\mu}^{DK}$
\end{tabular} \right), \quad \Gamma_0 =
 \left( \begin{tabular}{cccc}
$0_4$ & &   \\
& $I_6$ & & \\
& & $0_6$ & \\
& & & $I_4$
\end{tabular} \right)
\label{eq758}
\end{align}
($\Gamma_{\mu}^{DK}$ – 10-мерные матрицы Даффина–Кеммера).

Введение в лагранжиан топологического члена \eqref{eq747} приводит к изменению вида матриц $\Gamma_\mu$. Подстановки \eqref{eq752} эквивалентны некоторому унитарному преобразованию базиса \eqref{eq757}, в результате чего матрицы $\Gamma_\mu$ принимают первоначальный вид \eqref{eq758}. Матрица же $\Gamma_0$ при этом видоизменяется следующим образом:
\begin{align}
\Gamma_0 \to
 \left( \begin{tabular}{cccc}
$m I_4$ & &  & \\
& $m I_6$ & & \\
& & $0_6$ & \\
& & & $0_4$
\end{tabular} \right) =
\left( \begin{tabular}{cc}
$m I_{10}$ & \\
& $0_{10}$
\end{tabular} \right).
\label{eq759}
\end{align}

Таким образом, получается РВУ \eqref{eq:zero_mass} с матрицами $\Gamma_\mu$ \eqref{eq758} и $\Gamma_0$ \eqref{eq759}, которое представляет собой прямую сумму уравнения Даффина–Кеммера для спина $1$ и безмассовый предел ($m \to 0$) этого уравнения. По сути, данный способ генерации массы с точки зрения теории РВУ сводится к перестановке нулевых и единичных блоков матрицы $\Gamma_0$. При этом число степеней свободы полевой системы (равное трем) остается прежним, но имеет место их перераспределение: нотоф как бы передает свою степень свободы фотону, что автоматически приводит к появлению частицы с ненулевой массой и спином $1$. Можно сказать, что происходит своего рода «аннигиляция» фотона и нотофа с образованием векторной частицы с ненулевой массой.

\section{Массивные калибровочно-инвариантные поля \\ в теории РВУ}

\hspace{1.25cm}
Как отмечалось в предыдущем пункте, одно из отличий в описании бозонов с ненулевой и нулевой массой состоит в том, что в безмассовом случае часть компонент волновой функции $\Psi$ являются ненаблюдаемыми (потенциалы), а часть~-- наблюдаемыми (напряженности). На потенциалах можно задать калибровочные преобразования (речь по-прежнему идет о преобразованиях второго рода) и наложить дополнительные условия, исключающие «лишние» компоненты функции $\Psi$. При описании же частиц с ненулевой массой указанное разграничение компонент волновой функции явным образом не имеет места. Поэтому понятие калибровочной инвариантности в вышеуказанном смысле применяется обычно по отношению к теориям безмассовых частиц (полей).

Вместе с тем, известны работы (смотри, напр., [55]), в которых в разных подходах обсуждаются так называемые массивные калибровочно-инвариантные теории. В связи с этим возникают вопросы: каков статус таких теорий в подходе, основанном на использовании матричной формы РВУ; по каким признакам различаются калибровочно-инвариантные РВУ для частиц с ненулевой и нулевой массой?

Рассмотрим набор неприводимых представлений группы Лоренца
\begin{align}
(0,0) \oplus (\frac12 , \frac12) \oplus (0,1) \oplus (1,0),
\label{eq81}
\end{align}
образующих схему зацеплений
\begin{align}
\begin{tabular}{ccccc}
& & $(0,0)$ & & \\
& & $|$ & & \\
$(0,1)$ & --- & $(\frac12, \frac12)$ & --- & $(1,0)$.
\end{tabular}
\label{eq82}
\end{align}
Схеме \eqref{eq82} соответствует в общем случае тензорная система уравнений первого порядка
\begin{align}
\alpha \partial_{\mu} \psi_{\mu} + a \psi_0 =0,
\label{eq83} \\
\beta^* \partial_{\nu} \psi_{[\mu \nu]} + \alpha^* \partial_{\mu} \psi_0 + b \psi_{\mu} =0,
\label{eq84} \\
\beta (- \partial_{\mu} \psi_{\nu}+\partial_{\nu} \psi_{\mu}) + c \psi_{[\mu \nu]} =0.
\label{eq85}
\end{align}

В случае, когда ни один из параметров в этой системе не равен нулю, она описывает микрообъект с набором спинов $0,1$ и двумя массами
\begin{align}
m_1 = \frac{\sqrt{ab}}{|\alpha |}, \quad m_2 = \frac{\sqrt{bc}}{|\beta |},
\label{eq86}
\end{align}
причем масса $m_1$ относится к спину $0$, а масса $m_2$ – к спину $1$. Если наложить на параметры системы \eqref{eq83}--\eqref{eq85} условие
\begin{align}
\frac{\sqrt{a}}{|\alpha |} = \frac{\sqrt{b}}{|\beta |} ,
\label{eq87}
\end{align}
получается РВУ для микрообъекта со спинами $0, 1$ и одним значением массы $m = m_1 = m_2$. При $\alpha = 0$ рассматриваемая система переходит в уравнение типа Даффина–Кеммера для частицы со спином $1$ и массой $m = m_2$
\begin{align}
\beta^* \partial_{\nu} \psi_{[\mu \nu]} + b \psi_{\mu} =0,
\label{eq88} \\
\beta (- \partial_{\mu} \psi_{\nu}+\partial_{\nu} \psi_{\mu}) + c \psi_{[\mu \nu]} =0.
\label{eq89}
\end{align}
При $\beta = 0$ система  \eqref{eq83}--\eqref{eq85} переходит в уравнение типа Даффина–Кеммера для частицы со спином $0$ и массой $m = m_1$
\begin{align}
\alpha \partial_{\mu} \psi_{\mu} + a \psi_0 =0 ,
\label{eq810} \\
\alpha^* \partial_{\mu} \psi_0 + b \psi_{\mu} =0.
\label{eq811}
\end{align}

Теперь перейдем к интересующим нас случаям. Положим в  \eqref{eq83} $a = 0$. Получим систему
\begin{align}
\partial_{\mu} \psi_{\mu} =0,
\label{eq812} \\
\beta^* \partial_{\nu} \psi_{[\mu \nu]} + \alpha^* \partial_{\mu} \psi_0 + b \psi_{\mu} =0,
\label{eq813} \\
\beta (- \partial_{\mu} \psi_{\nu} +  \partial_{\nu} \psi_{\mu}) + c \psi_{[\mu \nu]} =0.
\label{eq814}
\end{align}
Вводя обозначение
\begin{align}
\varphi_{\mu} = \psi_{\mu} + \frac{\alpha^*}{b} \partial_{\mu} \psi_0 ,
\label{eq815}
\end{align}
систему \eqref{eq812}--\eqref{eq814} можно привести к виду
\begin{align}
\beta^* \partial_{\nu} \psi_{[\mu \nu]} + b \varphi_\mu =0,
\label{eq816} \\
\beta (- \partial_{\mu} \varphi_{\nu} +  \partial_{\nu} \varphi_{\mu})+ c \psi_{[\mu \nu]} =0,
\label{eq817}
\end{align}
совпадающему с точностью до обозначений с системой \eqref{eq88}, \eqref{eq89}.

К аналогичному результату можно прийти иначе, используя соображения, связанные с калибровочной инвариантностью. Система \eqref{eq812}--\eqref{eq814} инвариантна относительно преобразований
\begin{align}
 \psi_0 \to \psi'_0 = \psi_0 - \frac{1}{\alpha^*} \Lambda, \quad  \psi_\mu \to \psi'_\mu = \psi_\mu + \frac{1}{b}  \partial_{\mu} \Lambda,
\label{eq818}
\end{align}
где произвол в выборе калибровочной функции $\Lambda$ ограничен условием
\begin{align}
\square \Lambda =0.
\label{eq819}
\end{align}
Но точно такое же уравнение нетрудно получить из системы \eqref{eq812}--\eqref{eq814} и для функции $\psi_0$. Отсюда следует, что функция $\psi_0$ в системе \eqref{eq812}--\eqref{eq814} может рассматриваться как калибровочная. На нее можно наложить дополнительное условие $\psi_0 = 0$, приводящее систему \eqref{eq812}--\eqref{eq814} к системе Даффина–Кеммера \eqref{eq88}, \eqref{eq89}.

Данный вариант калибровочно-инвариантной теории впервые был предложен в работе [56] и известен в литературе как подход Штюкльберга к описанию частицы со спином $1$ и ненулевой массой. Детальный анализ этого подхода и его преимуществ по сравнению с обычными подходами Даффина–Кеммера и Прока дан в [57].

Теперь положим в \eqref{eq83}--\eqref{eq85} $c = 0$. Получим систему уравнений
\begin{align}
\alpha \partial_{\mu} \psi_{\mu} + a \psi_0 =0,
\label{eq820} \\
\beta^* \partial_{\nu} \psi_{[\mu \nu]} + \alpha^* \partial_{\mu} \psi_0 + b \psi_{\mu} =0,
\label{eq821} \\
- \partial_{\mu} \psi_{\nu}+\partial_{\nu} \psi_{\mu} =0.
\label{eq822}
\end{align}
Из \eqref{eq820}--\eqref{eq822} вытекает уравнение второго порядка
\begin{align}
\square \psi_0 - \frac{a b}{|\alpha |^2} \psi_0 =0,
\label{eq823}
\end{align}
которое означает, что эта система описывает частицу с ненулевой массой и спином $s=0$.

Система \eqref{eq820}--\eqref{eq822} инвариантна относительно преобразований
\begin{align}
\psi_{[\mu \nu]} \to \psi'_{[\mu \nu]} = \psi_{[\mu \nu]}- \frac{1}{\beta^*} \Lambda_{[\mu \nu]}, \quad \psi_{\mu} \to \psi'_{\mu} = \psi_{\mu}+ \frac{1}{b} \partial_{\nu} \Lambda_{[\mu \nu]},
\label{eq824}
\end{align}
где произвол в выборе калибровочных функций $\Lambda_{[\mu \nu]} (x)$ ограничен условием
\begin{align}
\partial_{\alpha} \partial_{\nu} \Lambda_{[\mu \nu]} -\partial_{\mu} \partial_{\nu}  \Lambda_{[\alpha \nu]} =0.
\label{eq825}
\end{align}
С другой стороны, как следует из уравнений \eqref{eq821}, \eqref{eq822}, аналогичному условию удовлетворяет тензор $\psi_{[\mu \nu]}$:
\begin{align}
\partial_{\alpha} \partial_{\nu} \psi_{[\mu \nu]} -\partial_{\mu} \partial_{\nu}  \psi_{[\alpha \nu]} =0.
\label{eq826}
\end{align}
Другими словами, произвол в выборе калибровочных функций $\Lambda_{[\mu \nu]}$ достаточен, чтобы на $\psi_{[\mu \nu]}$ наложить дополнительное условие
\begin{align}
\partial_{\nu} \psi_{[\mu \nu]} =0 .
\label{eq827}
\end{align}
При этом исключаются состояния со спином $1$, и система  \eqref{eq820}--\eqref{eq822}  приводится к виду  \eqref{eq810}, \eqref{eq811}.

Данный вариант калибровочно-инвариантной теории является своего рода аналогом подхода Штюкльберга, но для описания частицы с ненулевой массой и спином $s=0$. Впервые он был предложен нами в работе [58]. Опять-таки заметим, что систему \eqref{eq820}--\eqref{eq822}  с помощью подстановки
\begin{align}
\varphi_\mu = \psi_\mu + \frac{\beta^*}{b} \partial_{\nu} \psi_{[\mu \nu]}
\label{eq828}
\end{align}
можно преобразовать к виду
\begin{align}
\alpha \partial_{\mu} \varphi_{\mu} + a \psi_0 = 0 , \quad \alpha^* \partial_{\mu} \psi_0 + b \varphi_{\mu} =0,
\label{eq829}
\end{align}
совпадающему с \eqref{eq810}, \eqref{eq811} с точностью до обозначений.

Рассмотрим, наконец, случай, когда в \eqref{eq83}--\eqref{eq85} $b = 0$. Получим систему
\begin{align}
\alpha \partial_{\mu} \psi_{\mu} + a \psi_0 =0,
\label{eq830} \\
\beta^* \partial_{\nu} \psi_{[\mu \nu]} + \alpha^* \partial_{\mu} \psi_0 =0,
\label{eq831} \\
\beta (- \partial_{\mu} \psi_{\nu} + \partial_{\nu} \psi_{\mu} ) +c \psi_{[\mu \nu]} =0,
\label{eq832}
\end{align}
инвариантную относительно преобразования
\begin{align}
\psi_\mu \to \psi'_\mu = \psi_{\mu} + \partial_{\mu} \Lambda,
\label{eq833}
\end{align}
где калибровочная функция $\Lambda$ удовлетворяет условию \eqref{eq819}. Такое же условие (уравнение) вытекает из системы \eqref{eq830}, \eqref{eq831} для скалярной функции $\psi_0$. Это означает, что функция $\psi_0$ является фактически калибровочной функцией и ее, не уменьшая общности, можно выбрать равной нулю. В результате система  \eqref{eq830}--\eqref{eq832}  трансформируется к виду
\begin{align}
\partial_{\mu} \psi_{\mu} =0,
\label{eq834} \\
\partial_{\nu} \psi_{[\mu \nu]} =0,
\label{eq835} \\
- \partial_{\mu} \psi_{\nu} + \partial_{\nu} \psi_{\mu} + \psi_{[\mu \nu]} =0.
\label{eq836}
\end{align}
По своему содержанию система \eqref{eq834}--\eqref{eq836} совпадает с уравнениями Максвелла с той лишь несущественной разницей, что в теории Максвелла уравнение \eqref{eq834} выступает в качестве дополнительного условия, а в системе \eqref{eq834}--\eqref{eq836} – в качестве независимого уравнения. В обоих случаях речь идет о безмассовой частице со спиральностью $\pm 1$. Очевидно, что и эквивалентная \eqref{eq834}--\eqref{eq836} система \eqref{eq830}--\eqref{eq832} имеет тот же физический смысл.

Рассмотренные тензорные системы, будучи записанными в матричной форме \eqref{eq:zero_mass}, характеризуются одинаковым видом матриц $\Gamma_\mu$. Различие состоит только в матрице $\Gamma_0$. В базисе, в котором волновая функция $\Psi$ имеет вид столбца
\begin{align}
\Psi = (\psi_0 , \psi_{\mu}, \psi_{[\mu \nu]})^T,
\label{eq837}
\end{align}
для $\Gamma_0$ имеют место выражения:

1) $a=0$, система \eqref{eq812}--\eqref{eq814}
\begin{align}
\Gamma_0 =
 \left( \begin{tabular}{ccc}
$0$ & &  \\
& $I_4$ & \\
& & $I_6$
\end{tabular} \right);
\label{eq838}
\end{align}

2) $c=0$, система \eqref{eq820}--\eqref{eq822}
\begin{align}
\Gamma_0 =
 \left( \begin{tabular}{ccc}
$1$ & &  \\
& $I_4$ & \\
& & $0_6$
\end{tabular} \right);
\label{eq839}
\end{align}

3) $b=0$, система \eqref{eq830}--\eqref{eq832}
\begin{align}
\Gamma_0 =
 \left( \begin{tabular}{ccc}
$1$ & &  \\
& $0_4$ & \\
& & $I_6$
\end{tabular} \right).
\label{eq840}
\end{align}

Сопоставляя выражения \eqref{eq837}--\eqref{eq840}, нетрудно заметить, что характер поля (массивное или безмассовое), описываемого РВУ  \eqref{eq:zero_mass}, при прочих равных условиях зависит от лоренцевской структуры члена $\Gamma_0 \Psi$. В том случае, когда этот член содержит набор лоренцевских ковариантов, минимально необходимый для построения уравнения вида \eqref{eq:nonzero_mass} для частицы (поля) с ненулевой массой, возможно построение калибровочно-инвариантной теории массивного поля на основе РВУ вида \eqref{eq:zero_mass}. В противном случае может идти речь только о теории безмассового поля. Так, система \eqref{eq812}--\eqref{eq814}, представленная в форме  \eqref{eq:zero_mass}, содержит член $\Gamma_0 \Psi$, в котором присутствуют коварианты $\psi_{\mu}$, $\psi_{[\mu \nu]}$. На основе последних можно построить РВУ вида \eqref{eq:nonzero_mass} для частицы с ненулевой массой –  $10$-компонентное уравнение Даффина–Кеммера (спин $1$). Соответственно $11$-компонентное калибровочно-инвариантное РВУ вида  \eqref{eq:zero_mass} с волновой функцией \eqref{eq837} и матрицей $\Gamma_0$ \eqref{eq838} также описывает векторную частицу с ненулевой массой. Калибровочно-инвариантным аналогом уравнения Даффина–Кеммера для спина $0$ является система \eqref{eq820}--\eqref{eq822}, или $11$-компонентное матричное РВУ  \eqref{eq:zero_mass} с матрицей $\Gamma_0$ \eqref{eq839}, в котором член $\Gamma_0 \Psi$ содержит коварианты $\psi_0$, $\psi_\mu$. В калибровочно-инвариантном РВУ с волновой функцией \eqref{eq837}, которое сопоставляется тензорной системе \eqref{eq830}--\eqref{eq832}, член $\Gamma_0 \Psi$ содержит коварианты $\psi_0$, $\psi_{[\mu \nu]}$. На основе последних уравнение первого порядка для частицы с ненулевой массой, как это вытекает из базовых положений теории РВУ, построить нельзя. Поэтому РВУ с матрицей $\Gamma_0$ \eqref{eq840} может описывать только безмассовое поле, что и было установлено нами выше непосредственным образом.

\section{Теория РВУ и электрослабое поле}

\hspace{1.25cm}
В стандартной теории РВУ предусматривается лишь возможность раздельного описания микрообъектов с ненулевой и нулевой массой. Однако в современной физике высоких энергий фигурируют поля, кванты которых (для данного поля) обладают как нулевой, так и ненулевой массой. Хорошо известным примером такого рода является электрослабое поле.

Если свободное электрослабое поле рассматривать как единый физический объект (а при достаточно высоких энергиях так оно и есть), то возникает вопрос, можно ли в рамках теории РВУ совместно описывать поля с ненулевой и нулевой массой (так называемые массивно-безмассовые поля). Напомним, что под совместным здесь по-прежнему понимается описание в рамках одной не распадающейся в релятивистски-инвариантном смысле системы уравнений.

Остановимся подробно на этом вопросе. Снова обратимся к схеме зацеплений \eqref{eq41}, где представление $(\frac12, \frac12)$ является векторным, а $(\frac12, \frac12)'$ – псевдовекторным. Если в наиболее общей тензорной формулировке релятивистски-инвариантной системы первого порядка \eqref{eq727}, соответствующей схеме \eqref{eq41}, положить
\begin{align}
a = 0, \quad b = c = m ,
\label{eq91}
\end{align}
получим систему
\begin{align}
& \partial_{\nu} \psi_{[\mu \nu]} =0,
\label{eq92} \\
\beta & ( \partial_{\mu} \psi_{[\nu \alpha]}+ \partial_{\alpha} \psi_{[\mu \nu]}+ \partial_{\nu} \psi_{[\alpha \mu]}) + m \psi_{[\mu \nu \alpha]} =0,
\label{eq93} \\
-& \partial_{\mu} \psi_{\nu }+ \partial_{\nu} \psi_{\mu }+ \beta^*  \partial_{\alpha} \psi_{[\mu \nu \alpha]} + m \psi_{[\mu \nu]} =0.
\label{eq94}
\end{align}
Из \eqref{eq92}--\eqref{eq94} нетрудно получить уравнения второго порядка
\begin{align}
& (\square - \frac{m^2}{|\beta |^2}) \psi_{[\mu \alpha \nu]} =0 ,
\label{eq95} \\
& \varepsilon_{\mu \nu \alpha \beta} \partial_{\beta} \psi_{[\mu \nu \alpha]} =0,
\label{eq96} \\
& \square \psi_{\mu}  - \partial_{\mu}  \partial_{\nu} \psi_{\nu} =0 .
\label{eq97}
\end{align}

Уравнения \eqref{eq95}, \eqref{eq96} означают, что система \eqref{eq92}--\eqref{eq94} содержит описание псевдовекторной частицы с ненулевой массой. Уравнение же \eqref{eq97} указывает на то, что данная система описывает еще и безмассовое векторное поле с потенциалом $\psi_\mu$. Последнее обстоятельство позволяет использовать калибровочное преобразование вида \eqref{eq710}, относительно которого инвариантны и система  \eqref{eq92}--\eqref{eq94} и уравнение \eqref{eq97}. Эта инвариантность означает, что указанное безмассовое поле является полем максвеловского типа со спиральностью $\pm 1$.

Таким образом, не распадающаяся по группе Лоренца тензорная система  \eqref{eq92}– \eqref{eq94} дает совместное описание псевдовекторной частицы с ненулевой массой и безмассового векторного поля электромагнитного типа.

Отметим также, что если в \eqref{eq727} выбрать
\begin{align}
b = 0, \quad a = c = m,
\label{eq98}
\end{align}
то мы придем к дуально симметричной теории по отношению к \eqref{eq92}--\eqref{eq94} системе в том смысле, что при выборе \eqref{eq98} система \eqref{eq727} приводит к совместному описанию массивной векторной частицы и безмассового поля максвелловского типа с псевдовекторным потенциалом.

Кроме того, как уже отмечалось в главе 4, на основе схемы \eqref{eq41} можно осуществить совместное описание микрообъекта с двумя различными ненулевыми массами. Для этого в \eqref{eq727} надо положить
\begin{align}
a = b = c = m.
\label{eq99}
\end{align}
В результате получим систему первого порядка
\begin{align}
& \alpha \partial_{\nu} \psi_{[\mu \nu]} + m \psi_{\mu} =0 ,
\label{eq910} \\
& \beta (\partial_{\mu} \psi_{[ \nu \alpha ]} + \partial_{\alpha} \psi_{[ \mu \nu  ]} + \partial_{\nu} \psi_{[ \alpha \mu]}) + m \psi_{[\mu \nu \alpha]} =0 ,
\label{eq911} \\
& \alpha^* (-\partial_{\nu} \psi_{\alpha}+ \partial_{\alpha} \psi_{\nu} ) + \beta^* \partial_{\mu} \psi_{[\mu \nu \alpha]} + m \psi_{[\nu \alpha]} =0 .
\label{eq912}
\end{align}
Из \eqref{eq910}--\eqref{eq912} вытекают уравнения второго порядка \eqref{eq95}, \eqref{eq96}, а также уравнения
\begin{align}
(\square - \frac{m^2}{|\alpha |^2}) \psi_{\mu } =0 , \quad \partial_{\mu} \psi_{\mu} =0 ,
\label{eq913}
\end{align}
что и подтверждает вышесказанное.

Теперь рассмотрим возможность совместного описания векторных частиц (полей), сопоставляемых системам  \eqref{eq92}--\eqref{eq94} и \eqref{eq910}--\eqref{eq912}. Очевидно, что простое механическое объединение этих систем не приведет к желаемому результату, поскольку полученная таким образом система уравнений распадается в релятивистски-инвариантном смысле и, следовательно, с точки зрения теории РВУ не может описывать единый физический объект. Соответствующая схема зацеплений также распадается на два самостоятельных независимых фрагмента вида \eqref{eq41}.

Самый простой и естественный способ построения нераспадающейся системы уравнений, обеспечивающей совместное описание вышеуказанных векторных полей, заключается во введении в рассмотрение дополнительного скалярного представления $(0, 0)$.
В результате можно получить, например, схему зацеплений
\begin{align}
& \hspace{7.45mm} \line(0,-1){10} \line(1,0){185} \line(0,-1){10} \notag  \\[-0.5em]
& \begin{tabular}{ccccccccccccc}
\vline & & $(0,1)$ & & & & \vline  & & &  & $(0,1)$ &  &     \\
\vline & $\diagup$  &  & $\diagdown$ &   & & \vline & & &  $\diagup$  &         & $\diagdown$  &  \\
$(\frac12,\frac12)'$ & & & & $(\frac12,\frac12)$  & --- & $(0,0)$ &  --- & $(\frac12,\frac12)'$   &  &   &  & $(\frac12,\frac12) $ . \\
 & $\diagdown$  & & $\diagup$ & & &  &   & &   $\diagdown$ &  & $\diagup$  &   \\
 & & $(1,0)$ & & & &   &   & &  & $ (1,0)$ &  &
\end{tabular}
\label{eq914}
\end{align}

Соответствующая схеме \eqref{eq914} система уравнений первого порядка имеет следующий наиболее общий вид:
\begin{align}
& \partial_{\nu} \psi_{[\mu \nu]} =0 ,
\label{eq915} \\
& \alpha (\partial_{\mu} \psi_{[\nu \alpha]}+ \partial_{\alpha} \psi_{[\mu \nu]}+ \partial_{\nu} \psi_{[\alpha \mu]}) + \beta \varepsilon_{\mu \nu \alpha \beta} \partial_{\beta} \psi_0 + m \psi_{[\mu \nu \alpha]} =0 ,
\label{eq916} \\
- & \partial_{\mu} \psi_{\nu} + \partial_{\nu} \psi_{\mu} + \alpha^* \partial_{\alpha} \psi_{[\mu \nu \alpha]}
+ m \psi_{[\mu \nu]} =0,
 \label{eq917} \\
 & \rho \partial_{\nu} \varphi_{[\mu \nu]} + \gamma \partial_{\mu} \psi_0 + m \varphi_{\mu} =0 ,
\label{eq918} \\
 & \delta (\partial_{\mu} \varphi_{[\nu \alpha ]} + \partial_{\alpha} \varphi_{[\mu \nu ]} +\partial_{\nu} \varphi_{[\alpha \mu]} ) + \sigma \varepsilon_{\mu \nu \alpha \beta} \partial_{\beta} \psi_0 +m \varphi_{[\mu \nu \alpha]} =0,
\label{eq919} \\
 & \rho^* (- \partial_{\mu} \varphi_{\nu} + \partial_{\nu} \varphi_{\mu}) + \delta^* \partial_{\alpha}  \varphi_{[\mu \nu \alpha]} + m  \varphi_{[\mu \nu ]} =0 ,
\label{eq920} \\
 & \frac{1}{3!} \beta^* \varepsilon_{\beta \mu \nu \alpha} \partial_{\beta} \psi_{[\mu \nu \alpha]} + \gamma^*
 \partial_{\mu} \psi_{\mu} + \frac{1}{3!} \sigma^* \varepsilon_{\beta \mu \nu \alpha} \partial_{\beta} \varphi_{[\mu \nu \alpha]} + m \psi_0 =0 .
\label{eq921}
\end{align}

Найдем уравнения второго порядка, к которым приводит система \eqref{eq915}– \eqref{eq921}.

Подействуем на \eqref{eq917} оператором $\partial_{\nu}$. С учетом \eqref{eq915} получим
\begin{align}
\square \psi_{\mu} + \partial_{\mu} \partial_{\nu} \psi_{\nu} =0.
\label{eq922}
\end{align}
Используя калибровочное преобразование \eqref{eq710}, уравнение \eqref{eq922} стандартным способом можно привести к виду
\begin{align}
\square \psi_{\mu} =0, \quad \partial_{\mu} \psi_{\mu} =0 .
\label{eq923}
\end{align}

Применение оператора $\partial_{\nu}$ к уравнению  \eqref{eq920} дает
\begin{align}
\rho^* (\square \varphi_{\mu} - \partial_{\mu} \partial_{\nu} \varphi_{\nu}) + m \partial_{\nu} \varphi_{[\mu \nu]}= 0.
\label{eq924}
\end{align}
Из \eqref{eq918} выразим член $\partial_{\nu} \varphi_{[\mu \nu]}$
\begin{align}
\partial_{\nu} \varphi_{[\mu \nu]} = - \frac{\gamma}{\rho} \partial_{\mu} \psi_0  - \frac{m}{\rho}  \varphi_{\mu}.
\label{eq925}
\end{align}
Отсюда следует
\begin{align}
 \partial_{\mu} \varphi_{\mu} = - \frac{\gamma}{m} \square \psi_0  .
\label{eq926}
\end{align}
Подставляя \eqref{eq925}, \eqref{eq926}  в \eqref{eq924}, будем иметь
\begin{align}
\square \varphi_{\mu} +  \frac{\gamma}{m} \partial_{\mu} \square \psi_0  - \frac{m \gamma}{|\rho |^2}  \partial_{\mu} \psi_0
- \frac{m^2}{|\rho |^2} \varphi_{\mu} =0.
\label{eq927}
\end{align}

Если ввести в рассмотрение вектор
\begin{align}
\Phi_{\mu} = \varphi_{\mu} +  \frac{\gamma}{m} \partial_{\mu} \psi_0 ,
\label{eq928}
\end{align}
уравнение \eqref{eq927} принимает вид
\begin{align}
\square \Phi_{\mu} - \frac{m^2}{|\rho |^2} \Phi_{\mu} =0.
\label{eq929}
\end{align}
При это вектор $\Phi_{\mu}$ с учетом \eqref{eq926}  удовлетворяет условию
\begin{align}
\partial_{\mu} \Phi_{\mu} =0.
\label{eq930}
\end{align}

Теперь применим к уравнению \eqref{eq917} оператор $\varepsilon_{\rho \mu \nu \sigma} \partial_{\sigma}$:
\begin{align}
\alpha^* \varepsilon_{\rho \mu \nu \sigma} \partial_{\sigma} \partial_{\alpha} \psi_{[ \mu \nu \alpha ]} + m \varepsilon_{\rho \mu \nu \sigma} \partial_{\sigma} \psi_{[ \mu \nu ]} =0.
\label{eq931}
\end{align}
С помощью непосредственно проверяемого тождества
\begin{align}
\varepsilon_{\rho \mu \nu \sigma} \partial_{\sigma} \partial_{\alpha} \psi_{[\mu \nu \alpha] } = \varepsilon_{\rho \mu \nu \alpha} \square \psi_{[\mu \nu \alpha] }- \partial_{\rho} \partial_{\beta} \varepsilon_{\beta \mu \nu \alpha}  \psi_{[\mu \nu \alpha] } \notag
\end{align}
уравнение \eqref{eq931} приводится к виду
\begin{align}
\alpha^* ( \varepsilon_{\rho \mu \nu \alpha} \square \psi_{[\mu \nu \alpha] }- \partial_{\rho} \partial_{\beta} \varepsilon_{\beta \mu \nu \alpha}  \psi_{[\mu \nu \alpha] } ) + m \varepsilon_{\rho \mu \nu \alpha} \partial_{\alpha} \psi_{[ \mu \nu ]} =0.
\label{eq932}
\end{align}
Из уравнения \eqref{eq916} вытекает
\begin{align}
\partial_{\mu} \psi_{[\nu \alpha]}+ \partial_{\alpha} \psi_{[\mu \nu]} + \partial_{\nu} \psi_{[ \alpha \mu]} = - \frac{\beta}{\alpha} \varepsilon_{\mu \nu \alpha  \beta} \partial_{\beta} \psi_0 - \frac{m}{\alpha} \psi_{[\mu \nu \alpha]} .
\label{eq933}
\end{align}
Применяя к  \eqref{eq916} оператор $\frac{1}{3!} \varepsilon_{\mu \nu \alpha \rho} \partial_{\rho}$, получим
\begin{align}
\frac{1}{3!} \varepsilon_{\mu \nu \alpha \rho} \partial_{\rho}  \psi_{[\mu \nu \alpha] } = \frac{\beta}{m} \square \psi_0 .
\label{eq934}
\end{align}
Комбинация уравнений \eqref{eq932}--\eqref{eq934} дает
\begin{align}
\square \psi_{[\mu \nu \alpha] } - \frac{m^2}{|\alpha |^2} \psi_{[\mu \nu \alpha] }+ \frac{\beta}{m}  \varepsilon_{\mu \nu \alpha \beta} \partial_{\beta} \square \psi_0 - \frac{\beta m}{|\alpha |^2}  \varepsilon_{\mu \nu \alpha \beta} \partial_{\beta} \psi_0 =0.
\label{eq935}
\end{align}
Посредством обозначения
\begin{align}
 \Psi_{[\mu \nu \alpha]}  =  \psi_{[\mu \nu \alpha]} + \frac{\beta}{m} \varepsilon_{\mu \nu \alpha \beta} \partial_{\beta} \psi_0
\label{eq936}
\end{align}
приведем уравнение  \eqref{eq935} к виду
\begin{align}
\square \Psi_{[\mu \nu \alpha]} - \frac{m^2}{| \alpha |^2} \Psi_{[\mu \nu \alpha]} =0.
\label{eq937}
\end{align}
Одновременно с \eqref{eq937} в соответствии с \eqref{eq934} имеем условие
\begin{align}
\varepsilon_{\mu \nu \alpha \beta} \partial_{\beta} \Psi_{[\mu \nu \alpha]} =0 .
\label{eq938}
\end{align}

Аналогичным образом из системы \eqref{eq915}--\eqref{eq921} можно получить уравнение второго порядка
\begin{align}
\square \Phi_{[\mu \nu \alpha]} - \frac{m^2}{| \delta |^2} \Phi_{[\mu \nu \alpha]} =0
\label{eq939}
\end{align}
с дополнительным условием
\begin{align}
\varepsilon_{\mu \nu \alpha \beta} \partial_{\beta} \Phi_{[\mu \nu \alpha]} =0,
\label{eq940}
\end{align}
где
\begin{align}
 \Phi_{[\mu \nu \alpha]}  =  \varphi_{[\mu \nu \alpha]} + \frac{\sigma}{m} \varepsilon_{\mu \nu \alpha \beta} \partial_{\beta} \psi_0.
\label{eq941}
\end{align}

Найдем, наконец, уравнение второго порядка для скаляра $\psi_0$. Для этого, применяя оператор $\frac{1}{3!} \varepsilon_{\mu \nu \alpha \rho} \partial_{\rho}$ к уравнению \eqref{eq919}, будем иметь
\begin{align}
\frac{1}{3!} \varepsilon_{\mu \nu \alpha \rho} \partial_{\rho} \varphi_{[\mu \nu \alpha]}  = \frac{\sigma}{m} \square \psi_0 .
\label{eq942}
\end{align}
Подставляя теперь \eqref{eq926}, \eqref{eq934} и \eqref{eq942} в \eqref{eq921}, получим
\begin{align}
\square \psi_0- \frac{m^2}{|\beta |^2 + |\gamma |^2 + |\sigma |^2} \psi_0  = 0.
\label{eq943}
\end{align}

Уравнения \eqref{eq923}, \eqref{eq929}, \eqref{eq930}, \eqref{eq937}--\eqref{eq940} и \eqref{eq943}  показывают, что в не
распадающейся по группе Лоренца системе первого порядка  \eqref{eq915}--\eqref{eq921} содержится описание четырех частиц (полей) со спином $1$, одна из которых имеет нулевую массу, а также скалярной частицы с ненулевой массой.

Тензорная система \eqref{eq915}--\eqref{eq921} может быть представлена в стандартной матричной форме РВУ \eqref{eq:zero_mass}  с особенной матрицей $\Gamma_0$ вида
\begin{align}
\Gamma_0 = \left(\begin{tabular}{cc}
$0_4$ & \\
  & $m I_{25}$
\end{tabular} \right).
\label{eq944}
\end{align}
Дадим формулировку этого РВУ в базисе Гельфанда–Яглома.

Введем, как обычно, нумерацию неприводимых представлений,  содержащихся с схеме зацеплений \eqref{eq914}, например:
\begin{align}
\begin{split}
& (0,0) \sim 1 \,\, (\psi_0), \quad (\frac12, \frac12) \sim 2 \,\, (\psi_\mu), \quad (\frac12, \frac12)' \sim 3 \,\, (\psi_{[\mu \nu \alpha]}), \\
& (0,1),(1,0) \sim 4,5 \,\, (\psi_{[\mu \nu]}), \quad (\frac12, \frac12) \sim 6 \,\, (\varphi_\mu), \\
& (\frac12, \frac12)' \sim 7\,\, (\varphi_{[\mu \nu \alpha]}), \quad (0,1), (1,0) \sim 8,9 \,\, (\varphi_{[\mu \nu ]}).
\end{split}
\label{eq945}
\end{align}
Тогда для спиновых блоков $C^0$, $C^1$ матрицы $\Gamma_4$ получим следующие общие выражения:
\begin{align}
C^0 &= \left(\begin{tabular}{ccccc}
$0$ & $0$ & $c^0_{13}$ & $c^0_{16}$ & $c^0_{17}$  \\
$0$ & $0$ & $0$ & $0$ & $0$  \\
$c^0_{31}$ & $0$ & $0$ & $0$ & $0$  \\
$c^0_{61}$ & $0$ & $0$ & $0$ & $0$  \\
$c^0_{71}$ & $0$ & $0$ & $0$ & $0$
\end{tabular} \right), \quad
C^1 = \left(\begin{tabular}{cc}
$(C^1)'$ & \\
  & $(C^1)''$
\end{tabular} \right),
\label{eq946} \\
(C^1)' &= \left(\begin{tabular}{cccc}
$0$ & $0$ & $c^1_{24}$ & $c^1_{25}$  \\
$0$ & $0$ & $c^1_{34}$ & $c^1_{35}$  \\
$c^1_{42}$ & $c^1_{43}$ & $0$ & $0$   \\
$c^1_{52}$ & $c^1_{53}$ & $0$ & $0$
\end{tabular} \right), \quad
(C^1)'' = \left(\begin{tabular}{cccc}
$0$ & $0$ & $c^1_{68}$ & $c^1_{69}$  \\
$0$ & $0$ & $c^1_{78}$ & $c^1_{79}$  \\
$c^1_{86}$ & $c^1_{87}$ & $0$ & $0$   \\
$c^1_{96}$ & $c^1_{97}$ & $0$ & $0$
\end{tabular} \right).
\label{eq947}
\end{align}

Условия инвариантности  \eqref{c_cond} рассматриваемого РВУ относительно преобразований собственной группы Лоренца никаких ограничений на элементы $c^0_{ij}$, $c^1_{ij}$ здесь не накладывает. Требование $P$-инвариантности теории применительно к электрослабому полю не является актуальным. Условие  \eqref{c_eta_cond} возможности лагранжевой формулировки теории приводит к соотношениям
\begin{align}
\begin{split}
c^0_{31} &= f (c^0_{13})^* , \quad c^0_{61} = g (c^0_{16})^* , \quad c^0_{71} = h (c^0_{17})^*, \\
c^1_{42} &= p (c^1_{25})^* , \quad c^1_{52} = p (c^1_{24})^* , \quad c^1_{43} = q (c^1_{35})^*, \quad c^1_{53} = q (c^1_{34})^*, \\
c^1_{86} &= r (c^1_{69})^* , \quad c^1_{96} = r (c^1_{68})^* , \quad c^1_{87} = s (c^1_{79})^*, \quad c^1_{97} = s (c^1_{78})^*, \\
\end{split}
\label{eq948}
\end{align}
где
\begin{align}
\begin{split}
f &=\frac{\eta^0_{33}}{\eta^0_{11}}, \quad g = \frac{\eta^0_{66}}{\eta^0_{11}}, \quad h=\frac{\eta^0_{77}}{\eta^0_{11}} ,\\
p &= \frac{\eta^1_{45}}{\eta^1_{22}} , \quad q = \frac{\eta^1_{45}}{\eta^1_{33}} , \quad r= \frac{\eta^1_{89}}{\eta^1_{66}} , \quad s= \frac{\eta^1_{89}}{\eta^1_{77}}.
\end{split}
\label{eq949}
\end{align}

Вводя также для удобства обозначения
\begin{align}
\begin{split}
c^0_{13} &= \lambda_1 , \quad  c^0_{16} = \lambda_2 ,\quad c^0_{17} = \lambda_3,  \quad c^1_{24} = \lambda_4,  \quad c^1_{25} = \lambda_5, \\
c^1_{34} &= \lambda_6 , \quad  c^1_{35} = \lambda_7 , \quad c^1_{68} = \lambda_8,  \quad c^1_{69} = \lambda_9,  \quad c^1_{78} = \lambda_{10}, \quad c^1_{79} = \lambda_{11},
\end{split}
\label{eq950}
\end{align}
с учетом \eqref{eq948}, \eqref{eq949} для блоков $C^0$ \eqref{eq946}, $(C^1)'$, $(C^1)''$ \eqref{eq947} получим выражения
\begin{align}
C^0 &= \left(\begin{tabular}{ccccc}
$0$ & $0$ & $\lambda_1$ & $\lambda_2$ & $\lambda_3$  \\
$0$ & $0$ & $0$ & $0$ & $0$  \\
$f \lambda_1^*$ & $0$ & $0$ & $0$ & $0$  \\
$g \lambda_2^*$ & $0$ & $0$ & $0$ & $0$  \\
$h \lambda_3^*$ & $0$ & $0$ & $0$ & $0$
\end{tabular} \right), \quad
\label{eq951} \\
(C^1)' &= \left(\begin{tabular}{cccc}
$0$ & $0$ & $\lambda_4$ & $\lambda_5$  \\
$0$ & $0$ & $\lambda_6$ & $\lambda_7$  \\
$p \lambda_5^*$ & $q \lambda_7^*$ & $0$ & $0$   \\
$p \lambda_4^*$ & $q \lambda_6^*$ & $0$ & $0$
\end{tabular} \right), \quad
(C^1)'' = \left(\begin{tabular}{cccc}
$0$ & $0$ & $\lambda_8$ & $\lambda_9$  \\
$0$ & $0$ & $\lambda_{10}$ & $\lambda_{11}$  \\
$r \lambda_9^*$ & $s \lambda_{11}^*$ & $0$ & $0$   \\
$r \lambda_8^*$ & $s \lambda_{10}^*$ & $0$ & $0$
\end{tabular} \right).
\label{eq952}
\end{align}

Характеристические уравнения для блоков \eqref{eq951}, \eqref{eq952} имеют соответственно вид
\begin{align}
& \lambda^3 (\lambda^2 -  f |\lambda_1 |^2- g |\lambda_2 |^2- h |\lambda_3 |^2) =0,
\label{eq953} \\
\begin{split}
& \lambda^4 - \lambda^2 (p \lambda_4^* \lambda_5 + p  \lambda_4 \lambda_5^*  + q  \lambda_6^* \lambda_7  + q \lambda_6 \lambda_7^* ) + \\
& + pq (\lambda_4 \lambda_5^* \lambda_6^* \lambda_7  + \lambda_4^* \lambda_5 \lambda_6 \lambda_7^* - |\lambda_4 |^2 |\lambda_7 |^2 - |\lambda_5 |^2 |\lambda_6 |^2)=0 ,
\end{split}
\label{eq954} \\
\begin{split}
& \lambda^4 - \lambda^2 (r \lambda_8^* \lambda_9 + r  \lambda_8 \lambda_9^* + s  \lambda_{10}^* \lambda_{11}  + s \lambda_{10} \lambda_{11}^*) + \\
& + rs (\lambda_8 \lambda_9^* \lambda_{10}^* \lambda_{11}  + \lambda_8^* \lambda_9 \lambda_{10} \lambda_{11}^* - |\lambda_8 |^2 |\lambda_{11} |^2 - |\lambda_9 |^2 |\lambda_{10} |^2)=0 .
\end{split}
\label{eq955}
\end{align}
РВУ, эквивалентное тензорной системе \eqref{eq915}--\eqref{eq921} получится, если положить
\begin{align}
\begin{split}
& f |\lambda_1 |^2+ g |\lambda_2 |^2+ h |\lambda_3 |^2= |\beta |^2 + |\gamma |^2 + |\sigma |^2 , \\
& p \lambda_4^* \lambda_5 + p  \lambda_4 \lambda_5^*  + q  \lambda_6^* \lambda_7  + q \lambda_6 \lambda_7^* = |\alpha |^2 +1, \\
& p q ( \lambda_4 \lambda_5^* \lambda_6^* \lambda_7  + \lambda_4^* \lambda_5 \lambda_6 \lambda_7^* - |\lambda_4 |^2 |\lambda_7 |^2 - |\lambda_5 |^2 |\lambda_6 |^2) = |\alpha |^2, \\
& r \lambda_8^* \lambda_9 + r  \lambda_8 \lambda_9^* + s  \lambda_{10}^* \lambda_{11}  + s \lambda_{10} \lambda_{11}^* = |\rho |^2 +|\delta |^2 , \\
& r s (\lambda_8 \lambda_9^* \lambda_{10}^* \lambda_{11}  + \lambda_8^* \lambda_9 \lambda_{10} \lambda_{11}^* - |\lambda_8 |^2 |\lambda_{11} |^2 - |\lambda_9 |^2 |\lambda_{10} |^2)=  |\rho |^2 |\delta |^2 .
\end{split}
\label{eq956}
\end{align}
Соотношениям  \eqref{eq956} можно удовлетворить, выбрав, например:
\begin{align}
\begin{split}
\lambda_1 &= \beta, \quad \lambda_2 = \sqrt{2 (|\gamma |^2 + |\sigma |^2)} , \quad \lambda_3 = \sqrt{|\gamma |^2 + |\sigma |^2}, \\
\lambda_4 &= - \lambda_5 = \frac{i |\alpha|}{\sqrt{2}}, \quad \lambda_6 =  \lambda_7 = \frac{1}{\sqrt{2}}, \quad  \lambda_8 =  -\lambda_9 = \frac{i |\rho|}{\sqrt{2}}, \quad  \lambda_{10} =  \lambda_{11} = \frac{|\delta|}{\sqrt{2}},
\end{split}
\label{eq957} \\
f &= g = -h = -p = q =-r=s=1.
\label{eq958}
\end{align}
Равенства \eqref{eq958}, в свою очередь, приводят к следующим значениям элементов $\eta_{\tau \tau'}^s$ матрицы лоренц-инвариантной билинейной формы $\eta$:
\begin{align}
\eta^0_{11} = \eta^0_{33} =\eta^0_{66} =-\eta^0_{77} =\eta^1_{22} = -\eta^1_{33} = -\eta^1_{45} = - \eta^1_{66} = \eta^1_{77} = \eta^1_{89} = 1.
\label{eq959}
\end{align}

Спиновые блоки $C^0$ \eqref{eq951}, $(C^1)'$, $(C^1)''$ \eqref{eq952} в соответствии с \eqref{eq957}, \eqref{eq958} принимают вид
\begin{align}
C^0 &= \left(\begin{tabular}{ccccc}
$0$ & $0$ & $\beta$ & $\sqrt{2 (|\gamma |^2 + |\sigma |^2)}$ & $\sqrt{|\gamma |^2 + |\sigma |^2}$  \\
$0$ & $0$ & $0$ & $0$ & $0$  \\
$\beta^*$ & $0$ & $0$ & $0$ & $0$  \\
$\sqrt{2 (|\gamma |^2 + |\sigma |^2)}$ & $0$ & $0$ & $0$ & $0$  \\
$-\sqrt{|\gamma |^2 + |\sigma |^2}$ & $0$ & $0$ & $0$ & $0$
\end{tabular} \right), \quad
\label{eq960} \\
\begin{split}
& (C^1)' = \frac{1}{\sqrt{2}} \left(\begin{tabular}{cccc}
$0$ & $0$ & $i |\alpha |$ & $-i |\alpha |$  \\
$0$ & $0$ & $1$ & $1$  \\
$-i |\alpha |$ & $1$ & $0$ & $0$   \\
$i |\alpha |$ & $1$ & $0$ & $0$
\end{tabular} \right), \\
& (C^1)'' =  \frac{1}{\sqrt{2}}  \left(\begin{tabular}{cccc}
$0$ & $0$ & $i | \rho | $ & $- i | \rho |$  \\
$0$ & $0$ & $ | \delta | $ & $ |\delta |$  \\
$- i |\rho |$ & $|\delta |$ & $0$ & $0$   \\
$i |\rho |$  & $| \delta |$ & $0$ & $0$
\end{tabular} \right).
\end{split}
\label{eq961}
\end{align}
Вид блоков $\eta^0$, $\eta^1$ матрицы $\eta$ вытекает из \eqref{eq959}.

Спиновый блок $C^0$ \eqref{eq960} имеет один (с точностью до знака) ненулевой корень
\begin{align}
\pm \sqrt{|\beta |^2 +|\gamma |^2 +|\sigma |^2  },
\label{eq962}
\end{align}
который соответствует массе скалярного бозона
\begin{align}
m^{(0)} = \frac{m}{ \sqrt{|\beta |^2 +|\gamma |^2 +|\sigma |^2  }}.
\label{eq963}
\end{align}
Блок $(C^1)'$ имеет корни $\pm 1$, $\pm |\alpha|$. Первый из них в силу проективности матрицы $\Gamma_0$ \eqref{eq944} относится к безмассовому векторному полю максвелловского типа, второй – к векторной частице с массой
\begin{align}
m_1^{(1)} = \frac{m}{ |\alpha|}.
\label{eq964}
\end{align}
Корни $\pm |\delta|$, $\pm |\rho|$ блока $(C^1)''$ соответствуют массам
\begin{align}
m_2^{(1)} = \frac{m}{ |\delta|}, \quad m_3^{(1)} = \frac{m}{ |\rho|}
\label{eq965}
\end{align}
двух других векторных частиц, описание которых также содержится в обсуждаемом РВУ и эквивалентной ему тензорной системе.

Итак, схема зацеплений \eqref{eq914} позволяет построить релятивистское волновое уравнение, описывающее векторное поле с четырьмя типами квантов – одним безмассовым  и тремя массивными. При этом в теории с необходимостью возникает
скалярная частица с ненулевой массой, которая собственно и обеспечивает единство компонент указанного векторного поля. Векторная составляющая последнего может интерпретироваться «электрослабое» поле, а скалярное – как линейный аналог бозона Хиггса. Линейный характер уравнения \eqref{eq943}, описывающего скалярную частицу, связан с тем, что вопрос о происхождении массы не затрагивается в теории РВУ, рамками которой мы ограничиваемся. Наличие или отсутствие массы является в данной теории заданным фактом, поэтому нет необходимости вводить в \eqref{eq943} нелинейные члены. Таким образом, предлагаемая нами модель, не вступая в противоречие с общепринятым хиггсовским механизмом генерации массы, позволяет указать на еще одну возможную причину появления скалярной составляющей в теории векторного электрослабого поля. Однако если вместо хиггсовского механизма рассматривать иной способ генерации массы Кальба--Рамонда, описанный в главе 7, то мы придем к альтернативной модели электрослабого взаимодействия, в которой скалярный бозон играет исключительно роль связующей компоненты для векторных составляющих электрослабого поля.

\newpage
\section*{Заключение}
\addcontentsline{toc}{chapter}{ \bf{Заключение} }

\hspace{1.25cm}
Перечислим ещё раз  в сжатой форме представленные результаты. На основе использования расширенных наборов неприводимых  представлений группы Лоренца дано:
\begin{itemize}
  \item полуфеноменологическое описание внутренней структуры микрообъектов с низшими спинами;
  \item описание изоспиновых степеней свободы,  в частности киральности массивных микрообъектов, посредством нераспадающегося по группе Лоренца РВУ, обладающего  внутренней симметрией геометрического происхождения;
  \item совместное описание безмассовых полей со спиральностями $0$, $\pm 1$ как единого физического объекта, на этой основе показана возможность  полуфеноменологического описания взаимодействия струн и мембран в пространстве Минковского;
  \item матричная интерпретация механизма генерации массы векторных  полей, который отличается от общеизвестного механизма Хиггса и не приводит  к появлению дополнительных  скалярных или каких-либо иных  массивных  частиц;
 матричное истолкование массивных калибровочно-инвариантных полей в подходе теории РВУ.
  \item наконец предложено нераспадающееся по группе Лоренца РВУ, которое описывает массивно-безмассовое векторное поле с тремя типами массивных и одним безмассовым квантами. Данное поле вполне может быть интерпретировано как электрослабое поле. При этом необходимость появления скалярного массивного поля обосновывается совершенно по-новому. Оказывается, в подходе теории  РВУ указанное векторное поле может существовать только в "связке" с массивным скалярным полем, образуя вместе с ним единый единый физический объект. В противном случае свободные массивное и безмассовое векторные поля выступают как независимые, т.е.  описываются незацепляющимимся в релятивистски-инвариантном смысле уравнениями.
\end{itemize}

Новизна и возможность  применения полученных результатов заключается в следующем. Глобальные унитарные симметрии, которые используются в современных калибровочных моделях фундаментальных частиц и их взаимодействий, имеют в свой основе негеометрическое происхождение. Другими словами, в этих моделях в качестве исходного берется простейшее уравнение Дирака, на волновую функцию которого "навешивается"\vspace{0.3cm} свободный нелоренцевский индекс. Тем самым взаимосвязь  между  свойствами пространства-времени и материального мира проявляется лишь после локализации указанных симметрий. Наше предложение состоит в том, чтобы в локально-калибровочных  моделях опираться на внутренние симметрии, которые уже в исходном глобальном варианте имеют геометрическое происхождение, т.е. присущи уравнениям, нераспадающимися по полной группе Лоренца. Возможным кандидатом на роль таких РВУ могут служить рассмотренные в Главах 5, 6 уравнение Дирака--Кэлера и его алгебраические обобщения.

Данный подход, на наш взгляд, обеспечивает более тесную связь между про\-странством-временем и материальным миром. Кроме того, расширение класса базовых РВУ должно привести к получению новых физических  следствий и,  возможно, устранить  некоторые из  тех  трудностей, которые имеют  место в Стандартной Модели и теории суперструн.

% даём указание на включение данного место в оглавление как секции (\section)